\documentclass[twoside,12pt]{article}

\usepackage{epsfig}
\usepackage{hyperref}
\usepackage{cite}
\usepackage{amsmath}
\usepackage{amssymb}
\usepackage{common-defs}
\usepackage{caption}
\usepackage{color}
\usepackage{subfig}

\definecolor{mygreen}{rgb}{0,0.7,0}
\newcommand{\zT}{\mathbf{0}_T}
\newcommand{\wT}{\mathbold{w}_T}

\newcommand{\kT}{\mathbold{k}_T}
\newcommand{\T}{\mathbold{T}}
\newcommand{\Spbold}{\mathbold{Sp}}
\newcommand{\Dbold}{\mathbold{D}}
\newcommand{\Vbold}{\mathbold{V}}

\newcommand{\myred}{\color{red}}
\newcommand{\mygreen}{\color{mygreen}}
\newcommand{\distsq}{\Delta_{12}^2}
\newcommand{\dist}{\Delta_{12}}
\newcommand{\la}{\langle}
\newcommand{\ra}{\rangle}
\newcommand{\mean}[1]{\left\la\smash{#1}\right\ra}

\newcommand{\caesar}{\textsc{Caesar}\xspace}

\newcommand{\sherpa}{\textsc{Sherpa}\xspace}

\newcommand{\alpgen}{\textsc{Alpgen}\xspace}
\newcommand{\fxfx}{\textsc{FxFx}\xspace}
\newcommand{\pythia}{\textsc{Pythia}\xspace}
\newcommand{\pythiasix}{\textsc{Pythia6}\xspace}
\newcommand{\pythiaeight}{\textsc{Pythia8}\xspace}
\newcommand{\herwig}{\textsc{Herwig}\xspace}
\newcommand{\herwigpp}{\textsc{Herwig7}\xspace}

\newcommand{\nlojet}{\textsc{NLOjet++}\xspace}
\newcommand{\powheg}{\textsc{Powheg}\xspace}
\newcommand{\powhegbox}{\textsc{PowhegBox}\xspace}
\newcommand{\mcatnlo}{\textsc{Mc@NLO}\xspace}
\newcommand{\blackhat}{\textsc{BlackHat}\xspace}
\newcommand{\njet}{\textsc{Njet}\xspace}
\newcommand{\hej}{\textsc{Hej\xspace}}
\newcommand{\ariadne}{\textsc{Ariadne\xspace}}
\newcommand{\loopsim}{\textsc{LoopSim}\xspace}
\newcommand{\cascade}{\textsc{Cascade}\xspace}
\newcommand{\vbfnlo}{\textsc{Vbfnlo}\xspace}
\newcommand{\mcfm}{\textsc{Mcfm}\xspace}
\newcommand{\openloops}{\textsc{OpenLoops}\xspace}
\newcommand{\gosam}{\textsc{GoSam}\xspace}
\newcommand{\madgraphnlo}{\textsc{MadGraph5\_aMC@NLO}\xspace}
\newcommand{\helacnlo}{\textsc{Helac-NLO}\xspace}
\newcommand{\deductor}{\textsc{Deductor}\xspace}
\newcommand{\stripper}{\textsc{Stripper}\xspace}
\newcommand{\mepsatnlo}{\textsc{MePs@NLO}\xspace}
\newcommand{\tmdlib}{\textsc{TMDlib}\xspace}
\newcommand{\tmdplotter}{\textsc{TMDplotter}\xspace}
\newcommand{\fastjet}{\textsc{FastJet}\xspace}
\newcommand{\jaxodraw}{\textsc{JaxoDraw 2.0}\xspace}
\numberwithin{equation}{section}

\newcommand{\be}{\begin{equation}}
\newcommand{\ee}{\end{equation}}
\newcommand{\bea}{\begin{eqnarray}}
\newcommand{\eea}{\end{eqnarray}}

\begin{document}

%-----------------------------------------------------------------------------
\title{ \vspace{5em} 
QCD and Jets at Hadron Colliders~\footnote{Review article published in {\it
Prog. Part. Nucl. Phys. 89 (2016) 1-55}.}}

%-----------------------------------------------------------------------------
\author{Sebastian Sapeta  \\ \\
{\it \normalsize CERN PH-TH, CH-1211, Geneva 23, Switzerland }\\
{\it \normalsize and} \\
{\it \normalsize Institute of Nuclear Physics, Polish Academy of Sciences,} \\
{\it \normalsize ul.\ Radzikowskiego 152, 31-342 Krak\'ow, Poland }
}
\date{}

\maketitle

%--- Preprint numbers --------------------------------------------
\vspace{-29em}
\begin{flushright}
  CERN-PH-TH-2015-281\\
  IFJPAN-IV-2015-19
\end{flushright}
\vspace{20em}

%-----------------------------------------------------------------
\begin{abstract} 

We review various aspects of jet physics in the context of hadron colliders.
We start by discussing the definitions and properties of jets and recent
development in this area.
We then consider the question of factorization for processes with jets, in
particular for cases in which jets are produced in special configurations, like
for example in the region of forward rapidities. 
We review numerous perturbative methods for calculating predictions for jet
processes, including the fixed-order calculations as well as various matching
and merging techniques. 
We also discuss the questions related to non-perturbative effects and the
role they play in precision jet studies.
We describe the status of calculations for processes with jet vetoes
and we also elaborate on production  of jets in forward direction.
Throughout the article, we present selected comparisons between state-of-the-art
theoretical predictions and the data from the~LHC.

\end{abstract}

\setcounter{tocdepth}{3}
\tableofcontents

%-----------------------------------------------------------------------------
%-----------------------------------------------------------------------------
%-----------------------------------------------------------------------------
\section{Introduction}

In the era of the Large Hadron Collider~(LHC), as in the times of all precedent
hadron colliders, jets remain fundamental objects of interest.
They manifest themselves in detectors as collimated streams of charged
particles in the tracker, or as concentrated energy depositions in the
calorimeter. 

Jets measured in experiments are build of hadrons, hence, bound
states characterized by low energy scales of the order of a GeV or
less. However, their existence is a proof of violent phenomena happening at much
higher energies, from tens of GeV to half of the total initial energy of
the colliding particles. 
Such highly-energetic phenomena occur only in a tiny fraction of
hadron-hadron collisions, but, due to the large center-of-mass energy and
high luminosity, jet processes are extremely common at the LHC.
 
Because jets form signatures of large momentum transfers at short
distances, they belong primarily to the perturbative domain of Quantum
Chromodynamics~(QCD).
Predictions for processes involving jets are therefore computed at the level of
partonic degrees of freedom.
The relation between jets of hadrons, measured in experiments, and jets of
partons, for which theoretical results are obtained, is ambiguous. One source of
this ambiguity comes from the parton-to-hadron transitions (hadronization),
which are genuinely non-perturbative, and therefore cannot be controlled
precisely in theoretical calculations. The other reason is that jets at hadron
colliders are always produced in a very busy environment and full theoretical
control over the radiation prior to, or following, the hard scattering is
practically impossible.
 
As the ambiguity cannot be removed, continuous efforts have been made over the
years to formulate jet definitions that permit for precise studies of
short-distance phenomena being at the same time robust with respect to
hadronization or incoherent radiation from other parts of the event.
Such definitions are currently widely adopted and they allow for a fully
controlled comparisons between the theory and experiment.
This, in turn, opens innumerable possibilities for the use of jets.

Since they are genuinely QCD objects, jets can, first of all, be employed for
tests of Quantum Chromodynamics, and the Standard Model~(SM) at large.
Many high-precision studies of jets were performed at
Tevatron~\cite{Ellis:2007ib, Aaltonen:2008eq, Abazov:2011vi} and at the
LHC~\cite{Aad:2010ad, Aad:2011fc, Chatrchyan:2011qta, CMS:2011ab,
Chatrchyan:2012gwa, Chatrchyan:2012bja} finding so far no need for extensions
of the theoretical descriptions beyond the Standard Model~(BSM).
Jets are used for studies of various properties of the strong interactions, such
as measurements of the strong coupling~\cite{Malaescu:2012ts, Rojo:2014kta},
studies of the flavour sector of QCD~\cite{Voutilainen:2015lqa}, as well as
determination of the parton distribution functions (PDFs)~\cite{Rojo:2014kta}.
The modern PDF sets, such as  NNPDF3.0~\cite{Ball:2014uwa},
CT14~\cite{Dulat:2015mca} and MMHT14~\cite{Harland-Lang:2014zoa}, profit from a
great variety of jet data, including those from the LHC, which are crucial in
reduction of the gluon PDF uncertainties at large $x$.
Jet processes are also crucial for such fundamental questions as a validity of
factorization between the short- and the long-distance dynamics in
QCD~\cite{Sterman:2014nua} as well as the existence of a non-linear domain of
the strong interactions~\cite{Gelis:2010nm}. 
They are also instrumental in reaching out to extreme regions of QCD phase space
where theoretical modelling becomes challenging~\cite{Campanelli:2015oaa}.

The importance of jets extends however far beyond the strict domain of physics
of the strong interactions, where they are used as representatives of partons
participating in a hard process. 
This is because jets may have origins different than a short-distance
interaction between quarks and gluons.
They may, for example, also arise from hadronic decays of heavy objects such as
the Higgs boson or the vector bosons, which decay into a pair of jets, or the
top quark decaying into three jets.
Similarly, jets may be produced as decay products of new particles such a
hypothetical $Z'$ resonance,  which can show up as a peak in the tail of a dijet
mass spectrum~\cite{Chatrchyan:2013qha, Aad:2014aqa}, or a variety of SUSY
particles, which would readily decay into many-jet final states.
Even the dark matter and extra dimensions are looked for in events where a
monojet recoils against the missing energy~\cite{Aad:2011xw,
Khachatryan:2014rra}.
Many other jet processes are used to set limits on new
physics~\cite{Aad:2014vwa}.
But jets appear not only in the potential signals of BSM phenomena but they also
contribute to countless backgrounds to processes within and beyond the Standard
Model.
Just to give one example for each category: Higgs analyses divide events in samples with different jet
multiplicities for more efficient background subtractions~\cite{Aad:2012tfa, Chatrchyan:2012xdj},
while gluino production can be mimicked by a $W$+4 jets
process~\cite{Chatrchyan:2013qha, ATLAS:2012pu}.
Finally, jets are extensively used in heavy ion physics. The classic example is
the study of a dense medium created in collisions of large nuclei, which leads
to the asymmetry in dijet events~ \cite{Aad:2010bu}.

The above, long, yet still incomplete, list of applications motivates
considerable, multi-pronged efforts that are being made to develop better
control over jet processes. 
One direction of research focuses on improvements of our understanding of the
properties of jets, as well as the strengths and weaknesses of different jet
definitions and jet-related observables.
Another important area aims at establishing a solid theoretical basis for
the perturbative calculations by studying regions of validity and limitations of
various types of QCD factorization.
Yet another group of activities centres at systematic improvements of the
accuracy of perturbative predictions for all relevant processes with jets.

The aim of this review is to present selected topics from the theory behind
jets, and the phenomenology of jets produced in hadron-hadron collisions.
As jets have been discussed in the literature for nearly four
decades~\cite{Sterman:1977wj}, we will not attempt to fully cover the immense
field of jet physics. Instead, we shall focus on several chosen aspects of QCD
and jet production at hadron colliders and will refer the Reader to
the literature for complementary information.

We shall start from an overview of jet definitions and properties.
Jets turn out to be greatly diverse and rich objects. 
They vary in hardness, shape, mass, susceptibility to soft radiation, 
hadronization corrections and other aspects related to their internal structure. 
We shall elaborate on all of these issues in
Section~\ref{sec:jet-definitions-and-properties}.
 
But reliable QCD predictions for jet processes require not only that jets are
properly defined, but also that the short-distance physics, which we intend to
probe with jets, factorizes from the long-distance dynamics, which is then
parametrized in the form of the parton distribution functions.
This topic is discussed in Section~\ref{sec:factorizaion}.
QCD factorization becomes particularly delicate when one is interested in
using jets to stretch tests of the strong interactions to corners of phase space
where their current understanding is limited.
This often requires developments that go beyond the standard framework of
collinear factorization.

In the last part, which is presented in Section~\ref{sec:jet-production}, we
turn to the discussion of process with jet production in hadron-hadron collisions.
There, we start from elaborating on the factors that limit the precision of the
QCD predictions for jet processes, such as non-perturbative effects and
dependence of the result on jet definitions.
Then we turn to the state-of-the-art perturbative calculations for the processes
involving jets and show selected comparisons to the LHC data.
Those include both the next-to-leading order~(NLO) and the
next-to-next-to-leading order~(NNLO) results in QCD, as well as a variety of
methods for merging the NLO predictions with different jet multiplicities and
matching them to the parton shower~(PS).
Final subsections are devoted to the special cases of event selections, namely
those in which jet radiation is vetoed or where the jets are required to be
produced in forward direction.

Many topics had to be skipped or could only be mentioned briefly because of
space limits.
In particular, we do not provide a complete list of jets techniques and tools.
Many details on defining jets and understanding their properties can be found in
Refs.~\cite{Ellis:2007ib,Salam:2009jx, Altheimer:2013yza, Adams:2015hiv}.
We also do not cover all uses of jets. For those we refer to the recent
summaries devoted to jet physics at the LHC~\cite{Rojo:2014kta,
Voutilainen:2015lqa, Campanelli:2015oaa, Francavilla:2015yxa, Kokkas:2015gfa,
Cacciari:2015jwa}.
Finally, jets in heavy ions are mentioned only briefly in the context of forward
jet production. For complementary information, we refer to
Refs.~\cite{Abreu:2007kv, Mehtar-Tani:2013pia, Armesto:2015ioy, Qin:2015srf, Cacciari:2010te}.

%-----------------------------------------------------------------------------
%-----------------------------------------------------------------------------
%-----------------------------------------------------------------------------
\section{Jet definitions and properties}
\label{sec:jet-definitions-and-properties}

Jets of partons arise in QCD due to the fact that the collinear gluon emissions
are enhanced and the large-angle emissions are rare. 
Because of the former, most of the final state particles cluster into collimated
bunches. If such a bunch carries large transverse momentum, $p_T$, it
is referred to as a jet and its transverse momentum is associated with that of
the original parton that participated in the hard scattering.
Because of the latter, jets are the signatures of large momentum transfer
through local interactions and they form direct evidence of processes taking
places at distances $\sim 1/p_T$~\cite{Sterman:2014nua}.

We see that the concept of a jet is quite intuitive and structures of
collimated streams of particles can be indeed easily found on detector event
displays.
However, in order to relate the jets of hadrons, which are registered by
detectors, to the jets of partons, which can be computed within perturbative
QCD, one needs a precise and robust \emph{jet definition}. Only then, one is
able to meaningfully compare the experimental data with theoretical predictions
and fully exploit the information about the hard interaction carried by jets.

Before embarking on jets in hadron-hadron collisions, which is the main focus of
this review, it is appropriate to introduce the concept of a jet using the
historically first jet definition proposed by Sterman and
Weinberg~\cite{Sterman:1977wj} in the context of $\epem$ collisions. 
This definition
says
that a final state is classified as a 2-jet event if at least a fraction
$1-\epsilon$ of the total available energy is contained in a pair of cones of
half-angle~$\delta$. Hence, the definition depends on two parameters, $\epsilon$
and $\delta$, and it implies that jets take shape of a cone.
This simple definition can be used to compute fractions of 2- and 3-jet events.
At leading order we have $\epem \to q\qbar$, and all events fall into the 2-jet
class. At next to leading order, if the gluon emissions is sufficiently
large-angled and carries more that the fraction $\epsilon$ of the total energy,
the event corresponds to a 3-jet configuration.
The exact 3-jet fraction at NLO is given by 
$f_3 = \frac{g_s^2}{3\pi^2} \left(3 \ln \delta + 4 \ln \delta \ln 2\epsilon +
\frac{\pi^2}{3}-\frac74\right)$~\cite{Sterman:1977wj}. 
As expected, for $\delta, \epsilon \ll 1$, the 3-jet fraction increases with
the decreasing cone size, $\delta$, and with the increasing energy fraction
outside of the two hardest cones, $1-\epsilon$.

%-----------------------------------------------------------------------------
\subsection{Jets at hadron colliders}

As one moves to hadron-hadron collisions, jet definition has to be reformulated
since there is no special direction around which the first two cones could be
placed and the total energy of the final state particles cannot be determined.
It is therefore much more natural to define jets with a bottom-up approach,
starting to cluster the particles which are closest according to some distance
measure~\cite{Catani:1993hr, Ellis:1993tq, Dokshitzer:1997in, Wobisch:1998wt}.
This \emph{sequential-recombination} procedure was for a long time
believed to be very slow, with the time needed to cluster $N$ particles scaling
as $N^3$.
That led to developments of various 
\emph{cone-type algorithms}, which were more
practical in terms of the time required to cluster large numbers of particles,
since they were scaling as $N^2\ln N$.
The cone algorithms were widely used at Tevatron and we refer to
Refs.~\cite{Ellis:2007ib, Salam:2009jx} for further details.
However, because of the reasons just mentioned, there was no simple way to
introduce cones and that always came at the price of violating collinear and
infrared safety of a jet definition. This problem has eventually been solved
with the SISCone algorithm~\cite{Salam:2007xv}. 
Around the same time, the sequential recombination algorithms were optimized
and developed such that they needed only $\order{N\ln N}$~\cite{Cacciari:2005hq}
or $\order{N^{3/2}}$~\cite{Cacciari:2008gp} time to cluster $N$ particles.
Those modern jet algorithms are used for virtually all jet-related measurements
at the LHC and we shall discuss them in detail in Section~\ref{sec:jet-def}. 

In addition to the speed of an algorithm, the main concern is always the
\emph{infrared} and \emph{collinear}~(IRC) \emph{safety} of a jet definition.
This important problem will be explained in the next subsection.
 
Other problems specific to jet clustering in events with two incoming hadrons
have to do with the \emph{underlying event}~(UE) and \emph{pileup}~(PU).
The first is defined as a soft or moderately hard radiation accompanying the
production of hard objects, such as jets or vector bosons. The second stems from
multiple simultaneous hadron-hadron collisions per bunch crossing. We discuss the
issues of UE and PU in the context of jet physics in Section~\ref{sec:ue-pu}.
Finally, the hadronization of partons into hadrons has a potential impact on jet
properties and we elaborate on this topic in Section~\ref{sec:hadronization}.

%-----------------------------------------------------------------------------
\subsection{Infrared and collinear safety}
\label{sec:irc-safety}

\begin{figure}[t]
  \begin{center}
    %%% IRC safe jet definition
    \begin{minipage}{3.0cm}
     \begin{center} 
      \includegraphics[height=80pt]{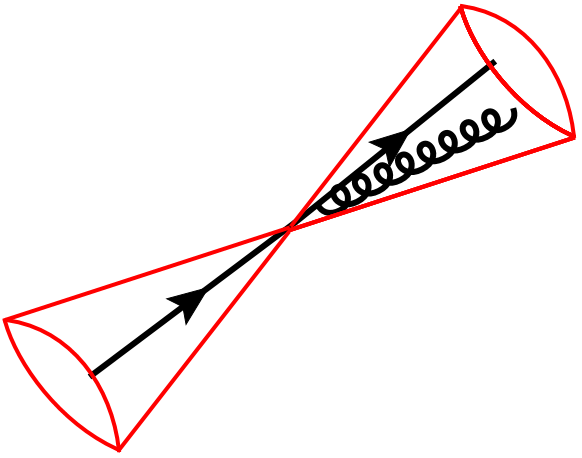}
     \end{center} 
    \end{minipage}
    \begin{minipage}{0.5cm}
      \centering
      \ \ +
    \end{minipage}
    \begin{minipage}{3.0cm}
      \includegraphics[height=80pt]{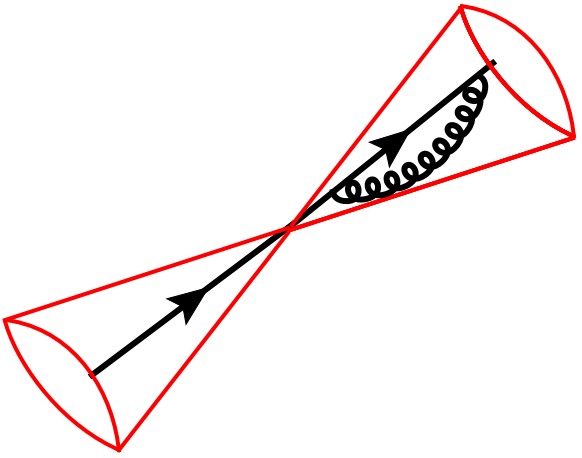}
    \end{minipage}
    \begin{minipage}{7cm}
     \begin{flushright}
      $ = \displaystyle \frac{1}{\epsilon}J^{(2)}_\text{IRC-safe}
        - \displaystyle \frac{1}{\epsilon}J^{(2)}_\text{IRC-safe} =
	{\mygreen \text{finite}}$
     \end{flushright}
    \end{minipage}
    \vspace{10pt}

    %%% IRC unsafe jet definition
    \begin{minipage}{3.0cm}
     \begin{center} 
      \includegraphics[height=80pt]{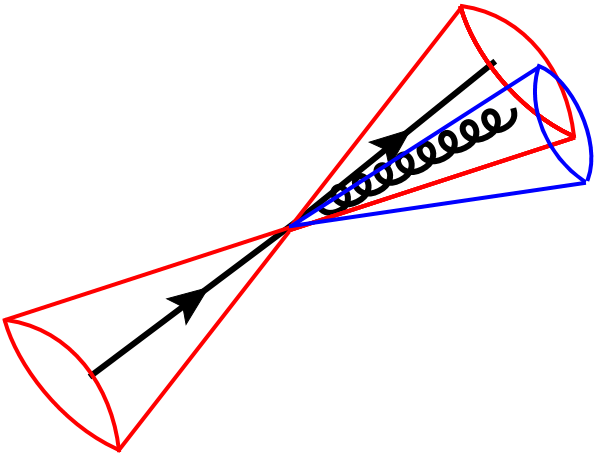}
     \end{center} 
    \end{minipage}
    \begin{minipage}{0.5cm}
      \centering
      \ \ +
    \end{minipage}
    \begin{minipage}{3.0cm}
      \includegraphics[height=80pt]{ee-2jets-cone-safe-virt.png}
    \end{minipage}
    \begin{minipage}{7cm}
      \begin{flushright}
      $ = \displaystyle \frac{1}{\epsilon}J^{(3)}_\text{IRC-unsafe}
        - \displaystyle \frac{1}{\epsilon}J^{(2)}_\text{IRC-unsafe} = 
	{\myred \infty}$
      \end{flushright}
    \end{minipage}
  \end{center}
  \caption{
    Importance of the infrared and collinear safety of a jet algorithm.  In the
    top row, an IRC-safe jet definition is used and the corresponding
    cross section is finite at any perturbative order. 
    In the bottom row, an IRC-unsafe jet definition brakes singularity
    cancellation between the real and virtual diagrams and leads to an infinite
    cross section.
  }
  \label{fig:irc-safety}
\end{figure}

Jets are meant to be proxies of the hard partons which participated in the
short-distance interaction at early times of a hadron-hadron collision. These
hard partons carry large transverse energy, which is subsequently released by
consecutive splittings. Because of the soft and collinear enhancements of the
QCD branchings, in the majority of cases, the series of emissions does not
change direction of the energy flow.

The cross sections in QCD diverge when the angle of emission or the energy of
the emitted gluon go to zero. In the perturbative regime, each emission
corresponds to the real part of a higher order correction and comes with a power
of the strong coupling, $\as$.
Hence, $n$ emissions contribute to $\order{\as^n}$ correction. However, the
complete $\order{\as^n}$ result requires also diagrams with up to $n$ loops. And
these diagrams come with divergences that match exactly those of the real
emissions. Once the real and the virtual contributions at the order
$\as^n$ are added together, the cross section becomes finite up to this
order.
This intuitively natural results stems from unitarity and was formally proved by
Kinoshita, Lee and Nauenberg~\cite{Kinoshita:1962ur, Lee:1964is}.
The above theoretical mechanism of singularity cancellations is also realised in
experiment thanks to the finite energy and angle resolution, which makes the events with ultra-soft
or collinear emissions indistinguishable from those with no emissions, 
the latter corresponding to virtual corrections.

The above mechanism of cancellation of the singularities may not work with
a bad choice of an observable and, in our context, a jet definition.
The problem is schematically illustrated in Fig.~\ref{fig:irc-safety}. In the
top row we see the real and virtual corrections to the dijet production. 
Each of them is separately divergent, which is denoted by the $1/\epsilon$ pole
of dimensional regularization on the right hand side. The red cones represent a
jet definition. We see that both the real and virtual diagram are classified as
2-jet configurations, hence the poles are multiplied by the same jet function
$J^{(2)}$ (which, in practice, is a definition of an observable) and the
divergent terms cancel in the sum leading to a finite result.

This is to be contrasted with the situation depicted in the bottom row of
Fig.~\ref{fig:irc-safety}, where a different jet definition was applied to the
very same real and virtual diagrams. As we see, now, the real diagram is
classified as a 3-jet event while the virtual diagram is still a 2-jet event.
This has severe consequences because the poles are now multiplied by
different jet functions, respectively $J^{(3)}$ and $J^{(2)}$. Thus,
the cancellation of singular terms does not occur, and the final result is infinite.
Infinities cannot of course appear in real experimental situations, where they
are always regularized by a finite granularity of a detector. Hence, the fact
that we obtain a nonsensical theoretical result in the above example comes from
the bad choice of a jet definition.

The situation from the top row of Fig.~\ref{fig:irc-safety} corresponds to the
\emph{infrared} and \emph{collinear} (IRC) \emph{safe} jet algorithm, which has
a property that the set of hard jets cannot be modified by an arbitrarily
collinear or soft emission (either of perturbative origin or coming from
non-perturbative dynamics at scales below $\Lambda_\qcd$).
In general, an IRC-safe observable forms a sum over all states with similar
energy flow into the same final state~\cite{Sterman:2014nua}.
On the contrary, the jet algorithm used in the bottom row of
Fig.~\ref{fig:irc-safety} is IRC-unsafe, as an arbitrary collinear emission is
capable of changing the set of hard jets.

It is clear from the above examples that the IRC-safety of a jet definition is
a crucial requirement if we are not to waste the results for higher order
corrections to process with jets. Many algorithms used in the past had problems
with IRC safety, which were appearing at different levels of perturbative
expansion (see~\cite{Ellis:2007ib, Salam:2009jx} for detailed discussions).
All modern jet algorithms used at the LHC fully comply with the
IRC safety requirement. Hence, they can be used for calculations at arbitrary
precision, which then can be meaningfully compared to the experimental results.

%-----------------------------------------------------------------------------
\subsection{Modern jet algorithms}
\label{sec:jet-def}

A comprehensive discussion of all the modern jet algorithms can be found
in~\cite{Salam:2009jx} as well as in the original articles~\cite{Catani:1993hr, Ellis:1993tq, Dokshitzer:1997in, Wobisch:1998wt,Salam:2007xv, Cacciari:2005hq,  Cacciari:2008gp}.
In order to make our review self-contained, below, we provide a brief summary of
the jet algorithms which became standard choices at the LHC. 

A complete jet definition consists of the following elements:

\begin{center}
  \it
  Jet definition = jet algorithm +  parameters  + recombination scheme.
\end{center}

As already mentioned, jet algorithms fall into two classes: the \emph{cone
algorithms} and the \emph{sequential-recombination algorithms}. Each jet
algorithm comes with at least one free parameter.
Recombination scheme specifies how the two 4-momenta of particles $i, j$ combine
into a 4-momentum of a particle $k$. Currently, one uses almost exclusively the
so-called $E$-scheme, where the 4-momenta of $i$ and $j$ are simply added,
hence, $p_k = p_i + p_j$.

The \emph{cone algorithms} represent a top-down approach to jet finding.
They were historically first, with the Sterman-Weinberg
algorithm~\cite{Sterman:1977wj} for $\epem$, and they were later extensively
used at hadron colliders, especially the Tevatron~\cite{Ellis:2007ib}.
Most of them were however plagued with the issues of the IRC
unsafety~\cite{Salam:2009jx}.
The problems originated from the need to define seeds in order to start an
iterative procedure to search for stable cones. Those seed were identified with 
final state particles. Such procedure is manifestly IRC-unsafe, as an emission
of a soft or collinear parton changes the set of initial seeds, which in turn,
for a non-negligible fraction of events, leads to a different set of the
final-state jets.
Resolution of this long-standing problem came with the Seedless Infrared-Safe
Cone jet algorithm (SISCone)~\cite{Salam:2007xv}, where an efficient procedure
for finding stable cones, without introducing initial seeds, was proposed.

The \emph{sequential recombination algorithms} dominate almost exclusively in
the jet measurements at the LHC. They represent a bottom-up approach by starting
to combine the closest particles, according to a distance measure which can be
generally written as
\begin{equation}
    d_{ij} = \min(p_{Ti}^{2p}, p_{Tj}^{2p})\frac{\Delta R_{ij}^2}{R^2}\,,
    \qquad \qquad \qquad
    d_{iB} = p_{Ti}^{2p}\,,
  \label{eq:seq-rec-dist}
\end{equation}
where $d_{ij}$ is a distance between the particles $i$ and $j$ and $d_{iB}$ is a
distance between the particle $i$ and the beam. The parameter $R$ is called
the \emph{jet radius} and $\Delta R_{ij}^2 = (y_i-y_j)^2 + (\phi_i-\phi_j)^2$
is the geometric distance between the particles $i$ and $j$ in the
rapidity-azimuthal angle plane.
The value of the parameter $p$ defines specific algorithm from the
sequential-recombination family: 
$p=1$ for the $k_T$ algorithm~\cite{Catani:1993hr, Ellis:1993tq}, 
$p=0$ for the Cambridge/Aachen~(C/A) 
algorithm~\cite{Dokshitzer:1997in, Wobisch:1998wt}, 
and $p=-1$ for the anti-$k_T$~\cite{Cacciari:2008gp} algorithm.

Given a set of the final-state particles, each procedure of finding jets with
the sequential-recombination algorithm consists of the following steps:
\begin{enumerate}
  \item
  Compute distances between all pairs of final-state particles, $d_{ij}$, as
  well as the particle-beam distances, $d_{iB}$, using the measure from
  Eq.~(\ref{eq:seq-rec-dist}).
  \item
  Find the smallest $d_{ij}$ and the smallest $d_{iB}$ in the sets of distances
  obtained above.
  \begin{itemize}
    \item
    If $d_{ij} < d_{iB}$, recombine the two particles, remove them from the list
    of final-state particles, and add the particle ${ij}$ to that list.
    \item
    If $d_{iB} < d_{ij}$, call the particle $i$ a jet and remove it from the
    list of particles.
  \end{itemize}
  \item
  Repeat the above procedure until there is no particles left.
\end{enumerate}

In spite of the fact that the distance measure of the three algorithms can be
written as a single formula~(\ref{eq:seq-rec-dist}), because of the different
values of the power $p$, each of them exhibits a different behaviour. The $k_T$
algorithm starts from clustering together the low-$p_T$ objects and it
successively accumulates particles around them.
The C/A algorithm is insensitive to the transverse momenta of particles and it
builds up jets by merging particles closest in the $y-\phi$ plane.
The anti-$k_T$ algorithm starts from accumulating particles around high-$p_T$
objects, just opposite to the behaviour of the $k_T$ algorithm. In the
anti-$k_T$ algorithm, the clustering stops when there is nothing within radius
$R$ around the hard center. For that reason, anti-$k_T$ leads to jets that take
circular shapes in the $y-\phi$ plane.
This last feature makes 
the anti-$k_T$ algorithm particularly attractive from the experimental point of
view. The reason is that jets with regular shapes allow for 
reliable interpolation between detector regions separated by dead zones. 
That is why the anti-$k_T$ algorithm became a default choice at the LHC.

All the algorithms discussed in this section are available within the
\fastjet package~\cite{Cacciari:2011ma}.

%-----------------------------------------------------------------------------
\subsection{Jet mass}

Amongst a number of properties of a jet, its mass turns out to be especially
important in numerous contexts. In the approximation of massless QCD partons,
the jet mass arises due to its substructure. In pure QCD, the substructure comes
from radiation of gluons and quarks.
However, in processes involving a hadronic decay of a heavy object of mass $m$
and $p_T \gg m$, the decay products will also end up in a single jet building up
its mass.
 
If a jet $J_{12}$ is obtained from clustering of the two subjets $J_1$ and
$J_2$, its exact mass is given by~\cite{QuirogaArias:2012nj}
\begin{equation}
  \label{eq:mjexact}
  m_{12}^2 = 
  2 m_{T1} m_{T2}\cosh(y_{2}-y_{1}) -
  2 p_{T1} p_{T2}\cos(\phi_{2}-\phi_{1})
  +m^2_{1}+m^2_{2} \,,
\end{equation}
where $m_{Ti}  = \sqrt{m_{i}^2 + p_{Ti}^2}$ are the
transverse masses of the subjets, while $p_{T1}$, $p_{T2}$ and $y_1$, 
$y_2$ are, respectively, the subjets' transverse momenta and rapidities.
In the limit of $m_{i} \ll p_{Ti}$ and $R \ll 1$, the above formula
reduces to
\begin{equation}
  m^2_{{12}}  \simeq m^2_{1}+m^2_{2}+ z (1-z)\, p_{T {12}}^2 \distsq\,,
  %\label{eq:}
\end{equation}
where $p_{T {12}}$ is the transverse momentum of the jet formed by recombination
of particles 1 and 2, while $\dist$ and $z$ are given by
\begin{equation}
   \dist = \sqrt{(y_1-y_2)^2+(\phi_1-\phi_2)^2} \equiv x R \,, \qquad \qquad
   z = \frac{\min(p_{T1}, p_{T2})}{p_{T1}+p_{T2}}\,.
  \label{eq:x-z-def}
\end{equation}

Jet mass is an infrared and collinear safe quantity that can be calculated order
by order in perturbation theory.
Because of the soft and collinear singularities of the QCD matrix element for
gluon emission, the distribution of masses, $m_J$, of the QCD jets receives
strong enhancement at low values of $m_J$. At the lowest,  non-trivial order,
the approximate result for the mass distributions of QCD jets is given by 
$\frac{d\sigma}{dp_{TJ}dm_J} \propto \alpha_s(p_{TJ}) \frac{4 C_i}{\pi m_J} \ln
\left(\frac{R\, p_{TJ}}{m_J}\right)$~\cite{Almeida:2008tp}, where $C_i$ is the
colour factor of the initiating parton and $R$, $p_{TJ}$, $m_{J}$ are the jet's
radius, transverse momentum and mass, respectively.
The higher order terms are enhanced by further powers of 
$\ln \frac{R\, p_{TJ}}{m_J}$. 
 
Contrary to the case of QCD, the distribution of jets coming from a decay of a
heavy object is flat in $z$ and therefore, the mass distribution of such jets is
peaked around the mass of the heavy object which originated them. 
This will be discussed further in Section~\ref{sec:substructure}.

%-----------------------------------------------------------------------------
\subsection{Jet area}
\label{sec:jet-area}

It is intuitive to think that the larger the jet, the more its transverse
momentum is susceptible to contamination from soft radiation, such as UE or PU.
This is just because the jets will capture the incoherent radiation
proportionally to their area, hence, larger jets will be more affected (in
absolute terms).
The naive geometrical expectation for the area of a jet with radius $R$ is $\pi
R^2$. A closer investigation reveals that the actual area of a jet is in most
cases different and that there is some freedom in its definition.
 
A quantitative discussion of jet areas started with the work
of Ref.~\cite{Cacciari:2008gn} were two types, the \emph{passive} and the
\emph{active} area, were introduced. They both use the concept of \emph{ghosts},
$\{g_i\}$, \ie infinitely soft particles which are added to the set of the
final state particles $\{p_k\}$. If the whole ensemble $\{p_k, g_i\}$ is
clustered with an IRC-safe algorithm, the resulting set of jets $\{J_n\}$ will
be identical to that from clustering just the physical particles $\{p_k\}$.

The scalar \emph{passive area} of the jet $J$ is defined 
as the area of the region in the $y-\phi$ plane in which the single ghost
particle, $g$, is clustered with $J$ 
\begin{equation}
  \label{eq:passive-area-def}
  a(J) \equiv \int dy\, d\phi\, f(g (y,\phi),J),
  \hspace{30pt}
  f (g, J) = 
  \bigg \{
  \begin{array}{cl}
    1  & {\rm for \ } g {\rm \ clustered\  with\ } J       \\
    0  & {\rm for \ } g {\rm \ not \ clustered\  with\ } J
  \end{array}\,.
\end{equation}
A 4-vector version of the passive area is introduced in a similar
way~\cite{Cacciari:2008gn}.

The passive area (\ref{eq:passive-area-def}) provides a measure of the
susceptibility of the jet to soft radiation in the limit in which this radiation
is pointlike.
For a 1-particle jet $J_1$,  $a(J_1) = \pi R^2$, for all four jet
clustering algorithms: $k_T$, C/A, anti-$k_T$ and SISCone.
For a 2-particle jet, the passive area starts to depend on the jet definition
and the geometrical distance between particles. The analytic results for
strongly ordered transverse momenta were obtained in
Ref.~\cite{Cacciari:2008gn}, and, in Ref.~\cite{Sapeta:2010uk}, they were
generalized to the case with arbitrary transverse momenta. 

The \emph{active area} has more physical relevance and it is defined with a
dense coverage of ghosts, randomly distributed in the $y-\phi$ plane.
If the number of ghosts from a particular ghosts ensemble~$\{g_i\}$ clustered
with the jet $J$ is ${\cal N}_{\{g_i\}}(J)$, and the number of ghosts from this
ensemble per unit area is $\nu_{\{g_i\}}$, then the active scalar area is given
by
\begin{equation}
    \label{eq:active-area-def}
    A(J) \equiv 
    \lim_{\nu_{g} \to \infty}
    \left\langle A(J\,|\,\{g_i\}) \right\rangle_g\,,
    \qquad
    A(J\,|\,\{g_i\}) = 
    \frac{{\cal N}_{\{g_i\}}(J)}{\nu_{\{g_i\}}}\,,
\end{equation}
where $\langle \ldots \rangle_g$ denotes the average over many ensembles
of ghosts with the number of ghosts in each ensemble $\nu_g \to \infty$.
Similarly, the 4-vector active area may be defined~\cite{Cacciari:2008gn}.

\begin{figure}[t]
    \centering
    \includegraphics[width=0.32\textwidth]
        {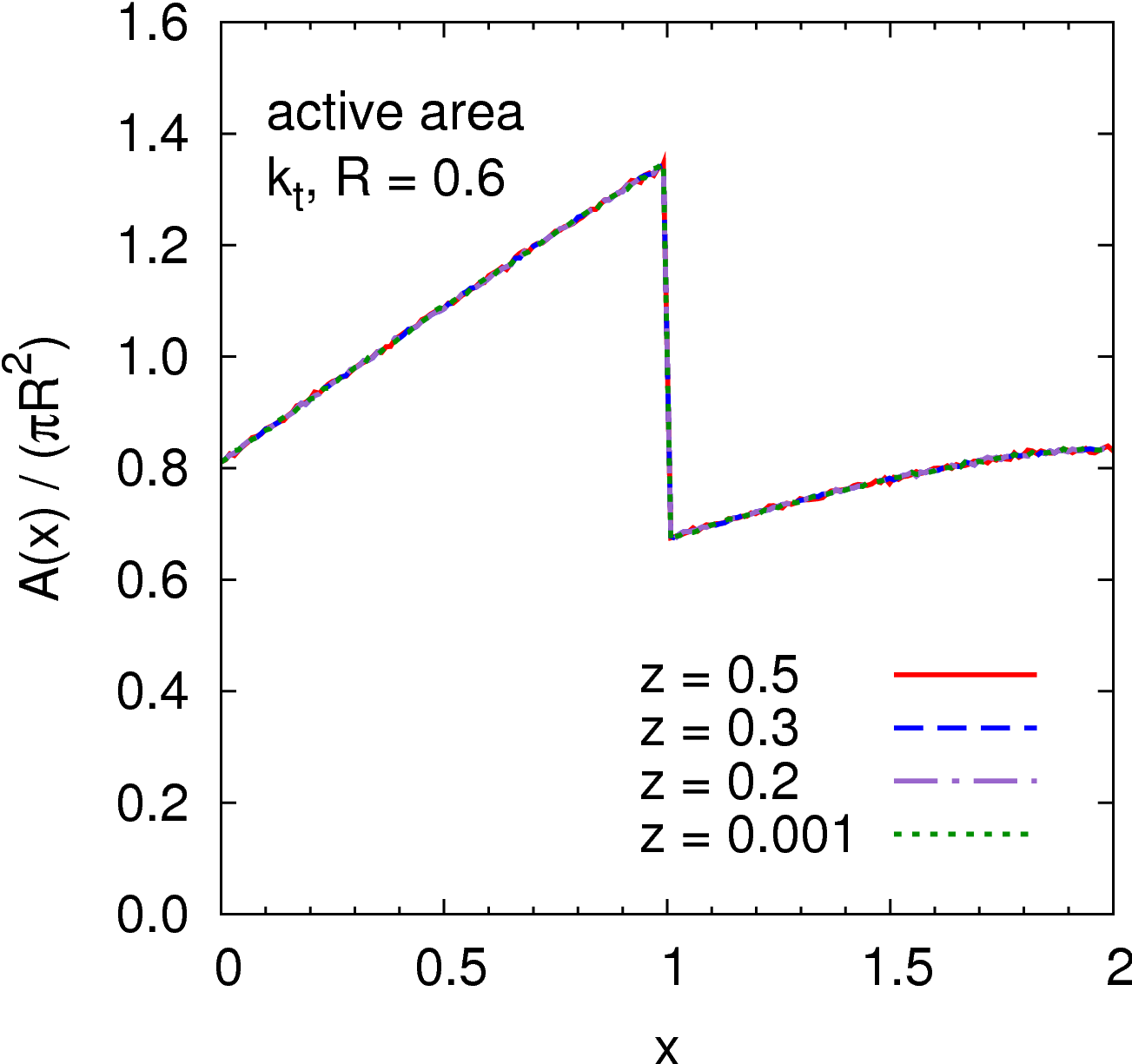}
    \hfill
    \includegraphics[width=0.32\textwidth]
        {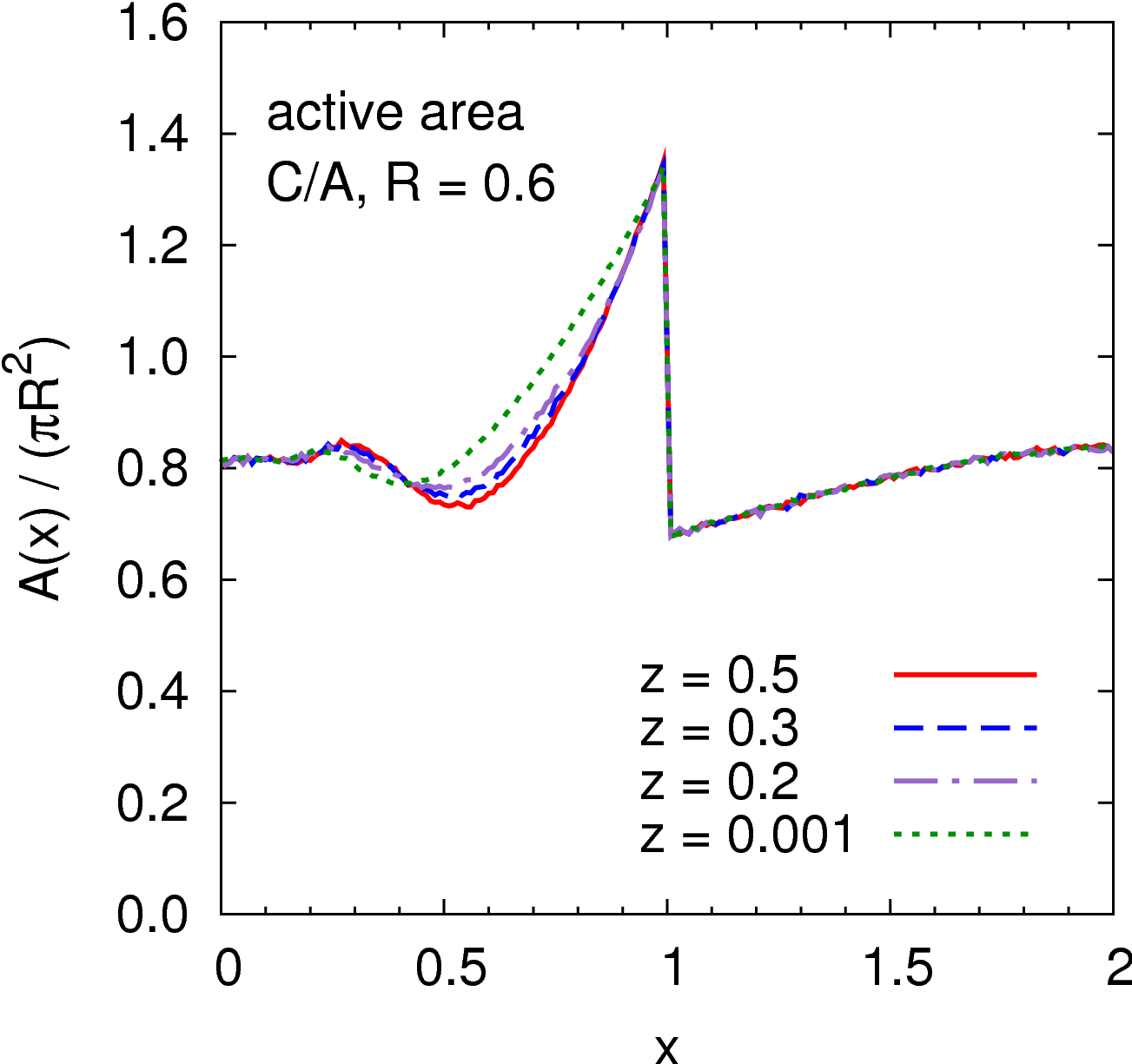}
    \hfill
    \includegraphics[width=0.32\textwidth]
        {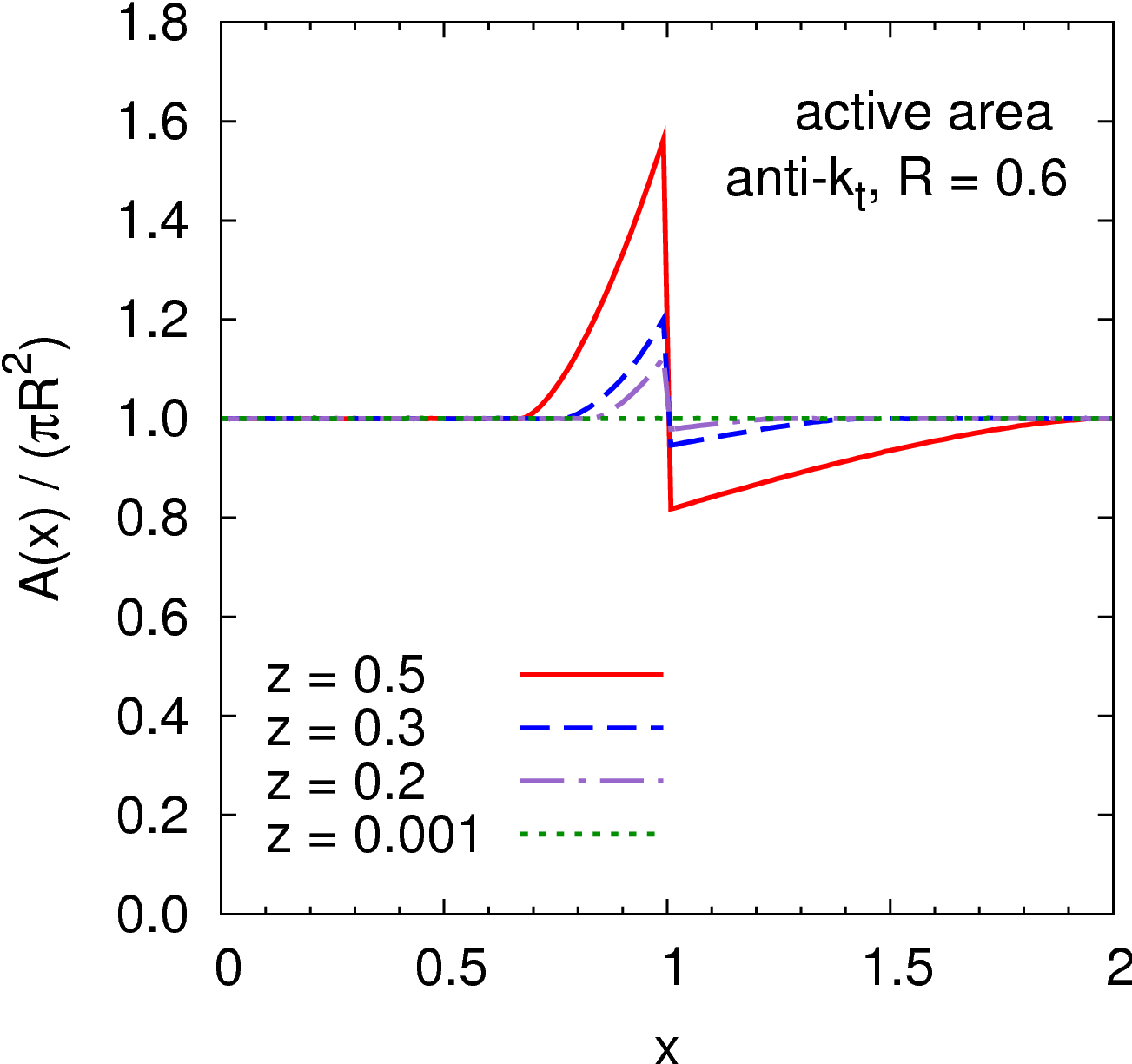}
    \caption{
    Active areas of the hardest jet for the system of two particles separated by
    the distance $x R$, defined in Eq.~(\ref{eq:x-z-def}) together with the
    asymmetry parameter $z$.
    Figure from Ref.~\cite{Sapeta:2010uk}.
    }
    \label{fig:active-area2p}
\end{figure}

Fig.~\ref{fig:active-area2p} shows the active area of the hardest jet in
a 2-particle system with
the geometrical separation between the particles, $\Delta_{12} \equiv x R$,
and the relative transverse momenta of the
constituents, $z$, given in Eq.~(\ref{eq:x-z-def}).
We see that the ``non-conical algorithms'', \ie $k_T$, which start from
clustering ghosts among themselves, and C/A, 
%with jets of highly
%irregular areas, 
exhibit virtually no dependence on the $z$ parameter, which
measures how much of the total jet's transverse momentum is taken by the softer
particles.
On the contrary, the active area of the ``conical'', anti-$k_T$ algorithm
depends quite strongly on the $p_T$ asymmetry of the constituents when those
constituents are separated by $\simeq R$.

In all three cases shown in Fig.~\ref{fig:active-area2p}, the areas increase
with the constituent separation for $x<1$ ($\dist < R$). 
In this region, the two particles form a single jet.
For $x>1$, each particle is clustered into a separate, 1-particle jet but
because the distance between the particles is smaller than $2R$, the area of the
hardest jet is smaller than that of a 1-particle jet in a single-particle event.
As $x\to 2 $, however, the hardest jet area tends to the result for 1-particle
active area.

%-----------------------------------------------------------------------------
\subsection{Jet mass area}
\label{sec:jet-mass-area}

\begin{figure}[t]
    \centering
    \includegraphics[width=0.32\textwidth]
        {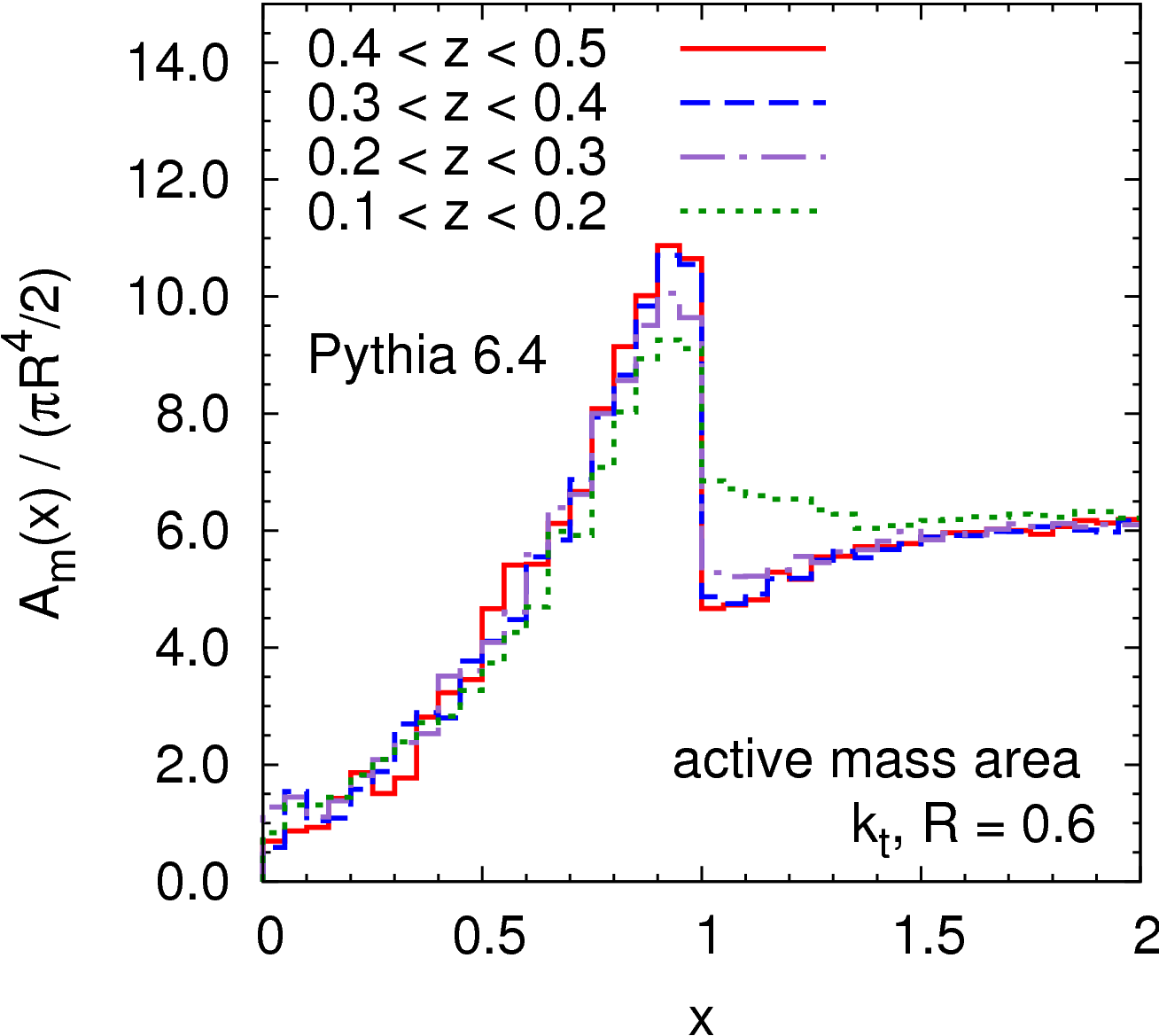}
    \hfill
    \includegraphics[width=0.32\textwidth]
        {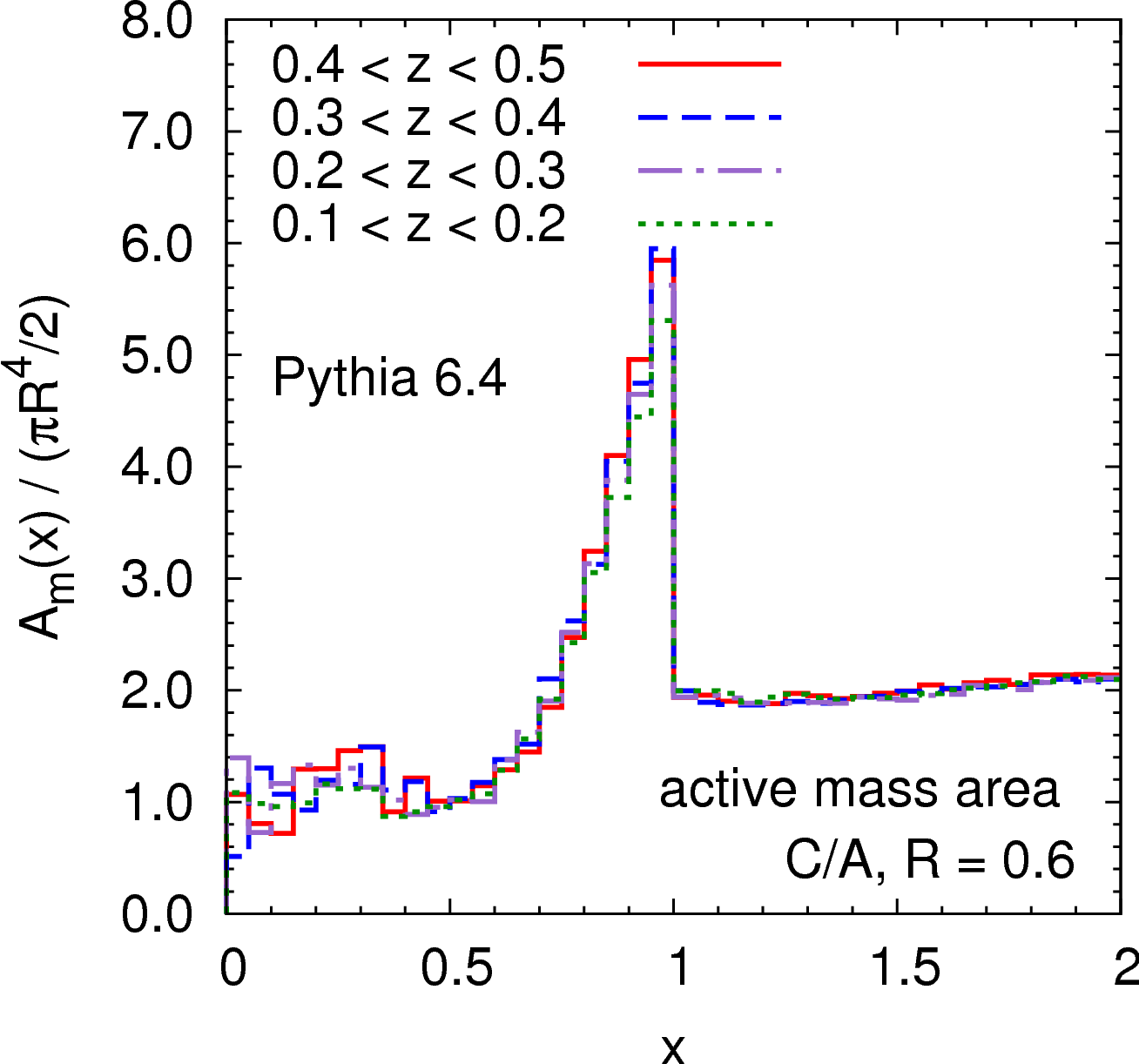}
    \hfill
    \includegraphics[width=0.32\textwidth]
        {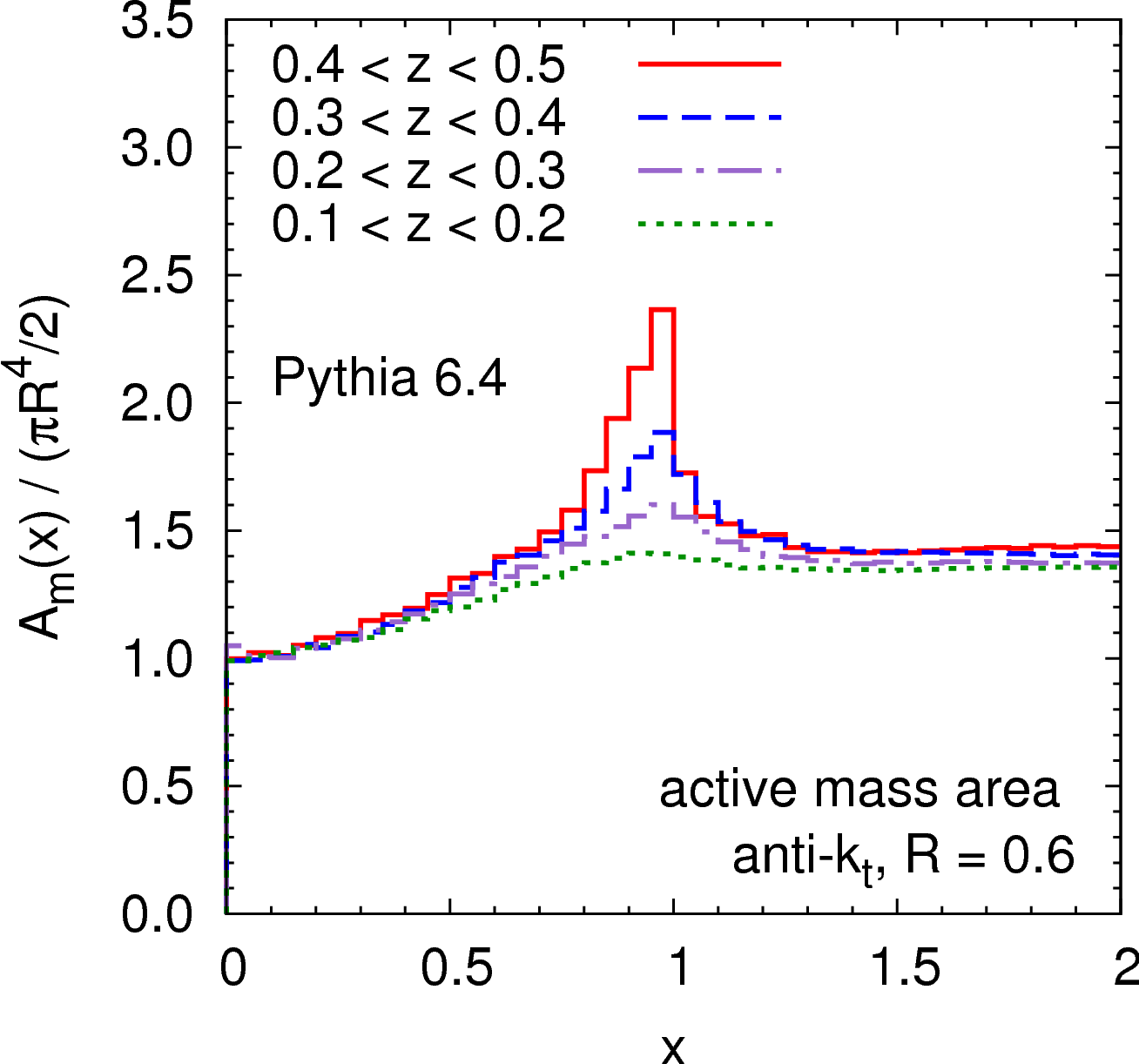}
    \caption{
    Active mass areas of the hardest jet from dijets events simulated with
    \pythia. The distance~$x$ and the asymmetry
    parameter~$z$, defined in Eq.~(\ref{eq:x-z-def}),  were obtained using the
    two hardest subjets.
    Figure from Ref.~\cite{Sapeta:2010uk}.
    }
    \label{fig:active-massarea}
\end{figure}

Just like the value of the jet area specifies susceptibility of jet's transverse
momentum to incoherent radiation, the value of the \emph{mass area} 
specifies how much that radiation affects the jet mass. It can be also defined
in the passive and active variants, of which the latter has more relevance for
jets produced at hadron colliders.

The \emph{active mass area} is defined as~\cite{Sapeta:2010uk} 
\begin{equation}
    \label{eq:active-ma-def}
    A_m(J) \equiv 
    \lim_{\nu_{\{g_i\}} \to \infty}
    \left\langle 
      \frac{m^2_{J{\{g_i\}}}-m^2_{J}}{\nu_{\{g_i\}}\mean{p_{tg}}p_{TJ\{g_i\}}} 
    \right\rangle_g
    =
    \frac{2}{p_{TJ}}\, p^\mu_{J} A_{\mu}(J)\,,
\end{equation}
where $m_{J}$ is a mass of the pure jet $J$ and $m_{J{\{g_i\}}}$ is a mass of
the jet consisting of $J$ and a dense coverage of ghosts from the random
ensemble  $\{g_i\}$. 
The last equation holds when all jet constituents are massless, and $A_\mu(J)$
is a 4-vector, active jet area.
 
Analytic study of the active jet areas for the 2-particle jets, as well as their
passive analogues, can be found in Ref.~\cite{Sapeta:2010uk}.
The 1-particle jet passive mass area is equal to $\pi R^4/2$ and this is how all
the active and passive mass areas scale.
The qualitative behaviour of the mass areas of the 2-particle jets is very
similar to that found for jet areas. The $x$ dependence follows that of 
Fig.~\ref{fig:active-area2p} and $z$ dependence is again very weak for $k_T$ and
C/A and fairly sizable for anti-$k_T$.

In Fig.~\ref{fig:active-massarea} we show the analogue of
Fig.~\ref{fig:active-area2p} for the active mass areas but, this time, the
hardest jet comes from full \pythiasix~\cite{Sjostrand:2006za} simulation of
dijets events at hadron level, with the underlying event switched off. 
The results from Fig.~\ref{fig:active-massarea} are in qualitative agreement
with those for the active mass areas of 2-particle jets and follow the shapes of
the jet areas of Fig.~\ref{fig:active-area2p}.
The general pattern of growth of the mass area with $x$ and then a drop at
$x\simeq 1$ is well observed. Also, the sensitivity to the $z$ value is similar
to that found in Fig.~\ref{fig:active-area2p}:
low for $k_T$ and C/A while noticeable for anti-$k_T$.

Jets from full simulation are of course much more complex, which leads to some
differences with respect to a simple 2-particle picture. We see, for example,
that the mass areas from the $k_T$ algorithm are significantly larger than those
from C/A and anti-$k_T$. This is related to a larger scaling violation in the
case of $k_T$, which means that even collinear emissions can lead to a
significant increase of the mass area~\cite{Sapeta:2010uk}.
Another complication with respect to the simple 2-particle jets is that, in the
latter case, the hardest jet area and mass area always return to the 1-particle
jet result as $x\to 1$. In real-life jets they stay much bigger since, with
many particles in the final state, widely separated subjets develop their own
substructure and cannot be any longer approximated by a single, massless
particle.

%-----------------------------------------------------------------------------
\subsection{Jet substructure}
\label{sec:substructure}

It is apparent from our discussion so far that jets have a very rich
substructure. Patterns of radiation found inside a jet carry important information about its origins.
The studies of jet substructure have become a separate, and by now well
developed, field of research. It is therefore impossible to do justice to all
important results that appeared over the last yeas as that alone would require a
dedicated review article.
We shall however briefly sketch the main ideas behind the studies of jet
substructure and describe the most important techniques.

The main motivation behind the studies of a structure of radiation inside a jet
is a potential enhancement of the signal vs background discrimination. Suppose
that we find a jet with certain mass and transverse momentum and with two well
defined subjets.  Those subjets can be, to a first approximation, modelled as
QCD partons that originated from a single vertex. Depending on the nature of
this vertex, the outgoing partons share the momentum of the incoming particle in
a very different way. In the collinear limit we have: $P(z) \propto 1$ for the
decay of a heavy boson $V \to q\qbar$ but 
$P(z) \propto \frac{1+z^2}{1-z}$ for the QCD splitting $q \to qg$. Here, $P(z)$
is a splitting probability with the momentum shared 
between the outgoing partons
in fractions $z$ and $1-z$.
This simple observation opens a possibility of discriminating between jets
coming from decays of heavy, colourless objects, and those originating from pure
QCD branchings. In the first case, the two subjets will share the momenta of a
jet symmetrically, whereas in the second case, one of the subjets will be much
harder than the other.
Similar considerations apply to the angle between the two subjets. In majority
of cases, this angle will be larger for $V \to q\qbar$ than for $q \to qg$, as
the latter splitting is collinearly enhanced.

The angle and momentum enter the distance measure of the $k_T$ algorithm, \cf
Eq.~(\ref{eq:seq-rec-dist}). Hence, by taking one step back in the clustering of
a jet, we obtain two subjets which we can treated as proxies of the partons
originating from the relevant splitting (\ie the one that builds up most of the
jet mass~\cite{QuirogaArias:2012nj}).
This procedure was first used in Ref.~\cite{Seymour:1993mx} in the study of $H
\to W W$ followed by one $W$ decaying leptonically and the other hadronically.
If the $p_T$ of a lepton from one $W$ is large, the hadronic decay products from
the other $W$ will end up in a single, fat jet. The $k_T$ measure distance
between the two hardest subjets, $d_{12}$, will be on average substantially
larger if those jet come from the $W \to jj$ decay than if they come from a QCD
splitting.  Therefore, by rejecting the events with  $d_{12} < d_\cut$, a
procedure generally called \emph{tagging}, one can significantly improve the
signal to background ratios.
Similar techniques were proposed later to study the $WW \to jj + \l
\nu$ process~\cite{Butterworth:2002tt} and to enhance SUSY signals~\cite{Butterworth:2007ke}.

The above ideas have been developed and refined in the BDRS study of
Ref.~\cite{Butterworth:2008iy} focused on suppressing large backgrounds to the
associated Higgs production with a subsequent decay to the $b\bbar$ pair, 
$pp \to Z H \to l^+ l^- b \bbar$. 
Here, the C/A algorithm has been used to study the substructure of the fat jet,
$J$, containing the two bottom quarks, expected to enter the two subjets, $J_1$
and $J_2$, of which $J_1$ is heavier. 

Contrary to the $k_T$ algorithm, C/A
does not necessarily end the clustering with the relevant splitting.
Therefore, to identify the latter, the \emph{mass drop} condition has been
introduced in addition to the \emph{asymmetry cut}. The two conditions read
\begin{equation}
  \frac{m_{J_1}}{m_J} < \mu_\text{cut} \, , 
  \qquad 
  \text{and }
  \qquad
  \frac{\min \left(p_{TJ_1}^2, p_{TJ_2}^2\right)}{m_J^2}\Delta^2_{12} 
  > y_{\cut}\,,
  \label{eq:bdrs}
\end{equation}
with $\Delta_{12}$ defined as in Eq.~(\ref{eq:x-z-def}).
If both conditions are met, \ie the splitting $J \to J_1 J_2$ is not too
asymmetric and the unclustering leads to a significant mass drop, than the
branching is identified as the $H \to b\bbar$ decay, with each of the $b$
quarks entering the subjet $J_1$ and $J_2$, respectively. Otherwise, $J$ is
redefined as $J_1$ and the whole procedure is iterated.

The advantage of using the C/A algorithm with the mass drop is that it already
cleans a jet from incoherent radiation as it makes its way to the relevant
splitting. 
To further improve its performance, the BDRS procedure has been supplemented with
one additional step, dubbed \emph{filtering}, in which the jet is reclustered
with a much smaller radius, $R_\text{filt}$, and only the $n$ hardest subjets
are taken for mass reconstruction. This helps to remove even more of the
unwanted contamination from the underlying event while keeping the most
important perturbative radiation from the Higgs decay products.
The original analysis of Ref.~\cite{Butterworth:2008iy} used $n=3$ but 
the filtering procedure can be defined with an arbitrary number of subjet taken
for mass reconstruction~\cite{Rubin:2010fc}.

In Ref.~\cite{Dasgupta:2013ihk}, a modification of the BDRS tagger has been
proposed, in which, in the case when the conditions from Eq.~(\ref{eq:bdrs}) are
not met,  the $J$ is redefined not as $J_1$ but as this of the two subjets,
$J_1$ and $J_2$, whose transverse mass,  
$m_{TJ_i}^2 = m^2+ p_{TJ_i}^2$, is the largest.
Such a tagger was called the \emph{modified mass drop tagger}~(mMDT).
This modified version turns out to perform better on a special class of
configurations in which a massless parton emits a soft gluon that subsequently
splits collinearly into a $q\qbar$ pair. Because the first parton is massless,
the BDRS tagger would choose the $g\to q\qbar$ branch for further iterations,
even though this branch comes from a soft gluon.
The modification made in mMDT fixes the above feature by elimination of
sensitivity to soft divergences and renders the tagger that is better-behaved
from the point of view theoretical calculations~\cite{Dasgupta:2013ihk}.
 
In general, procedures aimed at cleaning the incoherent radiation from a jet are
called \emph{grooming} techniques. 
Other, by now well established, examples include
\emph{pruning}~\cite{Ellis:2009su,Ellis:2009me},
\emph{trimming}~\cite{Krohn:2009th} and
\emph{$N$-subjettiness}~\cite{Thaler:2010tr,Thaler:2011gf}.

\emph{Pruning}~\cite{Ellis:2009su,Ellis:2009me} was designed to identify 
signal events with heavy objects decaying hadronically and to clean them from
incoherent radiation. 
The procedure modifies jet substructure in order to reduce the systematic
effects that obscure the reconstruction of hadronic heavy objects. It takes the
constituents of a jet and puts them through a new clustering procedure in which
each of the branchings is requested to pass a pair of cuts on kinematic
variables. If the cuts are not passed, then the recombination is vetoed and one
of the two branches is discarded. The conditions for each recombination 
$ij\to k$ are
\begin{equation}
  \frac{\min(p_{Ti},p_{Tj})}{p_{Tk}} > z_\text{cut} 
  \qquad
  {\rm and}
  \qquad
  \Delta_{ij} < D_{\cut} \,, 
  \label{eq:pruning}
\end{equation}
where $D_\cut$ is chosen dynamically according to
$D_\text{cut} = 2D\frac{m_J}{p_{TJ}}$ and the parameters $z_\cut$ and $D$ are
optimized based on Monte Carlo simulations.
If both conditions given in Eq.~(\ref{eq:pruning}) are satisfied, the merging
takes place, otherwise, the softer branch is discarded. 

\emph{Trimming} procedure~\cite{Krohn:2009th} takes a jet obtained with the
original definition which used the size $R$ and reclusters its constituents into
subjets employing an algorithm with the smaller jet radius $R_{\text{sub}}$.
In the next step, only the subjets whose transverse momenta that satisfy the
condition
\begin{equation}
  p_{Ti} > f_{\text{cut}}\, \Lambda_{\text{hard}} \,,
  \label{eq:trimm}
\end{equation}
are kept and they are subsequently recombined into the new, trimmed jet.
In the condition of Eq.~(\ref{eq:trimm}), $f_{\text{cut}}$ is a dimensionless
parameter, optimized based on simulations, and $\Lambda_{\text{hard}}$ is a hard scale characteristic to a given
process. 

\emph{N-subjettiness}~\cite{Thaler:2010tr} exploits the fact that
the pattern of the hadronic decay of a heavy object is characterised by
the presence of concentrated energy depositions corresponding to the decay
products.
On the contrary, a QCD jet represents a more uniformly spread energy
configuration.
The inclusive jet shape, $N$-subjettiness, is
defined, in its generalized version derived in~\cite{Thaler:2011gf}, as 
\begin{equation}
  \label{eq:nsub}
  \tau_N=
  \frac{1}{d_0}\sum_k p_{Tk}
  \min\left((\Delta R_{1k})^{\beta},...,(\Delta R_{Nk})^{\beta}\right)\,, 
\end{equation}
where $k$ runs over the constituent particles in the jet, $\Delta
R_{jk}=\sqrt{(\Delta y)^2+(\Delta \phi)^2}$ is a distance between the
constituent $k$ and the subjet $j$, $\beta$ corresponds to an adjustable
parameter called the angular
weighting exponent, and the normalization factor reads $d_0 = \sum_k
p_{Tk}R^{\beta}$, with $R$ being the jet radius.
 
The variable $\tau_N$ approaches 0 when the constituents of the jet are aligned
along $N$ directions. The latter correspond to $N$ subjets. On the
opposite end, large values of $\tau_N$ signal that the number of distinct
subjets is greater than $N$.
Hence, the ratio $\tau_N/\tau_{N-1}$ turns out to be a useful discriminating
variable.
For example, in the case of two-prong hadronic decays, $\tau_2/\tau_1$ is on
average smaller for the signal than for the background events.

Many different taggers and groomers appeared since the first proposals briefly
described above. Because of space limits, we cannot even mention all of them
here. Instead, we refer to the recent reports following the topical BOOST
conference~\cite{Altheimer:2013yza, Adams:2015hiv}, where the Reader can find
most of the essential information and references.
 
Let us conclude by noting that, after the initial stage characterized by
developments of new taggers and groomers, and essentially Monte Carlo-based
optimization of their parameters,
the efforts of the community turned
into comparisons between different tools~\cite{Soper:2010xk,
QuirogaArias:2012nj} and into gaining their analytic
understanding~\cite{Feige:2012vc,Dasgupta:2013ihk,Dasgupta:2013via,Dasgupta:2015yua, Larkoski:2015kga}.

%-----------------------------------------------------------------------------
\section{Factorization in hadroproduction of jets}
\label{sec:factorizaion}

Jets are defined as collimated streams of particles carrying sizable energy in
the transverse direction, hence, they are genuinely hard objects, which can be
treated by perturbative QCD. 
However, even though jets originate from collisions of highly energetic hadrons,
the hadrons themselves are characterized by low transverse scales, related to
their masses, only up to a few times larger than $\lqcd$ . 
This introduces a hierarchy of scales from $\lqcd$ to
$p_{T,\text{jet}}$, with the latter varying between tens of GeV and
several TeV.

Such a large span of scales creates a very rich dynamics, which poses serious
calculational challenge. Moreover, the low-scale dynamics of hadron binding
cannot be approached by perturbative QCD.
It turns out however, that the problem can be handled by extracting dominant
contributions at various scales, the so called, \emph{regions}, and by
factorizing the cross sections into the short- and long-distance pieces, which
can then be calculated separately. 
 
The \emph{factorization}, which is the subject of this section, is at
heart of all calculations at hadron colliders. Even though it has been intensely
studied since the very beginnings of QCD, with the seminal works
of Refs.~\cite{Ellis:1978ty, Bodwin:1981fv, Collins:1981uw, Collins:1981ta,
Collins:1985ue, Collins:1988ig, Collins:1989gx}, it is rigorously proven only
for a handful of processes and for the cases of sufficiently inclusive
observables.

%------------------------------------
\subsection{Collinear factorization}
\label{sec:coll-fac}

\begin{figure}[t]
  \begin{center}
    \begin{minipage}[t]{0.48\textwidth}
    \centering
    \includegraphics[width=0.95\textwidth]{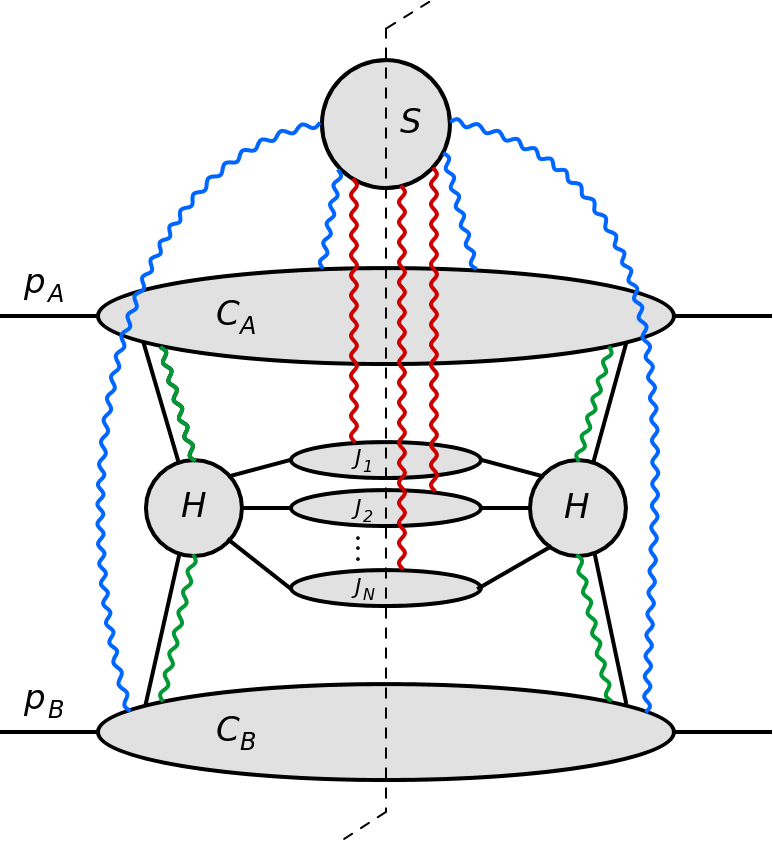}
    \caption{
    The most general leading pinch singular surface diagram for hadron-hadron
    collision.
    }
    \label{fig:general-LPSS}
    \end{minipage}
    \hfill
    \begin{minipage}[t]{0.48\textwidth}
    \centering
    \includegraphics[width=0.95\textwidth]{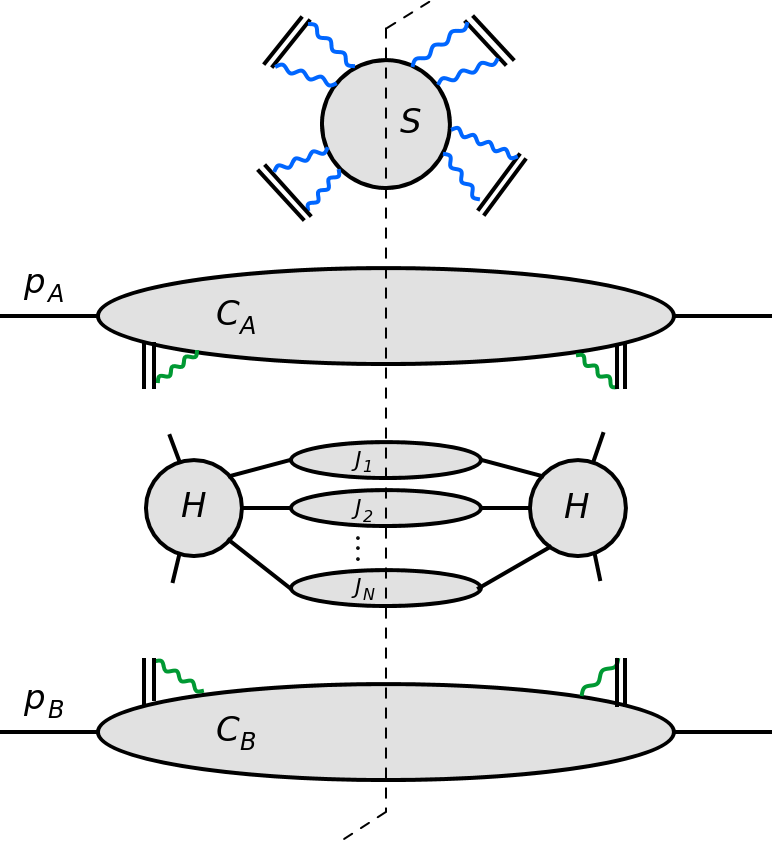}
    \caption{
    Factorized form of the cross section in hadron-hadron collision.
    }
    \label{fig:factorized-LPSS}
    \end{minipage}
  \end{center}
\end{figure}

A general cut diagram for the hadron-hadron scattering is shown in
Fig.~\ref{fig:general-LPSS}. 
It involves a low scale $ m \sim \lqcd-1\, \GeV$, which is in a typical range of
hadron masses, and the hard scale,~$Q$, which is of the order of jet's
transverse momentum. 
We denote the ratio of the two scales by $\lambda = m/Q$. 
In general, one is interested in the expression for the cross section
corresponding to the diagram of Fig.~\ref{fig:general-LPSS} in the limit
$\lambda \to 0$.
 
It turns out that, in the above limit, the integral over loop momenta 
is divergent.
Some of those divergences are only apparent and can be removed by deformation of
the integration contours into the regions of the complex plane where the
integrals are finite.
There is however an entire class of singularities that, in the limit $\lambda
\to 0$, pinch the integration contour such that it cannot be deformed. Those
integration surfaces, called the \emph{leading pinch singular surfaces} (LPSS),
give the dominant contribution to the cross section, the so-called \emph{leading
twist}, and all the other terms are suppressed as powers of $m/Q$ and hence
contribute at higher twist.

Fig.~\ref{fig:general-LPSS} shows the most general LPSS diagram divided into
several classes of subdiagrams, each of which corresponds to a different
leading region:
\begin{itemize}
  \item
  $C_A$ and $C_B$ are the \emph{collinear subgraphs} and characterize the
  incoming partons and the corresponding beam remnants after the hard reaction.
  The particles in the $C_{A,B}$ subgroups belong to the collinear region, which
  means that their momenta, in notation $p_\mu = (p_+, p_-, p_T)$, scale as
  \begin{equation}
    p_{C_A, \mu} \sim Q\, (1,\lambda^2,\lambda)\,,
    \qquad \qquad \qquad 
    p_{C_B,\mu} \sim Q\, (\lambda^2,1,\lambda)\,.
    \label{eq:collinear-region}
  \end{equation}
  \item
  $H$ is the \emph{hard subgraph}, with the interactions happening at the hard
  scale $Q$ and the momentum scaling
  \begin{equation}
    p_{H,\mu} \sim Q\, (1,1,1)\,.
    \label{eq:hard-region}
  \end{equation}
  \item
  $J_1,J_2, \ldots, J_N$ are the \emph{final-state jets} created after the hard
  interaction that happened in $H$. The jets are build-up from particles
  collinear to the momenta going out of the hard part $H$.
  \item
  $S$ is the \emph{soft subgraph} connecting to other subdiagrams with soft
  gluons whose momenta scale as
  \begin{equation}
    q_{S,\mu} \sim Q\, (\lambda, \lambda, \lambda)\,.
    \label{eq:soft-region}
  \end{equation}
\end{itemize}

The above subgraphs are in general connected by quark and gluon lines.
One can show, however that all connections except those involving
one collinear and an arbitrary number of longitudinally polarized gluons between
the collinear and the hard subgraphs, or the
soft gluons between the soft and the collinear, or the soft and the final state
jet subgraphs, contribute only to the higher twist~\cite{collins:book}.
On the contrary, any number of the remaining gluons connections, denoted by wavy
lines in Fig.~\ref{fig:general-LPSS}, will survive at the leading twist and
therefore must be resummed.

In simple theories, like the asymptotically free, non-gauge $\phi^3$ theory in
5+1 dimensions, contrary to QCD, there are no soft gluon connections between the
leading subgraphs. Moreover, only one collinear parton from the collinear part
connects to the hard part.  Hence, the contribution to each leading diagram
automatically takes the form of the product of the leading subgraphs: the hard
part and the collinear parton distribution functions. This is what we call
\emph{topological factorization}~\cite{Collins:1985ue}.
As shown in Fig.~\ref{fig:general-LPSS}, topological factorization does not
occur in QCD because of the soft gluon connections between various parts of the
diagram, as well as the longitudinal gluons connecting the collinear and the hard
part.
All the diagrams with multiple gluonic connections contribute to the leading
twist and that is why factorazibility of QCD is a highly non-trivial feature:
In order to arrive at the factorization formula, one needs to
show that the effects of the soft and longitudinal gluons cancel or can be
resummed.

The hadronic cross section factorizes if it can be written in the form
\begin{equation}
  d\sigma =
  \sum_{a,b}
  f_{a/A} \otimes f_{b/B} \otimes H_{ab} \otimes  S_{ab} \otimes
  J_1 \otimes \ldots J_N + \text{p.s.c.}\,,
  \label{eq:coll-fac-dy}
\end{equation}
where  $f_{a/A}$ and $f_{b/B}$ are the collinear (integrated) \emph{parton
distributions functions} (PDFs), corresponding to the collinear subgraphs and
the rest of the notation follows that of Fig.~\ref{fig:general-LPSS}.
The formula sums over all parton species and the extra term "p.s.c" denotes the
power-suppressed corrections (of higher twist), which are multiplied by extra
powers of $1/Q$ with respect to the leading part.

The PDFs, $f_{a/A}$ and $f_{b/B}$, are process-independent, whereas the sets of
the soft, $S_{ab}$, and the hard, $H_{ab}$,  functions  are process-specific.
The latter is also referred to as the partonic cross section and, in the
perturbative regime, \ie for $Q \gg \lqcd$, it can be calculated order by order
in powers of the strong coupling
\begin{equation}
  H_{ab} = 
  H_{ab}^\LO  +
  \left(\frac{\as(\mu_R^2)}{2\pi}\right)   H_{ab}^\NLO  +
  \left(\frac{\as(\mu_R^2)}{2\pi}\right)^2 H_{ab}^\NNLO  +  \ldots \,.
  \label{eq:hard-cross-section}
\end{equation}

Let us now sketch the proof of Eq.~(\ref{eq:coll-fac-dy}).
In the general diagram of Fig.~\ref{fig:general-LPSS}, various sub-graphs,
corresponding to interactions happening at different energy scales, are
connected. The aim of the factorization program is to show that many of these
connections disappear at leading twist and the expressions corresponding to
Fig.~\ref{fig:general-LPSS} can be written as a simple convolution of the
hard, the soft and the collinear functions.
 
QCD factorization is an enormously vast subject with plenitude of subtelties.
Our discussion presented below can be only brief on some of those issues. We
shall however try to signal the most important ingredients that enter proofs of factorization. More details can be found in the references provided in the
remaining part of this section. In addition, we point the Reader to the nicely 
structured overview of the different steps taken in factorization proofs
presented in Ref.~\cite{Diehl:2015bca}, where many issues are clarified in
order to prove the factorization in the very non-trivial case of the double
Drell-Yan process.

In gauge theories, like QCD, there are essentially three challenges that need be
addressed in order to demonstrate the factorization property of the cross
section~(\ref{eq:coll-fac-dy}).
They correspond to the following connections, which can spoil the factorization
~\cite{Collins:1985ue}:

\label{page:fac-problems}
\begin{enumerate}
  \item
  Soft-gluon connections between the wide angle jets in the hard subdiagram
  (red in Fig.~\ref{fig:general-LPSS}).
  \item
  Soft-gluon connections between the collinear subgraphs $C_A$ and $C_B$ via the
  soft function $S$ (blue in Fig.~\ref{fig:general-LPSS}). These are the
  interactions between the spectator particles.
  \item
  Longitudinally-polarized-gluon connections between the collinear subgraphs
  $C_{A,B}$ and the hard part~$H$ (green in Fig.~\ref{fig:general-LPSS}).
\end{enumerate}

The proof of factorization for the soft gluon connections of the first and the
second type proceeds via deformation of the contour of integration in the
complex plane of the soft gluon momenta such that one of the longitudinal
components dominates along the entire integration path. Subsequently, one uses
the non-abelian Ward identities (hence, gauge invariance) to show that the
corresponding contribution either vanishes or factorizes.

However, in some cases, the above procedure encounters extra difficulty because
the connecting gluons are pinched in the so-called Glauber
region~\cite{Bodwin:1981fv, Collins:1985ue}, which makes the necessary contour
deformation impossible. Let us now take a closer look at this problematic issue.

%------------------------------------
\subsubsection{Glauber region}
\label{sec:galuber-region}

Consider the soft gluon with momentum $q$, \cf Eq.~(\ref{eq:soft-region}),
coupling to the massless quark line with momentum $p$ collinear to $p_A$, hence
scaling similar to Eq.~(\ref{eq:collinear-region}). 
This leads to appearance of the propagator with the following denominator
%
% cf. (19) of Boer et al, NPB 667
%
\begin{equation}
  (p + q)^2 + i\epsilon \simeq
  2 p^+ q^- + q^+q^- - q_T^2 + i\epsilon\,.
  \label{eq:prop-glauber}
\end{equation}
By studying the integral over $q^-$, we notice that gluons from the soft
region satisfy
\begin{equation}
  p^+ q^- \sim \lambda \quad \gg  \quad
  q^+ q^-,\, q_T^2 \sim \lambda^2\,.
  \label{eq:non-glauber}
\end{equation}
Hence, the minus component of the gluon's 4-momentum dominates and 
$(p + q)^2 \simeq 2 p^+ q^- + i\epsilon$.

If we now go to the rest frame of the incoming hadron $A$,
which travels along the plus direction, the polarization vector of the soft
gluon will lie, to a good approximation, in the minus direction, $\epsilon_- \gg
\epsilon_+, \epsilon_T$, and, because of Eqs.~(\ref{eq:soft-region}) and
(\ref{eq:non-glauber}), also the minus component of the gluon's 4-momentum will
dominate. Hence, we have $\epsilon_\mu \propto q_\mu$, which implies that the
gluon arises from a pure gauge potential. Such a contribution can be eliminated
by using gauge invariance.
 
However, a complete integral over $q^-$ requires us to go to the lower limit of
$q^-$, where the relations~(\ref{eq:soft-region}) and (\ref{eq:non-glauber})
are not satisfied but, instead, we have $ p^+ q^- \sim \lambda^2 \sim q_T^2$,
which corresponds to 
\begin{equation}
  q^+ q^- \ll q_T^2\,.
  \label{eq:glauber-region}
\end{equation}
This is the so-called \emph{Glauber region}~\cite{Bodwin:1981fv, Collins:1985ue}
or the \emph{Coulomb region}~\cite{Collins:1981ta, Catani:2011st, Forshaw:2012bi}
in which the components of the gluon's 4-momentum are not
comparable, as assumed in Eq.~(\ref{eq:soft-region}), but the longitudinal
components are much smaller than the transverse one.
The gluons from the Glauber region are important in various contexts.  For
example, as discussed in Ref.~\cite{Fleming:2014rea}, they give rise to the
Lipatov vertex and emergence of the BFKL equation in SCET.
The Glauber region is not displayed in Fig.~\ref{fig:general-LPSS} and, in fact,
the contributions from that region must vanish in order for the factorization
procedure to be successful.
 
This is because, in the Glauber region, the propagator (\ref{eq:prop-glauber})
has a pole at $q^- \sim q_T^2/(2p^+) - i\epsilon \sim m^2/(2Q) - i\epsilon$ and
the above gauge-invariance arguments cannot be used.
If all such poles lie in the same part of the imaginary plane, the integration
contour may be deformed into the opposite part such that it is moved out of the
Glauber region restoring the relation (\ref{eq:non-glauber}).  However, if the
poles are distributed on both the positive and the negative half of the
imaginary plane, the integration contour is pinched in the Glauber region and
the cross section does not factorize.

The collinear factorization is rescued from the soft gluons pinched in the
Glauber region by the sum over final states. It turns out that, after such a sum
is performed, the soft gluon connections between the final state jets $J_i$ (red
lines in Fig.~\ref{fig:general-LPSS}) vanish~\cite{Collins:1985ue, Collins:1989gx}.
This cancellations between graphs takes
place because of unitarity, as those interactions happen long after the hard
process. 
 
It is important to note that canceling of the final state interactions does not
happen only in the fully inclusive cross sections but it occurs also in the jet
production cross sections. This is because the hard final state jets are well
separated in space and they cannot meet again to produce another hard
scattering. Therefore, the interactions between the final state jets can occur
only via soft gluons. Hence, the same line of arguments can be applied to jet
final states and the soft divergencies will cancel in the
sum~\cite{Sterman:1978bi,Sterman:1978bj}.
Regarding the collinear emissions from the hard final state partons, they are
compensated by virtual corrections, for the IRC jet definitions, as discussed in
Section~\ref{sec:irc-safety}, a feature guaranteed by the KLN
theorem~\cite{Kinoshita:1962ur,Lee:1964is}.

As for the soft connections between the collinear subgraphs, $C_A$ and $C_B$,
\ie blue lines in Fig.~\ref{fig:general-LPSS}, the pinches, which prevented us
from deforming the contour away from the Glauber region, disappear after the sum
over final states and the deformation is possible.
That allows one for application of the non-abelian Ward identities, which then
turn the $S-C_{A,B}$ connections into connections between the soft function $S$ and the Wilson lines
and that part of the formula factorized from the rest. 
This is illustrated in Fig.~\ref{fig:factorized-LPSS} and can be understood by a
classical analogy of $C_A$ passing through a soft colour filed of $C_B$. Since
the gluons are very soft, they correspond to interactions between partons in the
subgraphs $C_A$ and $C_B$ long before the hard scattering. But the soft gluons
cannot resolve the partons inside the hadron, hence, the subgraph $C_A$ appears
as a colour singlet and the soft gluons cannot couple to it. This is a
manifestation of unitarity.
 
Thus, the first two problems from the list on page~\pageref{page:fac-problems}
are solved.
We are now left with the third problem, \ie the longitudinally polarized  gluons
connecting $H$ with $C_A$ and $C_B$, which we discuss in the next subsection.

%-----------------------------------------------------------------------------
\subsubsection{Parton distribution functions and gauge links}
\label{sec:pdfs-gauge-links}

The third issue is solved by using gauge invariance and absorbing the
longitudinal gluons into the parton distribution functions via the so called
\emph{gauge links}. 
More specifically, eikonal propagators can be used for gluons linking the
collinear and the hard part, \ie green gluons in Fig.~\ref{fig:general-LPSS} and
then the non-abelian Ward identities allow one to show that the gluons from 
$C_A$ and $C_B$ effectively connect to the Wilson line, represented as double
line in Fig.~\ref{fig:factorized-LPSS}, which then factorizes from the rest of
the diagram. 
Physically, that comes from the fact that collinear gluons cannot resolve any
transverse structure of $H$ and are only sensitive to its longitudinal
components. Hence, $H$ appears to those gluons as a Wilson line.

The above procedure leads to the following, gauge invariant definitions of the
parton distribution functions for a quark and a
gluon~\cite{Collins:1981uw,Curci:1980uw, Collins:1989gx}
%
% Collins book, Eqs. (7.40) and (7.43)
\begin{subequations}
  \begin{align}
    f(\xi)_{q/h} & =
    \int \frac{dw_-}{2\pi} e^{-i \xi P^+ w_-}
    \langle P |
    \overline{\Psi}_q(0,0,\zT)\,
    \calW^n(0;w_-) \frac{\gamma^+}{2}
    \Psi_q(0,w_-,\zT)
    |P \rangle\,,
    \\
    f(\xi)_{g/h} & = 
    \int \frac{dw_-}{2\pi\xi P^+} e^{-i \xi P^+ w_-}
    \langle P |
    F^{+j}_{a}(0,0,\zT)
    \calW^n_{ab}(0;w_-)
    F^{+j}_{b}(0,w_-,\zT)
    |P \rangle\,,
  \end{align}
  \label{eq:coll-pdfs-def}
\end{subequations}
where $\Psi_q$ is a Dirac field of a quark $q$ and $F^{ij}_a$ is a gluon
field-strength tensor.

The object $\calW^n(0;w_-)$ in the above expressions is called the
\emph{Wilson line} (double line in Fig.~\ref{fig:factorized-LPSS}). It
effectively resums all exchanges of the longitudinal gluons between the hard
part and the collinear part.
The general form of the Wilson line reads
\begin{equation}
  \calW^C(a;b) =  \calP \exp\left[-i g\int_C ds_\mu A^\mu\right]\,,
  \label{eq:}
\end{equation}
where $C$ is a path connecting the space-time points $a$ and $b$, while $\calP$
denotes path ordering of the exponent. 

In the particular case of collinear PDFs, the path $C$ runs along the minus
direction with the endpoints $(0,0,\zT)$ and $(0,w_-,\zT)$
\begin{equation}
  \calW^n(0;w_-) =  \calP 
  \exp\left[-i g\int_0^{w_-} d\lambda\, n_\mu A^\mu(\lambda n^\mu)\right]\,,
  \label{eq:LLWL}
\end{equation}
where $n_\mu = (0,1,0)$.
We shall also call it a Wilson line in the light-cone direction.
 
Insertion of the simple Wilson line $\calW(w_-;0)$ between the quark or the
gluon fields in Eq.~(\ref{eq:coll-pdfs-def}) guarantees gauge invariance of the
the above definition of the collinear PDFs. As we shall see in next section, for
more exclusive parton distribution functions one needs more complex objects,
called the \emph{gauge links}, which are composed of multiple Wilson lines.
Light-cone Wilson lines are often written as a product of two
Wilson lines that connect the endpoints via infinity
\begin{equation}
  \calW^n(0;w_-) =  \calW^n(0;\pm \infty) \calW^n(\pm \infty;w_-)\,.
  %\label{eq:}
\end{equation}

It is important to mention that the gauge links involved in definitions of
the parton distribution functions in deep inelastic scattering (DIS) and the
Drell-Yan process (DY) are identical. 
Hence, collinear PDFs are universal.  The universality property of the collinear
factorization (\ie process independence of collinear PDFs) is a powerful feature
that enables one to parametrize the collinear parton distributions by fitting
them to the experimental data of one process, \eg DIS, and then use those PDFs
as an input to predictions for other processes, like DY or jet production.

Parton distributions given in Eqs.~(\ref{eq:coll-pdfs-def}) are already well
defined and useful, as they are gauge invariant and universal. There are however
still two more issues that need to be dealt with. They both arise from higher order
corrections to the subgraphs $H$, $C_A$ and $C_B$, which are implicit in the
diagrams of Figs.~\ref{fig:general-LPSS} and \ref{fig:factorized-LPSS}.
If we think of the $H$ subgraph for the simplest case of Drell-Yan, then the loop
corrections to the Born diagram $q\qbar \to Z$ exhibit UV divergences. Those are
removed by renormalization. However, the same loop correction, as well as the
corresponding real correction $q\qbar \to Z g$ contain also soft and collinear
divergences arising from masslessness of the incoming partons. The soft
divergences cancel between the real and virtual contributions while the
collinear divergencies do not.

However, as the momentum of the gluon correcting the Born term in $H$ becomes
collinear to the incoming parton, that gluon does not belong any more to the
hard region, defined in Eq.~(\ref{eq:hard-region}), but rather to one of the
collinear regions of Eqs.~(\ref{eq:collinear-region}). Therefore, contribution
from the collinear gluons should be subtracted from the hard part $H$ and added
to the collinear part $C_A$ or $C_B$.
That is most easily done using dimensional regularization, \ie performing the
integrals in $d=4-2\epsilon$ dimensions, with $\epsilon < 0$. Then, the
contributions from collinear gluons appear as poles in $\epsilon$, which are
moved from $H$ to $C_A$ or $C_B$. Since the dimensional regularization
introduces an arbitrary mass scale $\mu_F$, both the hard part, and the PDFs,
acquire dependence on this scale, which can be interpreted as a transverse
momentum scale separating the hard and the collinear regions and it is called
the \emph{factorization scale}.
Requirement of cancellation of the $\mu_F$ dependence between the hard partonic
cross section and the PDFs up to a given order in $\as$ leads to DGLAP evolution
equations~\cite{Altarelli:1977zs, Gribov:1972ri, Dokshitzer:1977sg}.
The above subtraction procedure has some degree of arbitrariness, which means
that PDFs are defined always in a certain \emph{factorization scheme}. 
Detailed discussion of ambiguities arising from the choice of factorization
scheme will be given in Section~\ref{sec:factorization-scheme}.

To summarize the discussion so far, the soft and the collinear gluon connections
between different subgraphs in Fig.~\ref{fig:general-LPSS} are shown to factor out
one by one. That is possible due to contour deformations, sums of diagrams and
gauge invariance, and the proof holds to all orders~\cite{Collins:1988ig}.
We emphasize that the sum over final states was critical to remove contributions
from the Glauber gluons. The latter implies that collinear factorization
will hold only for sufficiently inclusive observables.

%-----------------------------------------------------------------------------
\subsubsection{Factorization breaking}

In cases of less inclusive observables, one can indeed expect
factorization-breaking effects, for space-like collinear emissions, at higher
orders in QCD.  It turns out that such effects start at three loops for pure QCD
processes and at two loops for certain electroweak
processes~\cite{Catani:2011st, Forshaw:2012bi}.

\begin{figure}[t]
  \begin{center}
    \includegraphics[width=0.75\textwidth]{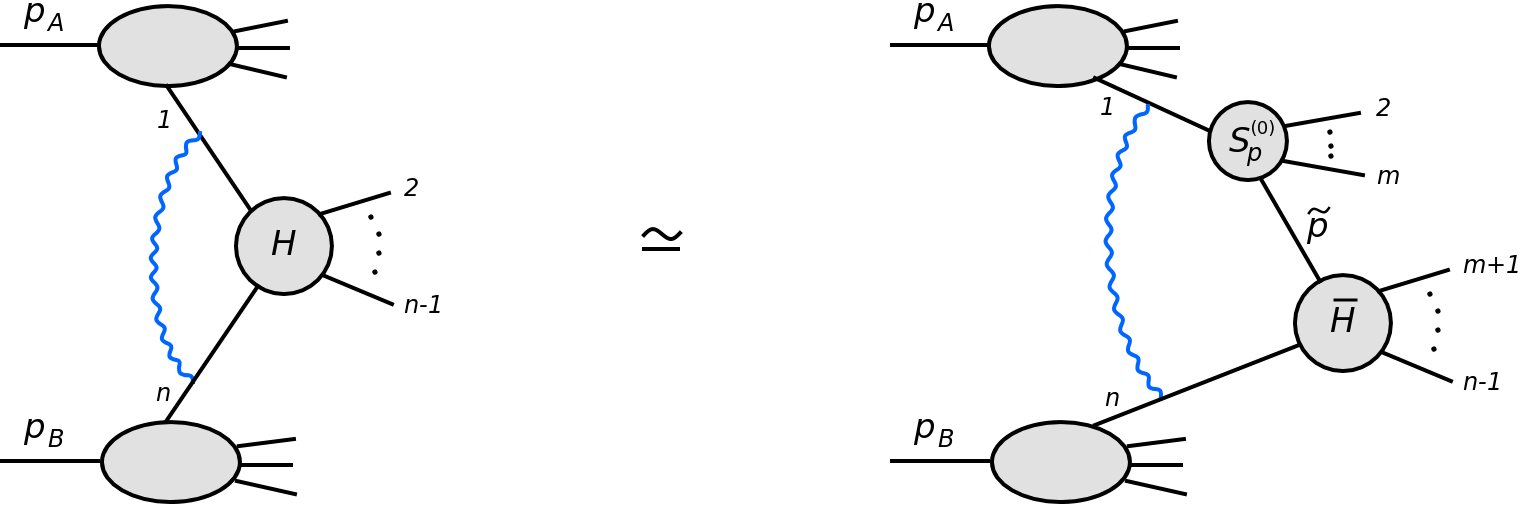}
  \end{center}
  \caption{
  One loop correction to the collinear factorization at the amplitude level.
  }
  \label{fig:coulomb-gluons}
\end{figure}

The origin of this breaking can be traced back to the Glauber/Coulomb gluons
discussed earlier, hence the gluons exchanged between two incoming partons long
before, or between two outgoing partons long after, the hard process.
Let us consider a simplified version of Fig.~\ref{fig:general-LPSS}, where we
concentrate just on the amplitude (part to the left from the cut) and the only
connection is that with a soft gluon between the two incoming partons. 
Left hand side of Fig.~\ref{fig:coulomb-gluons} represents the corresponding
one-loop amplitude.

Let us start from a tree level version of that diagram, \ie 
ignoring for a moment the gluon attached to the incoming partons $1$ and $n$.
In the limit where partons $1,\ldots, m$ become collinear, the amplitude can be
expressed as
\begin{equation}
  |\calM^{(0)}_H(1,\ldots, n) \rangle \simeq
  \Spbold^{(0)}(\tilde p, 1,\ldots, m) \,
  |\calM^{(0)}_{\bar H}(\tilde p, m+1,\ldots, n) \rangle\,,
  %\label{eq:}
\end{equation}
where the right hand side of the above equation corresponds to the right hand
side of Fig.~\ref{fig:coulomb-gluons}, again, ignoring the soft gluon
connection.
Hence, the tree level amplitude factorizes into an operator $\Spbold^{(0)}$,
which is a matrix in the colour+spin space and depends only on the collinear
momenta, and the reduced amplitude $|\calM^{(0)}_{\bar H}\rangle$, depending 
exclusively on the non-collinear partons.
 
This simple formula does not survive however at one-loop, hence the case
shown in Fig.~\ref{fig:coulomb-gluons}, as the $\Spbold^{(1)}$ operator
now has both a real and an imaginary part.
The real part arises from an integration over the eikonal gluons satisfying
Eq.~(\ref{eq:non-glauber}). The imaginary part corresponds to the Glauber
region of Eq.~(\ref{eq:glauber-region}) and it is proportional to 
$i\,\T_{m+1} \cdot (\T_1 - \T_{\tilde p})$, 
where $\T_i$ is a colour charge matrix of the particle $i$ and $\T_i
\cdot \T_k \equiv \sum_{c=1}^{N_c^2-1} T_i^c T_k^c$.
We see that the operator appearing in the imaginary part depends both on the collinear parton $1$, and on the
non-collinear parton $m+1$.  However, since the above operator is
anti-Hermitian, its contributions cancel in the NLO cross sections, which
involve interference terms between Born and one-loop diagrams. 

The corresponding operator becomes even more complex at two loops but, also
there, it can be shown that its contribution to the cross section 
vanishes~\cite{Catani:2011st, Forshaw:2012bi}.
It is only at three loops where the interplay between the eikonal and the
Coulomb gluons breaks the collinear factorization at the level of the cross
section.  Hence, the factorization breaking will have physical effects at this
order, which can manifest themselves \eg through \emph{super-leading
logarithms}~\cite{Forshaw:2006fk, Forshaw:2008cq}. The latter appear for
observables defined with non-trivial vetoes on the phase space that prevent
cancellations of the Coulomb gluon contributions.

The factorization breaking effects, discussed here for the amplitudes and
partonic cross sections, are expected to disappear when a sum over all
diagrams is performed, as explained earlier, hence the original result
of~\cite{Collins:1985ue, Collins:1989gx} remains valid for inclusive
observables. 
This can be understood in the context of the above discussion by noting that the
factorization breaking terms are associated with the colour factors and those
cancel each other when collisions of colourless hadrons are considered.

All order collinear factorization is a basic assumption in construction of
parton showers. There, the $n$-parton cross section is computed by applying a
sequence of $\Dbold_i$ and $\Vbold_i$ operators, which act on the $0$-parton
state $|M^{(0)}\rangle$. The operators $\Dbold_i$ correspond to real emissions
and the operators $\Vbold_i$ are the Sudakov factors that express non-emission
probabilities between two momentum scales.
 
Hence, the Sudakov operators resum virtual and unresolved real emissions on each
of the incoming or outgoing lines of the Born diagram. They do not, however,
include the virtual gluon exchanges \emph{between} two different external
particles.  Those would be exactly the Coulomb gluons depicted in
Fig.~\ref{fig:coulomb-gluons} and would contribute at subleading colour.
The study aimed at establishing whether the Coulomb gluons can be incorporated
in the currently assumed, factorized structure of the parton shower algorithms
has been presented in Ref.~\cite{Angeles-Martinez:2015rna}.
There it is found that, for the Drell-Yan process, in the first few orders of
perturbation theory, the Coulomb gluons can indeed be accommodated in the
probabilistic, $k_T$-ordered evolution algorithms.
Each individual diagram involving the virtual, Coulomb-gluon exchanges and one
or two real, eikonal-gluon emissions has different ordering conditions for those
gluons' transverse momenta. However, the sum of all diagrams at a given order
results in the final expression in which the Coulomb gluon's transverse momenta
are always ordered with respect to the transverse momenta of the emitted
(eikonal) gluons. Consequently, the factorized structure of the parton shower
emissions, realized through a sequence of the $\Dbold_i$ and $\Vbold_i$
operators, is preserved also after the inclusion of the Coulomb gluons.

%------------------------------------
\subsection{Factorization scheme}
\label{sec:factorization-scheme}

As discussed in Section~\ref{sec:pdfs-gauge-links}, factorization procedure
is not unique. The ambiguity seats in details of subtraction terms, as, while
all the terms divergent in the collinear limit need to be subtracted from the
hard part, in practice, one also subtracts a number of arbitrary finite terms. Differences
between the finite terms absorbed into PDFs is what differentiates between
factorization schemes.

In the context of hadron-hadron collisions, one uses almost exclusively the
$\msbar$ scheme~\cite{Bardeen:1978yd} and all the main PDF
sets~\cite{Butterworth:2015oua} are indeed the $\msbar$
PDFs. In the $\msbar$ factorization scheme, the PDFs absorb the $1/\epsilon$
terms as well as the constant $4\pi -\gamma_E$. All other finite terms are
left as part of the hard cross section $H$, \cf Eq.~(\ref{eq:hard-cross-section}).

However, some of the finite terms that arise from integration in $d$ dimensions
originate from large distances. This happens because the $1/\epsilon$ poles
are multiplied by the  dimensional-regularization-specific factor
$(k_T^2)^\epsilon = (k^2(1-z))^\epsilon \simeq 1+ \epsilon \ln k^2 + \epsilon
\ln(1-z)$~\cite{deOliveira:2013iya}, which results in 
$\order{\epsilon/\epsilon} = \order{1}$ contributions. The latter are not removed from the hard part by the
$\msbar$ subtraction scheme, though it would be more natural if they belonged to PDFs.
That is the motivation behind the efforts of Refs.~\cite{Jadach:2011cr,
Jadach:2015mza, deOliveira:2012qa,deOliveira:2013iya} to come up with a
factorization scheme that would better separate the short and the long distance
physics.
 
We repeat that all factorization schemes are equivalent at a given order of
perturbative expansion and the corresponding PDFs all satisfy the universality
(process-independence) property.
However, some finite terms may turn out to  be pathological, or just
inconvenient, and, therefore, switching to a different scheme  can offer
concrete advantages.

For example, the Monte Carlo (MC) scheme proposed in Refs.~\cite{Jadach:2011cr,
Jadach:2015mza} was motivated by construction of a matching procedure between
the NLO results and the parton shower, \cf Section~\ref{sec:matching}. 
The latter is implemented in practice as a Monte Carlo algorithm, hence the
emissions are effectively integrated in four dimensions using unitarity to
cancel the soft, and in the case of the final state radiation also the
collinear, divergences.
Thus, if we think of the parton shower as a procedure of unfolding PDFs, then,
those PDFs are definitely not of $\msbar$ type but they are defined by the MC
algorithm of the shower.
The NLO calculations, however, are performed in $d$~dimensions and they are
regularized by the $\msbar$ subtraction counter-terms.  Hence, they contain the
finite pieces originating from the $\epsilon/\epsilon$ cancellations, like for
example the term $\propto \ln(1-z)$, mentioned above.

Because of the non-compatibility of the factorization schemes used in the NLO
calculations and in parton showers, NLO+PS matching faces certain non-trivial
issues~\cite{Frixione:2002ik, Nason:2004rx}.
The way out, proposed in Refs.~\cite{Jadach:2011cr, Jadach:2015mza}, is to
consistently use a single factorization scheme that is compatible with the
parton shower. Such a scheme is therefore called the MC factorization scheme. 
In the MC scheme, the $z$-dependent terms related to the $\epsilon/\epsilon$
contributions that appear in the hard part calculated in $\msbar$, and come from
unphysical treatment of the phase space, are absorbed into PDFs. Hence, they do
not appear in the partonic cross section~(\ref{eq:hard-cross-section}). On top
of that, one subtracts contributions to the collinear space generated by the
shower, to avoid double counting (see Refs.~\cite{Jadach:2011cr, Jadach:2015mza}
for details). With that, one defines the MC PDFs by the following shift with
respect to the $\msbar$ PDFs
\begin{equation}
f_{q(\bar{q})}^{\rm MC}(x,\mu^2) = f_{q(\bar{q})}^{\overline{\rm MS}}(x,\mu^2)\; + \int_x^1 \frac{dz}{z}\,
f_{q(\bar{q})}^{\overline{\rm MS}}\left(\frac{x}{z},\mu^2\right) \Delta C_{2q}(z) \; + \int_x^1
\frac{dz}{z}\, f_g^{\overline{\rm MS}}\left(\frac{x}{z},\mu^2\right) \Delta C_{2g}(z)\,, 
\label{eq:PDFs-MC-quark}
\end{equation}
where
\begin{subequations}
  \label{eq:DeltaC}
  \begin{align}
  \Delta C_{2q}(z) & = 
  \frac{\alpha_s}{2\pi}\, C_F 
  \left[ \frac{1+z^2}{1-z}\ln\frac{(1-z)^2}{z} + 1-z \right]_+\,,
  \\
  \Delta C_{2g}(z) & = \frac{\alpha_s}{2\pi}\, T_R\,
  \left\{ \left[z^2+(1-z)^2\right] \ln\frac{(1-z)^2}{z} + 2z(1-z)\right\}\,.
  \end{align}
\end{subequations}
We recognize that the functions $\Delta C_{2q}$  and $\Delta C_{2g}$ contain the
well know pieces from the $\msbar$ coefficient function, \cf
Ref.~\cite{esw:book}, which in the MC scheme become parts of the MC PDFs.
Since the MC scheme was defined in the context of the Drell-Yan process, whose
Born level, $q\qbar \to Z$, does not involve an incoming gluon, the gluon PDF is
identical to that in the $\msbar$ scheme up to $\order{\as^2}$ corrections. 
However, an extension to processes with the initial gluon, \eg $gg \to H$, will
introduce a difference between the MC and the $\msbar$ schemes also for the
gluon PDF.

As we see from Eq.~(\ref{eq:PDFs-MC-quark}), the change of factorization scheme
can be regarded as a rotation in flavour space spanned by $(q_i, \qbar_i, g)$.
A similar rotation, with slightly different $\Delta C$ functions, has also been
used in definition of the \emph{physical scheme} in
Refs.~\cite{deOliveira:2012qa, deOliveira:2013iya}. By removing the $\msbar$
terms of $\epsilon/\epsilon$ origin, which are proportional to splitting
functions, the physical scheme avoids flavour mixing. The latter is a problem of
the $\msbar$ scheme in which, for example, the singlet-quark distribution gets
admixture of gluons.  This complicates calculations of heavy quark
effects~\cite{deOliveira:2013tya} as well as other non-inclusive
processes~\cite{deOliveira:2013iya}.

The differences between LO PDFs defined in the $\msbar$, the MC or the physical
scheme can be as large as 20\% or 
more~\cite{Jadach:2015mza, deOliveira:2013iya}. 
Hence, even though formally equivalent,
different choices for the factorization schemes lead to numerically
non-negligible differences for predictions of physical quantities.

Related efforts to account for the logarithmically enhanced threshold
contributions have been pursued in Ref.~\cite{Bonvini:2015ira}, where PDFs were
extracted at the NLO+NLL and NNLO+NNLL accuracy. Hence, these sets are in
principle the only consistent parton distributions to be used with
threshold-resummed matrix elements.

%------------------------------------
\subsection{TMD factorization}
\label{sec:tmd-factorization}

Collinear factorization assumes that components of the momentum of an incoming
parton emitted from the hadron $A$ moving in the plus direction satisfy
\begin{equation}
  k_+ \gg k_-, k_T\,,
  \label{eq:}
\end{equation}
hence only the $k_+$ component is kept while $k_-$ and $k_T$ are neglected,
\cf Eq.~(\ref{eq:collinear-region}).
This approximation is indeed valid in many situations as elaborated in the
Section~\ref{sec:coll-fac}.
However, there exist a class of observables which are directly sensitive to the
transverse component of the incoming parton's momentum.

Imagine for example measuring a distribution of the transverse momentum
imbalance, $q_T = |p_{T1}+p_{T2}|$ between the two final state leptons in the
Drell-Yan process or between two hardest jets in the hadronic production
of dijets with momenta $p_1$ and $p_2$.
In the limit where the two objects are oriented back-to-back in the transverse
plane, $q_T$ is very small and its value can be comparable to the transverse
momentum, $k_T$ of the incoming parton. In that case, neglecting $k_T$ will lead
to a significant modification of the $q_T$ distribution in the low-$q_T$ region.
In other words, observables like the $q_T$ spectrum or the distribution of the
azimuthal distance between final state leptons or jets are directly sensitive to
the transverse components of the 4-momenta of the incoming partons.
Hence, the corresponding parton distribution function should depend on both
$k_+$ and $k_T$. Such functions are known as \emph{transverse momentum dependent}~(TMD) parton distributions. 

Although TMDs are not the main focus of this review, we note in passing that
determination and modeling of the transverse momentum dependent distributions is
an active domain of research.
One of the classic approaches is due to Collins, Soper and
Sterman~(CSS)~\cite{Collins:1984kg}, where TMDs are expressed in terms of the
collinear PDFs convoluted with functions which resum the transverse emissions in
coordinate space.
For more details on evolutions, properties and parametrizations of TMDs
see Ref.~\cite{collins:book} as well as the recent reviews
~\cite{Angeles-Martinez:2015sea, Rogers:2015sqa}.
In order to facilitate usage and comparison between different fits and
parametrizations of TMDs, the project called \tmdlib (and a related 
\tmdplotter)~\cite{Hautmann:2014kza} has been started.
It provides a common interface to a wide range of distributions allowing for 
convenient phenomenology studies.

Understandably, TMDs call for an extension of the collinear factorization
formula to the transverse momentum dependent factorization.
As we shell see, this poses serious challenges and many questions are still
unanswered. The remaining part of this section aims at summarizing the status of
the TMD factorization across various processes, with special emphasis on jet
production, and recent progress in that domain.

%------------------------------------
\paragraph{Transverse gauge links and non-universality of TMDs}

\begin{figure}[t]
  \begin{center}
    \includegraphics[width=0.85\textwidth]{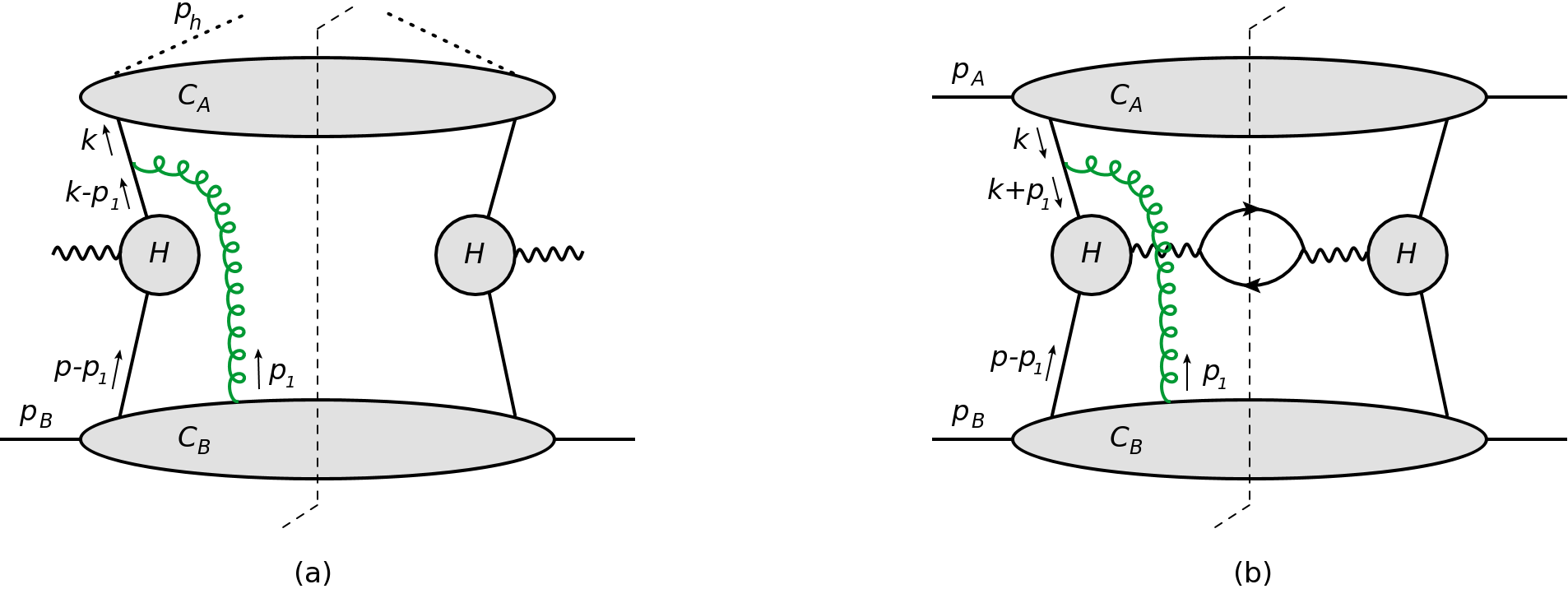}
  \end{center}
  \caption{
  Single gluon exchange in SIDIS (a) and DY (b).
  }
  \label{fig:sidis-dy}
\end{figure}

Let us consider a single-gluon exchange in the following two processes:
semi-inclusive deep inelastic scattering (SIDIS), shown in
Fig.~\ref{fig:sidis-dy}~(a) and Drell-Yan~(DY), depicted in
Fig.~\ref{fig:sidis-dy}~(b). 
The figures introduce necessary notation and 
it is understood that 
$p^- = p_1^- = k^+ = 0$, hence $k^- \approx q^- = Q/\sqrt{2}$, where $q$ is
the 4-momentum of the vector boson, while $Q$ denotes its virtuality.

The SIDIS diagram involves the following propagator~\cite{Boer:2003cm}
\begin{equation}
  \frac{\slashk - \slashp_1 + m}{(k-p_1)^2-m^2+i \epsilon} \approx
  \frac{(\slashk + m) -\slashn_+\,p_1^+  - \slashp_{1T}}
  {-p_1^+\,Q\sqrt{2} + (k_T-p_{1T})^2 -m^2 +i\epsilon}\,,
  \label{eq:prop-sidis}
\end{equation}
while, for DY, we have
\begin{equation}
  \frac{-\slashk - \slashp_1 + m}{(k+p_1)^2-m^2+i \epsilon} \approx
  \frac{-(\slashk - m) -\slashn_+\,p_1^+  - \slashp_{1T}}
  {p_1^+\,Q\sqrt{2} + (k_T+p_{1T})^2 -m^2 +i\epsilon}\,,
  \label{eq:prop-dy}
\end{equation}
where $m$ is a quark mass and $n_\pm$ are the light-like vectors satisfying $n_+
\cdot n_- = 1$.

The first thing we notice is that the two expressions have different signs in
front of $p_1^+\,Q\sqrt{2}$ in the denominators. 
This comes from different directions of the quark with momentum $k$ in
Fig.~\ref{fig:sidis-dy}. In the case of SIDIS, the gluon
connects the collinear subgraph $C_B$ with the outgoing quark,
whereas in DY, the gluon is attached to the incoming quark.
 
As is evident from Fig.~\ref{fig:sidis-dy}, the gluons connecting
the hard and the collinear parts break topological factorization graph by graph.
It turns out however that when the graphs are summed up for each process, one is
able to apply the Ward identities and consequently factor out all the 
gluon connections into Wilson lines and include them in definitions of
TMDs~\cite{Collins:1989gx}, very much like in the case of collinear
factorization discussed earlier.
There are however subtle differences. 
 
The diagrams in Fig.~\ref{fig:sidis-dy} and the expressions in
Eqs.~(\ref{eq:prop-sidis}) and (\ref{eq:prop-dy}) correspond to the first terms
in the expansion of the Wilson line (\ref{eq:LLWL}) for SIDIS and DY, with
$n^\mu \sim k^- \sim q^- = Q/\sqrt{s}$.
The term $p_1^+\,Q\sqrt{2}$ originates from eikonal vertex $\sim n\cdot p_1$
(see~\cite{Collins:1989gx} for details on eikonal Feynman rules).
Because of the sign
difference in the $p_1^+\,Q\sqrt{2}$ term, the gauge links resulting from
summation of multi-gluon exchanges will go along different paths for the
two processes.
Therefore, the factorization formula for SIDIS and DY will include \emph{different}
transverse momentum dependent parton distribution functions.
This difference disappears after integration over the transverse
momentum~\cite{Collins:1985ue, Collins:1988ig, Bodwin:1984hc, Bodwin:1984hcERR,
Aybat:2008ct}, however, at the level of the unintegrated TMDs, the strict
universality of parton distributions is lost.

Thus, we have discovered a very important feature of QCD factorization: the
parton distribution functions are universal at the level of the integral
(collinear PDFs) but not necessarily at the level of the integrand
(TMDs)~\cite{Bomhof:2007xt}.

The second subtlety concerning TMD factorization is related to the last term in
Eqs.~(\ref{eq:prop-sidis}) and (\ref{eq:prop-dy}), which is proportional to the
transverse component of gluon's momentum $p_{T1}$. 
As observed in Refs.~\cite{Belitsky:2002sm, Boer:2003cm}, this term will yield
leading twist contributions in the transverse direction which in general survive
at the light-cone infinity.  
More specifically, those contributions do not appear in covariant gauges but
they are non-vanishing for the light-cone gauges. 
Hence, unlike in the case of the collinear factorization, here, the
transverse components of the gauge links will not vanish.

Altogether, the gauge-invariant TMDs are defined as
%
% Collins book Eq. (8.44)
\begin{equation}
  f(\xi,k_T) = 
  \int \frac{d w^- d^2 \wT}{(2\pi)^3} 
  e^{-i \xi P^+ w^- + i \kT \cdot \wT}
  \langle P | 
  \phi^\dagger(0,0,0)\,
  \calU^{[\calC]}(0,w^-, \wT)\,
  \phi(0,w^-,\wT)
  | P \rangle\,,
  \label{eq:tmd-definition}
\end{equation}
where, this time, in order to assure gauge invariance, the object
$\calU^{[\calC]}(0,w^-, \wT)$ is a gauge link, which is in general  composed of
multiple Wilson lines in both the light-cone and the transverse directions
\begin{equation}
    \calW^{n}_{[a,b]}  =  
    \calP \exp\left[-ig \int_a^b dz\, n\cdot A(z) \right]\,,
    \qquad \qquad
    \calW^{T}_{[a,b]}  =  
    \calP \exp\left[-ig \int_a^b dz_T \cdot A_T(z) \right]\,.
\end{equation}

\begin{figure}
  \centering
  \includegraphics[width=0.3\textwidth]{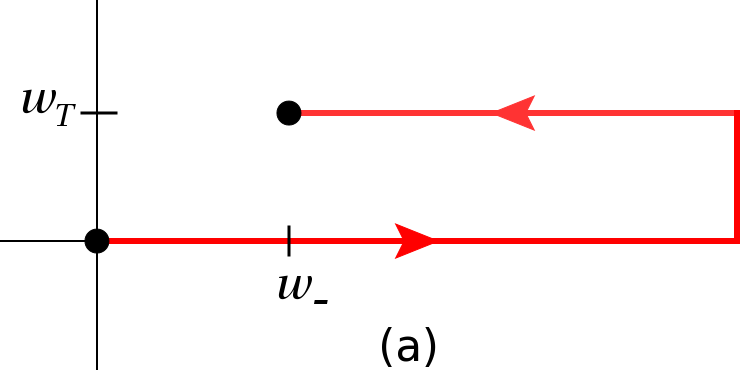}
  \hspace{80pt}
  \includegraphics[width=0.3\textwidth]{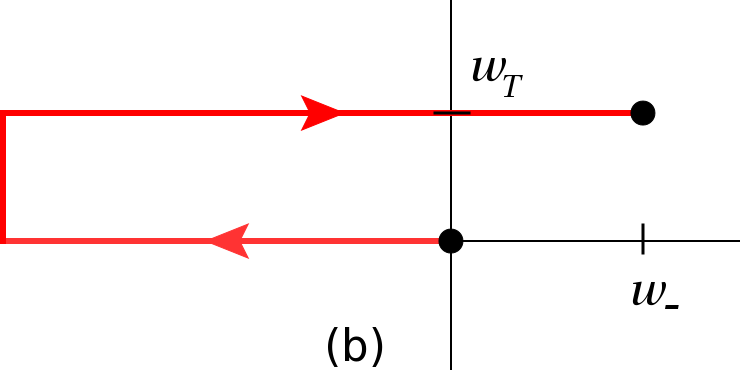}
  \caption{The gauge-link structure in the 
  correlator $\Phi$ in (a) SIDIS: $\mathcal U^{[+]}$ and 
  (b) DY: $\mathcal U^{[-]}$.
  }
  \label{fig:sidis-dy-gauge-links}
\end{figure}

In particular, for SIDIS, one has to use the gauge link $\calU^{[+]}$ involving
future-pointing Wilson lines, wheres for DY, gauge invariance of the TMD is
achieved with the link $\calU^{[-]}$, which involves past-pointing Wilson lines.
The above gauge links for SIDIS and DY are defined as
\begin{equation}
  \calU^{[\pm]}
  =
  \calW_{[(0^-,\zT); (\pm\infty^-,\zT)]}^n
  \calW_{[(\pm\infty^-,\zT);(\pm\infty^-,\mathbold \wT)]}^T
  \calW_{[(\pm\infty^-,\mathbold\wT);(w_-,\mathbold\wT)]}^n\,,
  \label{eq:plus-minus-links}
\end{equation}
and the corresponding paths are depicted in
Fig.~\ref{fig:sidis-dy-gauge-links}.

Hence, we have two different types of TMDs for SIDIS and DY and the strict
factorization property is lost due to the loss of universality of the TMDs.
However, it turns out that the SIDIS and DY TMDs are related by time-reversal
and differ only by the sign flip for two of the TMDs~\cite{Collins:2002kn}, the
so-called Sivers function~\cite{Sivers:1989cc} and the Boer-Mulders
function~\cite{Boer:1997nt}. Hence, the loss of strict
universality does not spoil predictive power. 

It is important to notice that the non-equality of TMDs in SIDIS and DY comes
from the differences in the colour flow, \cf Fig.~\ref{fig:sidis-dy}. The colour
running via an outgoing quark in SIDIS results in future pointing Wilson line
and the colour flowing via an incoming quark in DY leads to past pointing Wilson
line.
 
For the $k_T$-integrated cross sections, the transverse link does not
affect the result and one gets 
$\calU^{[+]}$ = $\calU^{[-]} =\calW_{[0,w^-]}$~\cite{Boer:2003cm}.
This can be seen directly in Eq.~(\ref{eq:tmd-definition}). Integration over
$\kT$ produces delta function~$\delta^{(2)}(\wT)$, which fixes the transverse
gauge link at $\calW_{[(0;\zT);(0;\zT)]}^T = 1$. As can be seen in
Fig.~\ref{fig:sidis-dy-gauge-links}, without the transverse separation, the
gauge links $\calU^{[+]}$ and $\calU^{[-]}$ both reduce to $\calW_{[0,w^-]}$.

However, in order to achieve full gauge invariance of the transverse momentum
dependent gluons, one has to include also transverse gauge links, even at
leading twist~\cite{Belitsky:2002sm, Boer:2003cm}.
Without them certain distributions, referred to as $T$-odd
functions~\cite{Sivers:1989cc, Collins:1992kk, Boer:1997nt}, would be
zero~\cite{Boer:2003cm}.
The need for the transverse gauge links, connecting the light-cone gauge-links
at infinity, is visible most notably the light-cone gauges which introduce
additional singularities~\cite{Belitsky:2002sm} leading to non-vanishing
transverse components at light-cone infinity.
Those modes can be though of as zero-light-cone-momentum, transverse gluons.
Hence, as opposed to the collinear factorization, the TMD factorization requires
in general contributions from transverse-gluon connections between the hadron
and the hard part, which are resummed into transverse Wilson lines.

\paragraph{Dijet production}

The non-universality of the transverse momentum dependent parton distributions
between SIDIS and DY is a minor problem as it reduces to a sign difference.
The situation becomes much more complicated for processes with two incoming and
two outgoing partons, as for example dijet production in hadron-hadron
collisions.
Here, the longitudinal (and at the light-cone infinity, also the transverse)
gluons connect both to the partons in the initial and in the final state.
As found in Refs.~\cite{Bomhof:2004aw, Bomhof:2006dp}, eikonalization of those
gluons is possible (at least to the order $g$) for an arbitrary hard process,
but the procedure leads to appearance of new gauge link structures, like for
example
\begin{equation}
  \calU^{\left[\square \right]} =\mathcal{U}^{\left[ +\right]
  }\mathcal{U}^{\left[ -\right]\dagger}=\mathcal{U}^{\left[ -\right]
  }\mathcal{U}^{\left[ +\right]\dagger}\,.
  \label{eq:square-link}
\end{equation}
As a consequence, dijet production in hadronic collisions requires new types of
TMDs, which are not reducible to those encountered in SIDIS or DY. As we shall
see, the differences appear even between different channels in dijet
production, hence not only that the TMDs are not universal but the cross
section for dijet production does not factorize in the strong sense.
We shall now discuss this question in more detail.

\begin{figure}[t]
  \begin{center}
    \includegraphics[width=0.7\textwidth]{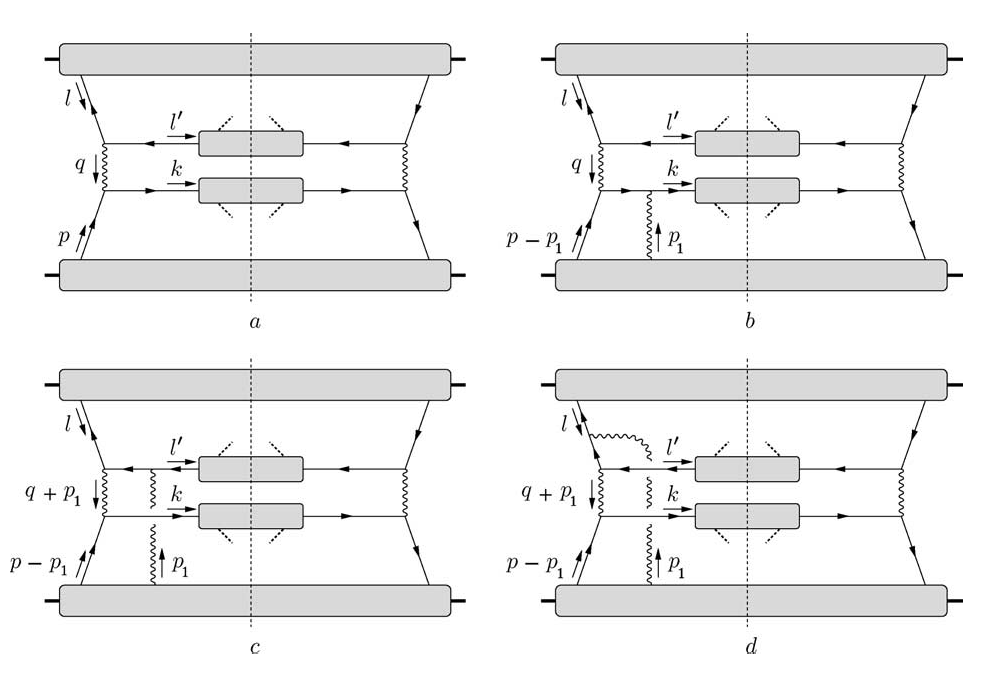}
  \end{center}
  \caption{
  Insertions of longitudinal gluons for $q\qbar\to q\qbar$ process. 
  Figure from Ref.~\cite{Bomhof:2004aw}.
  }
  \label{fig:qq2qq-long-gluon}
\end{figure}

In order to construct the gauge links for a process involving two incoming and
two outgoing partons, one needs to consider all gluon attachments between the
hadron and the hard part of the diagram. 
Fig.~\ref{fig:qq2qq-long-gluon} shows an example for the $q\qbar \to q\qbar$
process, where diagram~(a) is a tree level contribution, diagrams~(b) and~(c) correspond to attaching the gluon to the final-state partons, and, in
diagram~(d), the gluon is attached to the initial-state parton.
The result for the cross section comes from the sum of diagrams of
Fig.~\ref{fig:qq2qq-long-gluon} (a)-(d) and reads~\cite{Bomhof:2004aw}
\begin{eqnarray}
  \sigma&\propto&
  \int \mathrm{d}^4 p\,\mathrm{d}^4 k\,\mathrm{d}^4 l\,\mathrm{d}^4 l^\prime\ 
  \delta^4(p+q-k)\ \Big(\frac{1}{q^2}\Big)^2\ \bigg\{
  \tr\big[\Phi(p)\gamma^\mu\Delta(k)\gamma^\nu\big]
  \tr\big[\,\overline\Phi(l)\gamma_\nu\overline\Delta(l^\prime)\gamma_\mu\big] \nonumber \\
  &&\ -g_2\int \mathrm{d}^4 p_1 \frac{1}{-p_1^+ + i \epsilon}
  \tr\big[\Phi_A^\alpha(p,p-p_1) \gamma^\mu\Delta(k)\gamma^\nu\big]
  \tr\big[\,\overline\Phi(l)\gamma_\nu\overline\Delta(l^\prime)\gamma_\mu\big]
  \nonumber \\
  &&\ +g_1\int \mathrm{d}^4 p_1 \frac{1}{-p_1^+ + i \epsilon}
  \tr\big[\Phi_A^\alpha(p,p-p_1) \gamma^\mu\Delta(k)\gamma^\nu\big]
  \tr\big[\,\overline\Phi(l)\gamma_\nu\overline\Delta(l^\prime)\gamma_\mu\big]
  \nonumber \\
  &&\ +g_1\int \mathrm{d}^4 p_1 \frac{1}{p_1^+ + i \epsilon}
  \tr\big[\Phi_A^\alpha(p,p-p_1) \gamma^\mu\Delta(k)\gamma^\nu\big]
  \tr\big[\,\overline\Phi(l)\gamma_\nu\overline\Delta(l^\prime)\gamma_\mu\big]\,
  \bigg\}\,,
  \label{eq:linkfirstorder}
\end{eqnarray}
where the couplings $g_1$ and $g_2$ were introduced to distinguish between the
connections to the final and to the initial state particles, respectively. The
transition: hadron $\to$ quark is described by the correlator $\Phi(p)$, while
the transition: hadron $\to$ quark + gluon by $\Phi^\alpha_A(p,p-p_1)$.
The functions $\Delta(k)$ and $\Delta(l')$ represent quark~$\to$~hadron
transitions, hence these are fragmentation functions.

The three $\order{g}$ terms in Eq.~(\ref{eq:linkfirstorder}) correspond to the
first terms in the expansion of the gauge links
$U_{g_2}(\infty^-,\wT;w^-,\wT)$, 
$U_{g_1}(w^-,\wT;\infty^-,\wT)$, and
$U_{g_1}(-\infty^-,\wT;w^-,\wT)$, respectively.
Hence, at least to the order $g$, it is possible to account for the
gluon connections by reweighting the non-local bi-products of
operators, which define the TMDs, by the gauge links, as in
Eq.~(\ref{eq:tmd-definition}).  The above implies that, even in the complex
processes like the dijet production, it is possible to
define TMDs in a gauge invariant manner.

As shown in Ref.~\cite{Bomhof:2006dp}, similar procedure can applied to
subprocesses with gluons. The resulting TMDs involve all
types of gauge links, $\calU^{\left[ +\right]}$, $\calU^{\left[ -\right]}$ and
$\calU^{\left[\square \right]}$, multiplying each other and combined with 
several different colour factors. The complete set of definitions of the TMDs
appearing in the dijet process can be found in Ref.~\cite{Bomhof:2006dp}.

We see that  the dijet production in hadronic collisions requires very
complicated, subprocess-dependent TMDs. 
Hence, the strict factorization property does not hold.  It survives however in
a generalized form since the differential cross section for the process of
hadroproduction of two coloured partons, 
$h_1 + h_2 \mapsto a + b \to c + d$ can be written as
\begin{equation}
  d\sigma = \sum_i 
  \mathcal{F}_{a/h_1}^{(i)}(x_1,k_T) \otimes
  \mathcal{F}_{b/h_2}^{(i)}(x_2,k_T) \otimes
  H_{ab\to cd}^{(i)}\,.
  \label{eq:tmd-generalized}
\end{equation}
In the above, $h_1$ and $h_2$ denote the incoming hadrons. Each of them provides
a QCD parton, here, respectively $a$ and $b$.
$\mathcal{F}_{p/h}^{(i)}$ is a TMD of $i$-th type involving the parton $p$ and
the hadron $h$. 
Those TMDs are convoluted with the hard factors $H_{ab\to
cd}^{(i)}$, where $c$ and $d$ correspond to the outgoing partons. The hard
factors are gauge invariant combinations of the cut diagrams
contributing in a given channel, as explained in Refs.~\cite{Dominguez:2011wm,
Kotko:2015ura}.
Each term of the above sum has the same form as the strict factorization
formula. However, the gauge links appearing in the
generalized TMDs, $\mathcal{F}_{p/h}^{(i)}$, depend on the hard
subprocess.
 
Let us stress that we have deduced Eq.~(\ref{eq:tmd-generalized}) based on the
complete analysis with just a single gluon attachment, following
Refs.~\cite{Bomhof:2004aw, Bomhof:2006dp}. Partial contributions from two-gluon
attachments have also been studied~\cite{Bomhof:2006dp} but, for now, it is only
a conjecture that Eq.~(\ref{eq:tmd-generalized}) will hold at higher orders.
 
Differences in the link structures and the resulting process dependence can be
understood by looking again at Fig.~\ref{fig:qq2qq-long-gluon}.
As discussed earlier, the gluons connecting TMDs with the hard part can
be absorbed into a given quark or gluon correlator. This achieves two goals: it
renders the correlator gauge invariant and it allows one to factorize it from
the rest of the diagram.
 
However, this procedure requires pulling the gluons through the hard part, which
can be done using colour flow identities like Fierz identity.
The hard part,  in general, has a non-trivial colour structure.
Hence, the result of this procedure
depends on the type of partons participating in the hard scattering, as well
as whether the gluon which is pulled through the hard part
was attached to the initial or to the final state parton.
The situation is particularly complex for diagrams involving the $gg\to gg$ hard
process, where six different colour structures appear in the cut diagrams
(see Refs. \cite{Bomhof:2006dp,Kotko:2015ura} for details).
It is quite clear that the procedure of pulling the gluons into the
correlator introduces dependence on the sub-process and, as a consequence, leads
to process-dependence of the resulting TMDs.
Similar problems occur in proofs of the collinear factorization but, there, the
differences amongst the gauge links cancel between graphs after integration over
the transverse momentum~\cite{Aybat:2008ct}.

Construction of the gauge links necessary for gauge invariant definitions of
process-dependent TMDs, which allowed for a conjecture of
Eq.~(\ref{eq:tmd-generalized}) has been performed with up to two longitudinal
gluon attachments (thus up to the order $g^2$ in expansion of the gauge
links)~\cite{Bomhof:2004aw, Bomhof:2006dp}.
What is more important however, is that the two gluons were 
always taken from the same hadron.
This, in general, is not enough for QCD in processes involving four hadrons.
Indeed, as shown in Ref.~\cite{Rogers:2010dm}, even the generalized
factorization of Eq.~(\ref{eq:tmd-generalized}) breaks down in QCD at the level
of two gluons. The terms responsible for this effect correspond to diagrams in
which each of the two gluons is attached to different hadron.

\begin{figure}[t]
  \begin{center}
    \begin{minipage}[t]{0.48\textwidth}
    \centering
    \includegraphics[width=0.9\textwidth]{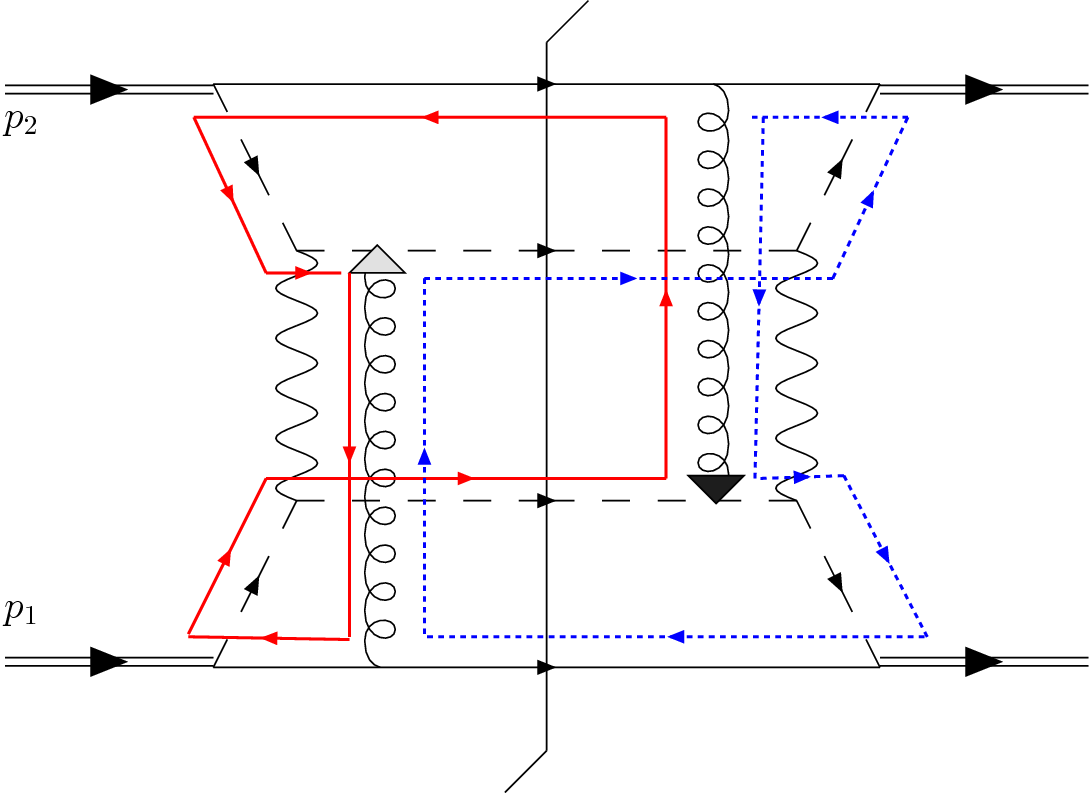}
    \caption{
      Example of a graph contribution to violation of generalized TMD
      factorization in a non-Abelian theory. The red an the blue lines
      illustrate colour flow. Figure from Ref.~\cite{Rogers:2010dm}.
    }
    \label{fig:genTMD-violation}
    \end{minipage}
    \hfill
    \begin{minipage}[t]{0.48\textwidth}
    \centering
    \includegraphics[width=0.75\textwidth]{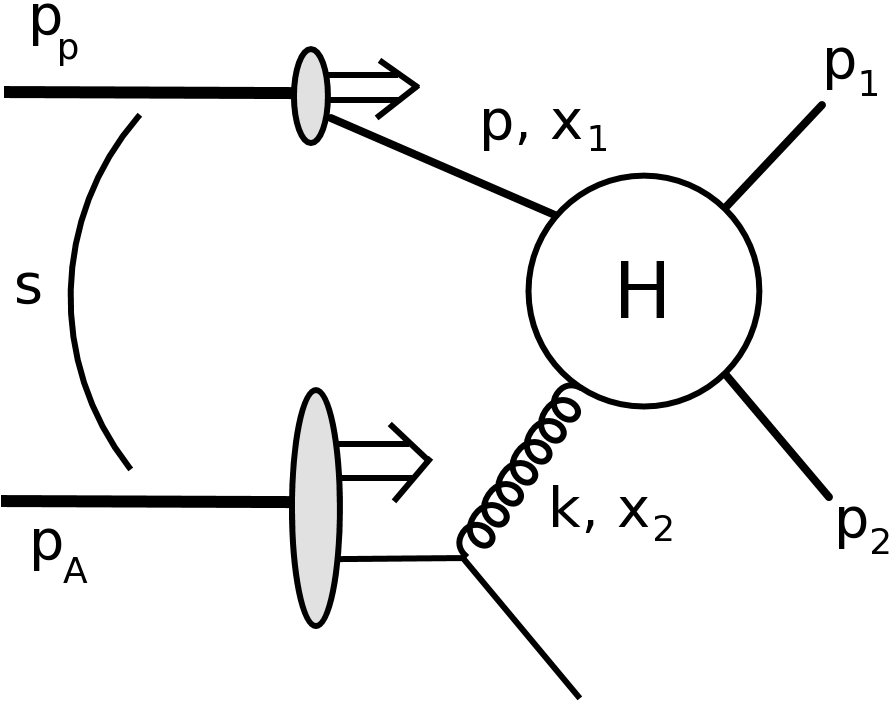}
    \caption{
      Inclusive dijet production in p+A collision. $H$ represents hard
      scattering. The solid lines coming out of $H$ can
      be either quarks or gluons.
      Figure from Ref.~\cite{Kotko:2015ura}.
    }
    \label{fig:dijets-pA}
    \end{minipage}
  \end{center}
\end{figure}

Ref.~\cite{Rogers:2010dm} verifies the assumption of the generalized TMD
factorization by finding an explicit counter-example, in which
Eq.~(\ref{eq:tmd-generalized}) is not valid 
for the double spin asymmetry (DSA) in a non-Abelian theory.
In order to facilitate the proof, Ref.~\cite{Rogers:2010dm} uses a simplified
model, where the complicated structure of multiple sub-processes and colour
flows of QCD is reduced to just a single sub-process with one colour flow.
The particle content of the model is shown in Fig.~\ref{fig:genTMD-violation}.
It consists of the incoming Dirac fields, neutral under the $SU(N_c)$ gauge
group (double line), spectator diquark (solid line), active, scalar ``quark''
(dashed line), the gluon (curly line) and a massive, colour-neutral boson (wavy
line) exchanged in the hard interaction. The latter allows for great
simplification of the colour flow in the hard process as compared to the
standard QCD. In fact, as seen in Fig.~\ref{fig:genTMD-violation}, the colour
(denoted by red and blue lines) is not exchanged in the hard process but only
through eikonal gluons.

As discussed earlier, the standard TMD factorization is violated in QCD at
the level of one gluon. In the simplified non-Abelian
theory~\cite{Rogers:2010dm}, similar breaking arises at the level of two gluons when
both connect the same hadron to a parton on the opposite side of the hard
process. These diagrams cannot be absorbed into standard Wilson lines. Hence,
new types of gauge links need to be introduced, which results in
non-universality of the TMDs.

At the level of two gluons there are however other diagrams, which are not
included even in the modified Wilson lines. These are the two-gluon diagrams
with one gluon radiated from each hadron attaching to the parton on the opposite
side of the hard process. An example of such diagram is shown in
Fig.~\ref{fig:genTMD-violation}.

The reason why the diagram of Fig.~\ref{fig:genTMD-violation} cannot be absorbed
by the modified Wilson line is that the latter includes a trace over
colour, hence, its $\order{g}$ expansion is proportional $\tr[t^a]=0$.
Therefore, the diagram shown in Fig.~\ref{fig:genTMD-violation} cannot be
generated by a Wilson loop expansion, yet it turns out to be non-zero because of
a non-trivial colour flow, impossible to accommodate by a product of
$\order{g}$ terms from two Wilson lines.
The diagram of Fig.~\ref{fig:genTMD-violation}
requires exchanges of non-singlet gluons. However, due to the
condition $\tr[t^a]=0$, only colour singlet gluon can be exchanged between a
spectator and the Wilson loop at the single-gluon level.
Therefore, we are left with a diagram that goes beyond what can be generated
by the Wilson lines and the factorization is broken.

The above direct counter-example, found for the double
spin asymmetry observable, shows that the general proof of TMD factorization
does not exist in a simplified model. Similar breaking is expected to happen
also for unpolarized cross sections in real QCD at the level of three gluon
exchanges (a diagram similar to Fig.~\ref{fig:genTMD-violation} with an extra
copy of the gluon on one side of the cut)~\cite{Rogers:2010dm} .

%-----------------------------------------------------------------------------
\subsection{Factorization in forward jet production}

We have seen that the general formula for the TMD factorization for the dijet
process is complex and its theoretical status is a subject of current
debate.
In this section, we shall discuss a special sub-class of dijet processes in
which both jets are produced at large rapidities.
This case corresponds to an asymmetric situation, in which the two colliding
objects are probed in very different momentum ranges.

The process is shown schematically in Fig.~\ref{fig:dijets-pA}.
The fractions of the longitudinal momenta of the incoming parton from the
projectile, $x_1$, and from the target, $x_2$, can be expressed in
terms of the rapidities and the transverse momenta of the produced jets as
\begin{equation}
  x_1  = \frac{1}{\sqrt{s}} \left(p_{T1}
  e^{y_1}+p_{T2} e^{y_2}\right)\,, 
  \quad
  \quad
  x_2  = \frac{1}{\sqrt{s}} \left(p_{T1} 
  e^{-y_1}+p_{T2} e^{-y_2}\right)\,.
  \label{eq:2to2kinematics}
\end{equation}
In the limit $y_1, y_2 \gg 0$, we have: $x_1 \sim 1$ and $x_2 \ll 1$. 
As we know, the number of gluons grows quickly with decreasing momentum
fraction. Therefore, the forward dijet production corresponds to dilute-dense
collisions, where the projectile, probed at a high momentum fraction, hence
appearing as dilute, can be described in terms of a collinear parton
distribution. On the other hand, one has to use an unintegrated parton
distribution on the dense target side, as the latter, probed at low $x$, will
have much bigger fraction of partons with high transverse momenta. Since those
partons will be mostly gluons, we limit our discussion to subprocesses in
which the parton from the target is always an off-shell (meaning $k^2 \neq 0$)
gluon, as illustrated in Fig.~\ref{fig:dijets-pA}.

%-----------------------------------------------------------------------------
\subsubsection{Generalized TMD factorization for forward jets}
\label{sec:generalized-tmd}

Ref.~\cite{Dominguez:2011wm} exploited the results of Ref.~\cite{Bomhof:2006dp}
to derive the exact form of the effective TMD factorization, given schematically
in Eq.~(\ref{eq:tmd-generalized}), for a range of processes studied in the
small-$x$ limit. In the case of dijet production in dilute-dense, hadron-hadron
collisions, one obtains
\begin{equation}
  \frac{d\sigma^{pA\rightarrow {\rm dijets}+X}}{dy_1dy_2d^2p_{T1}d^2p_{T2}} =
  \frac{\alpha _{s}^{2}}{(x_1x_2s)^{2}} \sum_{a,c,d} x_1 f_{a/p}(x_1,\mu^2) \sum_i H_{ag\to cd}^{(i)} \mathcal{F}_{ag}^{(i)}(x_2,k_T) 
  \frac{1}{1+\delta_{cd}}\,,
  \label{eq:tmd-main}
\end{equation}
where the notation is similar to that of Eq.~(\ref{eq:tmd-generalized}), but,
now, the incoming parton $b$ is always a gluon, $g$, and it comes from the dense
object $A$, while the parton from the dilute initial hadron, $p$, denoted as
$a$, can be either a (anti-)quark or a gluon.  The factor $1/(1+\delta_{cd})$ is
a symmetry factor needed in the case the $gg$ final state. (This factor is not
explicit in Eq.~(\ref{eq:tmd-generalized}) as the latter is only schematic.) 
 
As found in Ref.~\cite{Dominguez:2011wm}, the above, most complicated case
involves eight different gluon TMDs, which in the limit of large $N_c$ reduce to
five that can be written down as convolutions of two fundamental distributions:
the so-called \emph{dipole distribution} and the \emph{Weiz\"asacker-Williams
distribution}~(WW).
We recall that the dipole distribution enters directly into various inclusive
observables like the structure function in DIS, total cross section in DY and
SIDIS, as well as hadron production in DIS and $pA$ collisions and photon-jet
production in $pA$. None of the above involves the WW gluon. The latter can be
however determined from dijets production in DIS, as it is the only distribution
appearing in that process. Finally, dijets production in $pA$ involves both the
dipole and the Weiz\"asacker-Williams gluon.
Though appearing in factorization formulae for many processes, the dipole
distribution does not have a clear partonic interpretation. On the other hand,
the WW distribution corresponds to the number density of gluons, in the
light-cone gauge, inside a hadron.
 
All the accompanying hard factors, $H_{ag\to cd}^{(i)}$, appearing in
Eq.~(\ref{eq:tmd-main}), have been also calculated in the limit of large $N_c$
and ignoring the transverse momentum dependence (hence taking
$k_T=0$)~\cite{Dominguez:2011wm}. 
This
last approximation restricted the result to the case $p_{T1}, p_{T2} \gg k_T
\sim Q_s$, which, following the relation
\begin{equation}
  k_{T}^2 = |\mathbold{p_{T1}}+\mathbold{p_{T2}}|^2 = 
  p_{T2}^2 + p_{T2}^2 + 2p_{T1} p_{T2} \cos\Delta\phi\,,
  \label{eq:ktglue}
\end{equation}
corresponds to the production of two jets which are nearly back-to-back in the
transverse plane.
The~$Q_s$ above denotes the \emph{saturation scale}, \ie a momentum scale below
which non-linear effects become very relevant. Its typical value is estimated at
around a few GeV, depending on $x$, as well as the number of nucleons in the
target.

It was also found in Ref.~\cite{Dominguez:2011wm} that the same factorization
formula, Eq.~(\ref{eq:tmd-main}), involving identical gluon distributions, can
be derived from the \emph{colour glass condensate}
formalism~(CGC)~\cite{Gelis:2010nm} in the correlation limit (\ie for nearly
back-to-back dijet configurations).
In CGC, the gluons inside a dense, colour-neutral object are treated
semi-classically, which is justified at high densities. This leads to an
effective theory with colour sources, $\rho$, forming the static and large-$x$,
and the gauge fields, forming the dynamical and small-$x$, degrees of freedom.
Then, renormalization group equation is used for evolution of the sources in
$x$. The distribution of the sources, $W[\rho]$, describes properties of
saturated gluons at small~$x$.
Given the differences between the standard QCD framework and that of CGC, the
fact that the two lead to identical results (in the correlation limit) should be
regarded as a highly non-trivial check of the generalized TMD factorization
formula of Eq.~\ref{eq:tmd-main}.

We note that, since Ref.~\cite{Dominguez:2011wm} relies on gauge link
classification obtained rigorously at the level of a single-gluon exchange
~\cite{Bomhof:2006dp}, it may in principle suffer from the same issues that
were raised in Ref.~\cite{Rogers:2010dm} and discussed above in
Section~\ref{sec:tmd-factorization}.
Hence, the all-order validity of the generalized TMD factorization formula of
Eq.~(\ref{eq:tmd-main}) still needs to be demonstrated.
What is special about the dilute-dense limit is the fact that one of the
incoming hadrons, which we call a projectile, is described with an integrated
parton distribution. This opens a possibility that the TMD-factorization-breaking contributions, like the one illustrated in
Fig.~\ref{fig:genTMD-violation}, will cancel after integration, just like they
do in the case of the collinear factorization.
However, rigorous study is still needed to settle this point.

%-----------------------------------------------------------------------------
\subsubsection{High energy factorization}
\label{sec:hef-framework}

We note that the production of a dijet system moving in the forward direction is
a multi-scale problem. The highest scale is the collision energy $\sqrt{s}$,
then the jets' transverse momenta $p_{T1}, p_{T2}$ and, finally, the dijet
imbalance (or equivalently, the transverse momentum of the off-shell gluon).
As we see in Eq.~(\ref{eq:ktglue}), this last scale can in principle be anywhere
below the transverse momenta of the individual jets. 
The case with $k_T \sim p_{T1}, p_{T2}$, corresponds to a very small angle,
$\Delta\phi$, between the two forward jets, while if $k_T \ll p_{T1}, p_{T2}$,
the two jets are almost back-to-back. The latter is the domain of application of
the generalized (effective) TMD factorization, just discussed in
Section~\ref{sec:generalized-tmd}. 
In the former case, however, another type of factorization emerges.
This is the so-called \emph{high energy factorization} (HEF),
often also referred to as the \emph{$k_T$ factorization}.
 
The high energy factorization has been proposed in the context of heavy quark
production~\cite{Catani:1990xk, Catani:1990eg} (which is a multi-scale problem
as well, with the mass of a heavy quark playing a role similar to jet's
transverse momentum in the case discussed here).
 
The key observation is that, at high energies, the dominant contribution to the
cross section comes from exchanges of longitudinal gluons, whereas other terms
are sub-dominant. Hence, the cross section formula can be factorized into an
unintegrated parton distribution function (TMD), which emits an off-shell (\ie
$k_T \neq 0$) gluon, and an off-shell matrix element.
Ref.~\cite{Catani:1990eg} derives an effective procedure which guarantees gauge
invariance of the off-shell amplitudes within a subclass of axial gauges (see
also Ref.~\cite{Kotko:2015ura} for details).
Recently, fully general methods for calculating the off-shell amplitudes, which
can be used in the framework of the high energy factorization, have been
developed~\cite{vanHameren:2012uj, vanHameren:2012if}.

In the context of the forward dijet production, the HEF formula takes again a
hybrid form, as we only need to consider the off-shell gluon effects in one of
the colliding hadrons~\cite{Deak:2009xt, Kutak:2012rf}
\begin{equation}
  \frac{d\sigma^{pA\rightarrow {\rm dijets}+X}}{dy_1dy_2d^2p_{T1}d^2p_{T2}} 
  =
  \frac{1}{16\pi^3 (x_1x_2 s)^2}
  \sum_{a,c,d} 
  x_1 f_{a/p}(x_1,\mu^2)\,
  |\overline{{\cal M}_{ag^*\to cd}}|^2
  {\cal F}_{g/A}(x_2,k_T)\frac{1}{1+\delta_{cd}}\,.
  \label{eq:hef-formula}
\end{equation}
Note that the above cross section needs only a single TMD, in contrast to the
effective TMD factorization discussed earlier. 
The ${\cal F}_{g/A}$ parton distribution is the dipole distribution, usually
determined from fits to DIS data.
Another difference with respect to Eq.~(\ref{eq:tmd-main}) is that, here, the
hard factor (which, because we have only one TMD, is just a matrix element
squared) depends on the transverse momentum, hence, we shall refer to it as an
off-shell matrix element.
The complete set of the off-shell matrix elements needed for hadroproduction of
forward jets were calculated and analyzed in Refs.~\cite{Deak:2009xt,
Deak:2010gk}.
We shall further elaborate on the relation between Eqs.~(\ref{eq:tmd-main}) and
(\ref{eq:hef-formula}) in the following subsection.
 
As demonstrated in Ref.~\cite{Kotko:2015ura}, the factorization
formula~(\ref{eq:hef-formula}) can be also derived for all channels from the CGC
approach in the kinematic window $p_{T1}, p_{T2} \sim k_T \gg Q_s$. This
limit corresponds to the dilute target approximation, hence, it should not be
employed to study non-linear effects in dense systems. It can however be used in
the, so-called, geometric scaling region, where the linear approximation is
still valid but saturation effects can be felt~\cite{Kutak:2012rf,
vanHameren:2014lna, vanHameren:2014ala}.
However, the HEF formula is not applicable in situations 
corresponding to $k_T \sim Q_s$. This deficiency is fixed by the improved TMD
factorization framework, which we discuss next.

%-----------------------------------------------------------------------------
\subsubsection{Improved TMD factorization}
\label{sec:improvedTMD}

A framework unifying the HEF formalism (applicable when $k_T \sim p_{T1},
p_{T2}$) and the generalized TMD formalism (applicable for $k_T \ll p_{T1},
p_{T2}$) was proposed in Ref.~\cite{Kotko:2015ura}. It can be regarded as a
generalization of Ref.~\cite{Dominguez:2011wm} to the case in which the $k_T$
dependence is kept also in the hard factors. The latter were computed with two
independent methods: the original procedure of Refs.~\cite{Catani:1990xk,
Catani:1990eg} as well as the colour ordered amplitudes
approach~\cite{Mangano:1990by}. The improved factorization formula reads
\begin{equation}
  \frac{d\sigma^{pA\rightarrow {\rm dijets}+X}}{d^{2}P_{t}d^{2}k_{t}dy_{1}dy_{2}}=\frac{\alpha_{s}^{2}}{(x_1 x_2 s)^{2}}
  \sum_{a,c,d} x_{1}f_{a/p}(x_{1}, \mu^2)\sum_{i=1}^{2}K_{ag^*\to cd}^{(i)}
  \Phi_{ag\rightarrow cd}^{(i)} (x_2, k_T)\ \frac{1}{1+\delta_{cd}}\ ,
  \label{eq:gg2gg-mod}
\end{equation}
where $K_{ag^*\to cd}^{(i)}$ and $\Phi_{ag\rightarrow cd}^{(i)}$ are the new
hard factors and the new TMDs, replacing, respectively, 
$H_{ag\to cd}^{(i)}$ and $\mathcal{F}_{ag}^{(i)}$ from Eq.~(\ref{eq:tmd-main}).
As we see, $K_{ag^*\to cd}^{(i)}$  is a hard factor for an off-shell, hence
$k_T$-dependent, incoming gluon. We also note that the new formula has two TMDs
per each channel, thus, six altogether. This is fewer than the eight TMDs
appearing 
in Eq.~(\ref{eq:tmd-main}). The reason is that the generalized TMD factorization
formula (\ref{eq:tmd-main}) was found to have some redundancy, which has been
removed in Eq.~(\ref{eq:gg2gg-mod}). 
That is most readily seen in the colour ordered amplitude formalism.
The last improvement of Ref.~\cite{Kotko:2015ura} was a restoration of the full
$N_c$ dependence in the hard factors. We refer to the original article for
explicit expressions for the new hard factors and TMDs.

The improved TMD factorization (\ref{eq:gg2gg-mod}) is valid in the limit 
$p_{T1}, p_{T2} \gg Q_s$ for an arbitrary value of~$k_T$, hence it
provides a robust framework for studies of the non-linear domain of QCD
with hard jets.
Preliminary results indicate differences with respect to the HEF
formalism~\cite{Kotko:2016num} but the full potential of this new
framework for phenomenology of forward jet production is yet to be explored.

We note that theoretical status of the improved TMD factorization formula is
similar to that of the generalized TMD factorization discussed in
Section.~\ref{sec:generalized-tmd}. In particular, it remains to be shown that
large logarithms generated by higher order corrections can be absorbed into TMDs
and the remaining, factorization-breaking, contributions vanish.
As explained in Section~\ref{sec:tmd-factorization}, TMD factorization for dijet
production appears to be broken, even in its generalized form.
What is special about the case considered here is that, in the
formula~(\ref{eq:gg2gg-mod}), only one hadron is described by the TMDs. The
other incoming hadron is taken in the collinear approximation and, because we
discuss jet production, the outgoing partons do not involve fragmentation
functions.
%
%Hence, many of the potential factorization-breaking corrections will not appear
%in forward dijet production, \cf Ref.~\cite{Buffing:2011mj}.

We emphasize that Eq.~(\ref{eq:gg2gg-mod}) has limits, which are solid
results of  QCD. 
In the case $k_T \ll p_{T1}, p_{T2}$, the $k_T$ dependence can be neglected
in the hard factors. Then, when Eq.~(\ref{eq:gg2gg-mod}) is integrated over
$k_T$, one effectively integrates only the TMDs,  which, by the same principle
as discussed for SIDIS and DY in Section~\ref{sec:tmd-factorization}, will all
(except one, which vanishes) turn into a single, integrated PDF and the standard
collinear factorization formula is recovered.
On the opposite end, when $k_T \lesssim p_{T1}, p_{T2}$, the transverse part of the Fourier transform of Eq.~(\ref{eq:tmd-definition})
becomes quickly varied and the $\wT$ dependence of the gauge links can be
neglected. Then, again, all the gauge links become identical and all the six $\Phi_{ag\rightarrow cd}^{(i)}$ TMDs collapse to a
single function, which can be identified with $\mathcal{F}_{g/A}$.
Hence the HEF factorization (\ref{eq:hef-formula}) is recovered as it should be.

%-----------------------------------------------------------------------------
\section{Jet production in hadron-hadron collisions}
\label{sec:jet-production}

After having discussed the definitions and properties of jets and the subtelties
related to QCD factorization, in this section, we turn to jet production
processes at hadron colliders. 
Aiming at precise QCD predictions, we shall first discuss potential limiting
factors, such as non-perturbative effects and dependence of the results on jet
definition choices.
Then, we will turn to the state-of-the-art perturbative calculations for
processes involving jets and show selected comparisons to the LHC data.
Finally, we shall devote separate subsections to the questions of vetoing jets
and to the forward jet production.

%-----------------------------------------------------------------------------
\subsection{Nonperturbative effects}
\label{sec:ue-pu}

Although jets can be, to a first approximation, regarded as the objects that 
point to the partons undergoing a hard scattering, the exact relation between
partons and jets is significantly more complex.
One type of corrections to this simple picture comes from higher order terms of
the perturbative series, as each quark or gluon leaving the hard scattering can
dress into real and virtual partons. The other type of correction is related to
the partons $\to$ hadrons transition, which has to happen at some low scale
before jets are registered in the detector. This \emph{hadronization} effects
are of genuinely non-perturbative nature. 
A third type of corrections to the partons $\leftrightarrow$ jets relation is
specific to hadron colliders and it arises from interactions between hadron
remnants or between other hadrons from the same bunches that cross during the
collision. Those result in additional radiation that is likely to contaminate
the jets initiated by the hard partons, thus obscuring the relation between the
two.
Such radiation, called the \emph{underlying event} and \emph{pileup} has mixed,
perturbative and non-perturbative nature.
We start our discussion from the corrections with non-perturbative components,
while in the following sections, we shall turn to the effects arising due to
higher order perturbative radiation.

%-----------------------------------------------------------------------------
\subsubsection{Underlying event and pileup}

As mentioned above, a hard jet in hadron-hadron collision is always produced on
top of a pedestal formed by the soft and semi-hard particles, referred to as the
\emph{underlying event} (UE). 
The underlying event, although consisting mostly of soft particles, is
different than the activity measured without a jet trigger, the so-called
\emph{minimum bias}~(MB). In particular, the overall transverse momentum
accumulated in UE will depend on the $p_T$ of the produced jets, as well as on
the azimuthal distance to the jet.

On top of the UE activity, each measurement at the LHC faces the problem of
\emph{pileup} (PU), that is the radiation coming from independent hadron-hadron
collisions occurring in the same bunch crossing. 
Typical numbers of such crossing varies between $10-100$ in the LHC environment.
Because pileup comes from incoherent proton-proton collisions, unlike UE, it is
expected to be independent of jet activity.

\begin{figure}
  \centering
  \includegraphics[width=\textwidth]{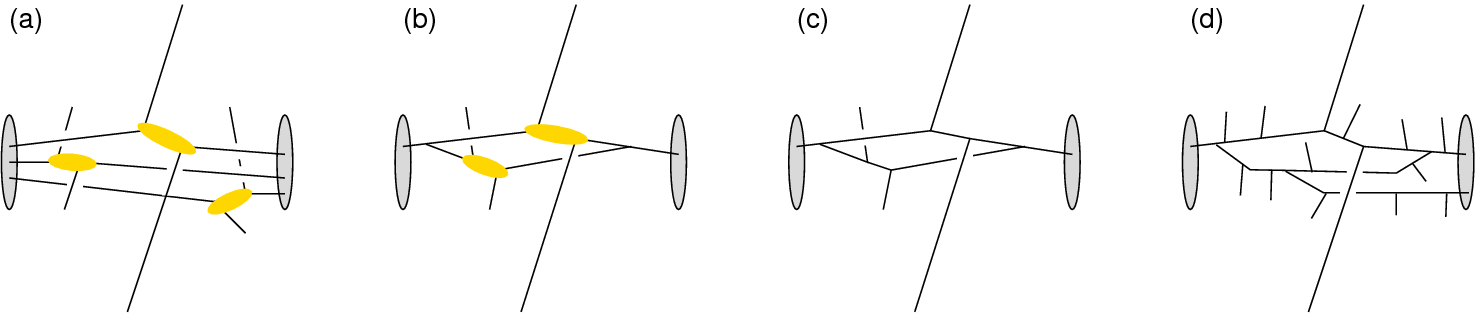}
  \caption{
   Contributions to the underlying event:
   (a)~multiple $2\to2$ interactions, 
   (b)~collinear splitting in the initial-state 
   followed by double $2\to2$ scattering, 
   (c)~perturbative, 1-loop, $2\to4$ diagram, and 
   (d)~BFKL-type emissions.
    }
  \label{fig:UE-model-pics}
\end{figure}

In order to understand why UE is different from MB, let us examine
Fig.~\ref{fig:UE-model-pics}, which depicts some of the sources of the
underlying event taking dijet production as an example. In
Fig.~\ref{fig:UE-model-pics}~(a) the hard interaction in which the dijet system is
produces, is supplemented by multiple, independent, \mbox{lower-$p_T$}, $2\to 2$
scatterings. 
Such mechanism, called the \emph{multi-parton interactions} (MPI) turns out
to be responsible for a large fraction of the underlying event.
Because the incoming particles all originate from the same proton, it is
natural to expect that the individual $2\to 2$ scatterings shall be correlated
both through the energy conservation or due to
colour connections. 
Moreover, MPIs can be also realized via an alternative graph, shown in
Fig.~\ref{fig:UE-model-pics}~(b), where both pairs of the incoming particles
have in fact a common origin and arise from the initial-state collinear
splitting.

However, the lowest order diagram corresponding to
Fig.~\ref{fig:UE-model-pics}~(b), shown in Fig.~\ref{fig:UE-model-pics}~(c), can
be regarded as a one-loop-double-real correction to the dijet process, which 
starts to be relevant at the order N$^3$LO.
Current modeling does not prevent us from double-counting between
multiple parton interaction and perturbative higher orders~\cite{Diehl:2011tt, Diehl:2011yj}.

The soft radiation that fills the event is not bound to come just from the
$2\to 2$ scatterings, but may also arise from configurations that involve
BFKL-like, multiple chains of low-$p_T$ emissions spread in rapidity, as 
shown in~Fig.~\ref{fig:UE-model-pics}~(d).
Finally, additional radiation emitted from the initial-state particles
(\emph{initial-state radiation}, ISR) and the beam remnants will also contribute
to the underlying event. 

We see that it is general not conceptually clear
how to separate the hard part, amenable to perturbative treatment, and the
underlying, low-$p_T$ activity.
One can turn the above statement around and say that because UE has a
significant non-perturbative component, which cannot be calculated from first
principles and can only be modeled, measurements of the underlying event are
important as they provide a way to constraint parameters of phenomenological
models.

To a first approximation, we expect that UE/PU modifies jet's $J$ transverse
momentum according to~\cite{Dasgupta:2007wa}
\begin{equation}
 \langle \delta p_T(J) \rangle_{\text{UE}/\text{PU}} =
 \frac{\rho}{2} 
 \left(
 R^2 - \frac{R^4}{8} + \order{R^6}
 \right)\,,
 \label{eq:ue-analytic}
\end{equation}
where $\rho$ is an average transverse momentum per unit rapidity
carried by the underlying event or pileup for a given event.
A fully realistic correction needs to take into account the fact that jets are
not exactly circular, but their areas depend on the algorithm and vary on the
jet-by-jet basis. Therefore, the exact correction from UE/PU
reads~\cite{Cacciari:2007fd}
\begin{equation}
 \langle \delta p_T(J) \rangle_{\text{UE}/\text{PU}} =
 \rho A(J) \pm \sigma \sqrt{A(J)} \,,
  %\label{eq:}
\end{equation}
where $A(J)$ is an active jet area introduced in Section~\ref{sec:jet-area} and
$\sigma$ is a standard deviation from the average UE/PU level, when measured
across sub-regions of the event.
Then, the transverse momentum of a jet can be corrected for the effects coming
from UE/PU according to
\begin{equation}
 p_{T}(J)^\text{sub} = p_T^\text{measured}(J) - \rho A(J)\,,
  \label{eq:pt-corr-jet-area}
\end{equation}
where $p_{T}(J)^\text{sub}$ is expected to correspond to 
genuine transverse momentum of the jet $J$.

%-----------------------------------------------------------------------------
\subsubsection*{Approaches to UE measurement}
\label{sec:overview-of-approaches}

Measurements of the underlying event are on one hand important for jet physics,
as they allow for corrections of jets' transverse momenta and masses. On the
other hand, jets themselves can be helpful in pining down the properties of UE.

\begin{figure}
  \centering
  \includegraphics[width=0.22\textwidth, trim=  0 -150 0 0]{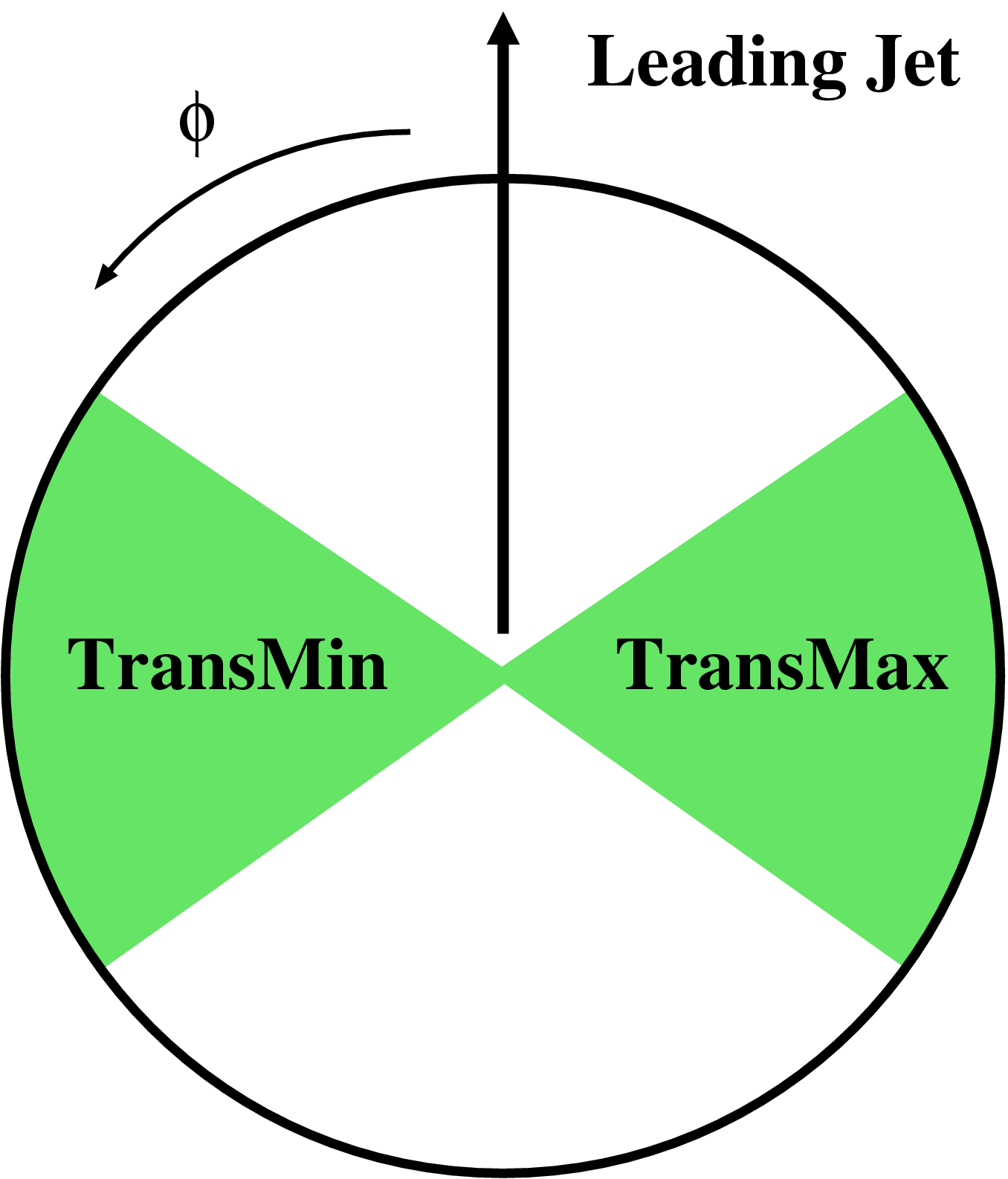}
  \hspace{5pt}
  \includegraphics[width=0.18\textwidth, trim = 0 -50 0 0]{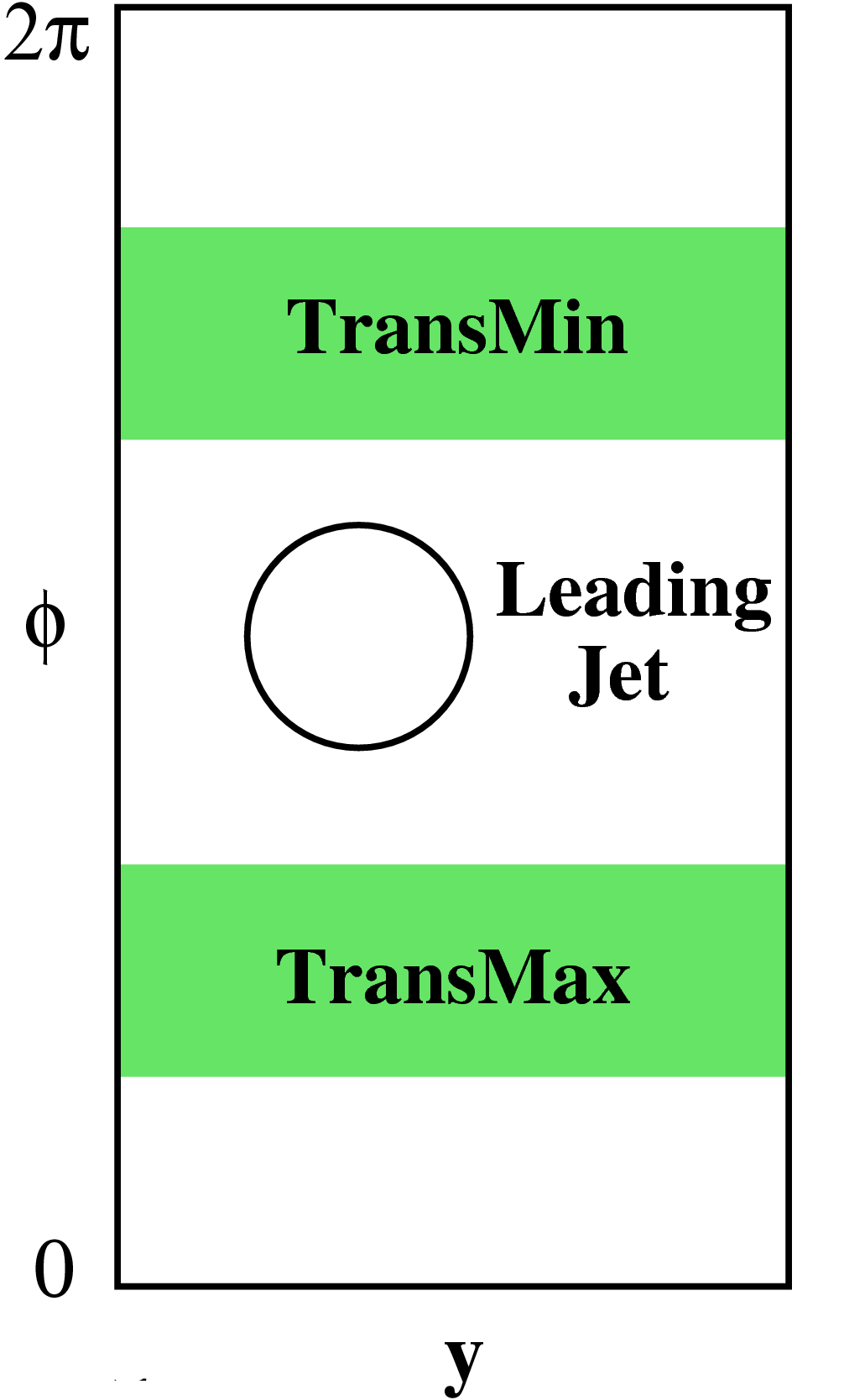}
  \hspace{50pt}
  \includegraphics[width=0.35\textwidth]{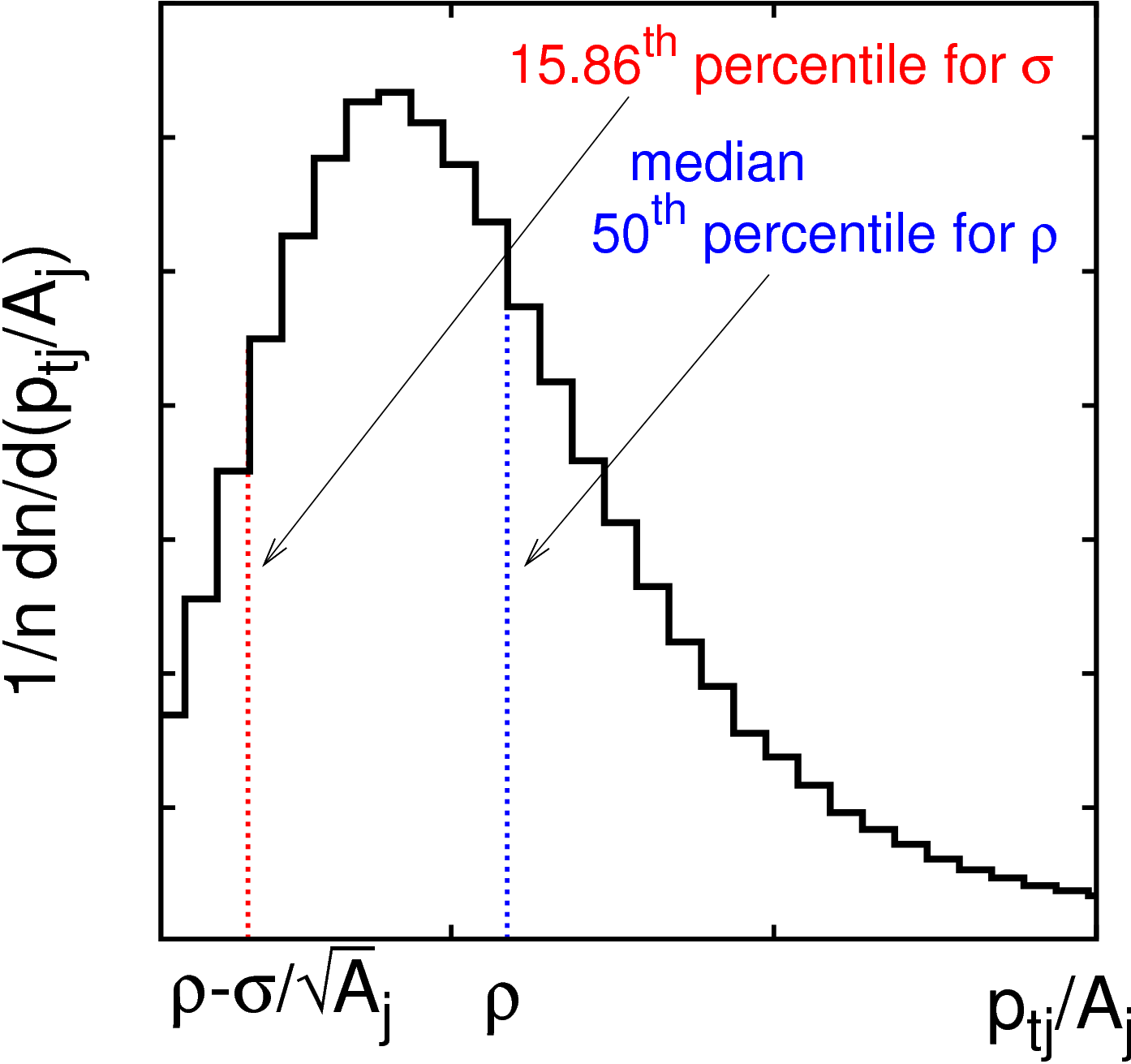}
  \caption{%
    Methods of UE measurements. 
    Left: traditional approach with the ``TransMin'' and ``TransMax'' regions 
    in the transverse plane.
    Right: area/median based approach in which distribution of the transverse
    momentum over area, $p_{tj}/A_j$ for the set of jets in a single event is
    used to determine the average level of UE, $\rho$, and its standard
    deviation, $\sigma$.
    Figures from Ref.~\cite{Cacciari:2009dp}.
  }
  \label{fig:approaches}
\end{figure}

%----------------------------------------------------------------------
\paragraph{Traditional approach}
\label{sec:trad-appraoch}

The most widely spread approach \cite{Albrow:2006rt} to measuring the
properties of the underlying event involves 
tagging events based on the
presence of a jet or a hard particle, and measuring the radiation separated from
the hard object by the distance $\phi$ in the transverse plane.
One defines four regions, as illustrated in Fig.~\ref{fig:approaches} (left):
the ``towards'' region, typically with $|\phi| < \pi/3$, the ``away'' region
with $2\pi/3 < |\phi| < \pi$, and two transverse regions, covering $\pi/3 <
|\phi| < 2\pi/3$.
 
The idea behind this division is that most of the radiation generated in the
hard scattering will occupy the towards and the away regions and, by measuring
the activity in the transverse regions, one gets access to the genuine
underlying event.
Hence, one measures the multiplicity of charged tracks above some
transverse-momentum threshold, as well as the total transverse momentum
contained in the charged tracks, and typically presents the result as an average
across many events and as a function of the $p_T$ of the leading jet. 

The method is usually further improved by labeling the two transverse regions as
``TransMin'' and ``TransMax'', respectively for the less and for the more active
of the two, where the ``activity'' is defined with the total transverse momentum
entering into the region.
This is motivated by the fact that there is still 
a probability of order $\order{\as}$ that one of the transverse regions receives
radiation from the hard scattering, which significantly affects the extracted
information about the average UE $p_T$ flow.
The largest component of the perturbative contamination should contribute
to the TransMax region, hence, by restricting the measurements to the
TransMin region one reduced the bias from the hard radiation.
The average over the two region is called ``TransAv''.

The method gives freedom as to where exactly place the transverse
regions, as well as to the choice of their size and shape.
The LHC experiments tend to divide the transverse plane exactly at $|\phi| =
\pi/3$ and $|\phi| = 2\pi/3$, and consider jets or other hard objects with
rapidities $|\eta| < 2.5$ for measurements based on charged
tracks~\cite{Aad:2010fh}.
UE measurements at Tevatron used the range o $|\eta| < 1.0$~\cite{Aaltonen:2010rm}.

%----------------------------------------------------------------------
\paragraph{Jet-area/median approach}
\label{sec:area-median-approach}

Another method to measure the underlying event and pileup, proposed
in Refs.~\cite{Cacciari:2007fd, Cacciari:2009dp}, is based on the concept of jet
areas~\cite{Cacciari:2008gn}.
In this approach, one first clusters the event with an IRC-safe jet algorithm,
like C/A or $k_T$ and then attributes the active jet area, $A_j$ to each jet
$j$, as described in Section~\ref{sec:jet-area}.
Notice that, because the active jet area is calculated by adding a large number
ghost particles to the event, some jets will be formed uniquely from ghosts.
Those are called the ``pure ghost jets'' and are considered to have $p_T=0$.

The key point of the method is to measure the transverse momentum density of per
unit area in an event by taking the median of the distribution of the
$p_{Tj}/A_{j}$ for the ensemble of jets in that event within certain rapidity
window
\begin{equation}
  \label{eq:median-y}
  \rho(y) = \mathop{\mathrm{median}}_{j \in \mathrm{jets},\;
    |y_j - y|< \delta y} \left[ \left\{ \frac{p_{Tj}}{A_j}\right\}\right]\,,
\end{equation}
as shown schematically in Fig.~\ref{fig:approaches} (right).
The motivation for the use of the median is that it is much less susceptible to
contamination from hard perturbative radiation, which is the main source of bias
in the traditional approach.
 
The UE level, $\rho$, is measured on the event-by-event basis.  In addition, one
can also determine the intra-event fluctuations of the UE, $\sigma$, 
defined such that a fraction $X/2$ of jets satisfy
$\rho - \sigma/\sqrt{\langle A_j \rangle} < p_{Tj}/A_j < \rho$, 
where $X$ is the fraction of a Gaussian
distribution within one standard deviation of the mean~\cite{Cacciari:2009dp}.
As shown in
Fig.~\ref{fig:approaches} (right), this definition is one-sided (\ie just
considering jets with $p_{Tj}/A_j < \rho$). This helps
 to limit contamination of $\sigma$ coming from the hard jets when the total
number of jets is small.

The jet area-based method is very robust when applied to both pileup and to UE
measurements. The latter case is more challenging since the UE is softer, and
has relative fluctuations that are larger than those in PU or MB. Hence, one
faces the trade-off between getting the most differential information, by taking
small rapidity windows in Eq.~(\ref{eq:median-y}) and sensitivity to the hard
contamination.
If the rapidity window is too small, one has a limited number of jets that enter
the formula (\ref{eq:median-y}) and the relative impact of the presence of a
hard jet in the region of interest is amplified by the small total number of
jets.
One way to help in such situations, when studying UE in processes with $n$ jets
is to remove the $n$ hardest jets from the list of jets used to compute the
median.

The jet area-based method has been used by CMS to study the underlying
event~\cite{Chatrchyan:2012tt}, where it has been also adjusted to cope with
scarcely populated events.
It has been also widely employed to the event-by-event UE subtraction for
processes with photons~\cite{Aad:2010sp,Aad:2014eha}.

%-----------------------------------------------------------------------------
\subsubsection*{Experimental results of UE measurements}

\begin{figure}[t]
  \begin{center}
    \begin{minipage}[t]{0.50\textwidth}
    \centering
    \includegraphics[width=0.95\textwidth]{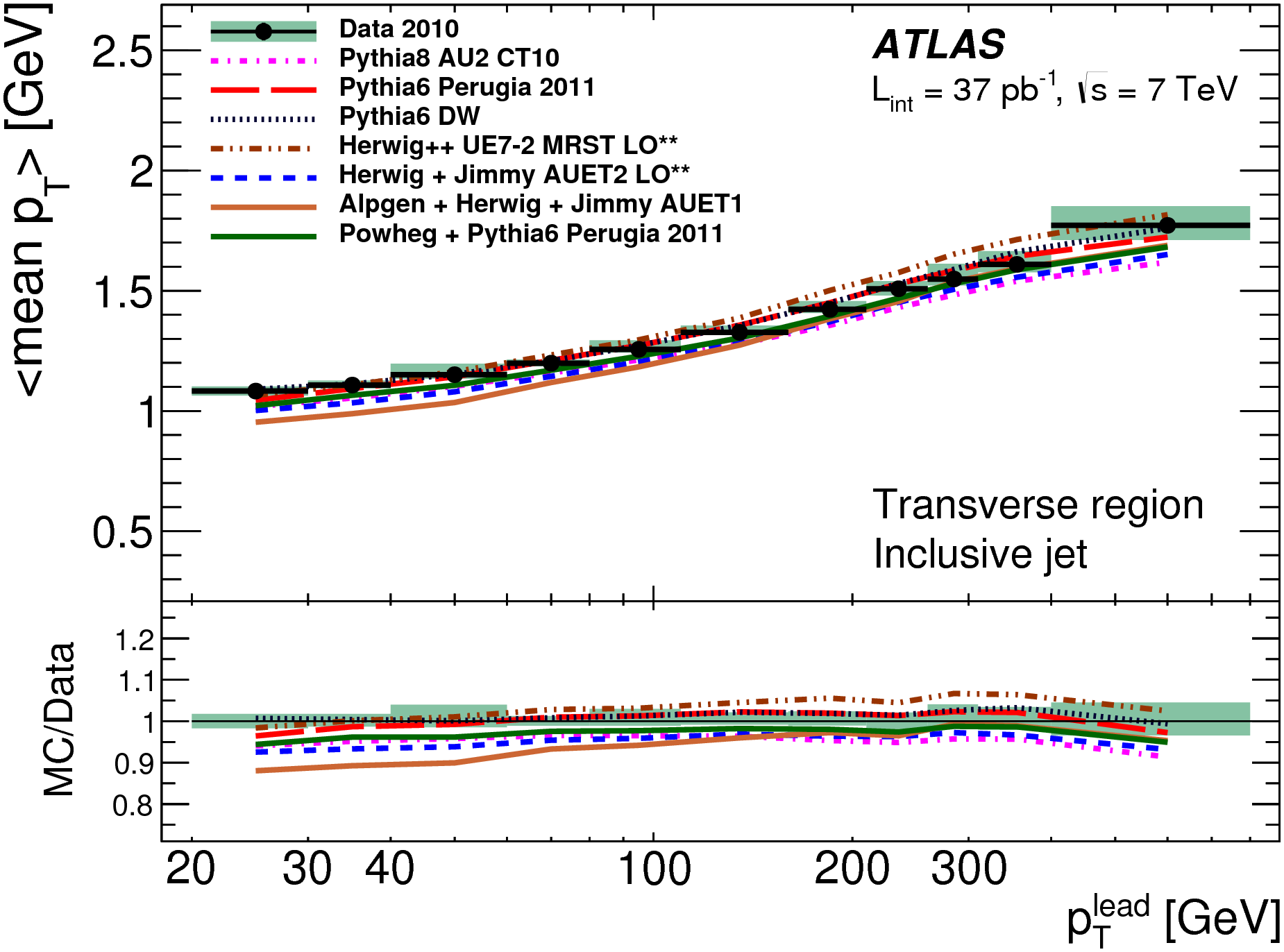}
    \caption{
    Transverse region profiles of the mean $p_T$ of charged particles for 
    inclusive selection against the leading-jet $p_T$ compared to a range of MC
    models.
    The shaded area shows the combined statistical and systematic uncertainty.
    Figure from Ref.~\cite{Aad:2014hia}.
    }
    \label{fig:ue-atlas-traditional}
    \end{minipage}
    \hfill
    \begin{minipage}[t]{0.46\textwidth}
    \centering
    \includegraphics[width=0.9\textwidth]{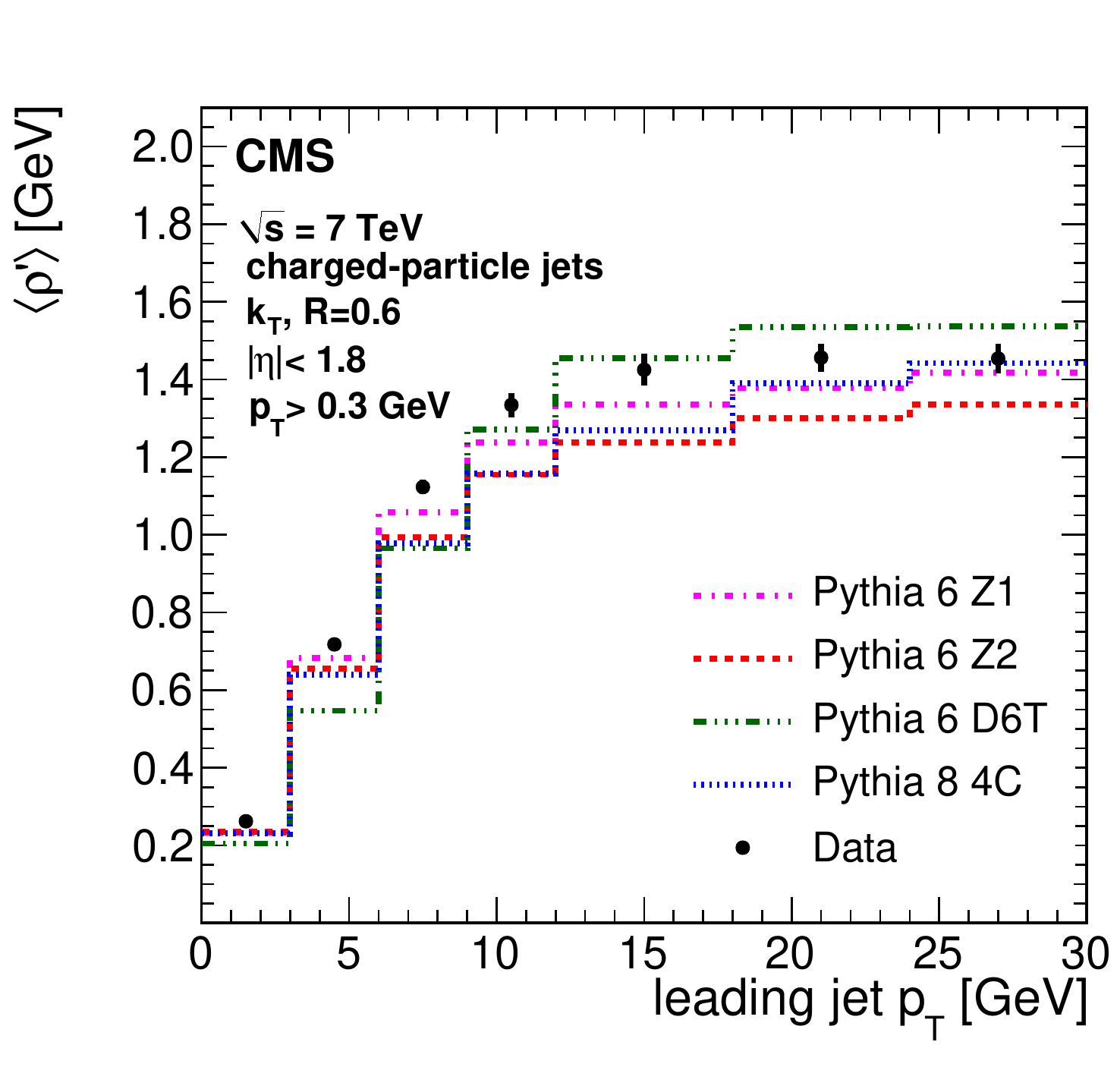}
    \caption{
      Mean values of $\rho'$ distributions versus leading charged-particle jet
      transverse momentum compared to different generator tunes.
      Figure from Ref.~\cite{Chatrchyan:2012tt}.
    }
    \label{fig:ue-cms-area}
   \end{minipage}
  \end{center}
\end{figure}

A range of measurements of the underlying event have been performed since the
start of the LHC and they have greatly helped to tune the MC models of soft and
semi-hard radiation in this new energy domain.

Most of the results were obtain with the traditional method and using charged
tracks as the leading hard objects\cite{Aad:2010fh, Khachatryan:2010pv,
Chatrchyan:2011id} 
Complementary analysis, based on calorimeter information only, has also been
performed~\cite{Aad:2011qe}.
This was an important cross-check as the neutral hadrons account for $\sim$ 40\%
of the produced particles. 
 
The measurements covered the energies from $\sqrts= 900\, \GeV$, through
2.76 TeV up to 7 TeV. Early UE measurements at 13 TeV start appearing
now~\cite{ATL-PHYS-PUB-2015-019}.
The qualitative features of the UE at higher energies remained the same as
those found at Tevatron. That is, the UE in the TransAv region retains fairly
strong dependence on the transverse momentum of the leading hard object, which
is reduced in the TransMin region. The latter confirms the expectation that the
TransMin region is less affected by hard radiation~\cite{Aad:2014hia}.
On the quantitative level, it turned out that none of the Monte Carlos tuned to
the pre-LHC data was able to correctly predict the UE activity. The predictions
for 7 TeV were lower than what has been found in the data, often up to 50\%.
In view of the above, the new UE data from ATLAS and CMS were used to retune
model parameters~\cite{Chatrchyan:2013ala, Aad:2014hia, Skands:2014pea}.
The current status is that most of the modern tunes allow for 20\% or
better accuracy in the description of the ensemble of the UE characteristics, as
shown in the example Fig.~\ref{fig:ue-atlas-traditional}

The jet area/median based methods has also been used to study UE at $\sqrts =
0.9$ and $7\, \TeV$~\cite{Chatrchyan:2012tt}. 
The original formula for $\rho$,
given in Eq.~(\ref{eq:median-y}), has been adjusted,
by taking only the physical jets in the set
used to determine the median, and the modified quantity was called $\rho'$.
This was motivated by the fact that, in the experimental situations with low
average charged-particles multiplicities, the number of ghost jets is so large
that $\rho=0$  in majority of cases.

Fig.~\ref{fig:ue-cms-area} shows the mean value of $\rho'$ measured by CMS, as a
function of the transverse momentum of the leading jet, together with
predictions from various MC tunes. The overall description is reasonably good,
with the discrepancies at the level of 20\% or lower. The observed sensitivity
to UE modeling shows usefulness of the jet area/median approach in constraining
models of soft radiation.

%-----------------------------------------------------------------------------
\subsubsection*{Further theoretical developments}

\begin{figure}[t]
  \begin{center}
    \begin{minipage}[t]{0.48\textwidth}
    \centering
    \includegraphics[width=0.88\textwidth]{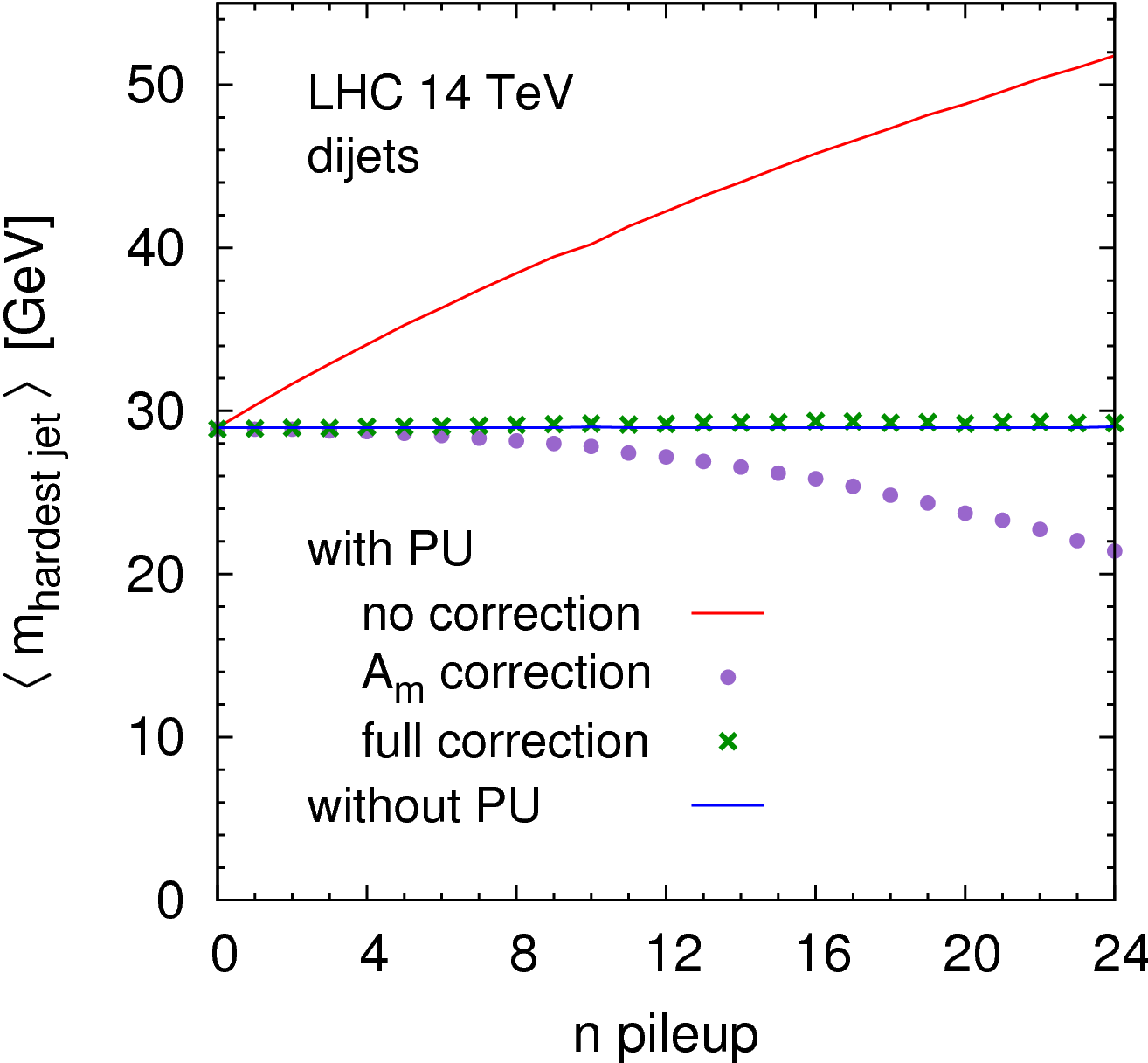}
    \caption{
    Average mass of the hardest jet in dijet events at the LHC. Jets were found
    with anti-$k_T$, $R=0.7$ and the cut $p_{T,\text{hardest}} > 150\, \GeV$ was
    imposed.
    Figure from Ref.~\cite{Sapeta:2010uk}.
    }
    \label{fig:mass-area}
    \end{minipage}
    \hfill
    \begin{minipage}[t]{0.48\textwidth}
    \centering
    \includegraphics[width=0.9\textwidth]{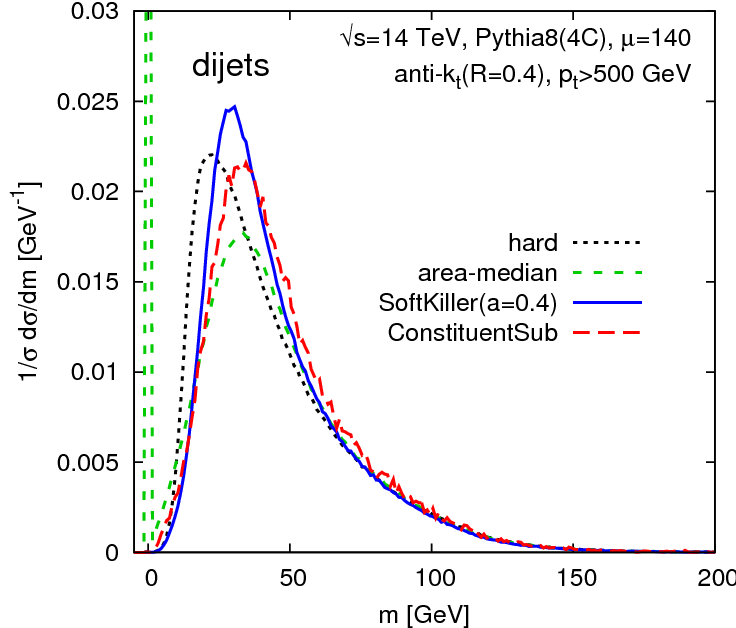}
    \caption{
    Performance of the Soft Killer and Constituent Subtraction methods as
    compared to the standard area/median subtraction for jet mass distribution.
    Figure from Ref.~\cite{Cacciari:2014gra}.
    }
    \label{fig:soft-killer}
   \end{minipage}
  \end{center}
\end{figure}

The method of Ref.~\cite{Cacciari:2007fd} for correcting jet transverse momenta for the contamination from
UE/PU by using jet areas has been extended to the case of jet masses by
introducing the concept of mass areas~\cite{Sapeta:2010uk} as described in
Section~\ref{sec:jet-mass-area}. The amount of mass coming from
pileup, which needs to be subtracted from the jet mass, is given
by~\cite{Sapeta:2010uk}
\begin{equation}
  \delta m^2 (J_\PU)=  p_{TJ_\PU}\,\rho\, A_m(J_\PU) - \rho^2 A_\mu^2(J_\PU)\,,
  \label{eq:mass-area-sub}
\end{equation}
where $\rho$ is the level of UE/PU, $A_m$ is the active mass area while $A_\mu$
is the active jet area. The mass area is computed from uncorrected jets and the
first term in Eq.~(\ref{eq:mass-area-sub}) provides the leading contribution to
the jet mass correction.
Fig.~\ref{fig:mass-area} shows how the above procedure fares for a dijet
production at the LHC. We see that, for events with pileup, the mass of a jet
increases with the number of simultaneous collisions. By subtracting the first
term of Eq.~(\ref{eq:mass-area-sub}), one recovers the correct jet mass as long
as the pileup is not too large. By adding the second term from
Eq.~(\ref{eq:mass-area-sub}), one obtains the correct jet mass even with large
pileup.

A somewhat related approach for measurement and subtraction of the incoherent
energy flow (\eg pileup) for massive jets has been proposed
in Ref.~\cite{Alon:2011xb}. It allows one to correct the jet mass, angularity
and planar flow.
The method is data-driven and it estimates the UE/PU activity, which should be
subtracted from the jet, by summing energy in cones of the area $a_0 = \pi R^2$
rotated by 90$^\circ$ in the transverse plane with respect to the leading jet.
The correction to the jet mass is found to have the form 
$\delta m \sim \sum_{i \in R^{90^\circ}} \frac{\delta m_i^2}{2 m_J}$, 
where $i$ runs over particles in
the cone perpendicular to the jet and the proportionality coefficient is found
from a fit. The above functional dependence has been confirmed by fit to
mass distributions measured by CDF.

A new jet shape variable, called the ``angular structure function''  was
proposed in~Ref.~\cite{Jankowiak:2012na}. Its average over an ensemble of jets
is defined as
\begin{equation}
  \langle \Delta {\cal G} (R) \rangle \equiv
  R \frac{\sum_{k=1}^N \sum_{i \neq j} p_{Tk,i} p_{Tk, j} \Delta R_{ij}^2 
          \delta_{dR} (R-\Delta R_{ij})}{\sum_{k=1}^N \sum_{i \neq j} 
	  p_{Tk,i} p_{Tk, j} \Delta R_{ij}^2 \Theta_{dR} (R-\Delta R_{ij})}\,,
  \label{eq:av-struct-func}
\end{equation}
where $k$ runs through jets in the ensemble, $i, j$ label the constituents of the
jet $k$ and $\Delta R^2_{ij} = (\eta_i - \eta_j)^2 + (\phi_i - \phi_j)^2$ is the
distance between two constituents in the transverse plane. $\delta_{dR}$
and $\Theta_{dR}$ are the Gaussian and error functions with the width $dR$. The
quantity defined in Eq.~(\ref{eq:av-struct-func}) is a logarithmic derivative
of the angular correlation function introduced in~ Ref.\cite{Jankowiak:2011qa}
and the latter has an interpretation of the fractional mass contribution from
constituents separated by the angular distance $R$ or less.
 
The angular structure function is formulated in terms of two-particle
correlations and hence provides information complementary to the usual jet
shapes. In particular, it receives contributions from uncorrelated radiation,
hence it is sensitive to UE modelling and can be used to determine the
level of the underlying event $\Lambda_\text{UE}$. As found
in Ref.~\cite{Jankowiak:2012na}, $\Lambda_\text{UE}$ comes out very different
for different Monte Carlo generators.
The quantity
$\langle \Delta {\cal G} (R) \rangle$ can be interpreted as an average scaling
exponent with $\langle \Delta {\cal G} (R) \rangle = 2$ for the leading order
perturbative result. 
Most of the UE model
give quasi-universal form of $\langle \Delta {\cal G} (R) \rangle \simeq 2$ at
small R, which follows from the perturbative $2\to 2$ processes used to build up
the MPI interactions. This feature can be directly compared with the
experimental data.

A method for correcting the jet transverse momenta for the contamination from
PU, called \emph{jet cleansing} was proposed in~Ref.~\cite{Krohn:2013lba}. It
proceeds by rescaling the 4-momentum of each subjet, $p^\tot_\mu$,  by a
factor determined from the constraint
\begin{equation}
 p_T^\text{tot} = \frac{p_T^{C,\text{PU}}}{\gamma_0} + 
                  \frac{p_T^{C,\text{LV}}}{\gamma_1}\,,
  %\label{eq:}
\end{equation}
where $p_T^{C, \text{PU}}$ and $p_T^{C, \text{LV}}$ are the transverse momenta
of the charged particles coming from the leading vertex and pileup, respectively,
while $p_T^\text{tot}$ is the total transverse momentum of a subjet. 
$\gamma_{0}  = p_T^{C,\text{PU}}/p_T^{\text{PU}}$ and
$\gamma_{1}  = p_T^{C,\text{LV}}/p_T^{\text{LV}}$ are the ratios of the charged
particles to all particles for the pileup and for the leading vertex. 
Hence, the method takes the input values for the charged particle transverse
momenta, $p_T^{C,\text{PU}}$ and $p_T^{C,\text{LV}}$, as well as the total
transverse momentum of a subjet, $p_T^\text{tot}$, to guess the ratio of 
$p^\text{LV}_T/p^\text{tot}_T$ and use it to rescale
the measured $p^\tot_\mu$ back to the original 4-momentum of a subjet formed
uniquely from the leading vertex particles, $p_\mu^{\text{LV}}$.
Depending
on the assumptions on $\gamma_0$ and $\gamma_1$, one defines different variants
of the cleansing procedure. The most sophisticated version, called ``Gaussian
cleansing'' requires input from simulations but it is then able to 
rescale the jet four-momenta such that they reproduce the true value at the
level of $\sim 98$\%~\cite{Krohn:2013lba}.

Several methods extending the original jet area/median-based approach have
been proposed in recent years. 
In Ref.~\cite{Soyez:2012hv} the subtraction formula of
Eq.~(\ref{eq:pt-corr-jet-area}) was generalized to any jet shape, $V$, and
to higher orders in the pileup level parameters $\rho$ and $\rho_m$
\begin{equation}
  V_{\jet,\text{sub}} = 
  V_\jet - \rho V_\jet^{(1,0)} - \rho_m V_\jet^{(0,1)}
  + \frac12 \rho^2 V_\jet^{(2,0)} + \frac12 \rho_m^2 V_\jet^{(0,2)} + 
  \rho \rho_m V_\jet^{(1,1)}\,,
  %\label{eq:}
\end{equation}
where $V_\jet^{(m,n)}$ denotes the $m^\text{th}$ derivative with respect to the
ghost transverse momentum, $p_{T,g}$ and $n^\text{th}$ derivative  with respect
to the component $m_{\delta, g} = \sqrt{m_g^2 + p_{T,g}^2}-p_{T,g}$. 
These derivatives can be determined numerically for a specific
jet~\cite{Soyez:2012hv}.
The PU levels are given by
\begin{equation}
 \rho = 
 \text{median} \left\{\frac{p_{T,\text{patch}}}{A_\text{patch}} \right\}\,,
 \qquad \qquad \qquad
 \rho_m = 
 \text{median} \left\{\frac{m_{\delta,\text{patch}}}{A_\text{patch}} \right\}\,,
  \label{eq:media-patches}
\end{equation}
where the patches can be defined in various ways and Ref.~\cite{Soyez:2012hv}
simply used the jets obtained with the~$k_T$~algorithm with $R=0.4$.
Applying the above general procedure results in the subtracted distribution
returning very close to their original shapes with the second derivative playing
a non-negligible role for certain processes like for example dijets. Also the
$\rho_m$ component is shown to be important in some cases, notably for the
filtered jet-mass distributions of fat jests in \ttbar\ events.

The \emph{constituent subtraction} technique~\cite{Berta:2014eza} proceeds by
applying the pileup correction to jet constituents rather then to the final jet
momenta. Each event is populated with ghosts with $p_T^g$ proportional to the
pileup density $\rho$. Then the transverse momenta of particles $i$ and ghosts
$k$ are corrected iteratively following the procedure: 
If $p_{Ti} \geq p_{Tk}^g$: 
$p_{Ti} \to p_{Ti} - p_{Tk}^g$, $p_{Tk}^g = 0$, otherwise
$p_{Ti} \to 0$,  $p_{Tk}^g \to p_{Tk}^g - p_{Ti}$, starting from the pair with
the lowest $R_{ik}$ distance. Similar procedure is used to correct for the
constituent masses.
The method has been applied to dijet production and to $Z'$ processes where it
was shown to result in better mass resolution than the area based method and
the shape expansion method~\cite{Soyez:2012hv}. An example of the performance is
presented in Fig.~\ref{fig:soft-killer}.

\emph{Soft Killer} method~\cite{Cacciari:2014gra} for pileup removal has been
proposed as a follow-up to the jet area based technique. It proceeds by
eliminating particles with the transverse momenta $p_T < p_T^\cut$ , where
$p_T^\cut$ is chosen as the minimal value that ensures that $\rho$ is zero, with
$\rho$ defined as in Eq.~(\ref{eq:media-patches}) only that squared patches of
size $a \times a$ are used instead of jets.
The Soft Killer method exhibits similar, in the case of jet transverse momentum,
or slightly larger, in the case of jet mass, bias than the area jet method,
where the bias is defined as the difference between the value of $p_T$ or mass
returned by the method and the true value.
At this small expense, it brings however a significant, in some case as big as
30\% improvement in the resolution of jet energy, mass and other jet shapes.
Example comparison for the jet mass is shown if Fig.~\ref{fig:soft-killer},
where the peak resulting from the Soft Killer procedure is clearly the most
pronounced.

Performance of the method is a function of the resolution parameter $a$. The
optimal value of this parameter depends on jet radius and was found to be 0.4
for $R=0.4$ and 0.8 for $R=1$.
Finally, Soft Killer  brings nearly two orders of magnitude speed improvement as
compared to the area/median method.

%-----------------------------------------------------------------------------
\subsubsection{Hadronization}
\label{sec:hadronization}

The other major non-perturbative effect relevant for jet production processes
comes from hadronization. Contrary to UE/PU, hadronization corrections appear
not only in the hadron-hadron collisions but also in DIS and $\epem$.

According to the result of Ref.~\cite{Dasgupta:2007wa}, hadronization changes
jet transverse momentum by the average amount
\begin{equation}
  \langle \delta p_T\rangle_\text{hadr} = 
  \frac{2}{\pi} 
  \left[
    -\frac{2 C_R}{R} + \order{R}
  \right]
  \calM\, \calA(\mu_I)\,,
  \label{eq:hadr-cor-pt}
\end{equation}
where $C_R$ is the Casimir factor, equal to $C_F$ for the quark and to $C_A$ for
the gluon jet, while $\calA(\mu_I)$ is an integral over a non-perturbative contribution to the
strong coupling up to some infrared matching scale $\mu_I$. It is calculated as
a difference between the average of the full coupling in the infrared region
 $\alpha_0(\mu_I) \equiv 1/\mu_I \int_0^{\mu_I}\as(k_T) dk_T$,
and the perturbative contribution between the scales $\mu_I$ and $p_T$.
The value of $\alpha_0(\mu_I)$ is obtained from fits to DIS and $\epem$ and it
comes out around 0.5 for $\mu_I = 2\, \GeV$.
The parameter $\calM$ is called the Milan factor and it is an
algorithm-dependent quantity with $\calM \simeq 1.49$ for the anti-$k_T$ and
$\calM \simeq 1.01$ for the $k_T$ algorithm~\cite{Dasgupta:2009tm}.

We see from Eq.~(\ref{eq:hadr-cor-pt}) that the hadronization corrections are
large and negative for small-size jets.
With similar methods, one obtains corrections to the jet mass
\begin{equation}
  \langle \delta m^2 \rangle_\text{hadr} = 
  \frac{4 C_R}{\pi} p_T
  \left[ R + \order{R^3} \right] \calM\,
  \calA(\mu_I)\,,
\end{equation}
whose dependence is not divergent in $R$, hence, the mass of small-$R$ jets is
not as strongly affected by hadronization as their transverse momentum.

\begin{figure}[t]
  \begin{center}
    \begin{minipage}[t]{0.48\textwidth}
    \centering
    \vspace{-185pt}
    \includegraphics[width=0.9\textwidth, angle=-90]{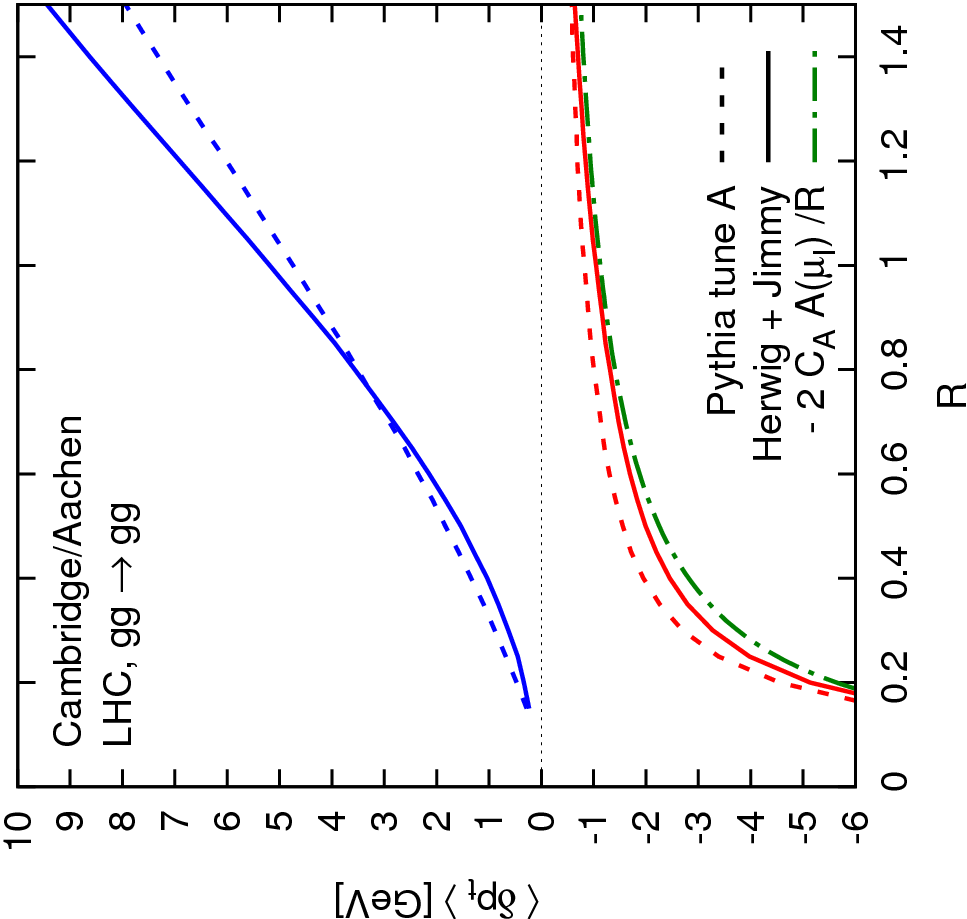}
    \caption{
    Modification of jet transverse momentum due to UE (upper curves) and
    hadronization (lower curves) for $qq \to qq$ scattering at the LHC, 14 TeV.
    Figure from Ref.~\cite{Dasgupta:2007wa}.
    }
    \label{fig:hadr-ue}
    \end{minipage}
    \hfill
    \begin{minipage}[t]{0.48\textwidth}
    \centering
      \includegraphics[width=0.92\textwidth]{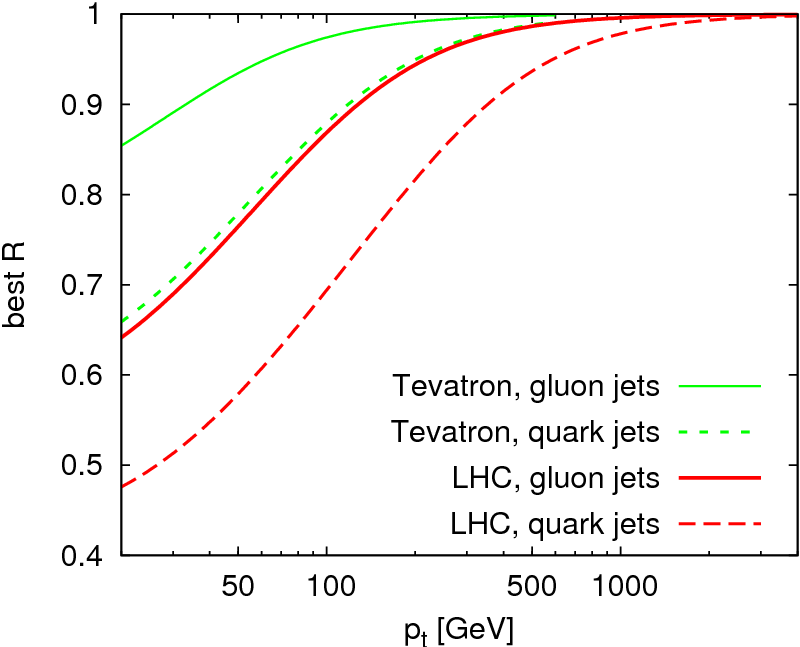}
    \caption{
    The value of jet radius, $R$, that minimises the sum of average, squared
    perturbative, hadronization and UE contributions as a function of jet
    transverse momentum. The level of UE per unit rapidity was taken at 
    $\rho =  4\, \GeV$ for Tevatron and $\rho =  10\, \GeV$ for LHC.
    Figure from Ref.~\cite{Dasgupta:2007wa}.
    }
    \label{fig:optimal-R}
   \end{minipage}
  \end{center}
\end{figure}

A summary of the nonperturbative effects coming from hadronization and 
underlying event is presented in Fig.~\ref{fig:hadr-ue}.
The upper curve corresponds to UE and we see that the change of jet's momentum
is positive and it becomes larger for large jets, which is qualitatively
consistent with Eq.~(\ref{eq:ue-analytic}).
On the contrary, the hadronization corrections are negative and their absolute
value grows for small jets. Hence, the two have a chance to balance each other
for moderate values of the jet radius $R\simeq 0.6$.
In Fig.~\ref{fig:hadr-ue}, the effects of hadronization are  compared between
the analytic calculations, taking just the leading order in $R$, as given in
Eq.~(\ref{eq:hadr-cor-pt}), and Monte Carlo results.
We see that the two are very close to each other.

Let us conclude this section by noting that experiments use their own
procedures to determine non-perturbative corrections from UE and hadronization
(see for example Refs.~\cite{Aad:2013tea, Aad:2014vwa, Aad:2014rma}).
The common practise is to rely on LO MC generators with LL showers.
The corrections are calculated as bin-by-bin ratios of
the MC cross sections obtained with and without modeling of hadronization and
underlying event. 
They are determined using several generators and UE tunes. Then, an
envelope of all correction factors is taken as a systematic hadronization+UE
uncertainty.
The final factors are subsequently used to multiply the parton level,
perturbative predictions for jet observables.

The non-perturbative corrections determined that way~\cite{Aad:2014vwa} reach up
to several percent and play an important role at low $p_T$ while significantly
decreasing at large $p_T$. They depend on the jet radius and are mostly negative
with the choice $R=0.4$ and positive with $R=0.6$.
This is exactly consistent with the picture of the interplay between the UE and
hadronization, and their respective dependence on jet radius, which we discussed
above in the context of Fig.~\ref{fig:hadr-ue}.

%-----------------------------------------------------------------------------
\subsection{Choice of jet radius}
\label{sec:jet-radius}

Jet radius is a parameter of jet definition, \cf Section~\ref{sec:jet-def}. 
As discussed in the preceding subsection, the corrections to jet transverse
momentum coming from UE/PU and those coming from
hadronization behave differently with $R$. 
In addition to the above, there is also a purely perturbative
mechanism that leads to further modification of jet's $p_T$.
 
For any jet of size $R$, there will be a fraction of partons that are emitted
outside of the jet area.
The approximate perturbative formula for the corresponding transverse momentum loss reads~\cite{Salam:2009jx}
\begin{equation}
  \frac{\langle p_{T,\jet} - p_{T,\text{parton}}\rangle}{p_{t,\text{parton}}}
  \simeq a\, \as \ln R\,,
  \label{eq:jet-pert-corr}
\end{equation}
where the prefactor $a=0.43$, if the parton that originated the jet  was a
quark, and $a=1.02$, if that parton was a gluon.
We notice that when the jet radius is small, the transverse momentum (or energy)
of the jet deviates significantly from that of the original parton. 

By comparing Eqs.~(\ref{eq:ue-analytic}), (\ref{eq:hadr-cor-pt}) and
(\ref{eq:jet-pert-corr}), we see that the choice of a jet
radius is a trade-off between different corrections, as the hadronization and leakage of perturbative radiation
point towards larger jets, while the underlying event and pileup prefer smaller
sizes.
There are further reasons to be considered while choosing the jet radius.
For example, if $R$ is large, than multiple, rather than a single
hard structures end up inside a jet. This is sometimes useful, especially in the
context of jet substructure analyses, \cf Section~\ref{sec:substructure},
however, in standard measurements, one is typically interested in resolving each
collimated energy flow into a single jet. 
 
The values of jet radius used at the LHC range from $R=0.4-0.7$ for the
proton-proton collisions~\cite{Aad:2011fc, CMS:2011ab} and are usually smaller,
around $R=0.2$, for the heavy-ion collisions~\cite{Aad:2014wha}. The latter is
motivated by very busy backgrounds.
 
Theoretical studies support the above choices adopted by the LHC
experiments.
Using the analytic or simulated results for the UE/PU contamination, hadronization effects and
perturbative leakage, one can define the optimal radius as, for example, a value
that minimizes the sum of the squared averages: 
$\langle \delta p_T\rangle^2_\text{pert} + \langle \delta
p_T\rangle^2_\text{hadr} +\langle \delta p_T\rangle^2_\text{UE}$.
The results, based on the analytic study performed in
Ref.~\cite{Dasgupta:2007wa}, as a function of jet $p_T$, are shown in
Fig.~\ref{fig:optimal-R}.
We see that the sizes which minimize the above sum depend on whether the jet
originated from a quark or a gluon, with the latter preferring substantially
larger jets.
The average values are however in the range around $R=0.6$ for 100 GeV jets.

A more realistic Monte Carlo studies,
with the so called \emph{quality measures}~\cite{Cacciari:2008gd}, confirm the above
pattern to a first approximation. 
Further details depend on the algorithm, type of the process, the parton
initiating the jet, and the jet transverse momentum. Therefore, ideally, the jet
radius should be chosen on the case-by-case basis.

Because the effects related to the choice of $R$ are non-trivial, they have 
been further studied both theoretically and experimentally.
In Ref.~\cite{Soyez:2011np} the ratios of the $p_T$ distributions in the inclusive jet
production with two different values of jet radius, 
$\calR(p_T; R_1, R_2) = \sigma^\NLO(pt, R_1)/\sigma^\NLO(pt, R_2)$,
have been computed in perturbative QCD
and supplemented with non-perturbative corrections from hadronization following
the prescription of Ref.~\cite{Dasgupta:2007wa}.
Note that the leading order contribution to the inclusive jet spectra does not
depend on $R$,  as it involves only two partons recoiling against each other.
For the same reason, the two-loop correction is also $R$-independent. 
That implies that the two-loop contribution, which is need for the NNLO jet
cross sections, does not appear in the ratio $\calR(p_T; R_1, R_2)$ and one is
able to obtain that quantity at the complete order $\as^2$ using just the NLO
distributions.
As found in Ref.~\cite{Soyez:2011np}, 
both the perturbative effects and the non-perturbative corrections are 
sizable at the LHC energies and the ratios $\calR(p_T;0.4,0.6)$ vary from 0.7 to 0.9 at low and high transverse momenta, respectively.

The measurement of the ratios of the inclusive jet cross sections with two
different jet radius values, $R=0.5$ and $R=0.7$, has been performed by
CMS~\cite{Chatrchyan:2014gia} and compared to the result of
Ref.~\cite{Soyez:2011np}.
It was found that the higher order and the non-perturbative correction are very
important in obtaining the right shapes of $\calR(p_T;0.5,0.7)$  as a function
of the transverse momentum in different rapidity bins. 
However, the NLO$\times$ non-perturbative corrections are still not entirely
sufficient to exactly describe the measurements and better agreement is
obtained with the matched calculations from \powheg+~\pythia.
This points to the relevance of higher multiplicities and resummation in
precise description of jet ratios.

All order resummation of small-$R$ effects have been recently performed at the
leading logarithmic accuracy, hence including the terms 
$(\as\ln R^2)^n$~\cite{Dasgupta:2014yra}.
The method exploits the angular ordering of the successive emissions in the
range between the jet size $R$ and 1.
Similar job is also done in practice by the standard MC parton showers, however,
there, the small-$R$ corrections are entangled with other physical effects.
The analytic approach of Ref.~\cite{Dasgupta:2014yra} allows for a
robust estimate of the size of the small-$R$ corrections. Such result can be
then combined with other calculations \eg jet veto resummation, \cf
Section~\ref{sec:jet-veto}.
The effects of the leading-logarithmic, all-order, small-radius resummation
turn out to be sizable.  For example, they can reduce the inclusive jet spectrum
by 30-50\% for jets defined with $R$ in the range 0.4-0.2, respectively.
The expansion in $(\as\ln R^2)^n$ usually converges well for moderate-size jets
but the all-order resummation is necessary for micro-jets of size $ R \lesssim
0.1$.  Such small jets are relevant in the context of the substructure studies,
\cf Section~\ref{sec:substructure}.

%-----------------------------------------------------------------------------
\subsection{Perturbative calculations}

We now turn to the perturbative calculations for jet processes. Those are
crucial both for tests of QCD and for searches of new phenomena. 
We shall start with 
the cleanest possible approach in which the order of the strong coupling is
fully controlled and fixed. Then, we will turn to
frameworks where those fixed-order results are merged with each other or
matched to a parton shower.
As we elaborate on the stat-of-the-art theoretical results, we shall also
discuss comparisons to selected distributions measured by the LHC experiments.

%-----------------------------------------------------------------------------
\subsubsection{Fixed order calculations}
\label{sec:fo-calculations}

All jet processes of interest have been calculated for hadron colliders up to
NLO in QCD. Several of them have been recently pushed to the NNLO accuracy,
although, in some cases only dominant contributions were included.
Hence, significant work is still needed to make a complete transition from the
current state-of-the-art of NLO to the desired NNLO accuracy.
Below, we list and discuss at some length the fixed-order perturbative results
for processes with jet production.
Because the NLO calculations for many processes  have been repeated several
times with different methods and tools, it is impossible to cite them all here.
Instead, we provide references to the first articles where the corresponding
corrections were calculated, and, at the end of the section, we also give a
list of modern tools which are used for phenomenological studies of jet
processes at NLO. The complete, fixed-order literature can be gathered from
there.

%------------------------------------
\paragraph{Pure jets}
We start from the pure QCD jets, where the single inclusive results at NLO were
obtained in Refs.~\cite{Aversa:1989xw, Ellis:1990ek}.
The most precise, complete theoretical calculations for dijets observables are
currently also known at the NLO accuracy~\cite{Ellis:1992en,
Giele:1994gf}. Similarly, the three-~\cite{Nagy:2001fj, Nagy:2003tz}, 
four-~\cite{Bern:2011ep} and five-jet~\cite{Badger:2013yda} productions were
calculated at NLO.

\begin{figure}[t]
  \centering
  \includegraphics[width=0.8\textwidth]{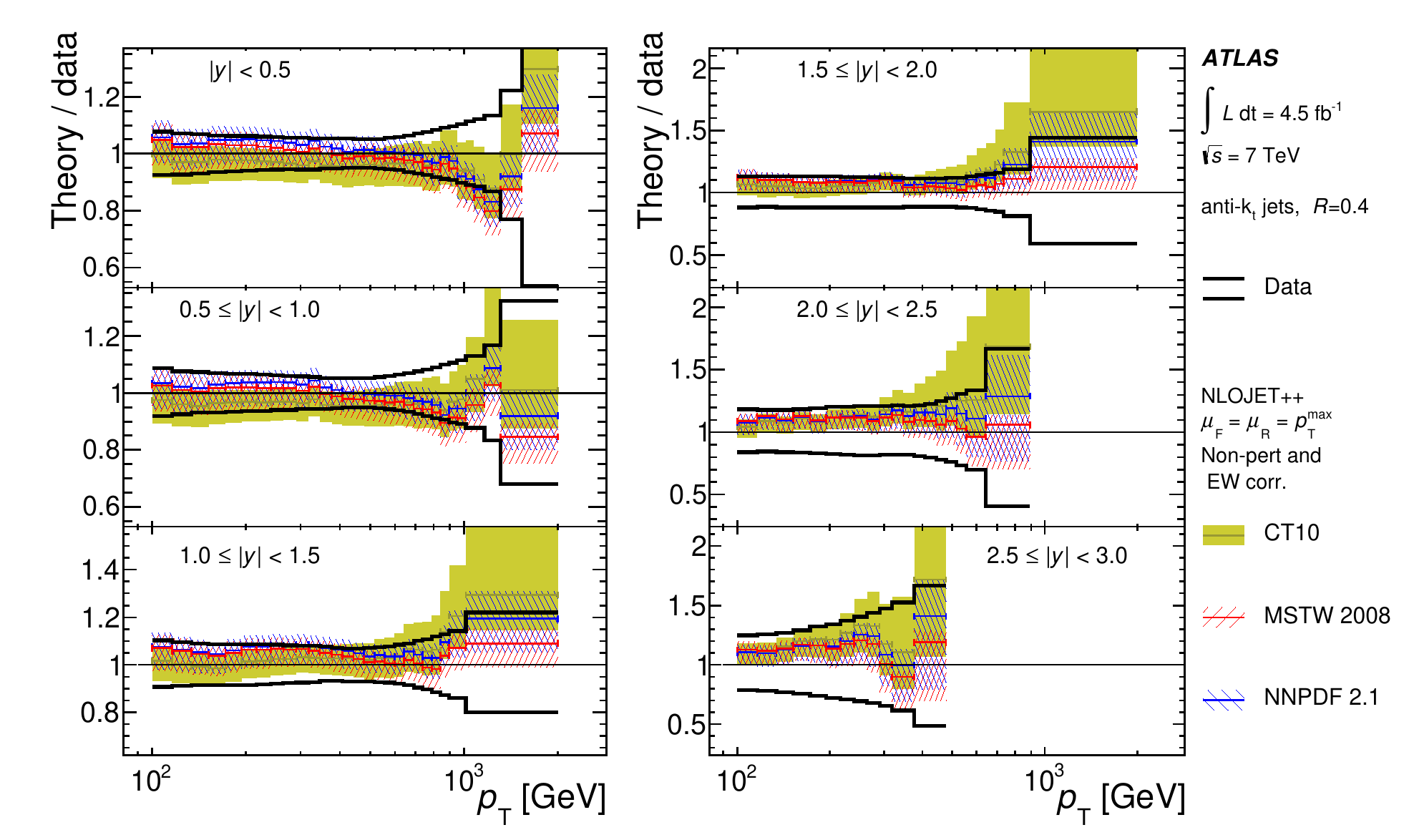}
  \caption{
  Ratio of NLO pQCD predictions to the measured inclusive
  jet transverse momentum distributions shown  in bins of the jet
  rapidity, for anti-$k_T$ jets with R=0.4. The predictions were calculated
  using \nlojet~\cite{Nagy:2003tz} with different NLO PDF sets. Theory
  uncertainties include scale variations and PDF uncertainties and are given by
  the bands.  The data lines show the total experimental uncertainty except the
  1.8\% uncertainty from the luminosity measurement. 
  Figure from Ref.~\cite{Aad:2014vwa}.
  }
  \label{fig:atlas-singl-incl}
\end{figure}

Single and multi-jet processes have been extensively studied at Run I of the
LHC.
In Fig.~\ref{fig:atlas-singl-incl}, we show the single inclusive jet spectra
measured by ATLAS~\cite{Aad:2014vwa} presented as ratios to the NLO calculations
from \nlojet~\cite{Nagy:2003tz} (see Ref.~\cite{Aad:2014vwa} for the
corresponding cross sections). Similar measurement has been performed by
CMS~\cite{Chatrchyan:2012bja}.
Overall good agreement, within, in some cases sizable, theoretical and
experimental uncertainties is observed. 
Disagreement in the tails of distributions shows the potential for using jest to
improve PDF fits.
The dominant systematic experimental uncertainty comes from jet calibration.
PDFs bring typically 10-20\% uncertainty, much larger at large transverse
momenta and masses.

The measurements of dijets cross sections and distributions
have also been performed by ATLAS~\cite{Aad:2014vwa} and CMS~\cite{Chatrchyan:2012bja}. 
Dijet invariant mass spectra have been compared to the fixed order NLO
calculations and the level of agreement is comparable to that found
in the single-inclusive distributions.
The conclusion from the studies of three-jet mass distributions at 7 TeV
~\cite{Aad:2014rma} is similar.
Finally, distributions for the inclusive production of four jets have also been
measured at the LHC at 7~\cite{Aad:2011tqa} and 8~TeV~\cite{Aad:2015nda}. 
Here, the LO $2\to$ ME+PS calculations describe the data poorly, while the
fixed-order NLO results provide good description for virtually all
distributions, bearing in mind that the experimental errors reach 30\% at low
momenta. 

\begin{figure}[t]
    \centering
    \includegraphics[width=0.60\textwidth]{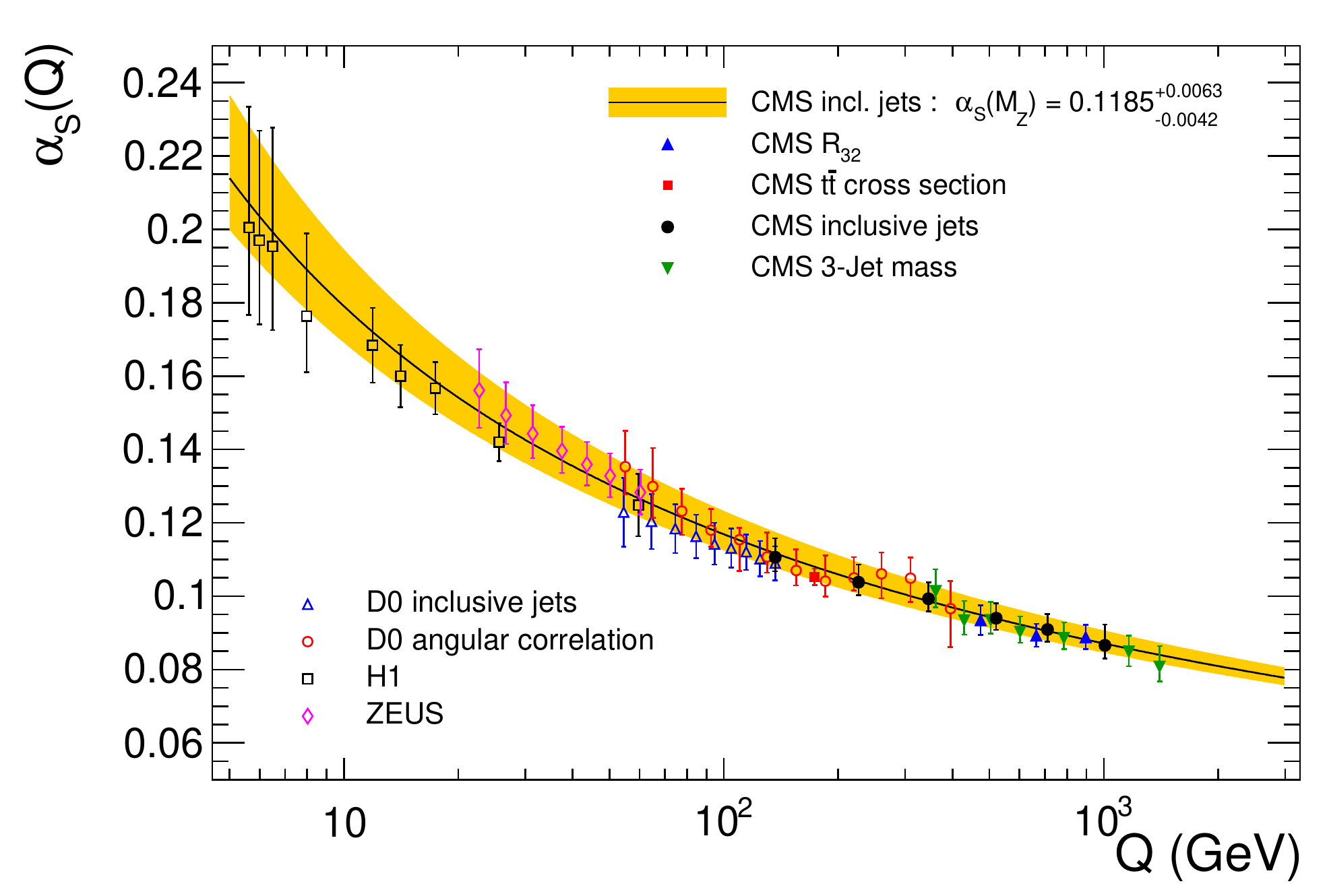}
    \caption{
    CMS measurements of the strong coupling in processes with jets at 7 TeV.
    Three different observables have been used: ratios of inclusive 3- to 2-jet
    distributions~\cite{Chatrchyan:2011wn}, single inclusive jet
    spectra~\cite{Khachatryan:2014waa} and 3-jet mass
    distributions~\cite{CMS:2014mna}. Figure from Ref.~\cite{Khachatryan:2014waa}.
    }
    \label{fig:cms-as}
\end{figure}

CMS experiment has used jet processes to measure the strong coupling in several 
different ways.
The first approach was based on studying the ratios $R_{32}$ of the inclusive
3-jet to the inclusive 2-jet cross sections at~7~TeV~\cite{Chatrchyan:2011wn}.
The ratio $R_{32}$ is proportional to $\as(Q)$, where $Q$ was taken as
the average transverse momentum of two hardest jets.
The measured ratio is compared to NLO predictions in which the $\as(m_Z)$ value
is varied and different PDF sets are used for calculations. The values of
$\as(m_Z)$ and $\as(Q)$, in various $Q$ bins, are determined by minimizing
$\chi^2$. Differences between PDFs contribute to the experimental uncertainty.
Another approach used the inclusive, 3-jet differential cross sections
at~7~TeV~\cite{CMS:2014mna}. With this approach, the strong coupling has been
determined for scales between 0.4 and 1.4 TeV by minimizing $\chi^2$ and a
combined fit for points above the 3-jet mass of $\sim 0.6$ TeV has been
performed to determine $\as(m_Z)$.
Finally, inclusive jet measurements at 7 TeV were also used to measure $\as$
by employing a very similar methodology~\cite{Khachatryan:2014waa}.

All of the above results of the strong coupling measurements are consistent with
the renormalization group evolution and the value $\as(m_Z)$ agrees with the
world average.
The CMS measurements of $\as$ are collected in Fig.~\ref{fig:cms-as}, together
with earlier measurements from Tevatron and HERA. As we see, jet-based
determinations of the strong coupling provide direct results for $\as(Q)$ at
scales from $\sim 0.1-2$~TeV.

\begin{figure}[t]
\centering
\begin{minipage}[b]{0.48\linewidth}
  \centering
  \raisebox{0.1\height}{
    \includegraphics[width=1.0\textwidth]{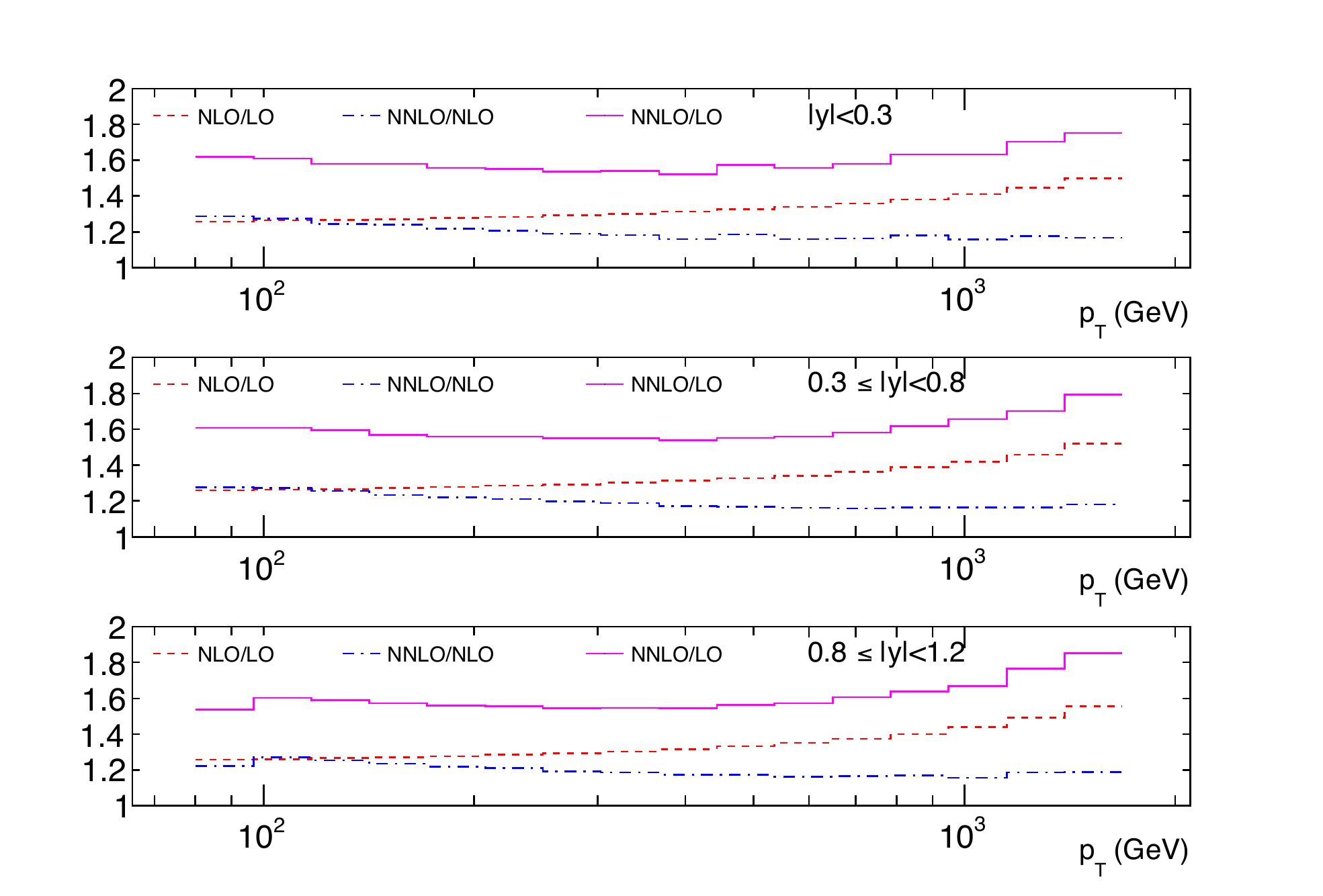}
  }
  \caption{
  Ratios of the NNLO, NLO and LO cross sections for inclusive jet
  production, $d^2\sigma/dp_T$, in three rapidity slices.  The results
  correspond to $\sqrt{s} = 8$~TeV and the anti-$k_T$ jet algorithm with
  $R=0.7$.
  Figure from Ref.~\cite{Currie:2013dwa}.
  }
  \label{fig:nnlo-incl-jets}
\end{minipage}
\hspace{20pt}
\begin{minipage}[b]{0.45\linewidth}
  \centering
  \includegraphics[width=1.0\textwidth]{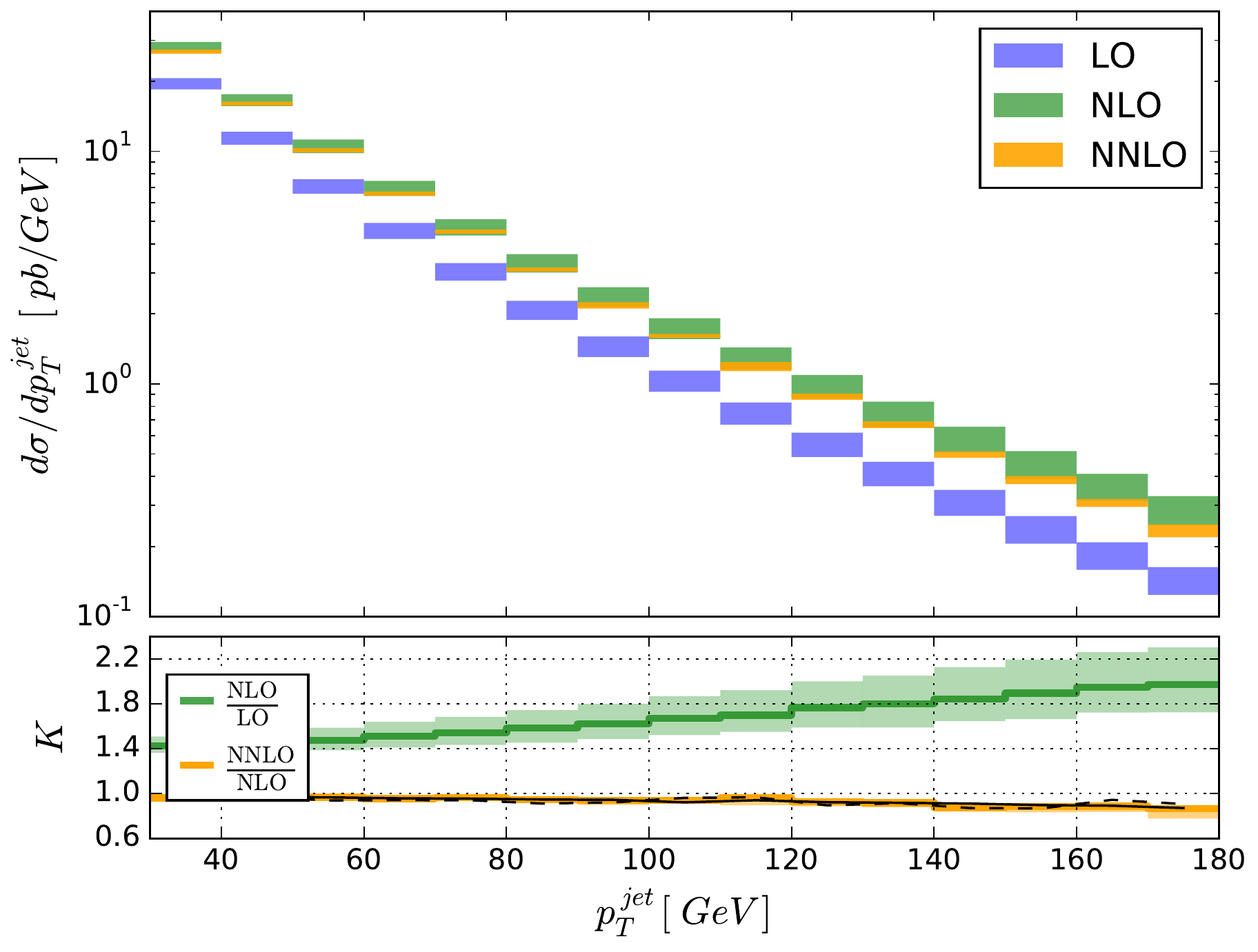}
  \caption{
  Transverse momentum spectra of the leading jet in $W$+jets process at LO, NLO
  and NNLO.
  The results correspond to $\sqrt{s} = 8$~TeV and the
  anti-$k_T$ algorithm with $R=0.5$.
  Figure from Ref.~\cite{Boughezal:2015dva}.
  }
  \label{fig:nnlo-wjets}
\end{minipage}
\end{figure}

Coming back to theoretical predictions, 
the NNLO result for dijet process has been recently obtained for the all-gluon
channel. 
The calculation utilized the \emph{antenna subtraction
method}~\cite{GehrmannDeRidder:2005cm} and was first performed in the
leading colour approximation~\cite{Ridder:2013mf} and later extended to the
subleading colour~\cite{Currie:2013dwa}. 
In the antenna method, the subtraction terms are constructed from the so-called
antenna functions, whose soft and collinear divergencies match those of the
double-real, real-virtual and double-virtual contributions to the NNLO cross
section. By adding the antenna functions and subtracting their integrated forms,
one constructs the finite contributions from the $n+2$, $n+1$ and $n$-particle
phase space, which are amenable to numerical integration.
The study presented in Ref.~\cite{Currie:2013dwa} focuses on the comparisons
of predictions at various orders of the perturbative expansion and between the
leading and the full-colour results.
A representative result is presented in Fig.~\ref{fig:nnlo-incl-jets}, which
shows the jet transverse momentum distributions for the single inclusive jet
production in various rapidity bins.
The NNLO/NLO K-factors turn out to be large, ranging between 16\% and
26\%.
The picture is similar in all rapidity bins.
The subleading colour terms contribute around 10\% to the final result, in
accordance with a naive power counting of colours, $ 1/N^2_c$.

%------------------------------------
\paragraph{Jets in association with electroweak bosons}

Single vector boson production in association with 
one-~\cite{Giele:1993dj}
two-~\cite{Campbell:2002tg}
three-~\cite{Berger:2009zg, KeithEllis:2009bu}
four-~\cite{Berger:2010zx}
and five-~\cite{Bern:2013gka} jets are known at the NLO accuracy in QCD.
Very recently, the first NNLO results for the inclusive $W$+1 jet~\cite{Boughezal:2015dva} and $Z$+1 jet~\cite{Ridder:2015dxa} productions have
been obtained.
 
The calculation for the $W$+ jets process has been performed with a new
subtraction scheme based on the $N$-jettiness~\cite{Stewart:2010tn} variable
$\tau_N$, which turns out to completely capture the singularity structure of QCD
amplitudes with final state parton.
This variable can be used as a resolution parameter to partition the phase space
into the $N+1$-jet region, where the NLO result for $N+1$ production is used,
and the $N$-jet region, where the missing NNLO contribution is obtained from
the all-order, small-$\tau_N$ factorization formula derived within SCET~\cite{Stewart:2009yx}.  

As shown in Fig.~\ref{fig:nnlo-wjets}, the NNLO corrections to the $W$+jet
process turn out to be small and almost flat as a function of the leading jet
transverse momentum.  The scale uncertainty for that distribution is reduced
from 20\% at NLO to a few percent at NNLO.
The total cross section decreases from NLO to NNLO by approximately 3\%.

\begin{figure}[t]
  \begin{center}
    \begin{minipage}[t]{0.46\textwidth}
    \centering
    \includegraphics[width=\textwidth]{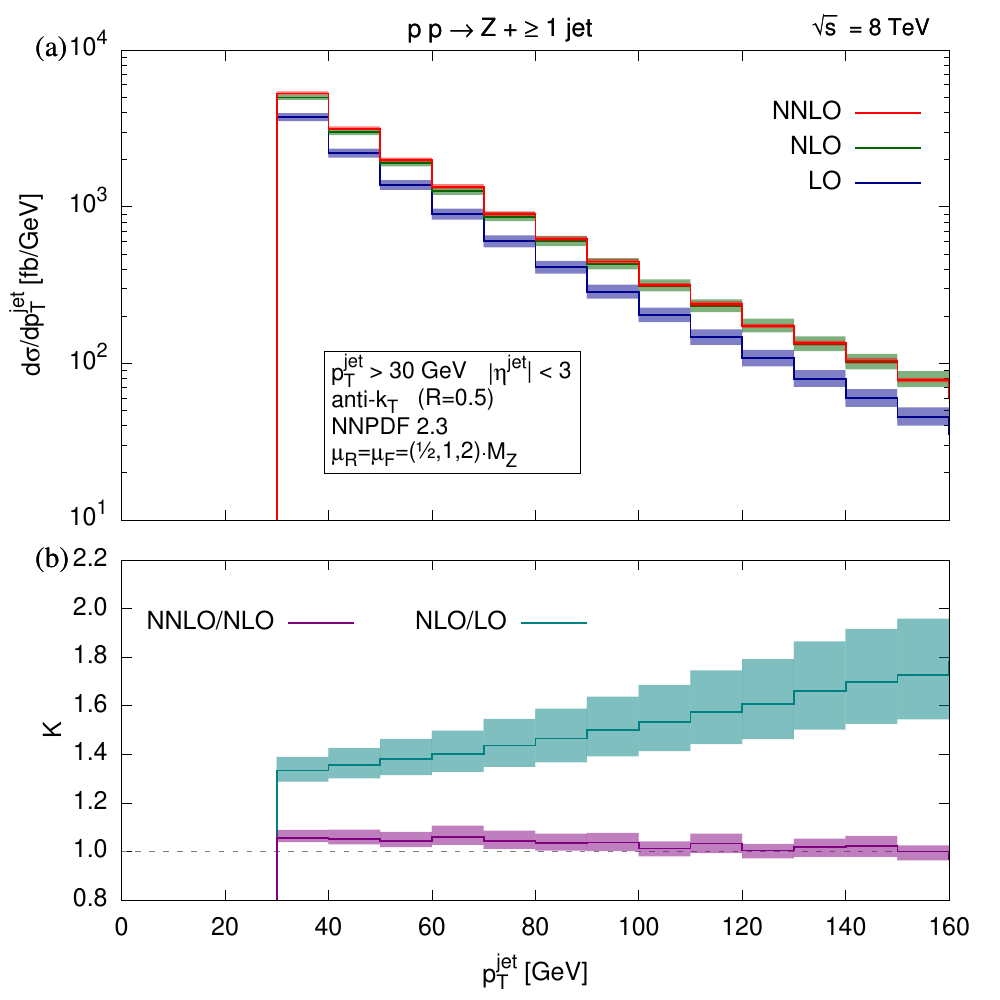}
    \caption{Distributions and corresponding K-factors of the leading jet
    transverse momentum in inclusive $Z$+jet production at $\sqrt{s} = 8\, \TeV$.
    Figure from Ref.~\cite{Ridder:2015dxa}.
    }
    \label{fig:nnlo-zjets}
    \end{minipage}
    \hfill
    \begin{minipage}[t]{0.50\textwidth}
    \centering
    \raisebox{0.3\height}{
      \includegraphics[width=\textwidth]{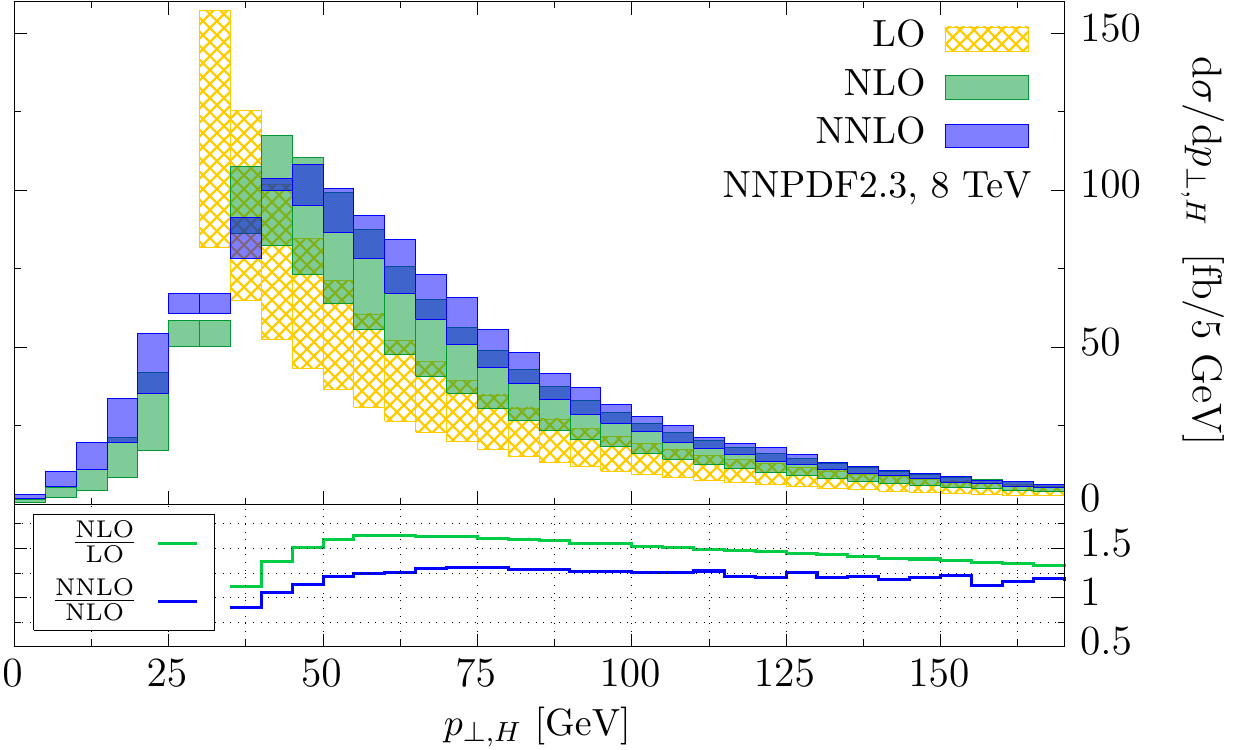}
    }
    \vspace{-45pt}
    \caption{
    Higgs boson transverse momentum distribution in $H$+jet process at 
    $\sqrt{s} = 8\, \TeV$. The bands correspond to renormalization and
    factorization scale variations around the central value taken at $\mu =
    m_H$. Jets are defined with the anti-$k_T$ algorithm with $R=0.5$.
    Figure from Ref.~\cite{Boughezal:2015dra}.
    }
    \label{fig:nnlo-hjets}
   \end{minipage}
  \end{center}
\end{figure}

The $Z$+jets calculation has been performed in the framework of the antenna subtraction
scheme~\cite{GehrmannDeRidder:2005cm} for the dominant $qg$, $\qbar g$, $q\qbar$
and $gg$ channels at leading colour.
The distributions of the leading jet transverse momentum are shown in
Fig.~\ref{fig:nnlo-zjets}, where we see that, similarly to the case of $W$+jets,
 the NNLO correction is small, within 6\% with respect to NLO, and the
scale uncertainty is greatly reduced. The inclusive cross section raises by a
similar amount.

Fully differential results are also available for the production of jets in
association with the Higgs boson.
The $H$+jet production is known at NLO~\cite{deFlorian:1999zd} and
NNLO~\cite{Boughezal:2013uia, Chen:2014gva, Boughezal:2015aha}.
The latter was calculated independently within three different approaches: the
antenna method, the $N$-jettiness formalism and the \stripper
method~\cite{Czakon:2010td}.
As an example, in Fig.~\ref{fig:nnlo-hjets}, we show the differential
distribution of transverse momentum of the Higgs. We see that the result
stabilizes at NNLO. The central value still moves upwards by $\sim20\%$ with
respect to NLO but it stays inside the NLO uncertainty band. The scale
uncertainty is significantly reduced at NNLO.
 
The $H$+2 jets process has been computed at NLO for the gluon
fusion~\cite{Campbell:2006xx} and for the vector boson
fusion~(VBF)~\cite{Figy:2003nv}.  The latter has been recently upgraded to the
NNLO accuracy~\cite{Cacciari:2015jma} with a new ``projection-to-Born'' approach
and it was found that the NNLO corrections to differential distributions can
reach up to 10-12\% with respect to NLO.
The $H$+3 jets production is known at NLO for both the gluon
fusion~\cite{Cullen:2013saa} and VBF~\cite{Figy:2007kv}.

As for diboson production in association with jets, $V_1 V_2 + n\, \jets$, where
$V_1, V_2 \in \left\{W^\pm, Z, H, \gamma \right\}$, essentially all processes
from this large group are known at NLO in QCD. This includes both the
QCD-initiated production and the vector boson fusion. The results are
differential in leptonic decay products and they account for spin correlations. 
The NLO corrections to those processes are  sizable, typically in the range
10-40\%, and they exhibit non-trivial phase space dependence.
Detailed discussion of theoretical results for this class of processes, as well
as the relevant references, can be found in Ref.~\cite{Campanario:2015vqa}.

It is clear from the above examples that the fixed-order calculations for the
processes with jets are of crucial importance as the corrections found at NLO
and NNLO are in many cases sizable. The current quest for the NNLO accuracy
is well motivated as it is only this order that effectively allows one to
reduce the theory uncertainties, coming from the unknown higher orders, down to
a few percent level.

All of the differential calculations at NLO have been implemented in efficient
numerical programs such as: 
\mcfm~\cite{Campbell:1999ah},
\vbfnlo~\cite{Arnold:2008rz, Baglio:2014uba}, 
\nlojet~\cite{Nagy:2003tz}, 
\sherpa~\cite{Gleisberg:2008ta}, 
\njet~\cite{Badger:2012pg}, 
\blackhat~\cite{Bern:2013gka},
\madgraphnlo~\cite{Alwall:2014hca},
\gosam~\cite{Cullen:2014yla} and
\openloops~\cite{Cascioli:2011va}.
In many cases, the calculations are fully automated.
The complete list of processes possible to study with those tools, and the
relevant references, can be found in the corresponding web pages.

In addition to the QCD higher orders, also the electroweak~(EW) NLO corrections
have been computed for several processes such as dijets~\cite{Dittmaier:2012kx},
$W$+jet~\cite{Denner:2009gj} and $Z$+2 jets~\cite{Denner:2014ina}.
The interplay between the radiative and the loop corrections has been analysed
in Ref.~\cite{Baur:2006sn}.
The calculations for $W$+1 and 2 jets have been recently repeated in a fully
automated setup and supplemented with that for the $W$+3 jets
production~\cite{Kallweit:2014xda}.
These predictions include the orders $\as^{n+1}\aew$ and
$\as^{n}\aew^2$ for the $W^+ + n\, \jets$ production processes.
Notice that separation of the NLO corrections into the QCD and EW type is in
general ambiguous. For example, $\order{\as^2 \aEW^2}$ can arise as a NLO
correction to the $W$+2 jets production obtained by inserting an electroweak
particle into the squared, tree-level diagram with an $s$-channel gluon
emission.
The same order, however, can be obtained by inserting a gluon into an
interference diagram between the tree level amplitudes with the gluon and
the electroweak boson, $s$-channel exchanges, respectively.
%
% -- COMMENT --
% first case:  |qq-g-qqZ|^2
% second case: |qq-g-qqZ|* |qq-z-qqZ|
 
The mixed QCD EW results depend in a nontrivial way on observables as
well as jet multiplicities. 
For example, the NLO QCD+EW corrections to the $W$+1 jet production at the LHC
at 13 TeV~\cite{Kallweit:2014xda} turn out to be negative for the distributions
of $p_{T,W}$ but positive for the $p_{T,\text{leading jet}}$ distributions.
The NLO QCD+EW corrections to the distributions of transverse momenta of the
vector boson as well as the leading and subleading jets come out negative for
$W$+2 and $W$+3 jets, with significantly smaller uncertainties than in the $W$+1
jet case.
As for the dijet production~\cite{Dittmaier:2012kx}, the pure weak loop
corrections, $\order{\as^2\aEW}$,  to the distributions of the leading and
subleading jets are negative, in the range of $-$12-16\% at high $p_T$. They
partially cancel with the LO EW $\order{\as\aEW, \aEW^2}$ contributions but the
degree of cancellation depends on observable and the net effect can still reach
$\sim 10\%$ in tails of distributions.

Processes of jet production in association with electroweak bosons have been
extensively studied at the LHC. The most recent results include
$Z$+jets~\cite{Aad:2013ysa, Khachatryan:2014zya} and
$W$+jets~\cite{Aad:2014qxa, Khachatryan:2014uva}.
We shall discussed them in the next subsection together with the matched and
merged predictions.

%-----------------------------------------------------------------------------
\subsubsection{Matching and merging for multi-jet processes}
\label{sec:matching}

The predictions for multi-jet processes calculated within the fixed-order
perturbative approach, described in the previous section, are invaluable.
However, the complexity of those calculations increases very quickly when moving
to higher orders and higher numbers of jets.
This means that, within fixed-order pQCD, we can effectively model our final
state with just a handful of quarks and gluons. Moreover, often only the
inclusive cross section is available, with no access to the final state
kinematics.
 
For those cases where the differential distributions are available, the
fixed-order approach works reliably only at the high transverse momentum and
fails as $p_T \to 0$. This is because of the large logarithms, $\ln p_T$, which
compensate the small coupling at low $p_T$ and yield $\as\ln p_T \sim 1$. Hence,
in the region of small $p_T$, each order contributes comparably and 
the logarithmic terms should resummed. 
One way to achieve the latter is provided by the \emph{parton shower}~(PS)
approach, which sums the dominant, leading-logarithmic~(LL) contributions, $(\as
\ln p_T)^n$, as well as a subset of sub-leading corrections, to all orders.
The events simulated with parton showers are fully exclusive and contain
abundance of final-state partons. The distributions are however computed in the
collinear (small-$p_T$) approximation, hence they differ from the exact results
at \mbox{high $p_T$.}

The complementary advantages of the NLO calculations and the parton shower can
be used simultaneously in a combined framework that goes under the name of
\emph{NLO+PS matching}.
For processes with $n$ tagged jets, matched result are NLO-accurate for $n$-jet
observables, LO accurate for $n+1$-jet observables and parton shower accurate
for $n+1, n+2,\ldots$-jet observables.
Such results can be further improved by correcting the vertices for the
2$^\text{nd}, 3^\text{rd},\ldots,  n^\text{th}$ emission with the exact NLO
results for higher multiplicities. This procedure, called \emph{NLO merging},
has to be, however, implemented carefully in order to remove all the double
counting between contributions from samples with different multiplicities and to
minimize dependence on merging parameters.

This section is devoted to discussion of the main methods for multi-jet matching
and merging at NLO as well as NNLO.
Examples of comparisons with LHC data are also discussed.

%-----------------------------------------------------------------------------
\paragraph{MC@NLO} was historically the first successful, general method of
NLO+PS matching~\cite{Frixione:2002ik} (for earlier proposals see
Refs.~\cite{Bengtsson:1986hr, Baer:1991caa}). 
The procedure starts by generating an
emission according to the cross section
\begin{equation}
  d\sigma =  
  \Big[
     B(\Phi_B) + \hat V(\Phi_B) + \int K^\MC(\Phi_B, \Phi_1) d\Phi_1 
  \Big] d\Phi_B
   + \Big[
   R(\Phi_B, \Phi_1) - K^\MC(\Phi_B, \Phi_1) 
   \Big] d\Phi_B d\Phi_1\,,
  \label{eq:mcatnlo-1st}
\end{equation}
where $B(\Phi_B)$ represents the Born contribution over the phase space
$\Phi_B$, $\hat V$ is a virtual term (before subtraction of soft and
collinear divergencies), $R(\Phi_B, \Phi_1)$ gives the real correction with
extra emission over the phase space $\Phi_1$, and $K^\MC(\Phi_B, \Phi_1)$ is a
kernel of the shower, which is  a PS-specific approximation to the real term,
hence, converging to $R(\Phi_B, \Phi_1)$ in the collinear limit. 
An event generated according to the cross section~(\ref{eq:mcatnlo-1st}) has
either $n$ or $n+1$ particles in the final state. The first case corresponds to
the Born or virtual configuration and it is represented by the first term in
Eq.~(\ref{eq:mcatnlo-1st}). These events are called $\mathbb{S}$-type. The
second case corresponds to a real, resolvable emission and it is given by the second
term of in Eq.~(\ref{eq:mcatnlo-1st}). Such events are called $\mathbb{H}$-type.
In either case, the event is passed to the parton shower, which generates
subsequent emissions in the collinear approximation.

With the integrated and unintegrated $K^\MC(\Phi_B, \Phi_1)$ functions,
Eq.~(\ref{eq:mcatnlo-1st}) has a structure of the NLO subtraction
formula~\cite{Catani:1996vz}. Hence, the NLO accuracy of MC@NLO is manifest.

Since the definitions of the MC counterterms depend on the shower, those terms
need to be calculated case by case.  Currently, the MC@NLO matching can work
with parton showers from the following generators:
\sherpa~\cite{Gleisberg:2008ta}, \herwigpp~\cite{Herwig7}, which both use
the dipole showers based on the Catani-Seymour subtraction scheme, hence, the NLO
subtraction terms are identical to the kernels of the shower, and
\madgraphnlo~\cite{Alwall:2014hca}, which is capable of performing the matching
with \herwig, \herwigpp, \pythiasix and \pythiaeight showers.

Recently, the MC@NLO method has been also worked out for the Nagy-Soper (NS)
shower~\cite{Nagy:2014mqa} in Ref.~\cite{Czakon:2015cla}. 
The potential advantage of the NS shower is that it includes soft
effects at subleading colour, as well as spin correlations.
The shower is currently implemented in \deductor~\cite{Nagy:2014mqa} 
in the so-called LC+ approximation. The above code has been used together with
\helacnlo~\cite{Bevilacqua:2011xh} to construct the matched samples.
The implementation has been validated by studying the inclusive $\ttbar+\jet$
production and comparing it to other matched results as well as to pure NLO.
For observables insensitive to the shower, the matched results recover fixed
order NLO distributions. For those sensitive to the soft and collinear
radiation, results differ between various showers. 
The differences between NG and LL showers are however similar to the
differences among various LL showers.

Together with \powheg, the \mcatnlo method has become a standard for matched
calculations in QCD. It has been applied to numerous processes with jets,
including the highly non-trivial cases of dijets~\cite{Hoche:2012wh} and W + up
to three jets~\cite{Hoeche:2012ft}. 
\mcatnlo is currently implemented in \madgraphnlo, \sherpa and \herwigpp, and we
point the Reader to those tools for further references and examples of
applications to phenomenology.

%-----------------------------------------------------------------------------
\paragraph{Powheg} matching method~\cite{Nason:2004rx} generates the first
emission according to

\begin{equation}
  d\sigma = 
  \bar B (\Phi_B)
  \left[
  \Delta_S(\mu_0,\mu) + \Delta_S(p_T, \mu) 
  \frac{R^S (\Phi_B, \Phi_1)}{B (\Phi_B)}d\Phi_1 \right] d\Phi_B + 
  R^F (\Phi_B, \Phi_1)\, d\Phi_R\,,
  \label{eq:powheg-1st}
\end{equation}
where
\begin{equation}
  \bar B = B + \hat V + \int R^S d\Phi_1\,, 
  \quad 
  R = R^S + R^F\,,
  \quad 
  \Delta_S(p_T, \mu) = \exp\left[-\int_{p_T}^{\mu} \frac{R^S}{B}d\Phi_1
  \Theta(k_T(\Phi_1)-p_T)\right]\,.
\end{equation}
The real contribution is split into the singular, $R^S$, and finite, $R^F$,
parts in the small $p_T$ limit, and only the former is exponentiated in the
Sudakov form factor $\Delta_S(p_T,\mu)$. The scale $\mu_0$ separates the
non-resolvable and resolvable emissions whereas $\mu$ is the uppers scale of the
shower.

The Powheg formula for the first emission~(\ref{eq:powheg-1st}) is essentially a
\emph{matrix element correction} (MEC)~\cite{Bengtsson:1986hr} with the
replacement $B \to \bar B$ and the splitting of the real part into $R^S$ and
$R^F$. The former achieves full NLO accuracy of the matched
result~\cite{Nason:2012pr} whereas the latter avoids generation of spurious
contributions from exponentiation of finite NLO terms that are unrelated to 
the all-order resummation of the collinear limit, which is effectively achieved
by the Sudakov form factor.
For further details on Powheg and MC@NLO, as well as for discussions of
differences between these two methods, we refer to~\cite{Nason:2012pr, Hoeche:2011fd}.
As in the case of MC@NLO, the first emission in Powheg, generated with
Eq.~(\ref{eq:powheg-1st}), is
passed to a parton shower, which supplements the event with subsequent emissions
in the leading logarithmic approximation.

The Powheg method is available in \powhegbox~\cite{Alioli:2010xd} and
\herwigpp~\cite{Herwig7}.
It has been used to calculate numerous predictions for processes with jets
including dijet production~\cite{Alioli:2010xa}, three-jet
production~\cite{Kardos:2014dua}, $Z$+jet~\cite{Alioli:2010qp},
Z/$W$+2jets~\cite{Re:2012zi, Campbell:2013vha}, $\ttbar$+
jet~\cite{Alioli:2011as} and $H$+1,2 jets~\cite{Campbell:2012am}.

%-----------------------------------------------------------------------------
\paragraph{KrkNLO} approach~\cite{Jadach:2011cr,Jadach:2015mza} was constructed
with the aim to maximally simplify the NLO+PS matching procedure.
As discussed in the original articles~\cite{Frixione:2002ik, Nason:2004rx}, both
MC@NLO and Powheg are designed to work with $\msbar$ PDFs. This is convenient in
the context of the fixed-order calculations, however, it leads to a class of purely
collinear terms in the partonic cross section, which need to be included in the
MC shower generators aiming at achieving the NLO accuracy.
That is problematic as the Monte Carlo produces particles in three, space dimensions
and forcing it to generate partons in the strictly collinear phase space is
not straightforward. 

The KrkNLO method circumvents this problem by departing from the $\msbar$ scheme
into a new factorization scheme, called the Monte Carlo (MC) scheme, discussed
at some length in Section~\ref{sec:factorization-scheme}. In the MC scheme, all
the problematic, collinear contributions are essentially moved to the parton
distribution functions that now become the new, MC PDFs. The latter can be
obtained from the $\msbar$ PDFs using Eqs.~(\ref{eq:PDFs-MC-quark}) and
(\ref{eq:DeltaC}).

The method can be used with any shower whose coverage of the phase space
of real emission is complete. Then, it boils down to correcting the hardest emission via
reweighing with $W_R = R/K$ (notation of Eq.~(\ref{eq:mcatnlo-1st})) and
multiplying the result by the  $(1+W_V)$. 
Here, $W_R$ is a weight related to the real radiation and it corrects the PS
matrix elements, $K$, valid in the collinear approximation, to the exact matrix
elements, $R$.  The virtual weight, $W_V$, is a constant in the KrkNLO approach,
as all the $z$-dependent terms are contained in the MC PDFs.
The KrkNLO technique has been so far implemented for the Drell-Yan
process~\cite{Jadach:2015mza} and the results are in general close, within $\sim
20\%$, to those from Powheg and MC@NLO.
They are also closer to the complete NNLO prediction than the pure NLO
and other matched calculations.

%-----------------------------------------------------------------------------
\paragraph{Geneva} method~\cite{Alioli:2013hqa, Alioli:2015toa} achieves
matching at the NNLO+PS accuracy. Moreover, for some observables, the NNLO
accuracy can be extended by resummation up to the order NNLL' from
SCET~\cite{Becher:2014oda}.
The key feature of this approach is a use $N$-jettiness~\cite{Stewart:2010tn},
$\tau_N$, as a resolution variable which measures the degree to which the final
state is a $N$-jet state.  The limit $\tau_N \to 0$ corresponds to events with
exactly $N$~infinitely narrow jets, while $\tau_N \gg 0$ occurs for the events
with additional hard radiation between $N$ jets. Hence, $N$-jettiness can be
used for phase space slicing, similarly to case of the fixed-order calculations
discussed earlier.

The starting point of the Geneva approach is the inclusive, $N$-jet cross
section given by 
\begin{equation}
  d\sigma_{\geq N} = 
  d\Phi_N \frac{d\sigma}{d\Phi_N} \left(\tau^\text{cut}_N\right) +
  d\Phi_{N+1}\frac{d\sigma}{d\Phi_{N+1}} \left(\tau_{N+1}^\text{cut} \right)
  \theta\left(\tau_N-\tau^\text{cut}_N\right)\,,
  \label{eq:geneva-main}
\end{equation}
with the first term bringing the exclusive $N$-jet cross section at a given
fixed-order accuracy + resummation (currently, NNLO+NNLL' for
DY~\cite{Alioli:2015toa}) while the second terms corresponds to the inclusive
$N+1$-jet cross section at an accuracy available for a given process (NLO+NNLL' for DY).

The fixed-order+resummation formula~(\ref{eq:geneva-main}) matches naturally to
a shower ordered in the $N$-jettiness variable. One simply needs to generate
the radiation via probabilistic shower algorithm and impose the condition $\tau_N <
\tau_N^\text{cut}$ for the first emission from the $N$-jet event and
$\tau_{N+1} < \tau_{N+1}^\text{cut}$ for the first emission from the $N+1$
event. Then, each subsequent emission needs to satisfy 
$\tau_{N+1} < \tau_{N+1}^\text{cut}$.
One has to be slightly more careful when interfacing Geneva to showers ordered
in variables other then $N$-jettiness. In practice, the first emission is done
within Geneva, with the $N$-jettiness-ordered shower, which guarantees that it
has the largest value of the jet resolution scale. Then, an event is passed a
standard parton shower, \eg from \pythiaeight, and a veto technique is used to
ensure that subsequent emissions have lower resolution scales.
As discussed in Ref.~\cite{Alioli:2015toa}, such a procedure preserves the
original accuracy of Eq.~(\ref{eq:geneva-main}).
As mentioned earlier, the Geneva method has been implemented and
validated for the Drell-Yan process at NNLO+NNLL' but the framework is capable
of handling NNLO+PS matching for processes with final states jets.

%-----------------------------------------------------------------------------
\paragraph{LoopSim} method~\cite{Rubin:2010xp} allows for consistent mergings of
NLO samples with different multiplicities.

The approach is based on
the unitarity requirement, which leads to the condition that, at each order in
$\alpha_s$, the singularities of the loop diagrams must be canceled by the
singularities of the integrated real diagrams. This allows one to extract the
singular structure of the missing loop contributions from the real diagrams with
higher multiplicities. 

The main ingredient of the method is a procedure for taking events with $n$
particles in the final state and supplementing them with all ($n-k$)--particle
events (or equivalently all $k$-loop events), where $n$ and $k$ depend on the
specific process and the order that we want to calculate. 
Because of unitarity, the sum of the weights of the full set of the above events
will be zero. However, due to reshuffling of the weights between different bins
and because of acceptance cuts, the contributions to the differential
distributions will be finite.
The LoopSim procedure consists of several steps in which the branching structure
and the underlying hard structure are assigned to the original event, with the
help of jet clustering algorithms with radius $R_\LS$. 
Then, the QCD partons that were not
identified as final state particles are made virtual by recombining them, in all
possible ways, with the emitters.

The jet radius $R_\LS$ is a parameter of the LoopSim method. The smaller the
value of $R_\LS$, the more likely the particles are recombined with the beam.
Reversely, the larger $R_\LS$, the more likely the particles are recombined
together. The value of $R_\LS$ is irrelevant for the collinear (or soft)
radiation.  It affects only the wide-angle (or hard) emissions, where mergings
between particles compete with the mergings with the beam.

The difference between the exact NNLO distribution for an observable $A$  and
the LoopSim approximation seats in the finite constant associated with the
two-loop diagrams. Since, the latter have LO topology, the difference becomes
very small for distributions with significant NLO K-factors and goes as
$\sim \as^2/K^{(A)}_\NNLO$,
where $K^{(A)}_\text{NNLO} = \sigma^{(A)}_{\NNLO}/\sigma^{(A)}_{\LO} >
K^{(A)}_\text{NLO} = \sigma^{(A)}_{\NLO}/\sigma^{(A)}_{\LO} \gg 1$.
Therefore, differential distributions sensitive to new channels and new,
kinematically enhanced configurations, which lead to large NLO K-factors, are 
very close to the full NNLO result. 

The method has been used for a variety of processes at hadron colliders
including DY~\cite{Rubin:2010xp}, $W/Z$+jets~\cite{Maitre:2013wha}, 
dijets~\cite{Rubin:2010xp} and dibosons~\cite{Campanario:2012fk,
Campanario:2013wta, Campanario:2015nha}.
In particular, in Ref.~\cite{Maitre:2013wha}, the predictions for the $W$+jets and
$Z$+jets production have been compared to the 7-TeV LHC data.
There, it was found that the description of the data for the distributions of
the leading jet transverse momentum with the LoopSim-simulated NNLO result is
comparable to that of NLO. However, the LoopSim predictions exhibit significant
reduction of scale uncertainties.
On the other side, the $H_{T}$-type distributions obtained with LoopSim agree
much better with the 7-TeV LHC data than NLO~\cite{Maitre:2013wha}.

%-----------------------------------------------------------------------------
\paragraph{MiNLO} method~\cite{Hamilton:2012np, Hamilton:2012rf} can be regarded
as NLO extension of the CKKW~\cite{Catani:2001cc} procedure.
The latter is an algorithm for merging tree level events with different
multiplicities  and matching them to a parton shower. This provides a unified
framework valid both in the wide-angle, multi-jet region (domain of the Born
cross sections) and in the collinear region (domain of the shower).

The MiNLO method for merging the NLO samples with $N+0, N+1, \ldots, N+n$ jets, 
where $N$ is the minimal number of jets required in the process of interest,
proceeds as follows:
Each event, calculated at the scale $Q$, is re-clustered
with the~$k_T$ algorithm in order to reconstruct the most probable branching
history. If the clustering results in $N$ jets, we call that configuration a
$N$-jet \emph{primary system}. 
The lowest order, tree level, $N$-parton contribution to the primary system is
in general proportional to $\as^{m}$, where $m\geq N$.
For example, $m=N=2$ for dijet production but $m = 2 > N = 0$ for the Higgs
production in gluon fusion.
Hence, the tree event with $N+n$ partons is proportional to $\as^{m+n}$, while
the real and virtual events are proportional to $\as^{m+n+1}$.
 
To each of the first $n$ vertices, $i=1,\ldots, n$, where the lower value
corresponds to earlier time in the branching history of forward evolution, one
assigns a scale $q_i$, equal to the relative transverse momentum at that
branching. 
In the case of real events, the first clustering is associated with the scale
$q_0$.
Hence, the scales are ordered according to: $q_0 < q_1< q_2 < \cdots < q_n$.
If $q_n>Q$, which may happen for example in the $Z$ production in association
with a hard jet, one sets $Q=q_n$.
 
$n$ powers of of the coupling, those corresponding to the unresolved branchings,
are evaluated at scales $q_i$ by reweighting the with the factors
$\as(q_i)/\as(Q)$. Another $N$ powers, those corresponding to the primary
system, are evaluated at the scale $Q$. 
The choice of the scale in the $(m+n+1)^\text{th}$ power of the
coupling is a delicate problem and prescriptions vary from case to case.
In order to prove the formal NLO accuracy of the MiNLO result for production
of a colour singlet, $V$, obtained by merging the $V$+0 and $V$+1 NLO
samples, one needs to set the scale in that power of $\as$ to
the value of the transverse momentum of $V$.
This case corresponds to $N=0$. MiNLO has also been successfully used for
processes with $N>0$ but, in those cases, no formal claim about the accuracy is
made.

After the assignments of scales in all the powers of the strong coupling, the
intermediate lines between the vertices are re-weighted with the Sudakov form
factors. For real events, the external lines that join at the first node are not
multiplied by the Sudakovs.
In addition, one performs a subtraction of part of the NLO contribution already
present in the CKKW Born term. For further details see Refs.~\cite{Hamilton:2012np, Hamilton:2012rf}.
It is also worth to note that the MiNLO procedure does not involve a merging
scale.

MiNLO has been used to study $W$, $Z$ and the Higgs boson productions in
association with 0, 1 and 2 jets~\cite{Hamilton:2012np, Hamilton:2012rf}. 
The results were found to be well behaved in the Sudakov region for a large
class of distributions, contrary to the pure NLO results, which are unstable in
that region. Away from the Sudakov region, MiNLO performs similarly to a regular
NLO.
The results for $W/Z$+2 jets production have been also extensively compared to
data in Ref.~\cite{Campbell:2013vha}, finding a generally good agreement.
Other interesting study focused on the $VH+0$ and $VH+1$-jet
merging~\cite{Luisoni:2013kna}.
As this is an example of the production of a colour-neutral object, the result
yields complete NLO accuracy.

The MiNLO method has been also used in conjunction with Powheg to
achieve NNLO+PS matching for the Higgs production~\cite{Hamilton:2013fea}
and for Drell-Yan~\cite{Karlberg:2014qua}.
We recall that each of these processes, matched with Powheg at the NLO+PS level,
is NLO accurate only for the inclusive observables, while the spectrum of the
associated jet has LO accuracy.
By using MiNLO to merge $V$@NLO+PS with $V$+jet@NLO+PS, one attains the NLO
accuracy for the inclusive quantities simultaneously for the $V$ and $V$+jet
processes, hence both the rapidity and the $p_T$ spectra of the boson are
NLO-accurate.
Such result can be subsequently reweighted in order to upgrade 
the inclusive boson production observables to NNLO, thus achieving the NNLO+PS
matched result. As shown in Refs.~\cite{Hamilton:2013fea,Karlberg:2014qua}, by
doing the above reweighting carefully, the NLO accuracy of the 1-jet observables
is preserved.

%-----------------------------------------------------------------------------
\paragraph{MEPS@NLO} is also a method designed to consistently combine the NLO
samples with different multiplicities and to match them to a parton
shower~\cite{Hoeche:2012yf}.
Hence, it can be regarded as a variant of an NLO extension of the CKKW-L
merging.
In MEPS@NLO, the merging is achieved by combining the MC@NLO samples for various
number of jets with help of a truncated parton shower.
A \sherpa flavour of the MC@NLO method~\cite{Hoeche:2011fd} is used to match
each of the $(n+k)$-parton, NLO matrix elements to the parton shower
individually.  Here, $n$ denotes the number of QCD partons in the process of
interest at the Born level.

A branching with the smallest hardness, characterized by the scale $Q_{n+k}$, in
an event with $n+k$ final state partons, is used to separate events into
different multiplicity classes. 
When $Q_{n+k} > Q_\text{cut}$, the event is called a $(n+k)$-jet event,
otherwise it is labeled as a $(n+k-1)$-jet event.
The parameter $Q_\text{cut}$ is called a merging scale.
The real NLO events with $n+k+1$ partons are reduced to $n+k$-parton events by
clustering.

If $\calO$ is an arbitrary, infrared-safe observable, say a cross section for
the
production of a boson in a certain bin of $p_T$, then its $(n+k)$-jet exclusive
expectation value to the order $\as$ is schematically given
by~\cite{Hoeche:2012yf} 
\begin{eqnarray}
  \langle \calO \rangle_{n+k}^\text{excl} 
  & \!\!\!\!\!=\!\!\!\!\! &
  \int \!\!
  d\Phi_{n+k} \Theta(Q_{n+k}-Q_\text{cut}) \tilde B
  \left[
  \Delta^{(A)}(t_c) \calO_{n+k}
  + \int \!\! d\Phi_1 \frac{D^{(A)}}{B} \Delta^{(A)}(t_{n+k+1}) \calO_{n+k+1}
  \Theta(Q_\text{cut}-Q_{n+k+1})
  \right]
  \nonumber \\
  & &
  + \int \!\! d\Phi_{n+k+1}
  \Theta(Q_{n+k}-Q_\text{cut})
  \Theta(Q_\text{cut}-Q_{n+k+1})
  H^{(A)}
  \Delta^{(\text{PS})}(Q_\text{cut}) \calO_{n+k}\,,
  \label{eq:mepsatnlo}
\end{eqnarray}
where $\Delta^{(A)}(t)$ and $\Delta^{(\text{PS})}(t)$ are the Sudakov factors,
hence the probabilities of no-emissions between the scale $t$ and the upper
scale $\mu_Q$, with $t_c$ being an infrared cutoff.
The two Sudakov functions differ by the kernel, which is provided by the
subtraction terms $D^{(A)}$, or by the parton shower, respectively.
The Sudakov factor of the second line assures no shower emissions in
the $(n+k)$-jet region.
The $\tilde B$ and $H^{(A)}$ functions are the analogues of the $\mathbb{S}$ and
$\mathbb{H}$ terms of MC@NLO defined in the context of
Eq.~(\ref{eq:mcatnlo-1st}).
The $\Theta$ functions guarantee that only the $(n+k)$-jet events contribute to
the observable $\calO$.

By combining the results obtained with Eq.~(\ref{eq:mepsatnlo}) for a range of
multiplicities between $n$ and $n+k$, we achieve the NLO-merged result matched
to a parton shower.
As demonstrated in Ref.~\cite{Hoeche:2012yf}, the method preserves both the NLO
accuracy of the fixed-order, $n,n+1,\ldots, n+k$-multiplicity results and the
logarithmic accuracy of the shower

The MEPS@NLO method has been used to study $W$+jets~\cite{Hoeche:2012yf},
four-lepton + 0 and 1 jets~\cite{Cascioli:2013gfa},
$H$+jets~\cite{Hoeche:2014lxa} with a special focus on the analysis of
uncertainties, as well as $\ttbar$ production with up to two
jets~\cite{Hoeche:2014qda} and to processes with multiple weak bosons
~\cite{Hoeche:2014rya}, the latter being relevant as a background to the
associated Higgs production.
 
In the case of $W$+jets, significant decrease of the scale uncertainties is
observed and an overall excellent agreement with ATLAS data is found.
Similarly, \mepsatnlo applied to $\ttbar$ + jets helps to reduce the
uncertainties down to the level of 20\%, compared to 50\% observed in
tree-level-merged distributions.
Also, the four-lepton + 0 and 1 jets predictions profit from \mepsatnlo merging
as compared to pure NLO or \mcatnlo, both in terms of reduction in the scale
uncertainties (in some cases even down to 5\%), as well in improved tails of
distributions and in the inclusive cross sections.
On the contrary, for $H$+jets, no significant improvement in terms of the scale
uncertainties in observed. This is attributed to  the $\as^2$-dependence of the
Born contribution as well as to the intrinsic parton-shower uncertainties and
could probably only be improved by enhancing the accuracy of the shower.

%-----------------------------------------------------------------------------
\paragraph{UNLOPS and UN$^2$LOOPS} methods~\cite{Lonnblad:2012ix,Platzer:2012bs}
not only upgrades the CKKW-L-type merging to NLO but they also restore unitarity
of the merging procedure, which is not exactly satisfied in the original CKKW-L
prescription.
 
When examined carefully~\cite{Lonnblad:2012ng}, the CKKW-L method turns out not
to preserve the inclusive cross section. This comes from a mismatch between the
exact, tree level, $n$-jet matrix element and its parton shower approximation,
with the latter entering through the Sudakov functions.
A parton shower (in backward evolution) is unitary because the contribution
corresponding to a jet being emitted at the scale $\rho$, integrated between
$\rho_0$ and $\rho_\text{max}$, is cancelled with the contributions for no jets
being emitted between the scales $\rho$ and $\rho_\text{max}$. This is however
not true in a CKKW-L-merged result since, there, the higher-multiplicity samples
are only added, while they should also be subtracted, in an integrated form,
from the lower multiplicity states. 

One of the consequences of unitarity violation is a residual dependence on the
merging parameter through sub-leading logarithms of the merging scale cut.  The
UMEPS procedure~\cite{Lonnblad:2012ng} restores unitarity of the CKKW-L merging
by supplementing it with the above, missing subtraction terms.
This is done by using the exact, higher multiplicity states (rather than their
parton shower approximation) to calculate the integrated contributions relevant
for the resummation in the lower multiplicity samples. Then, the parton shower
domain is pushed to multiplicities higher than what is provided by the exact
tree level samples.
And indeed, the above procedure allows for use of lower merging scales, which in
turn improves description of certain observables at small $p_T$.
 
The UMEPS method was updated to the NLO merging in UNLOPS~\cite{Lonnblad:2012ix}
by providing additional set of subtraction terms, related to the exact NLO
contributions from higher-multiplicity samples.  Around the same time, a similar
approach was proposed in~Ref.~\cite{Platzer:2012bs}.
The UMEPS/UNLOPS methods have tight relation to LoopSim~\cite{Rubin:2010xp},
with the integrated contributions in the former being analogous to the simulated
loop contributions in the latter.
 
The results of merging of the $W/H$+ 0, 1 jet samples with UNLOPS  show very
small dependence on the merging scale~\cite{Lonnblad:2012ix} The ATLAS data for
jet multiplicities varying from 0 to 4 are very well described by the UNLOPS
predictions. 
A general tendency of producing the leading jet $p_T$ spectra with tails 
harder than what is seen in data is found.

The above method has been recently extended to construct a NNLO+PS matching
procedure.  The technique, dubbed UN$^2$LOPS, was used for
Drell-Yan~\cite{Hoeche:2014aia} and Higgs production in gluon
fusion~\cite{Hoche:2014dla}.
The matching is performed with help of the $q_T$-subtraction
method~\cite{Catani:2007vq}, where the phase space is sliced according to the
value of the transverse momentum of the electroweak boson, $q_T$. All
configurations with $p_{T,V} < q_T$ correspond to the zero bin and the NNLO
cross section in that bin is most readily obtained within
SCET~\cite{Becher:2014oda}. Contributions from configurations above $q_T$
correspond to simple application of the NLO+PS merging procedure for the
$Z/H$+1 jet process.
After adding the above two contributions and subtracting the $\order{\as}$
emission from the shower (to avoid double counting), one achieves the NNLO+PS
matched result.
The latter is found to be generally consistent with NNLO and superior over
NLO+PS for a range of observables.

%-----------------------------------------------------------------------------
\paragraph{FxFx} procedure~\cite{Frederix:2012ps} was designed to merge MC@NLO
results with different jet multiplicities.

As illustrated in Eq.~(\ref{eq:mcatnlo-1st}), a given tree level matrix element
$M_n$ contributes both to $n$-parton samples (via $\mathbb{S}$-type events) and
to $n-1$ parton samples (via $\mathbb{H}$-type events). Hence, naive addition
of the MC@NLO results with different multiplicities would lead to a clear double
counting.
The above problem can be solved by carefully vetoing emissions above or below a
merging scale $Q_\text{MS}$. For example, in order to merge the 0- and
1-particle samples, one modifies the $\mathbb{S}$ and $\mathbb{H}$ contributions
from the original MC@NLO prescription, Eq.~(\ref{eq:mcatnlo-1st}),  according to
\begin{subequations}
  \label{eq:fxfx-main}
  \begin{align}
  \label{eq:fxfx0}
  d\sigma_{\Sbb,0} = B_{0} + \hat V_0 + B_0\, K^\MC \Theta(Q_\text{MS}-d_1)\,,
  & \qquad \quad
  d\sigma_{\Hbb,0} = \left[B_{1} -  B_0\, K^\MC \right]
  \Theta(Q_\text{MS}-d_1)\,,
  \\
  \label{eq:fxfx1}
  d\sigma_{\Sbb,1} = 
  \left[B_{1} + \hat V_1 + B_1\, K^\MC \right] \Theta(d_1-Q_\text{MS})\,,
  & \qquad \quad
  d\sigma_{\Hbb,1} = \left[B_{2} -  B_1\, K^\MC \right]
  \Theta(d_1-Q_\text{MS})\,.
  \end{align}
\end{subequations}
The above formulae are of course only schematic and the notation is similar to
that of Eq.~(\ref{eq:mcatnlo-1st}), with $B_i$ corresponding to the $i$-parton,
tree level contribution and $\hat V_i$ denoting the related (subtracted) virtual
contribution. The $\Theta$ functions guarantee that the 
$(i+1)^\text{th}$, real emission, with the
hardness~$d_1$, does not, for $\sigma_{\Sbb,0}$ and $\sigma_{\Hbb,0}$, or does,
for $\sigma_{\Sbb,1}$ and $\sigma_{\Hbb,1}$, contribute to the cross section. 
After such a slicing, the sum of the contributions from
Eq.~(\ref{eq:fxfx-main}) provides a correctly merged result.
Because of the $\Theta$ functions, the extra parton emitted on top of the
$n$-particle sample is either 
always unresolved, Eq.~(\ref{eq:fxfx0}), or 
always resolved, Eq.~(\ref{eq:fxfx1}).
This means that the combination of Eqs.~(\ref{eq:fxfx-main}) can be effectively
regarded as a LO merging, which implies treatment similar to that of CKKW, \ie
each sample is reweighted with the Sudakov form factors
(see Ref.~\cite{Frederix:2012ps} for detailed discussion).

The \fxfx method has been validated for the Higgs production, $e^+\nu_e$
production, and \ttbar production in hadron-hadron collisions, where the
spectra were also compared to \mcatnlo and \alpgen. Theoretical uncertainties
related to the merging were found to be small.
Detailed discussion on subtelties of the choice of merging scales and
further phenomenological studies for the dibosons and associated Higgs production can be found in Ref.~\cite{Alwall:2014hca}.

A comprehensive study of the predictions from the \fxfx method for multi-jet
production in association with a vector boson has been recently presented in
Ref.~\cite{Frederix:2015eii}. A range of differential distributions were
compared the 7-TeV LHC results for the $W$+jets and $Z$+jets processes, as well
as for the inclusive dilepton production. 
Theoretical results correspond to merging the NLO samples for $V$+1 jet
and $V$+ 2 jets and matching them to the parton shower from \herwigpp and
\pythiaeight.
The study finds a good agreement between the predictions and the data from ATLAS
and CMS. As expected, the merged NLO samples exhibit reduced scale uncertainties.
Similarly to the result found in Ref.~\cite{Maitre:2013wha}, also here, merging
of the 0- and 1-jet samples does not bring the prediction for the leading jet
$p_T$ distribution into full agreement with the data, although the experimental
errors cover for most the difference.
Description of the $H_T$ distributions, however, is greatly improved by the
merged result from \fxfx.

%-----------------------------------------------------------------------------
\paragraph{Electroweak merging}

As discussed in Section~\ref{sec:fo-calculations}, in addition to the NLO and
NNLO QCD corrections, one should in principle always study the effects
coming from the electroweak higher orders. 
This is also true in the case of parton showers.
Consider the $W$ + 2 jets production at LO. Configuration with a hard $W$ boson
recoiling against a quark or a gluon, while the other parton is soft, can be
considered as a QCD correction to the $W$ production. Hence, it is expected to
be well described by the standard merged results of the CKKW-L type for $W$
production + parton shower.
If, however, in the same process, the two partons are hard, recoiling against
each other, with one of them emitting a $W$ boson at small relative transverse
momentum, then, it is more natural to describe such a configuration as a EW
correction to dijet production. 
Moreover, because the $W$ boson is emitted at small angle, the above
configuration should profit from resummation of collinear  $W$ emissions.
Hence, the second situation is expected to be well described by an analogue
of the CKKW-L merging but for the dijet process + EW shower.
 
Ref.~\cite{Christiansen:2015jpa} proposed a prescription to merge the above two,
merged results into a common framework with the EW and QCD showers.
In such a result, the hard emissions of QCD partons, as well as the $W$ boson,
are generated with the fixed-order matrix elements, while the soft emissions of
partons and $W$s come from the QCD and EW showers, respectively.
The method is currently worked out for the tree level merging and it is
implemented in \pythiaeight.
Detailed numerical studies for the LHC show that the new scheme improves
most of the distributions as compared the standard CKKW-L prescription with
a QCD shower only.
The importance of the soft and collinear $W$ emissions grows with the center of
mass energy. This is because, the higher $\sqrt{s}$, the more frequent the
events with very hard partons. Therefore, the hierarchy of scales between the
partons' and the $W$ bosons' transverse momenta increases and the resulting EW
logarithms become more important.  
Indeed, as found in Ref.~\cite{Christiansen:2015jpa}, the weak Sudakov effects
for dijet production can be large and will become especially important at the
future, 100 TeV proton collider, where they can reach up to 25\% at $p_T$s of
the order of 20 TeV.

%-----------------------------------------------------------------------------
\paragraph{Confronting theory and experiment}

\begin{figure}[t]
  \begin{center}
    \includegraphics[width=0.43\textwidth]{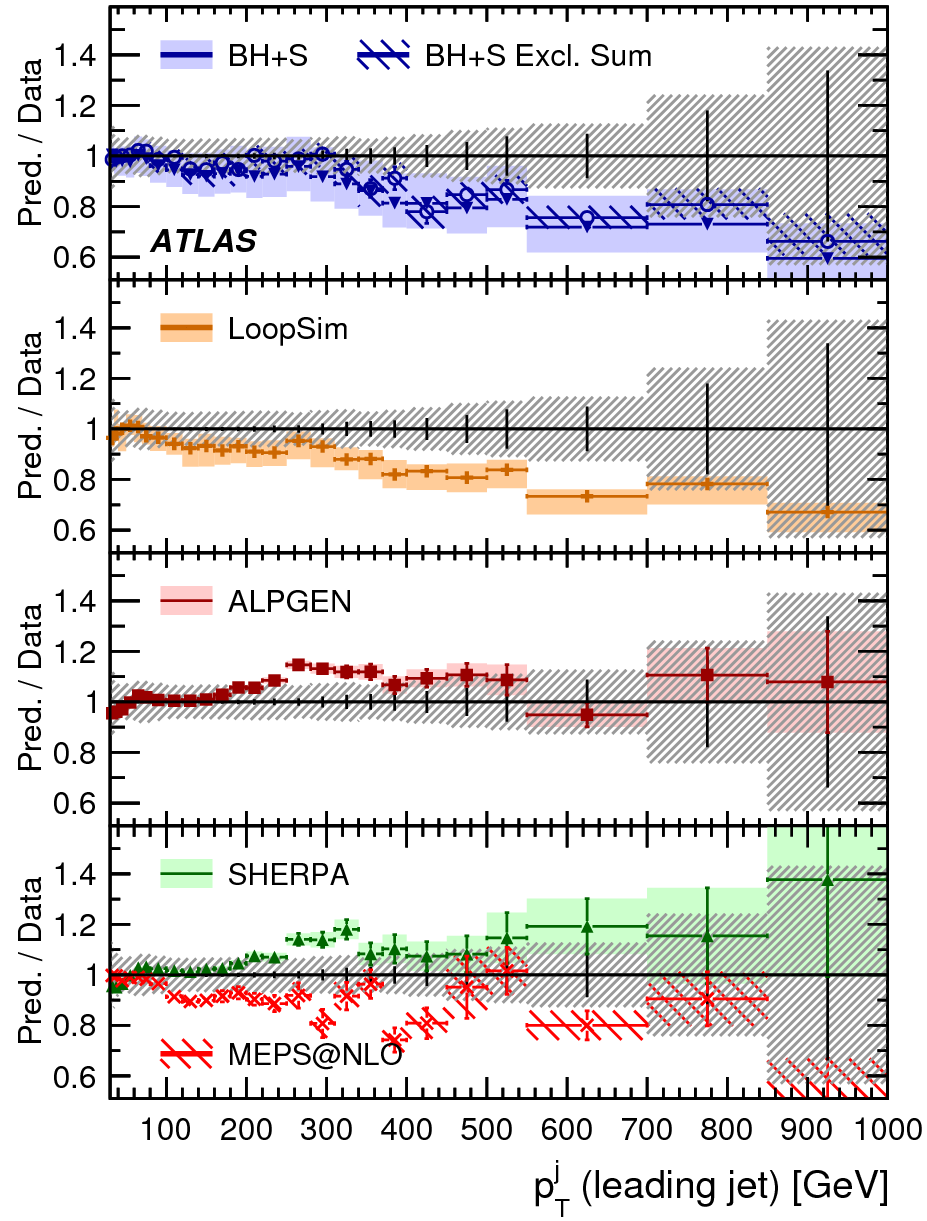}
    \hspace{25pt}
    \includegraphics[width=0.43\textwidth]{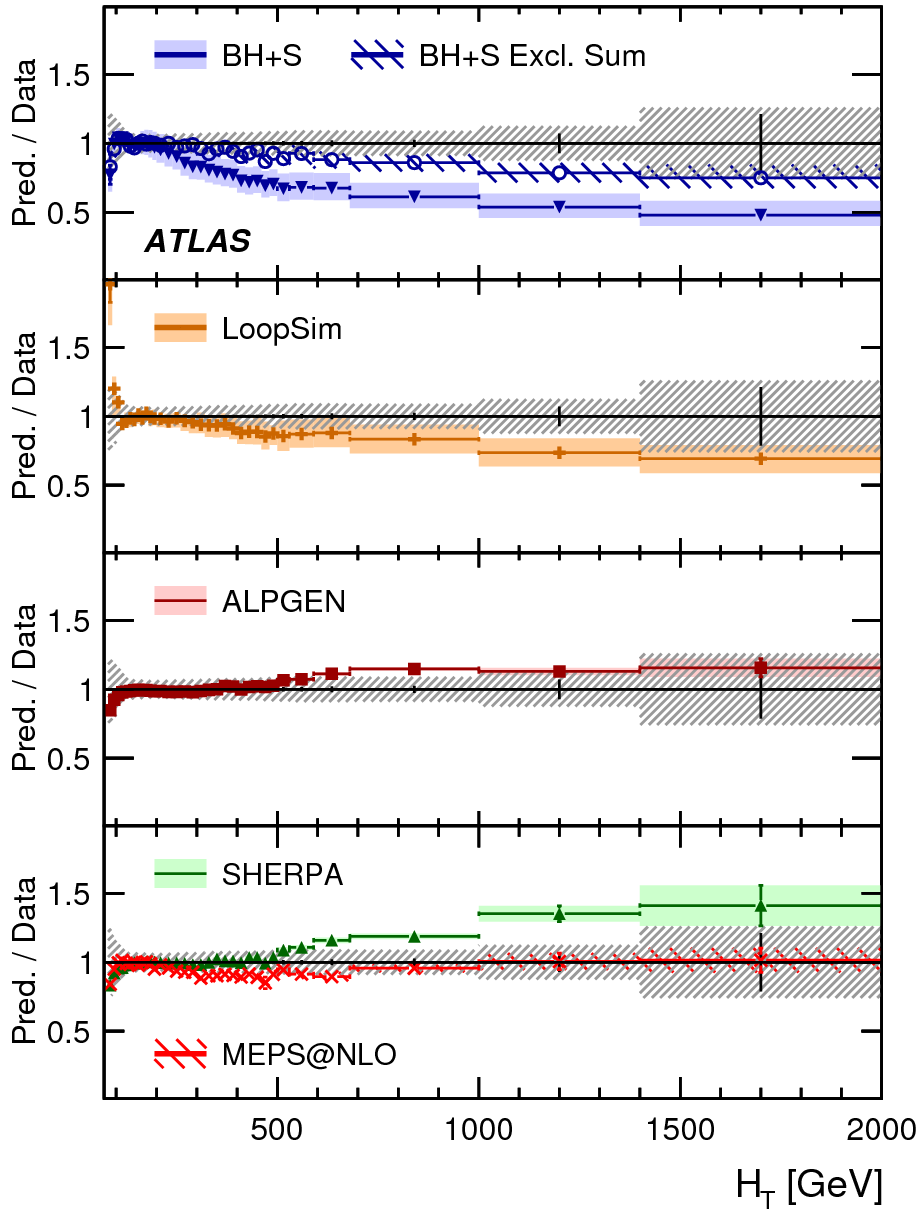}
  \end{center}
  \caption{
  Cross sections for the production of $W$+jets as a
  function of the leading-jet $p_T$ (left) or the $H_T$ variable (right).
  Statistical uncertainties of the data are shown as vertical bars while the
  hashed regions correspond to combined experimental uncertainties.
  The data are compared to various theoretical predictions, see text for
  discussion.
  Figure from Ref.~\cite{Aad:2014qxa}.
  }
  \label{fig:wj-atlas}
\end{figure}

We conclude this section by a brief discussion on how the perturbative
calculations discussed above  fare when confronted with the LHC data.
In Fig.~\ref{fig:wj-atlas}, we show a snapshot of the current status of the
theory vs experiment comparison for the $W$+jet production.
The ATLAS data are compared to predictions from the LO generator \sherpa~\cite{Gleisberg:2008ta}, LO
merged predictions from \alpgen~\cite{Mangano:2002ea}, fixed-order NLO results from
\blackhat+\sherpa~(BH+S)~\cite{Bern:2013gka, Gleisberg:2008ta} and merged
calculations from \loopsim~\cite{Rubin:2010xp, Maitre:2013wha},  
MEPS@NLO~\cite{Hoeche:2012yf} and the exclusive sums
approach~\cite{AlcarazMaestre:2012vp}. 

We see that the BH+S NLO calculations predict the leading-jet $p_T$
distribution, which is systematically lower than the data,
Fig.~\ref{fig:wj-atlas}~(left), and the discrepancy becomes bigger at large
$p_T$. This is not surprising as the NLO result for the inclusive $W$+1 jet
production has only the contributions for $W$+1 jet at NLO and $W$+2 jets at LO.
At high $p_T$, however, one expects higher multiplicities to become important
and those are not provided by the pure NLO result. They can be added by the
merging methods like the exclusive sums, LoopSim or MEPS@NLO, also shown in the
plot. Those methods supplement the NLO result for $W$+1 jet by the contributions
from $W$+2 jets at NLO, which, when taken inclusively, contain also $W$+3 jets
at LO. However, as we see in Fig.~\ref{fig:wj-atlas}~(left), merging does not
seem to improve the description of the data. Nevertheless, it certainly reduces
the theory uncertainties coming missing higher orders. 
Surprisingly, the best description of the ATLAS data is achieved with 
the LO results from \sherpa and \alpgen.

The situations looks different in the case of the $H_T$ variable, defined as a
scalar sum of $p_T$s of the jets, leptons and the missing energy.
Here, the NLO result suffers from a problem similar to the one discussed above
but, this time, the merged/matched results from the exclusive sums, LoopSim and
MEPS@NLO significantly improve the agreement with the data. 
This is related to the fact that the $H_T$ spectra receive giant corrections 
at NLO and NNLO, whose origin can be traced to extra real
radiation appearing at higher orders~\cite{Rubin:2010xp}. Significant part of
this contribution can be accounted for by means of merging the NLO samples with
higher jet multiplicities,  in this case, $W$+1 jet and $W$+ 2 jets at NLO.
 
As for the LO predictions, the \alpgen generator fares comparably to the
leading-jet $p_T$ case, while \sherpa overshoots the data for $H_T$
distributions

%-----------------------------------------------------------------------------
\subsection{Jet vetoes in single and diboson production}
\label{sec:jet-veto}

In studies of the Higgs boson and searches for new physics, the data is usually
divided into exclusive jet bin samples, as the background subtraction is
much more efficient if it is optimized separately to the events with 1, 2 or $n$
jets.
For example, in the case of the $W^+W^-$ production~\cite{Chatrchyan:2011tz}, a
huge background comes from the \ttbar process. In the latter however, the
$W^+W^-$ pair is in most cases accompanied by a hard jet. Hence, by rejecting
the events with jets above certain hardness, the procedure called \emph{jet
veto}, one is able to remove the \ttbar background almost entirely.

Jet vetoes pose a challenge to theoretical calculations as they introduce
logarithms of the type $\ln \left(\pTveto/Q\right)$, where $Q$ is a hard scale
of the process (\eg the  mass of the Higgs boson) and $\pTveto$ is the maximal
jet transverse momentum allowed for the event to be accepted.
Such logarithms can be large if $\pTveto \ll Q$ and that requires resummation of
the logarithmically enhanced terms to all orders.

\begin{figure}[t]
  \begin{center}
    \begin{minipage}[t]{0.48\textwidth}
    \centering
    \includegraphics[width=1.0\textwidth]{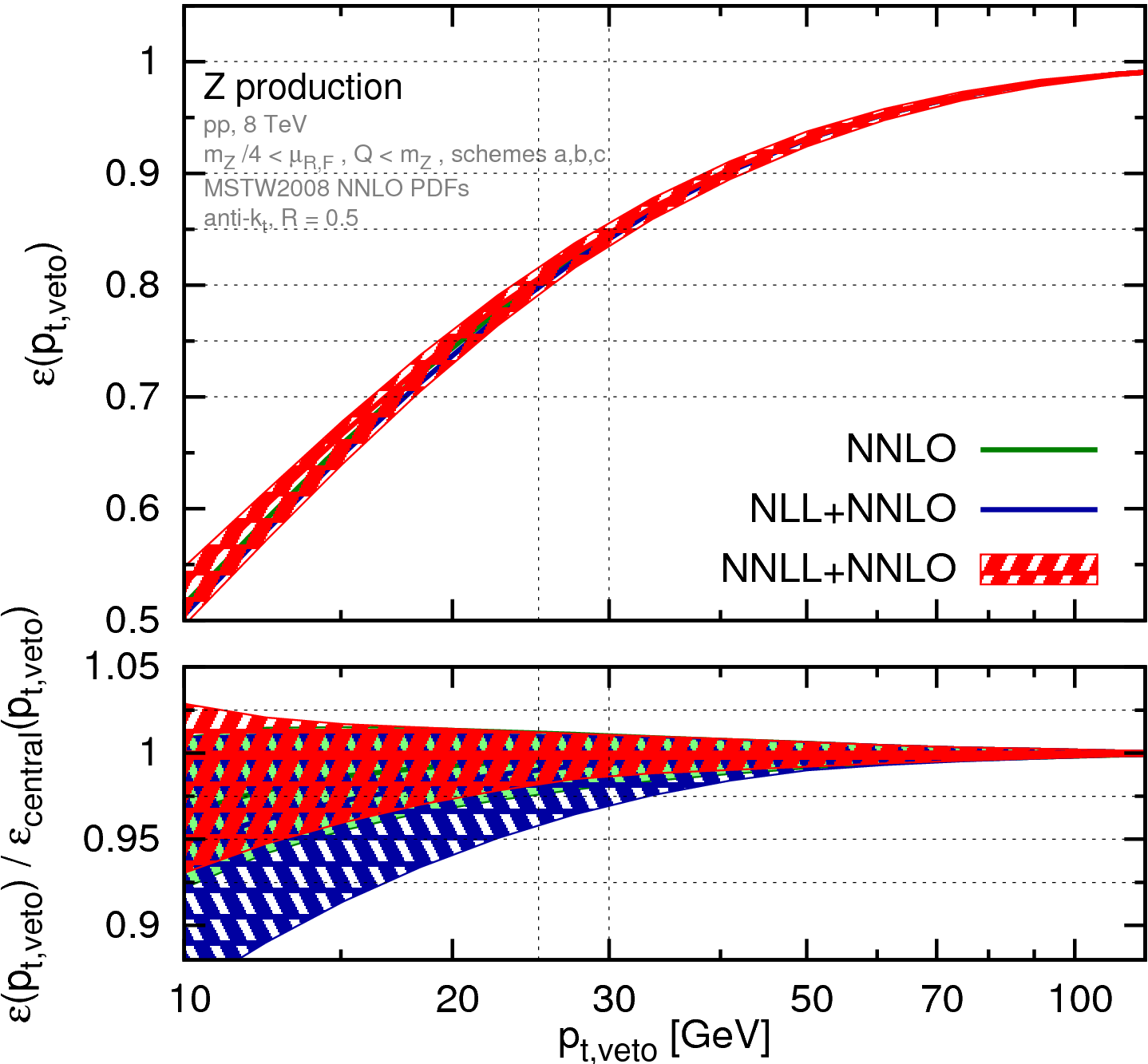}
    \caption{
    Jet-veto efficiency
    for $Z$-boson production at the 8 TeV LHC.  
    The lower panel shows distributions normalised to the
    NNLL+NNLO central value.
    Figure from Ref.~\cite{Banfi:2012jm}.
    }
    \label{fig:Z-jetvetoeff}
    \end{minipage}
    \hfill
    \begin{minipage}[t]{0.48\textwidth}
    \centering
    \includegraphics[width=1.0\textwidth]{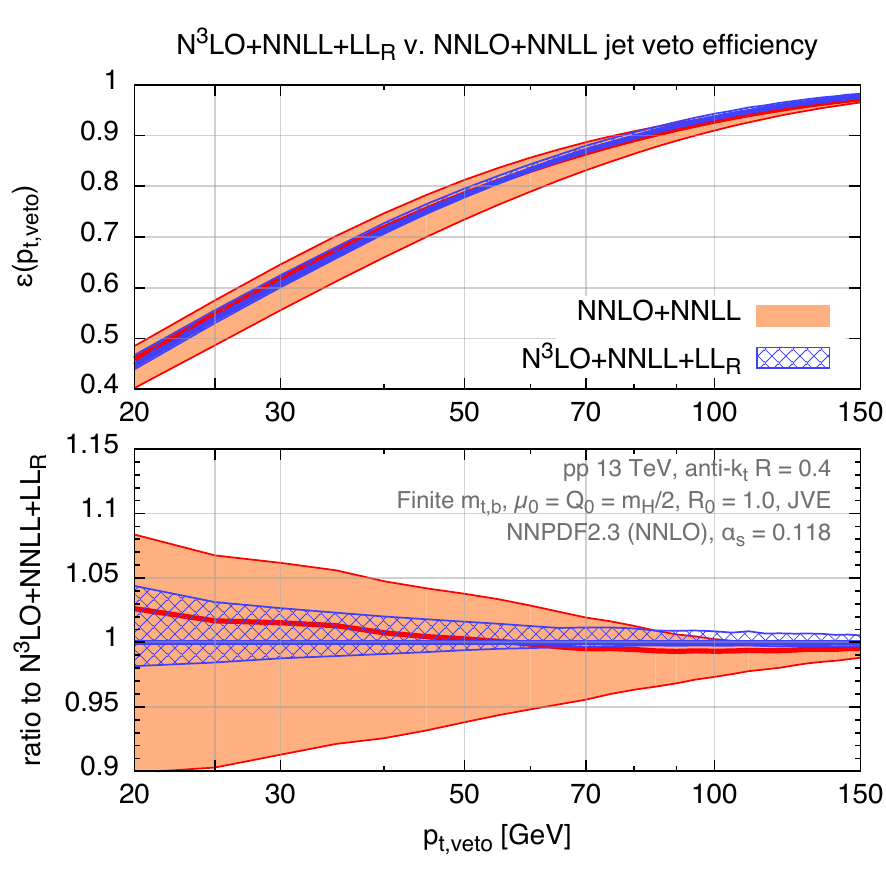}
    \caption{
    N$^3$LO+NNLL+LL$_R$, best prediction for jet veto efficiency at 13 TeV LHC
    compared to NNLO+NNLL. Figure from Ref.~\cite{Banfi:2015pju}.
    }
    \label{fig:H-jetvetoeff}
   \end{minipage}
  \end{center}
\end{figure}

In Ref.~\cite{Banfi:2012jm} such resummation has been performed at
the next-to-next-to-leading-logarithmic~(NNLL) accuracy, for
the Higgs and $Z$ boson production, and it has been matched to the NNLO
results~\cite{Anastasiou:2005qj,Catani:2007vq,Grazzini:2008tf,
Catani:2009sm}.
In Ref.~\cite{Banfi:2015pju}, the jet-veto resummed prediction for the Higgs
production in gluon fusion has been matched to the recent N$^3$LO
calculation~\cite{Anastasiou:2015ema} and further extended by including small-$R$
resummation~\cite{Dasgupta:2014yra} up to LL and finite mass effects up to
NLO~\cite{Banfi:2013eda}.

The above studies focused on computing the quantity
called the \emph{jet veto efficiency}, defined as~\cite{Banfi:2012yh}
\begin{equation}
  \epsilon(\pTveto) \equiv \frac{\Sigma(\pTveto)}{\sigma_\tot}
  \qquad
  \text{and}
  \qquad
  \Sigma(\pTveto) = 
  \sum_N \int d\Phi_N \frac{d\sigma_N}{d\Phi_N}
  \Theta(\pTveto-p_{T,j1}(p_1,\ldots, p_N))\,,
  \label{eq:jetveto-eff-def}
\end{equation}
where $\sigma_\tot$ is the total cross section, $p_{T,j1}$ is the transverse
momentum of the hardest jet and $d\sigma_N$ is a partonic cross section for
the production of the Higgs or $Z$ boson in association with $N$ partons.

The NNLL resummation in $\pTveto$ was performed using the \caesar
approach~\cite{Banfi:2004yd} and, up to the NLL accuracy, it could be
incorporated via a Sudakov form factor~\cite{Banfi:2012yh}. At that order, all
emissions can be treated as independent and one can assume that each emitted
parton forms a separate jet.
At the NNLL order, these assumptions do not hold and the resummation formula
needs to be supplemented with terms accounting for correlations between two
emissions, as well as corrections arising when two gluons are clustered into a
single jet~\cite{Banfi:2012jm}. These corrections, together with the previously
known NNLL pieces complete the NNLL result for the jet veto resummation.

The most precise, currently available predictions for
the jet veto efficiencies for the $Z$ boson 
production~\cite{Banfi:2012jm} at the LHC with $\sqrts = 8\, \TeV$ are shown in
Fig.~\ref{fig:Z-jetvetoeff}.
Because the efficiency (\ref{eq:jetveto-eff-def}) is defined as a ratio of two
cross sections, each of which has its own perturbative expansion, there exist
several ways in which the quantity $\epsilon$ can be computed. All of them are
equivalent up to the order NNLO and differ only by terms $\order{\as^3}$.
 
Refs.~\cite{Banfi:2012yh, Banfi:2012jm} adopted an envelope-based method to
asses uncertainties of the resummed predictions.
The motivation behind using envelopes is to avoid double counting between
uncertainties coming from various sources. Hence, at most one source is probed
at a time. In practice, one takes an envelope from the scale variation band in
one resummation scheme and the central values of the other
schemes~\cite{Banfi:2012yh}.

The above method of assessing the uncertainties has been further developed and
is now called the \emph{jet-veto efficiency} (JVE)
method~\cite{Andersen:2014efa}.  The efficiencies are effectively ratios of
$N$-jet exclusive to $N$-jet inclusive cross sections and their uncertainties
are driven by the Sudakov suppression terms. Therefore, one can treat the
uncertainties in the efficiencies and in the inclusive cross sections as largely
uncorrelated since the latter contains in addition uncertainties from genuine
higher order corrections. 
In the JVE method, the uncertainties of efficiencies are obtained as envelopes
and are combined with the uncertainties from inclusive cross sections using the
above assumption.

As we see in Fig.~\ref{fig:Z-jetvetoeff}, in the $Z$ boson case, the
uncertainties of the NNLL+NNLO result are visibly smaller than those of NLL+NNLO
and very similar to the fixed order NNLO result. 
In the case of Higgs production (see Ref.~\cite{Banfi:2012jm} for analogous
figure), the uncertainties of the NNLL+NNLO result are much smaller than those
of the NNLO result but comparable to NLL+NNLO.

The state-of-the-art predictions of the jet-veto efficiency for the Higgs boson
production in gluon fusion is shown in Fig.~\ref{fig:H-jetvetoeff}. The previous
results at NNLO+NNLL are compared to the updated prediction which includes the
complete order N$^3$LO as well as the small-$R$ and finite-mass effects.
The central value of the renormalization and  factorization scales was set at
$\mu_0 = m_H/2$.
We see that inclusion of the exact, three-loop result has a very strong effect
on the scale uncertainties, which are reduced from 10\% to less than 5\%, but 
change in the central value is small.
We note that the JVE prescription for estimation of the uncertainties has been
modified in Ref.~\cite{Banfi:2015pju} by limiting the number of schemes used
for envelope determination.
As further discussed in Ref.~\cite{Banfi:2015pju}, while the N$^3$LO corrections
are included, the effects of jet-veto resummation are very small, both in terms
of the central value and the uncertainties. Also the
finite-mass and small-$R$ effects are at a per-cent level.

Resummation of the jet-veto logarithms for the Higgs production cross sections
has been also performed  within SCET at the NNLL accuracy~\cite{Becher:2012qa},
as well as at a very good approximation to the N$^3$LL order, dubbed
N$^3$LL$_p$~\cite{Becher:2013xia}. In both cases, the results were matched to
the fixed order at the NNLO accuracy.
Unlike in Refs.~\cite{Banfi:2012yh, Banfi:2012jm, Banfi:2015pju}, which resum
logarithms in the jet-veto efficiency, \cf Eq.~(\ref{eq:jetveto-eff-def}), here
the resummation is performed directly for the Higgs cross section. 
A clear pattern of uncertainty reduction is observed when going from NLL,
through NNLL to the N$^3$LL$_p$ accuracy.
The components not captured by the N$^3$LL$_p$ result, one of which is the
$\order{\as^3}$  correction coming from the jet radius logarithms, were
estimated to be small. This was partially confirmed by a direct calculation in
Ref.~\cite{Alioli:2013hba}.

The jet-veto results from SCET are compatible with those obtained within
the standard QCD approach. In particular, the $\ln R$ dependence, known up
$\order{\as^2}$, agrees between the two frameworks~\cite{Becher:2013xia}.
Those jet-radius terms turn out to lead to sizable uncertainties for
small-$R$ jets, which provides further motivation for efforts to resum such
corrections, \cf Section~\ref{sec:jet-radius}.
 
Because the \caesar-based results were obtained for the jet veto
efficiencies while the SCET calculations were performed for the cross sections,
direct comparisons of the two is not unambiguous.
However, one can conclude that the numerical results from the two frameworks are
compatible, within theoretical uncertainties,  but the differences, which come
from incomplete $\order{\as^3}$ terms, are not negligible, \cf Fig.~12 of
Ref.~\cite{Becher:2013xia}.

\begin{figure}[t]
  \begin{center}
    \begin{minipage}[t]{0.48\textwidth}
    \centering
    \raisebox{0.2\height}{\includegraphics[width=0.5\textwidth]{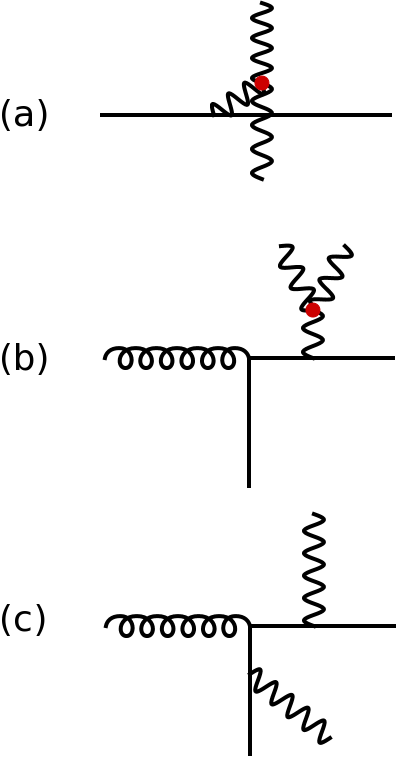}}
    \caption{
    Different configurations appearing in diboson process, $VV$: 
    (a) back-to-back production, 
    (b) $V$ recoiling against a jet, 
    (c) soft or collinear $V$ emission from a quark.
    }
    \label{fig:WZ-diagrams}
    \end{minipage}
    \hfill
    \begin{minipage}[t]{0.48\textwidth}
    \centering
    \includegraphics[width=0.85\textwidth]{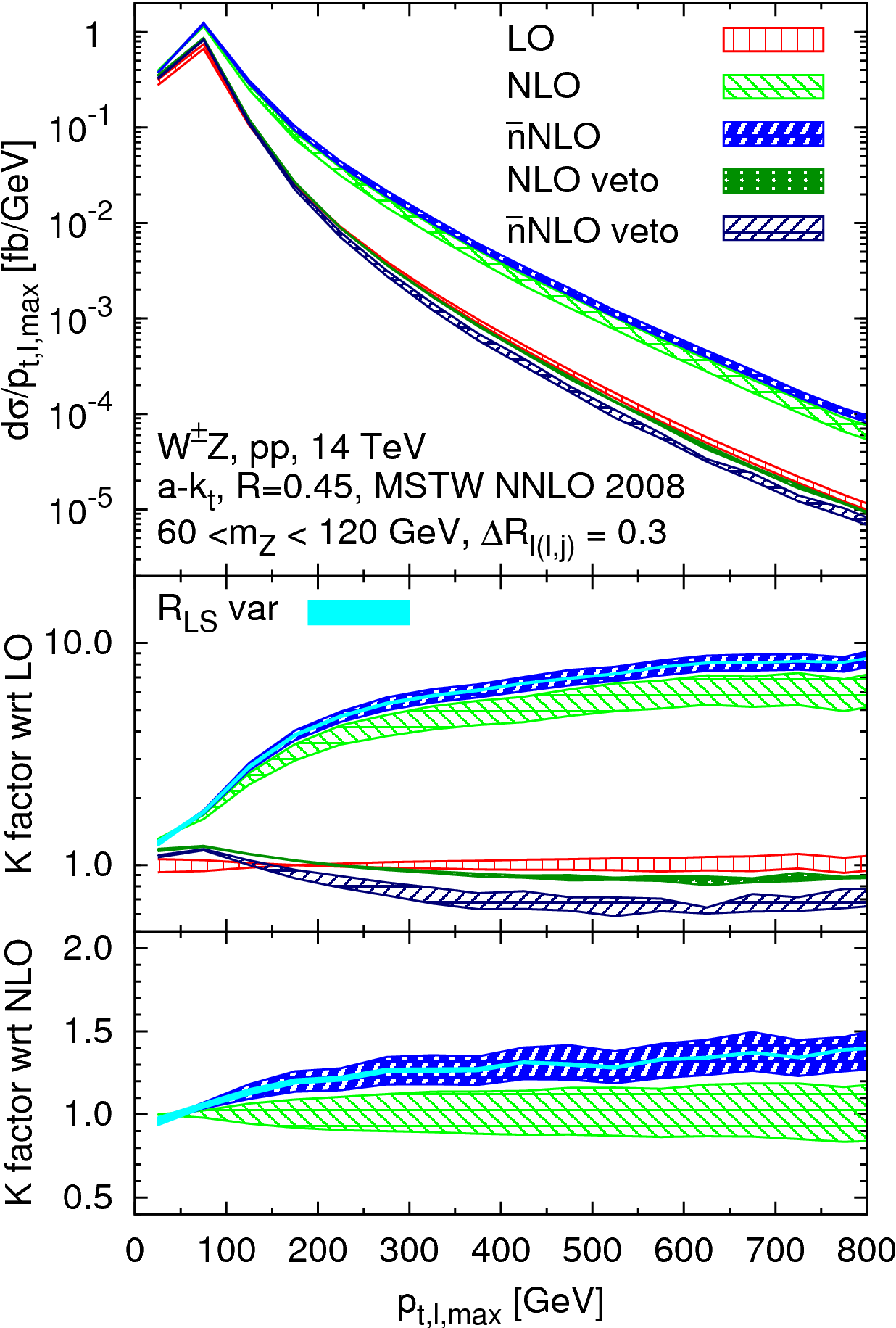}
    \caption{
    Differential cross sections and K factors for the $p_T$ of the hardest
    lepton in $WZ$ production at the LHC at $\sqrt{s}=14\, \text{TeV}$ with an
    without jet veto.
    Figure from Ref.~\cite{Campanario:2012fk}.
    }
    \label{fig:WZ-nNLO-veto}
   \end{minipage}
  \end{center}
\end{figure}

Another motivation for the jet veto procedure comes from searches for the
\emph{anomalous triple gauge boson couplings}~(aTGCs). These searches are
performed by looking at deviations in the cross sections for various
combinations of dibosons, like $WZ$, $WW$, $WH$, etc.
As shown in Ref.~\cite{Campanario:2010xn}, deviations in differential cross
sections are observed only if there is a significant momentum transfer through
the triple gauge boson vertex. This is achieved in configurations with the two
bosons going back to back, as depicted in Fig.~\ref{fig:WZ-diagrams} (a).
If one allows for a jet radiation produced in association with the diboson pair,
the amount of transverse momentum flowing through the triple gauge boson vertex
is reduced, as the hard jet will typically recoil against the system of the
collinear boson pair, \cf Fig.~\ref{fig:WZ-diagrams} (b).
Another way in which jets reduce sensitivity to aTGCs is via configurations
depicted in  Fig.~\ref{fig:WZ-diagrams} (c), where the bosons do not originate
from a single vertex but each of them is radiated separately from a quark
line. Triggering on large transverse momentum of one of the boson (or its decay
products) favours configuration in which the other boson is soft or collinear to
the energetic quark. Such diagrams are enhanced by the logarithmic factor 
$\ln \big({p_{T,\jet}^2}/{m_V^2}\big)$ and, since they do not involve the triple
boson vertex, sensitivity to the anomalous coupling effects is decreased.
 
By imposing a veto on the jet radiation, one can enhance configurations of the
type shown in Fig.~\ref{fig:WZ-diagrams}~(a), in which the two bosons are
produced back-to-back. Then, by requiring each of them to be sufficiently hard,
large transverse momentum flow through the triple gauge boson vertex is assured.

These effects have been studied in Refs.~\cite{Campanario:2012fk,
Campanario:2013wta, Campanario:2014lza} and Fig.~\ref{fig:WZ-nNLO-veto} shows an
example for the $WZ$ production.
The distributions of the hardest lepton transverse momentum were computed with
\vbfnlo~\cite{Arnold:2008rz, Baglio:2014uba} together with the
\loopsim~\cite{Rubin:2010xp, Maitre:2013wha} package, which allows one to
account for the dominant part of the NNLO corrections at high transverse
momentum of the leading lepton, $p_{T,l,\max}$.  The
renormalization and factorization scales were set to $ \mu_{F,R}= \frac12 \sum
p_{T,\text{partons}} + \frac12 \sqrt{p_{T,W}^2+m_W^2}+ \frac12
\sqrt{p_{T,Z}^2+m_Z^2} $
and the bands correspond to varying $\mu_F=\mu_R$ by factors 1/2 and 2 around
the central value. 
The cyan solid bands give the uncertainty related to the $R_\text{LS}$ parameter
of \loopsim, which is varied between 0.5 and 1.5. As we see, the factorization
and renormalisation scale uncertainties dominate above 100~GeV.

The large correction from LO to NLO for the inclusive sample (\ie with no
restriction on jet radiation) comes from the configurations of
Fig.~\ref{fig:WZ-diagrams} (c), with soft and collinear emissions of the vector
boson. 
We see that the approximate NNLO correction, labeled as \nNLO, is still
significant for the inclusive $WZ$ production, reaching up to
30\% of the NLO result at high $p_T$.
 
The situation is very different if the jet veto, forbidding the radiation with
$p_{T, \jet} > 50\, \GeV$, is imposed. This leads to suppression of the configurations
shown in Figs.~\ref{fig:WZ-diagrams} (b) and (c) and one is left with
contributions from events with the two hard vector bosons recoiling against each
other.

Fig.~\ref{fig:WZ-nNLO-veto} shows the generic problem of the standard jet veto
procedure. At NLO, the distribution seems to be very well behaved, with small
corrections with respect to LO and with the reduced scale uncertainty. The \nNLO result,
however, reveals further, significant corrections and the corresponding
uncertainty band is broader than that of NLO.
This effect comes from 
 Sudakov-type logarithms introduced by the veto procedure, which
forbid radiation in certain regions of phase space. These logarithms bring
negative corrections to the cross section at high $p_T$. 
Hence, the NLO/LO and \nNLO/LO K-factors rise a little, but as 
$p_{T,l,\max}$ increases, the restriction for additional
radiation leads to suppression and eventually fairly rapid drop of the
K-factors.
Of course, the full NNLO correction will also receive contribution from,
potentially non-negligible constant term of the two-loop diagrams.
However, the approximate \nNLO result for the vetoed case gives already an
indications what happens at $\order{\aEW^2\as^2}$. Moreover,
as shown in Fig.~\ref{fig:WZ-nNLO-veto}, as well as other distributions
discussed in Refs.~\cite{Campanario:2012fk, Campanario:2013wta,
Campanario:2014lza}, it demonstrates that the small scale uncertainties of many
of the NLO results with a jet veto are to a large extent accidental, as the
corresponding uncertainty at \nNLO comes out larger than that of NLO.
The misleadingly small scale uncertainties of the vetoed events come from
cancellations between large perturbative NLO corrections and terms involving
logarithms of the veto scale.

The problem of accidental cancellation occurring in events with jet vetoes has been
carefully addressed in Ref.~\cite{Stewart:2011cf}, where a method superior to a
simple scale variation in exclusive, fixed-order predictions has been proposed.
The exclusive, $N$-jet cross section can be defined as a difference of the inclusive
cross sections: $\sigma_N = \sigma_{\geq N} - \sigma_{\geq N+1}$. By varying a
scale in the exclusive  cross section, we implicitly assume 100\% correlation
between the uncertainties of the two inclusive cross sections.
However, there is \emph{a priori} no reason for that assumption. 
 
An inclusive, $N$-jet cross section is sensitive to the minimal jet transverse
momentum, $p_T^\cut$, through the logarithms $L=\ln\left(p_T^\cut/Q\right)$, where
$Q$ is the hard scale of the process, \eg mass of the Higgs boson.  Powers of
such logarithms multiply various powers of $\as$.
For experimentally relevant values of the transverse momenta, these logarithms
are large and the variation of $Q$, as well as the renormalization scale in
$\as$, provides a realistic assessment of the uncertainty due to missing higher
orders.
However, in the difference of the two inclusive cross sections, which gives the
exclusive cross section $\sigma_N$, the coefficients multiplying the powers of $\as$
vanish for certain value of $L$. This is a cancellation between the genuine
higher order corrections and the terms generated by the veto procedure and it
turns out to happen for the values of $p_T^\cut$ that are currently used in
analyses of the LHC data.

\begin{figure}[t]
  \begin{center}
    \includegraphics[width=0.48\textwidth]{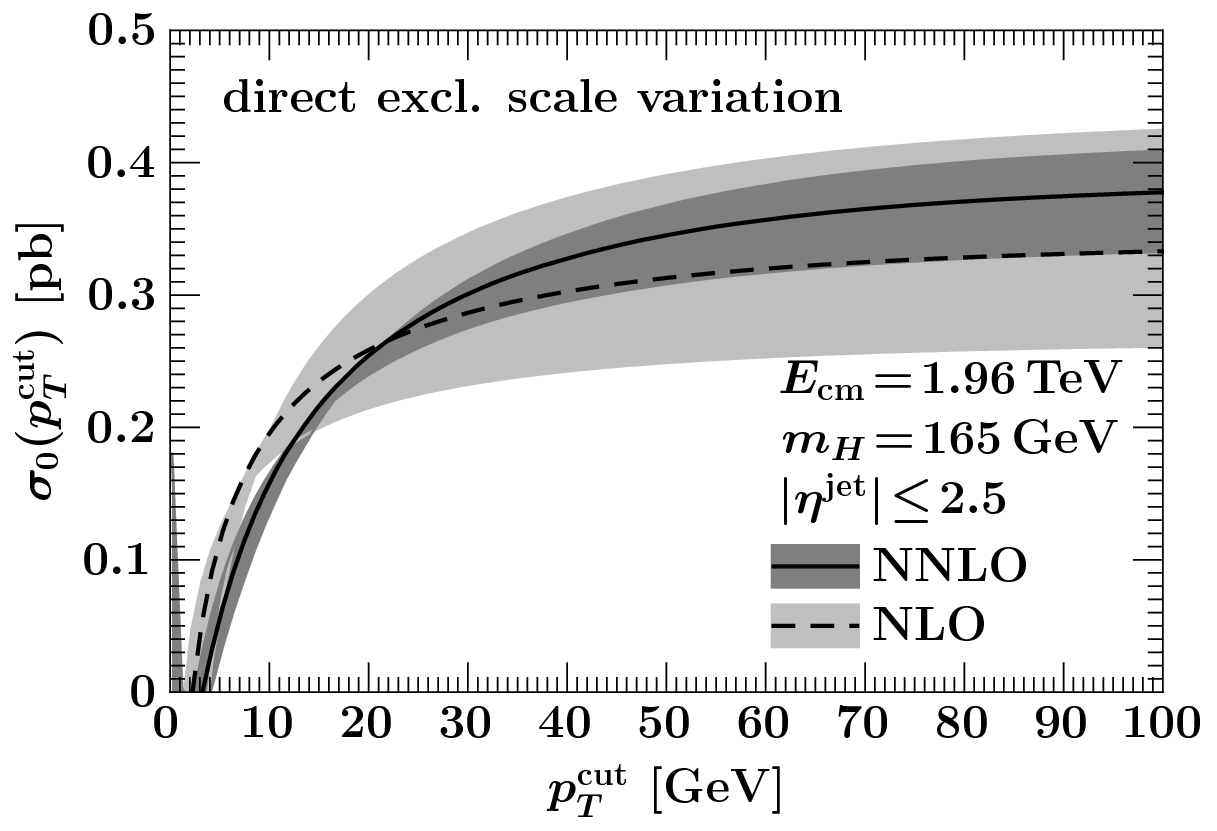}
    \hfill
    \includegraphics[width=0.48\textwidth]{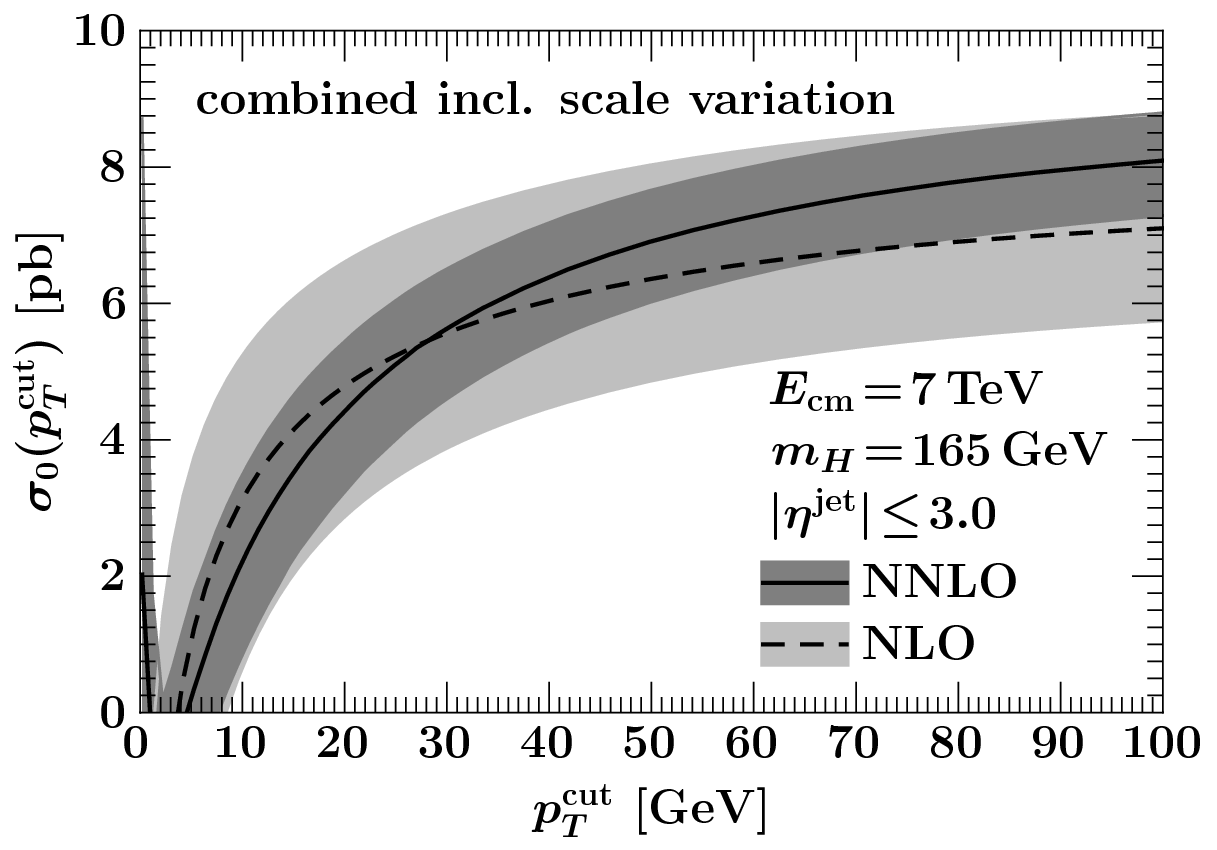}
  \end{center}
  \caption{
  Scale variation uncertainties for exclusive $gg \to H + 0\, \jets$ production
  at NLO and NNLO.  On the left, the bands are obtained by the standard scale
  variation in the exclusive cross section, while on the right, by combining the
  uncertainties of inclusive 0- and 1-jet cross sections in quadrature.
  Figure from Ref.~\cite{Stewart:2011cf}.
  }
  \label{fig:stewart-tackmann}
\end{figure}

Consider a simple case of the $0$-jet inclusive production. The inclusive cross
section reads:
$\sigma_{\geq 0} = \sigma_{0}(p_T^\cut) + \sigma_{\geq 1}(p_T^\cut)$.
Since the $0$-jet inclusive cross section cannot depend on $p_T^\cut$,
additional perturbative uncertainty induced by the cut (through the logarithms
mentioned above), $\Delta_\cut^2$, must be
100\% anti-correlated between $\sigma_{0}$ and $\sigma_{\geq 1}$.
Because the veto-induced terms are large, it is justified to assume that
$\Delta_{\geq 1}^2 \simeq \Delta_\cut^2$, \ie the uncertainty coming from the
veto is equal to the total uncertainty of the $1$-jet inclusive cross section. 
The uncertainty of the $0$-jet inclusive cross section, $\Delta_{\geq 0}^2$, is
independent of $\Delta_\cut^2$, it is therefore also independent of
$\Delta_{\geq 1}^2$. From the above, it follows that the uncertainty of the
$0$-jet exclusive cross section, which is what we want to calculate, is, to a
good approximation, given by
$\Delta_{\geq 0}^2 + \Delta_{\geq 1}^2$.
Hence, contrary to the standard procedure, in this approach, the $N$ and
$N+1$-inclusive cross sections are assumed to be 100\% uncorrelated.
 
As shown in Fig.~\ref{fig:stewart-tackmann}~(left), the standard procedure of
scale variation in the exclusive cross section for the Higgs production in gluon
fusion gives misleadingly small uncertainties at NLO. Hence, the NLO band does
not cover the NNLO distribution, which means that the uncertainty due to missing
higher order terms is not estimated correctly. On the contrary, the procedure of
combining the uncertainties of the inclusive cross sections, described above,
gives a much more realistic assessment of the theoretical errors, as shown in
Fig.~\ref{fig:stewart-tackmann}~(right).

Another approach to the problem of the large logarithmic corrections associated
with setting a sharp cut on jet radiation is to modify the veto procedure such
that the logarithmic enhancement is reduced.
Ref.~\cite{Campanario:2014lza} introduced the so-called \emph{dynamical jet
veto}, where, instead of using a fixed $p_T$ cut, the jet veto is applied on
the event-by-event basis with help of the variables
\begin{equation}
  x_\jet = \frac{\sum_\jets E_{T,i}}{\sum_\jets E_{T,i}+\sum_{V} E_{T,i}}\,,
  \qquad \qquad \qquad
  x_V = \frac{E_{T,V}}{\sum_\jets E_{T,i}+\sum_{V} E_{T,i}}\,.
\end{equation}

\begin{figure}[t]
  \begin{center}
  \includegraphics[width=0.45\textwidth]{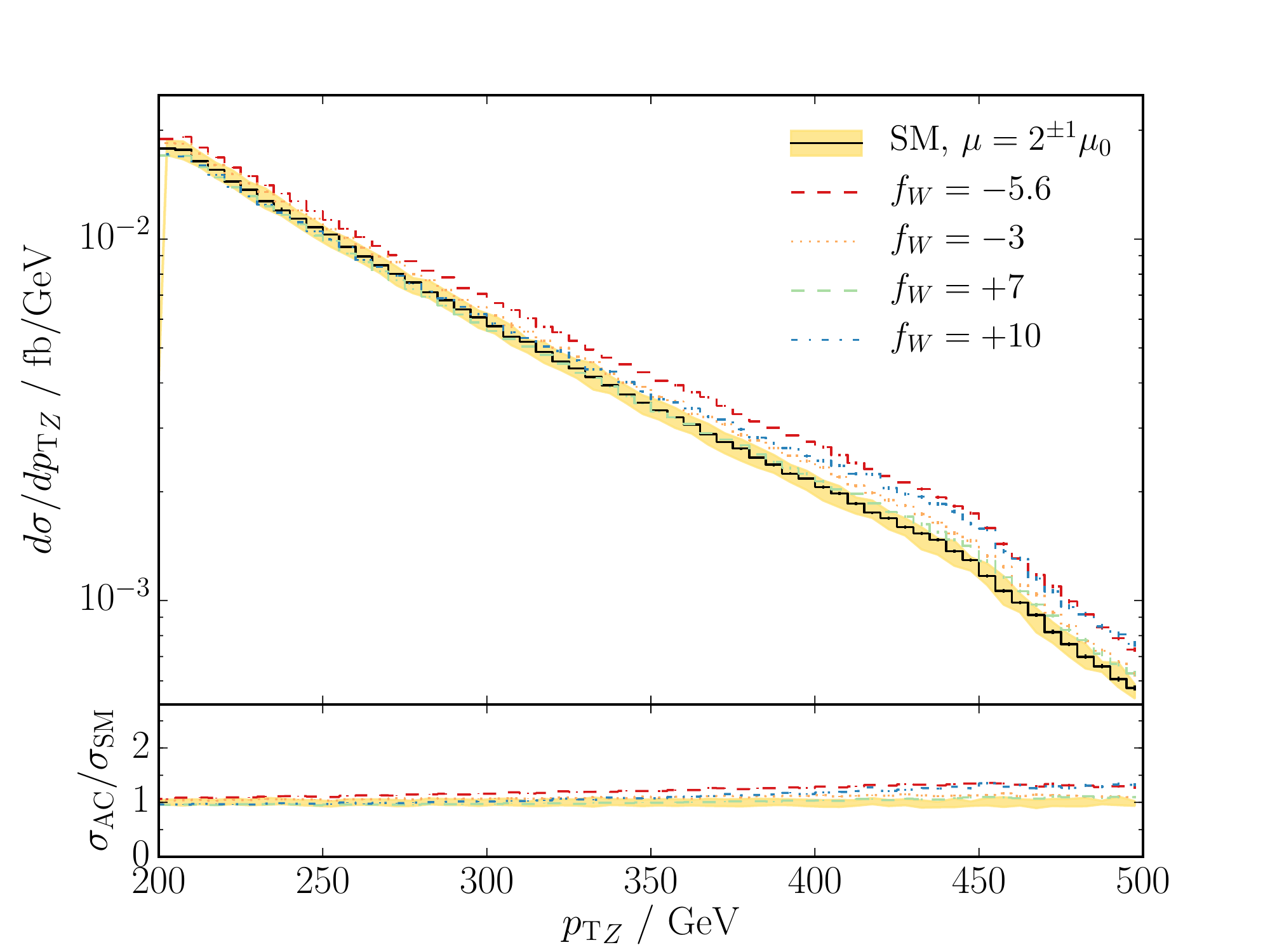}
  \includegraphics[width=0.45\textwidth]{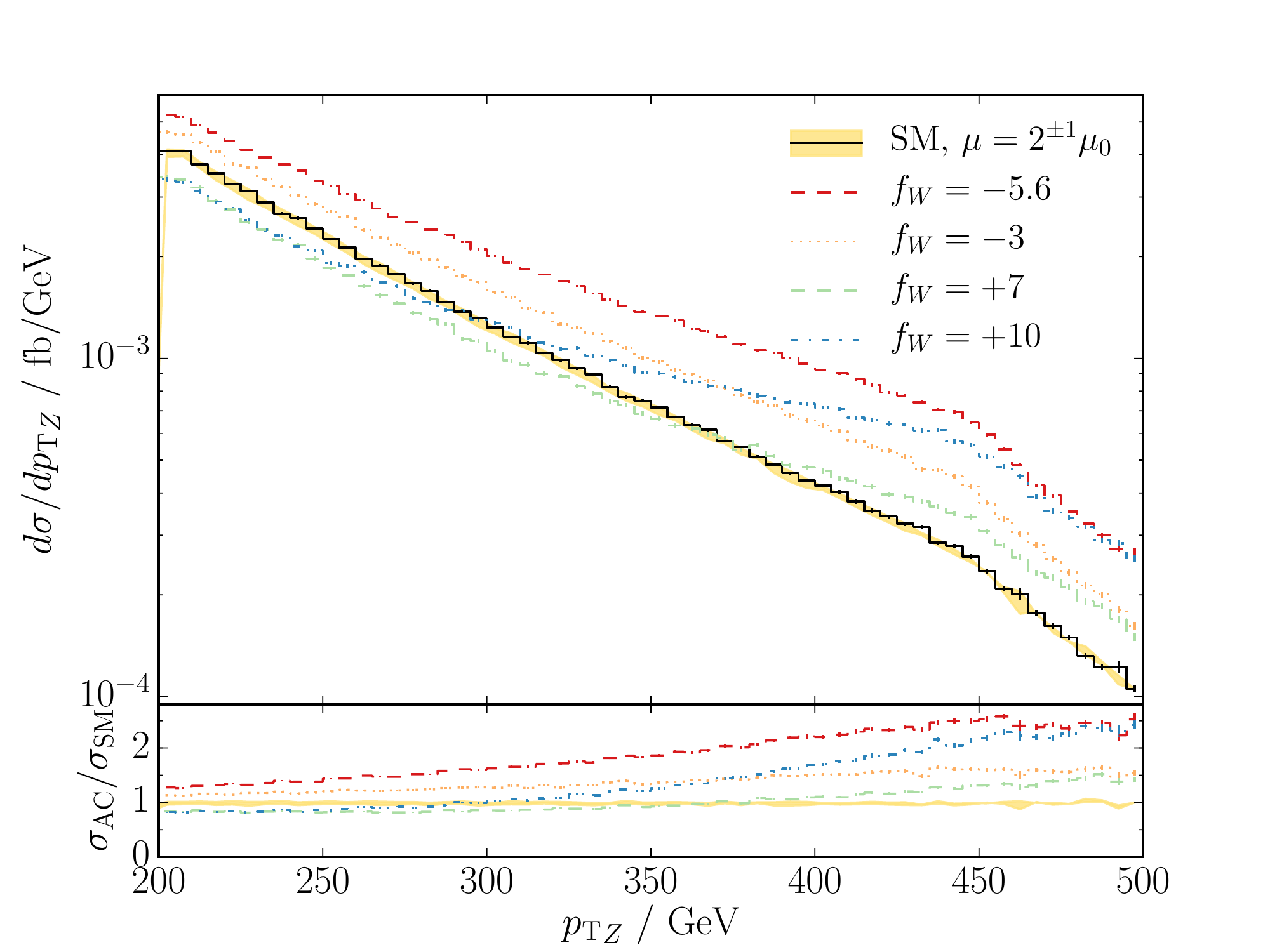}
  \end{center}
  \caption{
  Differential transverse momentum distribution of the $Z$
  boson in $WZ$ production at NLO for different values of the anomalous coupling
  $f_W$ without (left) and with (right) a dynamical jet veto.  
  Figure from Ref.~\cite{Campanario:2014lza}.
  }
  \label{fig:dynamical-veto-ptz}
\end{figure}

As shown in Ref.~\cite{Campanario:2014lza}, for the diboson production
processes, the sensitivity to aTGCs lies in the region of $x_\jet \lesssim
0.2-0.3$. 
At the same time, most of the events
have $x_\jet \simeq 0.4-0.5$. Hence, vetoing the events with $x_\jet>0.2$
will lead to a significant increase of the sensitivity to aTGC.
This is illustrated in Fig.~\ref{fig:dynamical-veto-ptz} which shows $p_{T,Z}$
distributions obtained with different values of the anomalous coupling $f_W$
(including the SM case of $f_W=0$). On the left hand side, no veto was used and
we see that the sensitivity to $f_W$ variation is limited.  The right plot shows
the effect of applying the dynamical veto procedure. We see that sensitivity to the
$f_W$ values increases significantly, especially at high $p_T$. At the same
time, the uncertainty band from scale variation (shown for SM only) decreases
only moderately unlike in the case of fixed-$p_T$ veto discussed earlier in the
context of Fig.~\ref{fig:WZ-nNLO-veto}, where it was nearly vanishing at NLO.
Stability of the result with the dynamical veto indicates that the effect of
Sudakov logarithms is mild.

%-----------------------------------------------------------------------------
\subsection{Forward jets}
\label{sec:forward-jets}

The main features of the kinematics of jet processes in hadron-hadron collisions
can be understood by analysing a simple $2\to 2$ partonic scattering.
The relation between the rapidities and the transverse momenta of the outgoing
partons (our proxies for jets) and the fraction of energies carried by the
incoming partons corresponds to Eq.~(\ref{eq:2to2kinematics}) introduced in
Section~\ref{sec:generalized-tmd}.

Most of the jets measured at the LHC are produced in the central rapidity
region, $y_\jet \sim 0$, hence,  
the energy fractions of the incoming partons are comparable, and
typically larger than $10^{-2}$. Therefore, the central dijet production
corresponds to the case $x_1 \sim x_2 \lesssim 1$. This region of phase-space is
amenable to standard treatment in the framework of the collinear factorization
discussed in Section~\ref{sec:coll-fac}, with PDFs evolved according to DGLAP
equations and combined with the collinear matrix elements. 

The above canonical framework is expected to break down when more extreme
corners of phase-space are probed. Those can be reached in 
processes with forward jets. In the case of dijet production, we can have both
of the jets going forward, one forward and one central jet, or one jet going 
in the forward and the other in the backward direction. The forward-forward case
corresponds to $x_1 \sim 1$, $x_2 \ll 1$, the forward-central to  $x_1 \sim 1$,
$x_2 \lesssim 1$ and the forward-backward to $x_1 \sim 1$, $x_2 \sim 1$.

Each of this regions poses challenges to the standard approaches based on the
collinear factorization. In the forward-forward case, one of the incoming
hadrons is probed at very low momentum fractions, which leads to appearance of
large logarithms, $\ln(1/x)$, from initial state emissions. These logarithms
should be resummed, which can be achieved by means of the
BFKL~\cite{Lipatov:1976zz, Kuraev:1977fs, Balitsky:1978ic} 
or CCFM~\cite{Ciafaloni:1987ur, Catani:1989sg, Catani:1989yc} equations.
Those formalisms go beyond simple collinear
factorization, as they result in parton distributions which are unintegrated in
the transverse momentum and therefore are only compatible with some form of the
TMD factorization, \cf Section~\ref{sec:tmd-factorization}.
 
Similar issues arise in the case of the forward-central dijet production, where,
in addition, a potentially large rapidity gap between the two jets opens a phase
for addition BFKL-type emissions from the final state.
Finally, in the case of the forward-backward dijet production, rapidity
separation between the jets is even larger and the process falls into the
category of the so-called Mueller-Navelet~(MN) jets~\cite{Mueller:1986ey}. In
this case, the use of collinear factorization is justified as both $x_1$ and
$x_2$ are large, but one needs to include BFKL resummation of the final state,
wide-angle radiation~\cite{Andersen:2009nu, Andersen:2009he}.

\begin{figure}[t]
  \begin{center}
    \raisebox{0.15\height}{
    \includegraphics[width=0.45\textwidth]{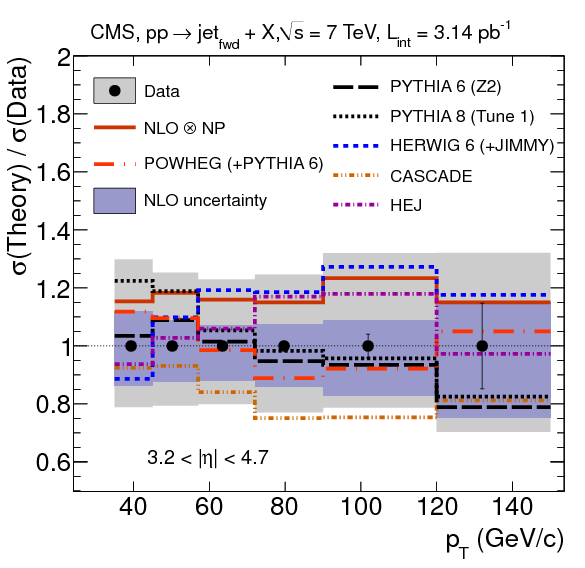}
    }
    \hfill
    \includegraphics[width=0.45\textwidth]{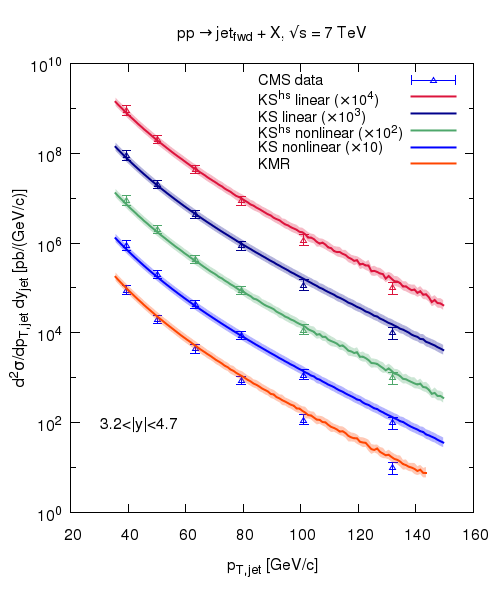}
  \end{center}
  \caption{
    Single inclusive production of forward jets.
    Comparison between CMS data and predictions from a range of theoretical
    approaches both within the collinear (left: \herwig, \pythia, \hej) and the
    high energy factorization frameworks (left: \cascade, right: HEF with KS and
    KMR gluons).
    Figures from Refs.~\cite{Chatrchyan:2012gwa} and \cite{Bury:2016cue},
    respectively.
  }
  \label{fig:cms-fwd-single}
\end{figure}

In the following, we shall briefly describe the existing theoretical approaches
to the forward jet processes and confront them with experimental data from the
LHC.

The \cascade~\cite{Jung:2010si} is a MC event generator based on the CCFM
evolution equation~\cite{Ciafaloni:1987ur, Catani:1989sg, Catani:1989yc}.  The
latter resummes both the $\ln \left(1/x\right)$ and $\ln Q^2$ terms, to all
orders in the leading logarithmic approximation, providing appropriate
unintegrated gluon distributions. 
In \cascade, the CFFM equation is used for probabilistic, backward generation of
gluon radiation, starting from the hard scattering process described by an
off-shell matrix element.
This procedure is, in principle, compatible with the high energy
factorization~\cite{Catani:1990xk, Catani:1990eg}. 
Final-state parton shower and hadronization are subsequently added via an
interface to \pythia.
The framework has been used to fit parameters of the unintegrated gluon
distributions to the HERA data~\cite{Jung:2010si}.

The \emph{hybrid high energy factorization} framework, introduced in
Section~\ref{sec:hef-framework} can be directly employed to study forward
dijets by integrating the formula~(\ref{eq:hef-formula}).
The calculations can be performed with the public codes of
Refs.~\cite{Bury:2015dla, Bury:2016cue, vanHameren:2015uia}.
Here, one can use a range of unintegrated gluons fitted to DIS data, like those
obtained in Refs.~\cite{Kutak:2012rf, Kimber:2001sc}. A generalized version of
HEF, the \emph{improved TMD
factorization}, \cf Section~\ref{sec:improvedTMD}, is in principle better,
as long as one is able to determine multiple gluon distributions entering
Eq.~(\ref{sec:improvedTMD}).

The \emph{high energy jets} (HEJ) framework~\cite{Andersen:2009nu,
Andersen:2009he, Andersen:2011hs}, implemented in the \hej program, provides
all-order resummation of the leading logarithmic contributions to the
wide-angle, hard QCD radiation. These contributions are dominant in the high
energy limit, where all invariants involving the outgoing partons are large and
their transverse momenta are fixed. In this limit, the scattering amplitudes
factorize into rapidity ordered  pieces which allows for efficient evaluation of
the matrix element.
This is analogous to what happens in the collinear limit, where multi-parton
cross sections factorize,  which in turn is a basis for construction of a
parton shower.
The HEJ approach can be thought of as being complementary to the parton shower
and the two have been indeed combined~\cite{Andersen:2011zd} providing
simultaneous description of the large- and small-angle, multi-particle
emissions.
Both pure $n$-jet processes, with $n\geq 2$, as well as dijets in association
with a vector boson or the Higgs boson~\cite{Andersen:2012gk}, are implemented
in the \hej program.
The approach is relevant for production of the central-forward and
Mueller-Navelet jets, as it capable of filling up the large rapidity gap between
the jets with gluon radiation.

\emph{BFKL-based approaches} are also used to study  Mueller-Navelet jets.
Differential cross sections are calculated in the framework of the high energy
factorization with the unintegrated gluons obtained from ordinary collinear PDFs
convoluted with jet vertices, currently known at the
next-to-leading-logarithmic~(NLL) accuracy~\cite{Bartels:2001ge,
Bartels:2002yj}. 
The off-shell matrix elements are provided by the BFKL kernel, also
known up to NLL.
As discussed in Ref.~\cite{Ducloue:2013wmi}, predictions for broad class of
observables, most notably the moments of distributions of the azimuthal distance
between the jets, are different in this framework, as compared to the standard
NLO.  The MN jets are therefore very suitable for searches of BFKL dynamics.
For further important studies of the Mueller-Navelet jets see
Refs.~\cite{Colferai:2010wu, Caporale:2011cc, Caporale:2013uva, Ducloue:2013bva,Caporale:2014gpa}.

\begin{figure}[t]
  \begin{center}
    \includegraphics[width=0.43\textwidth]{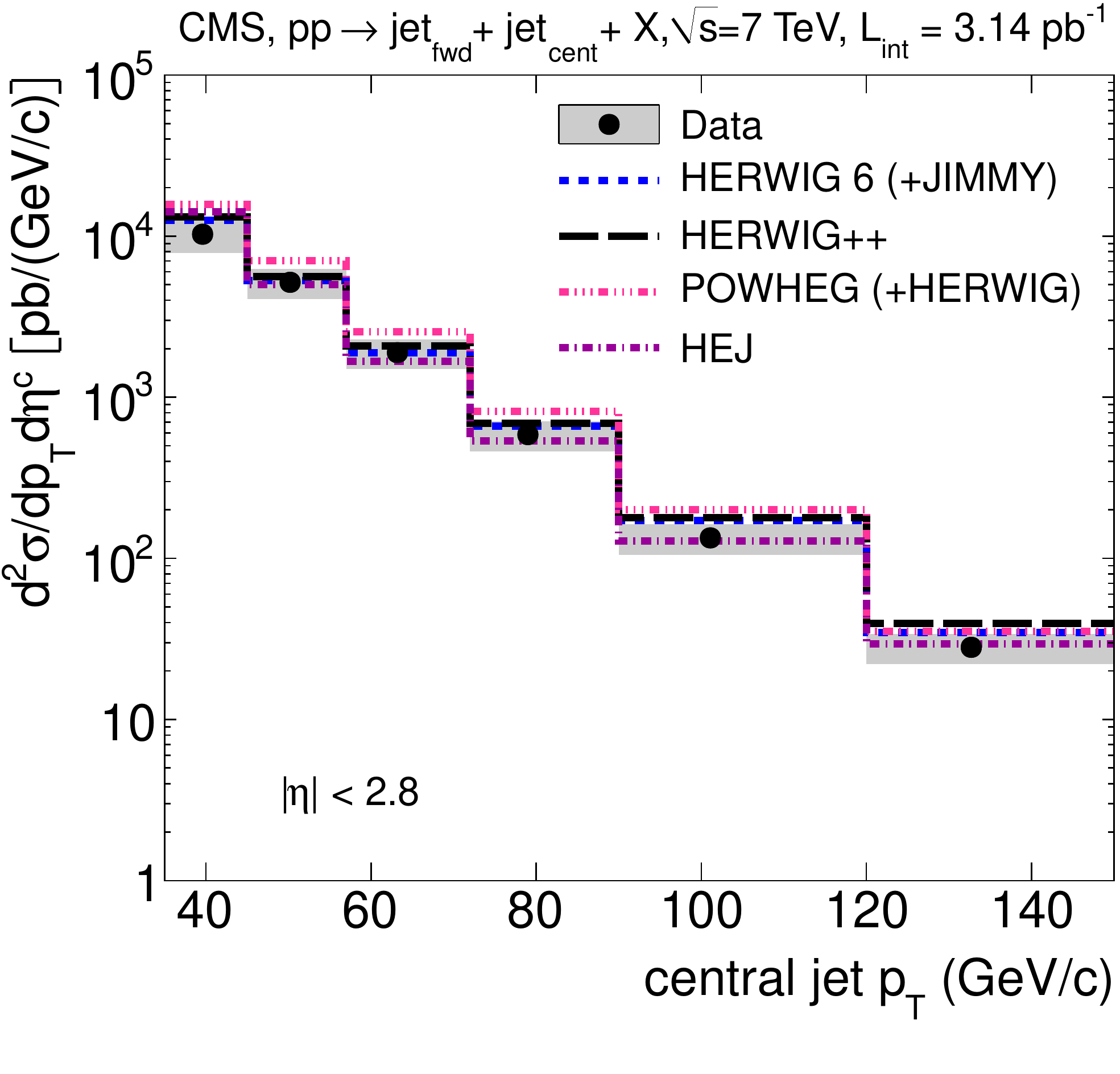}
    \hspace{40pt}
    \includegraphics[width=0.43\textwidth]{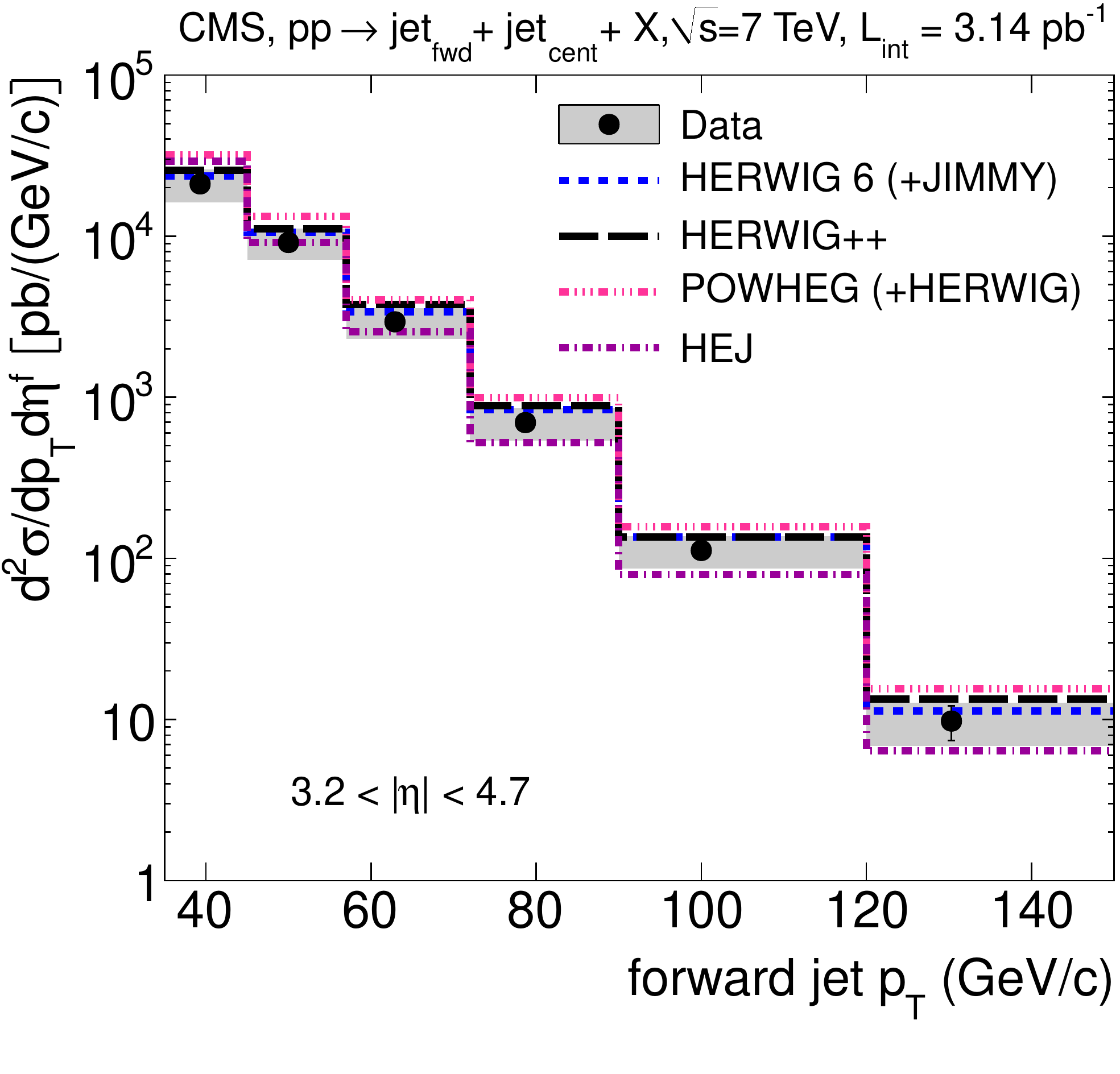}
  \end{center}
  \caption{
    Differential cross sections as a function of the central (left) and the
    forward (right) jet $p_T$ in the central-forward dijet production as
    measured by CMS in comparison with various theoretical predictions, all
    within the framework of collinear factorization.
    Figure from Ref.~\cite{Chatrchyan:2012gwa}.
  }
  \label{fig:cms-forward-central}
\end{figure}

The LHC experiments have performed many interesting  measurements with forward
jets. The inclusive jets and dijets production in the forward region have
been studied both by ATLAS and CMS.
In Fig.~\ref{fig:cms-fwd-single}, we show a
comparison between the CMS
data~\cite{Chatrchyan:2012gwa} for the jet's transverse momentum distributions
in the single inclusive production, and various theoretical calculations.
Overall agreement within 20\% experimental uncertainty band is found for most
predictions. We note that good description is obtained both within the collinear
factorization framework (NLO, \powheg), as well within the hybrid high energy
factorization with the KS~\cite{Kutak:2012rf, Kutak:2014wga} or
KMR~\cite{Kimber:2001sc} gluons.

Production of the central-forward dijets, with rapidities, $|y| < 2.8$ and $3.2
< |y| < 4.7$, respectively, was also studied by CMS.  As this measurement is
less inclusive than the single jet production, it is more challenging to
describe theoretically. Indeed, as found in Ref.~\cite{Chatrchyan:2012gwa}, the
results from \pythia, \herwig and \cascade show various levels of agreement with
the data, from 30\% to a factor of 2, depending on the distribution ($p_T$ of
the central or the forward jet) as well as a specific tune of the Monte Carlo.
The results from \herwig, \powheg and \hej describe the data better, within
30-50\%, as shown in Fig.~\ref{fig:cms-forward-central}.

Of particular interest is a distribution of the azimuthal separation between two
leading jets, $\Delta \phi$, the so-called \emph{azimuthal decorrelation}.
In the framework of the collinear factorization, azimuthal decorrelation is
sensitive to the initial state radiation (ISR), which boosts the dijet system in
the transverse direction. 
In the language of the high energy factorization, this ISR radiation is what
builds up the non-zero $k_T$ of the incoming gluon and it is effectively resumed
in the unintegrated gluon TMDs.

\begin{figure}[t]
  \begin{center}
    \begin{minipage}{0.48\textwidth}
    \centering
    \includegraphics[width=0.95\textwidth]{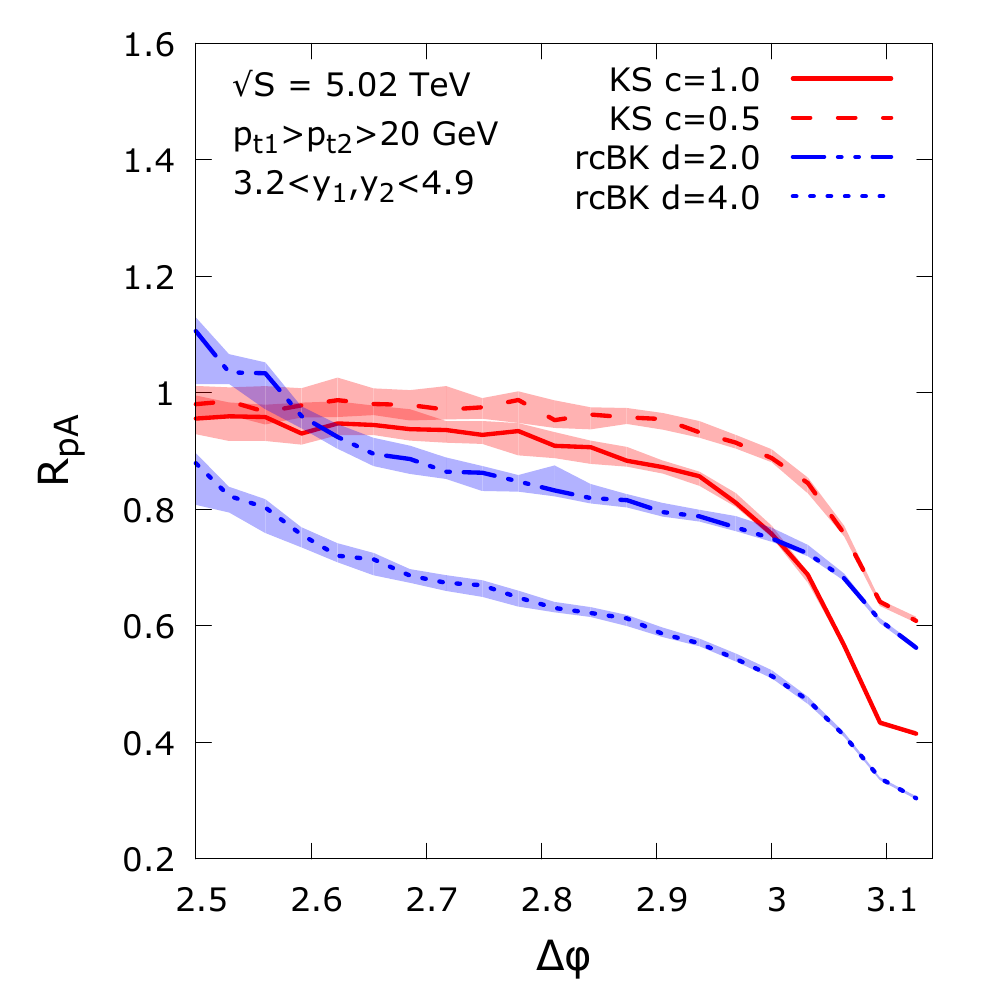}
    \caption{
    Nuclear modification ratios for azimuthal decorrelations in the
    forward-forward dijet production. The predictions were obtained within HEF.
    See text for details.  Figure from Ref.~\cite{vanHameren:2014lna}.
    }
    \label{fig:RpA-fwd-fwd}
    \end{minipage}
    \hfill
    \begin{minipage}{0.48\textwidth}
    \centering
      \includegraphics[width=0.95\textwidth]{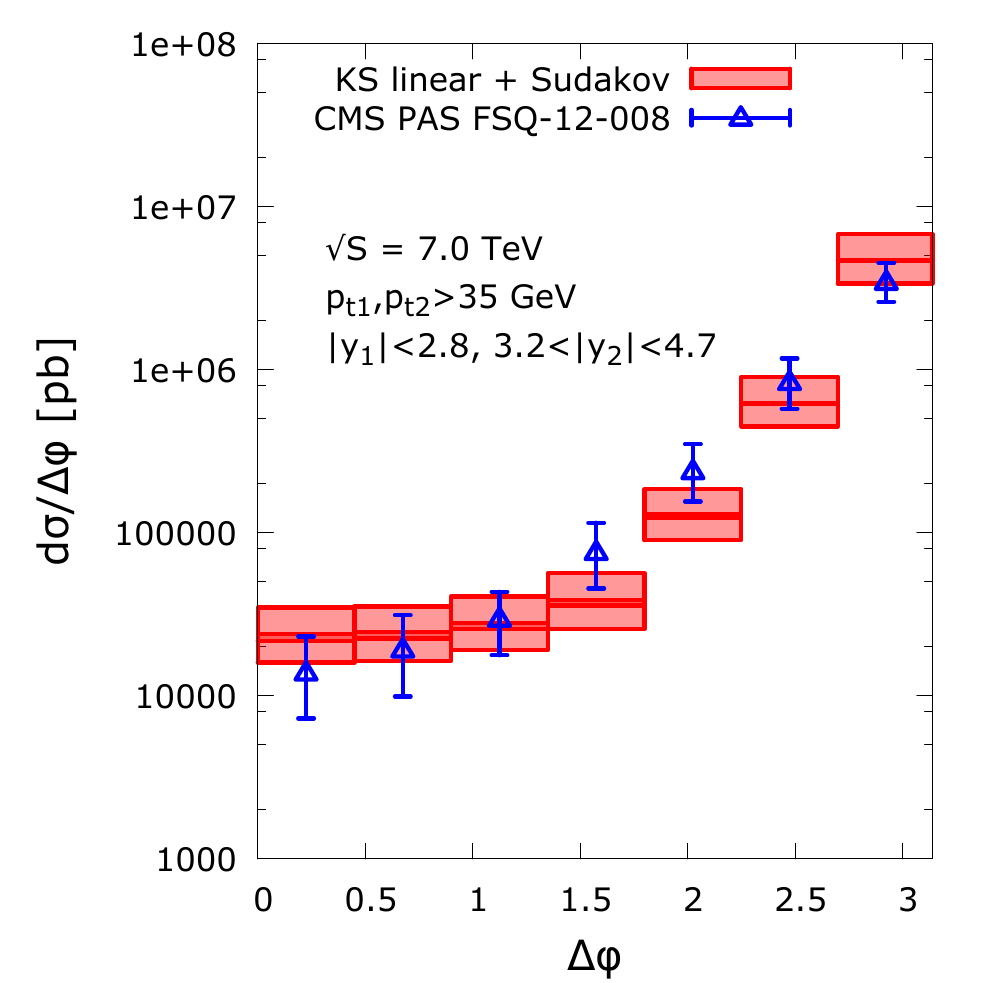}
    \caption{
    Azimuthal decorrelation in central-forward dijet production at the LHC, 7
    TeV, compared to theoretical predictions obtained within high energy
    factorization.
    Figure from Ref. \cite{vanHameren:2014ala}.
    }
    \label{fig:cms-ctr-fwd-decorr}
    \end{minipage}
  \end{center}
\end{figure}

According to Eq.~(\ref{eq:ktglue}), the region of $\Delta \phi \sim \pi$, which
corresponds to the nearly back-to-back dijet configurations, probes the gluon
distributions at very low transverse momenta. If, at the same time, the two jets
are produced in the forward direction, most of the contribution to the cross
section comes from the low-$x$ region. The corner of phase space with low $x$
and low $k_T$ is a domain of a non-linear QCD evolution and the related
phenomenon of saturation~\cite{Gribov:1984tu, Balitsky:1995ub,
Kovchegov:1999yj}. Hence, azimuthal decorrelation in the forward-forward dijet
production is a very promising observable to test this extreme regime of QCD.

Because the saturation scale  is proportional to $A^{1/3}$, where $A$ is the
atomic number of the target, the nonlinear regime of the gluon density is reached earlier in heavy
ions. By forming a ratio of the azimuthal decorrelations in the proton-lead and
proton-proton
collisions, we are able to obtain a robust signature of gluon saturation in
the forward dijet production.
Fig.~\ref{fig:RpA-fwd-fwd} shows such ratios, $R_{pA} = 1/A\,
(d\sigma^{pA}/d\Delta \phi)/(d\sigma^{pp}/d\Delta \phi)$, for several scenarios
with different unintegrated gluons (KS~\cite{Kutak:2012rf} or
rcBK~\cite{Albacete:2010sy}) and different parameters entering the modeling of
the transition from the proton, whose unintegrated PDFs were fitted to HERA
data, to lead.
We see that, regardless of the model, strong suppression of the $R_{pA}$ ratio
is predicted in the region $\Delta\phi \sim \pi$.

\begin{figure}[t]
  \begin{center}
    \includegraphics[width=0.48\textwidth]{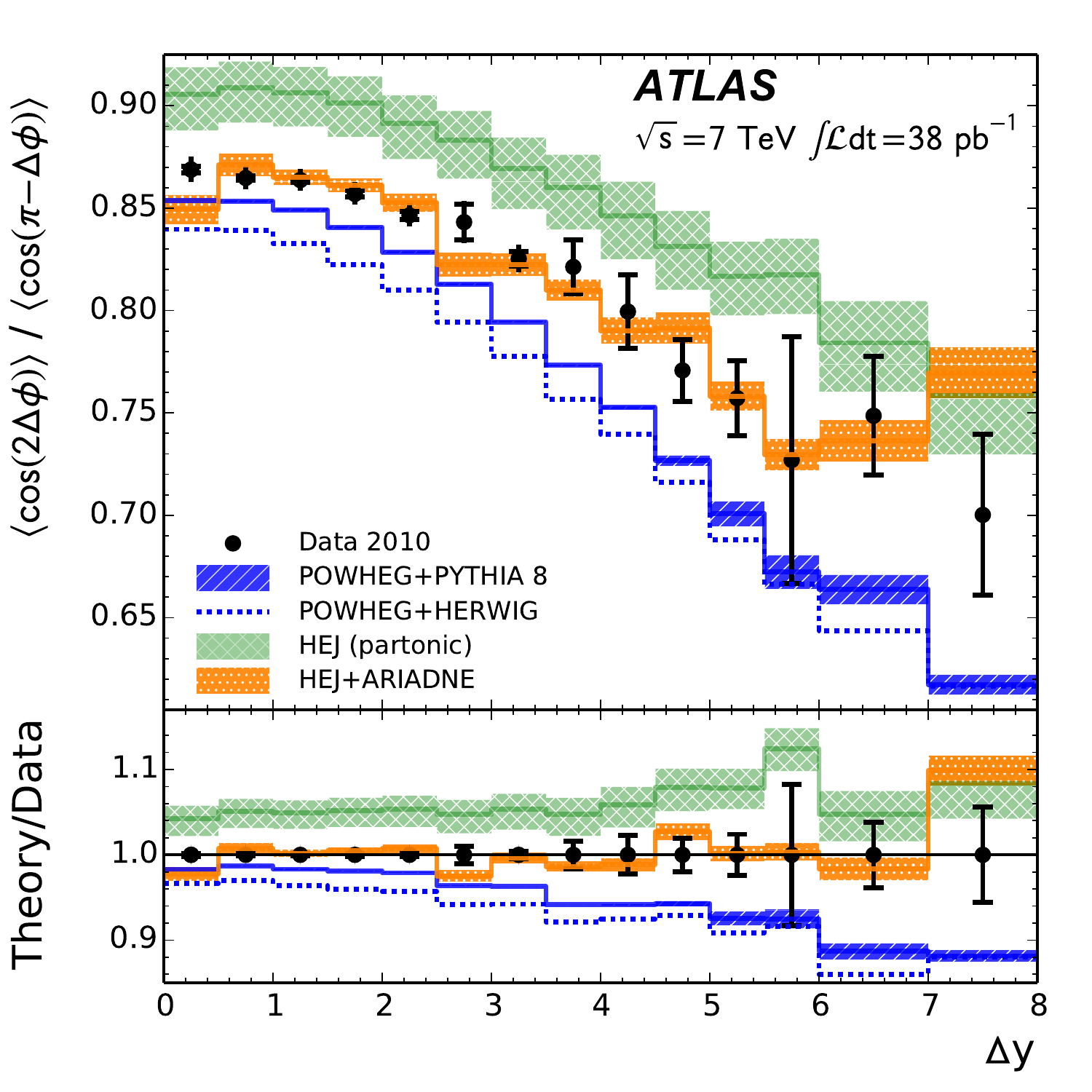}
    \hfill
    \includegraphics[width=0.48\textwidth]{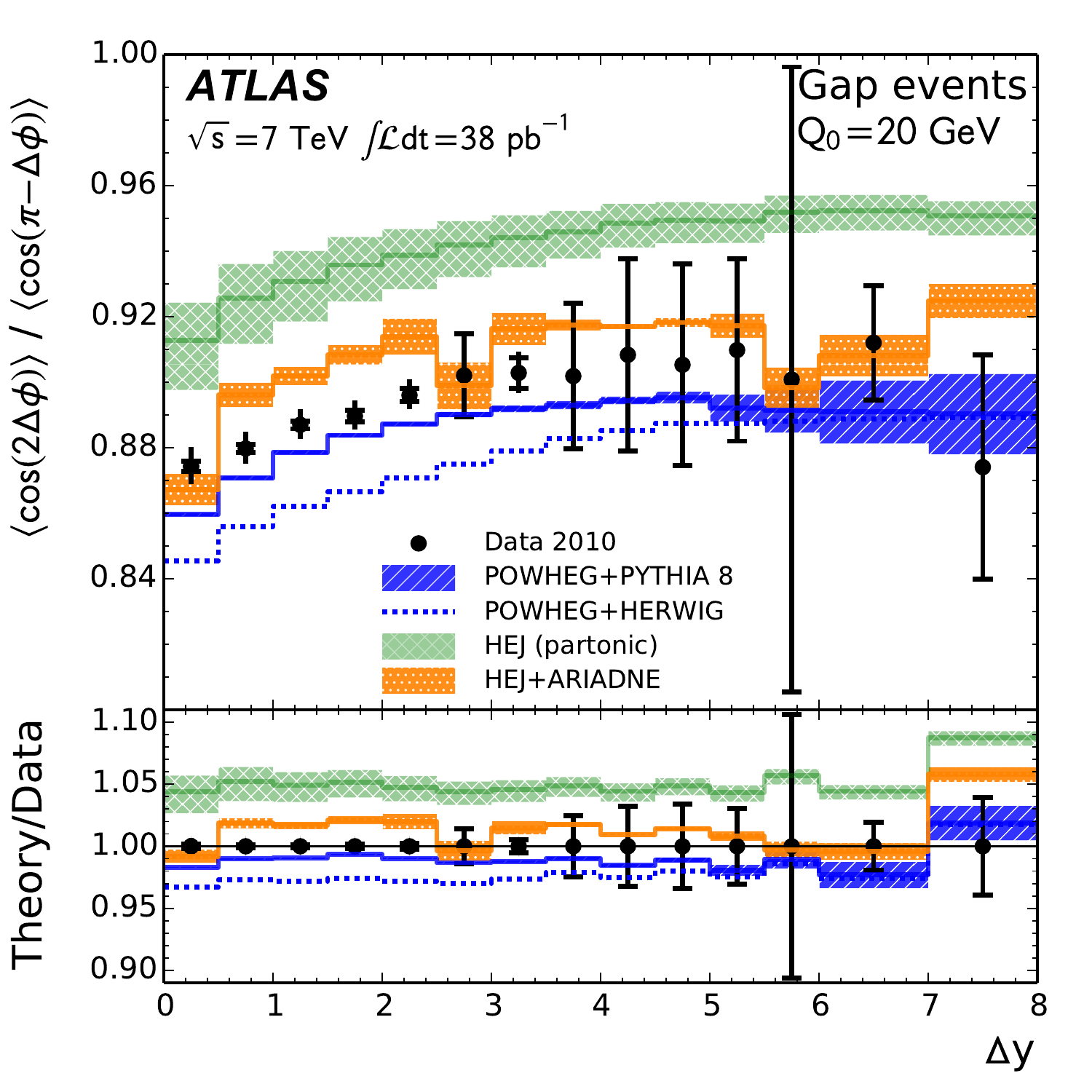}
  \end{center}
  \caption{
    Ratios of the second to the first moment of the azimuthal decorrelation
    distribution for the production of dijets separated by the rapidity
    distance $\Delta y$. The left plot corresponds to inclusive dijet production
    whereas on the right hand side no jet activity with $p_T$ above 20 GeV is
    allowed in the gap between the two leading jets.
    Figures from Ref.~\cite{Aad:2014pua}.
  }
  \label{fig:large-rapdity-jets}
\end{figure}

Azimuthal decorrelation has been studied so far by CMS in the central-forward
dijet production in Ref.~\cite{CMS:2014oma}. 
There, the data were compared to various theoretical predictions.
The latter were obtained with the standard Monte Carlo programs, where
decorrelation is generated by the parton shower, superimposed on top of the
$2\to 2$ hard scattering. The final-state partons are then hadronized and the
predictions include in addition contributions from the underlying event and
multi-parton interactions. The MC models tend to describe the data within
systematic uncertainties.
 
In Fig.~\ref{fig:cms-ctr-fwd-decorr}, the same data are
compared to predictions from the HEF formalism, with the KS gluon supplemented
by the hard scale dependence by means of reweighting the gluon distribution with
the Sudakov form factor~\cite{vanHameren:2014ala}. 
The predictions describe the CMS data comparably to the LL MCs studied in
Ref.~\cite{CMS:2014oma}.
This is an important observation as the two frameworks, HEF and LL MCs, are
constructed with very different sets of approximations. In HEF, the transverse
momentum of the dijet pair, $k_T$, is generated as a result of evolution of the
unintegrated gluon distribution function, and its value enters the off-shell
matrix element for the $2\to 2$ process.
However, the LL MCs are built in the framework of the collinear factorization,
hence the incoming partons have zero transverse momenta. In this case, the $k_T$
of the dijet pair comes from the subsequent, initial-state radiation~(ISR),
which is part of the hard process and not the parton distributions.
This radiation, generated in the parton shower approximation, effectively
provides parts of contributions coming from higher multiplicities.
At the end, however, the two frameworks, albeit through different means, realize
the same task of generation of the ISR emissions, hence they are supposed to
grasp the same physics. Therefore, it is not surprising that the same features
are seen in the distributions obtained in HEF and in the collinear factorization
supplemented with a parton shower. What is non-trivial in the context of
Fig.~\ref{fig:cms-ctr-fwd-decorr} (and its analogue, Fig.~4 of
Ref.~\cite{CMS:2014oma}) is that the two frameworks provide predictions which
are quantitatively consistent.

It is also interesting to analyse the azimuthal decorrelation as a function of
the rapidity separation between the jets in the dijet system. 
When this separation is large, patterns of radiation are expected to be very
different in the DGLAP- and BFKL-based approaches, as in the former, the
emissions are ordered in the transverse momentum while in the latter, in
rapidity.
The normalized cross section, $1/\sigma \, d\sigma/d\Delta \phi$, can be
expanded in Fourier series and the corresponding coefficients are given by the
moments $\langle\cos\left(n(\pi-\Delta \phi)\right)\rangle$. The ratios of the
second and the first moment are shown in Fig.~\ref{fig:large-rapdity-jets} for
the inclusive case of dijet production~(left) and for the case where the hard
radiation is forbidden in the space between the two leading jets~(right).
Here, none of the jets is explicitly required to be forward. However, when the
rapidity separation, $\Delta y$, is sizable, one or both are likely to be
produced at large (positive or negative) rapidities.
We see that the inclusive case is well described by \hej with the \ariadne
parton shower, while \powheg fails to describe the data. However, with the gap
requirement, hence forbidding radiation between the jets, both \hej+\ariadne and
\powheg predictions lie close to the experimental points. This can be
interpreted as an indication that, in the inclusive case shown in
Fig.~\ref{fig:large-rapdity-jets}~(left), the wide-angle, multi-jet radiation,
created abundantly in the space between two well-separated jets, is modeled
better by the approach based on BFKL evolution.

A related study has been performed by CMS~\cite{CMS:2013eda}, where  a
similar ratio is well described by analytic calculation within NLL
BFKL~\cite{Ducloue:2013wmi}. 
However, both in the ATLAS~\cite{Aad:2014pua} and
in the CMS~\cite{CMS:2013eda} studies, some aspects of the data are also well
reproduced by DGLAP-based approaches hence, no firm conclusion can be drawn at
this point.
A detailed analysis of dijet production with a veto on central jet
activity has been also published by ATLAS in Ref.~\cite{Aad:2011jz}, where the
predictions from \hej and \powheg are extensively tested against the data.

We conclude by noting that the range of theoretical frameworks used to describe
the forward jet production, and briefly mentioned above, differ considerably
both in the grasped aspects of physics of these class of processes, as well as
in the level of theoretical sophistication. Many of them are of LO or LO+LL
accuracy and, in some cases, jets are approximated only by a single, leading
particle. This certainly leaves a room for improvements, which, amongst other
things, would lead to more realistic modeling of the final states that could
be in turn analysed with the fully general jet finding techniques discussed in
Section~\ref{sec:jet-definitions-and-properties}.

%-----------------------------------------------------------------------------
\section{Summary and outlook}

Jets are omnipresent at hadron colliders and physics of jets is by now a
well-grounded and highly developed area of research.
That allows for precise and quantitative discussion and this review tried to
make an account of the concepts and results that are vital to studies of jet
processes.

The LHC opened a new chapter in jet physics through the exclusive use of
the infrared and collinear-safe jet definitions by all its experiments.
That laid a foundation for accurate measurements and
provided strong motivation for theoretical efforts aiming at calculating the
higher order corrections to jet processes.

We have discussed some of the advances in controlling jets and understanding
their properties.
We saw that individual jets produced in hadron-hadron collisions differ in many
ways and that diversity should be turned to our advantage. 
It is therefore advisable to choose jet definitions based on specificities of
the analyses one is interested to perform, such as: expected backgrounds,
final state cuts and potential contamination from incoherent radiation.
Recent years have seen many developments in the area of jet-based analyses
methods, especially those related to the substructure techniques, treated only
briefly in this review. These results allow for even broader use of jets in 
precise test of the Standard Model and searches for new physics.
A truly impressive progress has been also made in calculating the perturbative
predictions for the processes with jet production. We have discussed the highest
precision, fixed-order, results, which, in the recent years, have been pushed in
many cases to the NNLO accuracy or were supplemented by electro-weak
corrections. 
We have also extensively elaborated on techniques for matching the NLO and the
NNLO results to a parton shower or merging the NLO results with different
multiplicities.  Those methods allow one to achieve predictions of the highest
accuracy simultaneously for the inclusive and for a range of exclusive
observables.
All of the above advances went hand-in-hand with proposals of genuinely new
calculational techniques and developments of efficient numerical tools.

The overall description of the experimental results for jet processes with the
state-of-the-art theoretical predictions is very good. And it spans across
numerous distributions and many orders of magnitude.
However, there are still challenges and open questions that need to be
addressed in order to push the precision of jet physics to the next level.
Throughout the article, we have mentioned various sources that contribute to the
theoretical and experimental uncertainties. The latter type consists of errors
related to jet energy scale determination, luminosity measurements,
unfolding and pileup.
Theoretical uncertainties arise because of truncation of the perturbative
series, limited accuracy of the parton showers, missing electroweak
contributions, ambiguities in PDF determinations, including those related to the
choice of factorization scheme, as well as potential factorization breaking
effects.
Many of the above uncertainties are being dealt with by systematically adding
further terms with higher powers of the strong and electroweak couplings and
by improving logarithmic accuracy of the showers and resummations.
 
Finally, jets are also produced in special regions of phase space or with 
strong cuts on their momenta.
Such cases are very interesting, as they allow us to stretch tests of QCD to
extreme corners. 
At the same time, they are challenging to model theoretically because 
of inapplicability of the standard frameworks.
Therefore, we have discussed at length the delicate issue of factorization in
jet production processes, reporting on the recent developments and indicating 
promising future directions.
We have also pointed to potential measurements that could help to pin down the
effects going beyond the standard description based on the collinear
factorization and DGLAP evolution such as BFKL-type radiation and non-linear
effects inside the initial-state hadrons.

%-----------------------------------------------------------------------------
\section*{\normalsize Acknowledgments}

I am particularly indebted to all my collaborators, with whom some of the
results presented in this review have been obtained:
Marcin Bury, 
Matteo Cacciari, 
Francisco Campanario, 
Andreas van Hameren, 
Stanis\l{}aw Jadach,
Piotr Kotko, 
Krzysztof Kutak, 
Daniel Ma\^itre, 
Cyrille Marquet, 
Elena Petreska, 
Wies\l{}aw P\l{}aczek, 
Paloma Quiroga-Arias, 
Michael Rauch, 
Mathieu Rubin, 
Gavin Salam, 
Andrzej Si\'odmok,
Maciej Skrzypek
and Qi Cheng Zhang.
I acknowledge clarifying discussions with Ren\'e \'Angeles-Mart\'inez and
Piet Mulders.
I am grateful to
Simone Alioli,
Jeppe Andresen, 
Francisco Campanario, 
Stanis\l{}aw Jadach,
Krzysztof Kutak, 
Elena Petreska, 
Wies\l{}aw P\l{}aczek, 
Stefan Prestel,
Emanuele Re,
Robin Roth,
Marek Schoenherr
and Jennifer Smillie
for thorough reading of various parts of the manuscript and for numerous comments
that helped to improve the material presented in this review.
Finally, I would like to thank the authors and publishers who agreed to use
their figures in the body of this work.
Feynman diagrams were generated with \jaxodraw~\cite{Binosi:2008ig}.


\begin{thebibliography}{100}

\bibitem{Ellis:2007ib}
S.D. Ellis, J.~Huston, K.~Hatakeyama, P.~Loch, and M.~Tonnesmann.
\newblock {\em Prog.Part.Nucl.Phys.}, 60:484--551, 2008.

\bibitem{Aaltonen:2008eq}
T.~Aaltonen et~al.
\newblock {\em Phys. Rev.}, D78:052006, 2008.
\newblock [Erratum: Phys. Rev.D79,119902(2009)].

\bibitem{Abazov:2011vi}
Victor~Mukhamedovich Abazov et~al.
\newblock {\em Phys. Rev.}, D85:052006, 2012.

\bibitem{Aad:2010ad}
G.~Aad et~al.
\newblock {\em Eur.Phys.J.}, C71:1512, 2011.

\bibitem{Aad:2011fc}
Georges Aad et~al.
\newblock {\em Phys.Rev.}, D86:014022, 2012.

\bibitem{Chatrchyan:2011qta}
Serguei Chatrchyan et~al.
\newblock {\em Phys.Lett.}, B700:187--206, 2011.

\bibitem{CMS:2011ab}
Serguei Chatrchyan et~al.
\newblock {\em Phys.Rev.Lett.}, 107:132001, 2011.

\bibitem{Chatrchyan:2012gwa}
Serguei Chatrchyan et~al.
\newblock {\em JHEP}, 1206:036, 2012.

\bibitem{Chatrchyan:2012bja}
Serguei Chatrchyan et~al.
\newblock {\em Phys.Rev.}, D87(11):112002, 2013.

\bibitem{Malaescu:2012ts}
Bogdan Malaescu and Pavel Starovoitov.
\newblock {\em Eur.Phys.J.}, C72:2041, 2012.

\bibitem{Rojo:2014kta}
Juan Rojo.
\newblock {\em Int. J. Mod. Phys.}, A30:1546005, 2015.

\bibitem{Voutilainen:2015lqa}
Mikko Voutilainen.
\newblock {\em Int. J. Mod. Phys.}, A30(31):1546008, 2015.

\bibitem{Ball:2014uwa}
Richard~D. Ball et~al.
\newblock {\em JHEP}, 04:040, 2015.

\bibitem{Dulat:2015mca}
Sayipjamal Dulat, Tie-Jiun Hou, Jun Gao, Marco Guzzi, Joey Huston, Pavel
  Nadolsky, Jon Pumplin, Carl Schmidt, Daniel Stump, and C.~P. Yuan.
\newblock {\em Phys. Rev.}, D93(3):033006, 2016.

\bibitem{Harland-Lang:2014zoa}
L.~A. Harland-Lang, A.~D. Martin, P.~Motylinski, and R.~S. Thorne.
\newblock {\em Eur. Phys. J.}, C75(5):204, 2015.

\bibitem{Sterman:2014nua}
George Sterman.
\newblock {\em Acta Phys.Polon.}, B45:2205, 2014.

\bibitem{Gelis:2010nm}
Francois Gelis, Edmond Iancu, Jamal Jalilian-Marian, and Raju Venugopalan.
\newblock {\em Ann. Rev. Nucl. Part. Sci.}, 60:463--489, 2010.

\bibitem{Campanelli:2015oaa}
Mario Campanelli.
\newblock {\em Int. J. Mod. Phys.}, A30(31):1546006, 2015.

\bibitem{Chatrchyan:2013qha}
Serguei Chatrchyan et~al.
\newblock {\em Phys.Rev.}, D87(11):114015, 2013.

\bibitem{Aad:2014aqa}
Georges Aad et~al.
\newblock {\em Phys. Rev.}, D91(5):052007, 2015.

\bibitem{Aad:2011xw}
Georges Aad et~al.
\newblock {\em Phys. Lett.}, B705:294--312, 2013.

\bibitem{Khachatryan:2014rra}
Vardan Khachatryan et~al.
\newblock {\em Eur. Phys. J.}, C75(5):235, 2015.

\bibitem{Aad:2014vwa}
Georges Aad et~al.
\newblock {\em JHEP}, 1502:153, 2015.

\bibitem{Aad:2012tfa}
Georges Aad et~al.
\newblock {\em Phys. Lett.}, B716:1--29, 2013.

\bibitem{Chatrchyan:2012xdj}
Serguei Chatrchyan et~al.
\newblock {\em Phys. Lett.}, B716:30--61, 2012.

\bibitem{ATLAS:2012pu}
Georges Aad et~al.
\newblock {\em JHEP}, 1301:029, 2013.

\bibitem{Aad:2010bu}
Georges Aad et~al.
\newblock {\em Phys.Rev.Lett.}, 105:252303, 2010.

\bibitem{Sterman:1977wj}
George~F. Sterman and Steven Weinberg.
\newblock {\em Phys.Rev.Lett.}, 39:1436, 1977.

\bibitem{Salam:2009jx}
Gavin~P. Salam.
\newblock {\em Eur.Phys.J.}, C67:637--686, 2010.

\bibitem{Altheimer:2013yza}
A.~Altheimer et~al.
\newblock {\em Eur. Phys. J.}, C74(3):2792, 2014.

\bibitem{Adams:2015hiv}
D.~Adams et~al.
\newblock {\em Eur. Phys. J.}, C75(9):409, 2015.

\bibitem{Francavilla:2015yxa}
Paolo Francavilla.
\newblock {\em Submitted to: Int. J. Mod. Phys.}, 2015.

\bibitem{Kokkas:2015gfa}
Panagiotis Kokkas.
\newblock {\em Int. J. Mod. Phys.}, A30(31):1546004, 2015.

\bibitem{Cacciari:2015jwa}
Matteo Cacciari.
\newblock {\em Int. J. Mod. Phys.}, A30(31):1546001, 2015.

\bibitem{Abreu:2007kv}
N.~Armesto, N.~Borghini, S.~Jeon, U.~A. Wiedemann, S.~Abreu, V.~Akkelin,
  J.~Alam, J.~L. Albacete, A.~Andronic, D.~Antonov, et~al.
\newblock {\em J. Phys.}, G35:054001, 2008.

\bibitem{Mehtar-Tani:2013pia}
Yacine Mehtar-Tani, Jose~Guilherme Milhano, and Konrad Tywoniuk.
\newblock {\em Int. J. Mod. Phys.}, A28:1340013, 2013.

\bibitem{Armesto:2015ioy}
N.~Armesto and E.~Scomparin.
\newblock {\em Eur. Phys. J. Plus}, 131(3):52, 2016.

\bibitem{Qin:2015srf}
Guang-You Qin and Xin-Nian Wang.
\newblock {\em Int. J. Mod. Phys.}, E24(11):1530014, 2015.

\bibitem{Cacciari:2010te}
Matteo Cacciari, Juan Rojo, Gavin~P. Salam, and Gregory Soyez.
\newblock {\em Eur. Phys. J.}, C71:1539, 2011.

\bibitem{Catani:1993hr}
S.~Catani, Yuri~L. Dokshitzer, M.~H. Seymour, and B.~R. Webber.
\newblock {\em Nucl. Phys.}, B406:187--224, 1993.

\bibitem{Ellis:1993tq}
Stephen~D. Ellis and Davison~E. Soper.
\newblock {\em Phys. Rev.}, D48:3160--3166, 1993.

\bibitem{Dokshitzer:1997in}
Yuri~L. Dokshitzer, G.~D. Leder, S.~Moretti, and B.~R. Webber.
\newblock {\em JHEP}, 08:001, 1997.

\bibitem{Wobisch:1998wt}
M.~Wobisch and T.~Wengler.
\newblock In {\em {Monte Carlo generators for HERA physics. Proceedings,
  Workshop, Hamburg, Germany, 1998-1999}}, 1998.

\bibitem{Salam:2007xv}
Gavin~P. Salam and Gregory Soyez.
\newblock {\em JHEP}, 0705:086, 2007.

\bibitem{Cacciari:2005hq}
Matteo Cacciari and Gavin~P. Salam.
\newblock {\em Phys. Lett.}, B641:57--61, 2006.

\bibitem{Cacciari:2008gp}
Matteo Cacciari, Gavin~P. Salam, and Gregory Soyez.
\newblock {\em JHEP}, 04:063, 2008.

\bibitem{Kinoshita:1962ur}
T.~Kinoshita.
\newblock {\em J. Math. Phys.}, 3:650--677, 1962.

\bibitem{Lee:1964is}
T.~D. Lee and M.~Nauenberg.
\newblock {\em Phys. Rev.}, 133:B1549--B1562, 1964.

\bibitem{Cacciari:2011ma}
Matteo Cacciari, Gavin~P. Salam, and Gregory Soyez.
\newblock {\em Eur. Phys. J.}, C72:1896, 2012.

\bibitem{QuirogaArias:2012nj}
Paloma Quiroga-Arias and Sebastian Sapeta.
\newblock {\em Int. J. Mod. Phys.}, A28:1350087, 2013.

\bibitem{Almeida:2008tp}
Leandro~G. Almeida, Seung~J. Lee, Gilad Perez, Ilmo Sung, and Joseph Virzi.
\newblock {\em Phys. Rev.}, D79:074012, 2009.

\bibitem{Cacciari:2008gn}
Matteo Cacciari, Gavin~P. Salam, and Gregory Soyez.
\newblock {\em JHEP}, 0804:005, 2008.

\bibitem{Sapeta:2010uk}
Sebastian Sapeta and Qi~Cheng Zhang.
\newblock {\em JHEP}, 06:038, 2011.

\bibitem{Sjostrand:2006za}
Torbjorn Sjostrand, Stephen Mrenna, and Peter~Z. Skands.
\newblock {\em JHEP}, 05:026, 2006.

\bibitem{Seymour:1993mx}
Michael~H. Seymour.
\newblock {\em Z. Phys.}, C62:127--138, 1994.

\bibitem{Butterworth:2002tt}
J.~M. Butterworth, B.~E. Cox, and Jeffrey~R. Forshaw.
\newblock {\em Phys. Rev.}, D65:096014, 2002.

\bibitem{Butterworth:2007ke}
J.~M. Butterworth, John~R. Ellis, and A.~R. Raklev.
\newblock {\em JHEP}, 05:033, 2007.

\bibitem{Butterworth:2008iy}
Jonathan~M. Butterworth, Adam~R. Davison, Mathieu Rubin, and Gavin~P. Salam.
\newblock {\em Phys. Rev. Lett.}, 100:242001, 2008.

\bibitem{Rubin:2010fc}
Mathieu Rubin.
\newblock {\em JHEP}, 05:005, 2010.

\bibitem{Dasgupta:2013ihk}
Mrinal Dasgupta, Alessandro Fregoso, Simone Marzani, and Gavin~P. Salam.
\newblock {\em JHEP}, 09:029, 2013.

\bibitem{Ellis:2009su}
Stephen~D. Ellis, Christopher~K. Vermilion, and Jonathan~R. Walsh.
\newblock {\em Phys. Rev.}, D80:051501, 2009.

\bibitem{Ellis:2009me}
Stephen~D. Ellis, Christopher~K. Vermilion, and Jonathan~R. Walsh.
\newblock {\em Phys. Rev.}, D81:094023, 2010.

\bibitem{Krohn:2009th}
David Krohn, Jesse Thaler, and Lian-Tao Wang.
\newblock {\em JHEP}, 02:084, 2010.

\bibitem{Thaler:2010tr}
Jesse Thaler and Ken Van~Tilburg.
\newblock {\em JHEP}, 03:015, 2011.

\bibitem{Thaler:2011gf}
Jesse Thaler and Ken Van~Tilburg.
\newblock {\em JHEP}, 02:093, 2012.

\bibitem{Soper:2010xk}
Davison~E. Soper and Michael Spannowsky.
\newblock {\em JHEP}, 08:029, 2010.

\bibitem{Feige:2012vc}
Ilya Feige, Matthew~D. Schwartz, Iain~W. Stewart, and Jesse Thaler.
\newblock {\em Phys. Rev. Lett.}, 109:092001, 2012.

\bibitem{Dasgupta:2013via}
Mrinal Dasgupta, Alessandro Fregoso, Simone Marzani, and Alexander Powling.
\newblock {\em Eur. Phys. J.}, C73(11):2623, 2013.

\bibitem{Dasgupta:2015yua}
Mrinal Dasgupta, Alexander Powling, and Andrzej Siodmok.
\newblock {\em JHEP}, 08:079, 2015.

\bibitem{Larkoski:2015kga}
Andrew~J. Larkoski, Ian Moult, and Duff Neill.
\newblock {\em JHEP}, 05:117, 2016.

\bibitem{Ellis:1978ty}
R.~Keith Ellis, Howard Georgi, Marie Machacek, H.~David Politzer, and Graham~G.
  Ross.
\newblock {\em Nucl. Phys.}, B152:285, 1979.

\bibitem{Bodwin:1981fv}
Geoffrey~T. Bodwin, Stanley~J. Brodsky, and G.~Peter Lepage.
\newblock {\em Phys.Rev.Lett.}, 47:1799, 1981.

\bibitem{Collins:1981uw}
John~C. Collins and Davison~E. Soper.
\newblock {\em Nucl. Phys.}, B194:445, 1982.

\bibitem{Collins:1981ta}
John~C. Collins and George~F. Sterman.
\newblock {\em Nucl. Phys.}, B185:172, 1981.

\bibitem{Collins:1985ue}
John~C. Collins, Davison~E. Soper, and George~F. Sterman.
\newblock {\em Nucl.Phys.}, B261:104, 1985.

\bibitem{Collins:1988ig}
John~C. Collins, Davison~E. Soper, and George~F. Sterman.
\newblock {\em Nucl.Phys.}, B308:833, 1988.

\bibitem{Collins:1989gx}
John~C. Collins, Davison~E. Soper, and George~F. Sterman.
\newblock {\em Adv.Ser.Direct.High Energy Phys.}, 5:1--91, 1988.

\bibitem{collins:book}
John Collins.
\newblock {\em Foundations of Perturbative QCD}.
\newblock Cambridge University Press, 2011.

\bibitem{Diehl:2015bca}
Markus Diehl, Jonathan~R. Gaunt, Daniel Ostermeier, Peter Plößl, and Andreas
  Schäfer.
\newblock {\em JHEP}, 01:076, 2016.

\bibitem{Catani:2011st}
Stefano Catani, Daniel de~Florian, and German Rodrigo.
\newblock {\em JHEP}, 1207:026, 2012.

\bibitem{Forshaw:2012bi}
Jeffrey~R. Forshaw, Michael~H. Seymour, and Andrzej Siodmok.
\newblock {\em JHEP}, 11:066, 2012.

\bibitem{Fleming:2014rea}
Sean Fleming.
\newblock {\em Phys. Lett.}, B735:266--271, 2014.

\bibitem{Sterman:1978bi}
George~F. Sterman.
\newblock {\em Phys. Rev.}, D17:2773, 1978.

\bibitem{Sterman:1978bj}
George~F. Sterman.
\newblock {\em Phys. Rev.}, D17:2789, 1978.

\bibitem{Curci:1980uw}
G.~Curci, W.~Furmanski, and R.~Petronzio.
\newblock {\em Nucl. Phys.}, B175:27, 1980.

\bibitem{Altarelli:1977zs}
Guido Altarelli and G.~Parisi.
\newblock {\em Nucl.Phys.}, B126:298, 1977.

\bibitem{Gribov:1972ri}
V.N. Gribov and L.N. Lipatov.
\newblock {\em Sov.J.Nucl.Phys.}, 15:438--450, 1972.

\bibitem{Dokshitzer:1977sg}
Yuri~L. Dokshitzer.
\newblock {\em Sov.Phys.JETP}, 46:641--653, 1977.

\bibitem{Forshaw:2006fk}
Jeffrey~R. Forshaw, A.~Kyrieleis, and M.~H. Seymour.
\newblock {\em JHEP}, 08:059, 2006.

\bibitem{Forshaw:2008cq}
J.~R. Forshaw, A.~Kyrieleis, and M.~H. Seymour.
\newblock {\em JHEP}, 09:128, 2008.

\bibitem{Angeles-Martinez:2015rna}
René Ángeles Martínez, Jeffrey~R. Forshaw, and Michael~H. Seymour.
\newblock {\em JHEP}, 12:091, 2015.

\bibitem{Bardeen:1978yd}
William~A. Bardeen, A.~J. Buras, D.~W. Duke, and T.~Muta.
\newblock {\em Phys. Rev.}, D18:3998, 1978.

\bibitem{Butterworth:2015oua}
Jon Butterworth et~al.
\newblock {\em J. Phys.}, G43:023001, 2016.

\bibitem{deOliveira:2013iya}
E.~G. Oliveira, A.~D. Martin, and M.~G. Ryskin.
\newblock {\em JHEP}, 11:156, 2013.

\bibitem{Jadach:2011cr}
S.~Jadach, A.~Kusina, W.~Placzek, M.~Skrzypek, and M.~Slawinska.
\newblock {\em Phys. Rev.}, D87(3):034029, 2013.

\bibitem{Jadach:2015mza}
S.~Jadach, W.~Płaczek, S.~Sapeta, A.~Siódmok, and M.~Skrzypek.
\newblock {\em JHEP}, 10:052, 2015.

\bibitem{deOliveira:2012qa}
E.~G. de~Oliveira, A.~D. Martin, and M.~G. Ryskin.
\newblock {\em JHEP}, 02:060, 2013.

\bibitem{Frixione:2002ik}
Stefano Frixione and Bryan~R. Webber.
\newblock {\em JHEP}, 06:029, 2002.

\bibitem{Nason:2004rx}
Paolo Nason.
\newblock {\em JHEP}, 11:040, 2004.

\bibitem{esw:book}
R.~K. Ellis, W.~J. Stirling, and Webber~B. R.
\newblock {\em QCD and Collider Physics}.
\newblock Cambridge University Press, 2003.

\bibitem{deOliveira:2013tya}
E.~G. de~Oliveira, A.~D. Martin, M.~G. Ryskin, and A.~G. Shuvaev.
\newblock {\em Eur. Phys. J.}, C73(10):2616, 2013.

\bibitem{Bonvini:2015ira}
Marco Bonvini, Simone Marzani, Juan Rojo, Luca Rottoli, Maria Ubiali,
  Richard~D. Ball, Valerio Bertone, Stefano Carrazza, and Nathan~P. Hartland.
\newblock {\em JHEP}, 09:191, 2015.

\bibitem{Collins:1984kg}
John~C. Collins, Davison~E. Soper, and George~F. Sterman.
\newblock {\em Nucl. Phys.}, B250:199, 1985.

\bibitem{Angeles-Martinez:2015sea}
R.~Angeles-Martinez et~al.
\newblock {\em Acta Phys. Polon.}, B46(12):2501--2534, 2015.

\bibitem{Rogers:2015sqa}
Ted~C. Rogers.
\newblock {\em Eur. Phys. J.}, A52(6):153, 2016.

\bibitem{Hautmann:2014kza}
F.~Hautmann, H.~Jung, M.~Krämer, P.~J. Mulders, E.~R. Nocera, T.~C. Rogers,
  and A.~Signori.
\newblock {\em Eur. Phys. J.}, C74:3220, 2014.

\bibitem{Boer:2003cm}
Daniel Boer, P.J. Mulders, and F.~Pijlman.
\newblock {\em Nucl.Phys.}, B667:201--241, 2003.

\bibitem{Bodwin:1984hc}
Geoffrey~T. Bodwin.
\newblock {\em Phys.Rev.}, D31:2616, 1985.

\bibitem{Bodwin:1984hcERR}
Geoffrey~T. Bodwin.
\newblock {\em Phys.Rev.}, D34:3932, 1986.

\bibitem{Aybat:2008ct}
S.~Mert Aybat and George~F. Sterman.
\newblock {\em Phys. Lett.}, B671:46--50, 2009.

\bibitem{Bomhof:2007xt}
C.~J. Bomhof and Piet~J. Mulders.
\newblock {\em Nucl. Phys.}, B795:409--427, 2008.

\bibitem{Belitsky:2002sm}
Andrei~V. Belitsky, X.~Ji, and F.~Yuan.
\newblock {\em Nucl.Phys.}, B656:165--198, 2003.

\bibitem{Collins:2002kn}
John~C. Collins.
\newblock {\em Phys.Lett.}, B536:43--48, 2002.

\bibitem{Sivers:1989cc}
Dennis~W. Sivers.
\newblock {\em Phys. Rev.}, D41:83, 1990.

\bibitem{Boer:1997nt}
Daniel Boer and P.~J. Mulders.
\newblock {\em Phys. Rev.}, D57:5780--5786, 1998.

\bibitem{Collins:1992kk}
John~C. Collins.
\newblock {\em Nucl. Phys.}, B396:161--182, 1993.

\bibitem{Bomhof:2004aw}
C.J. Bomhof, P.J. Mulders, and F.~Pijlman.
\newblock {\em Phys.Lett.}, B596:277--286, 2004.

\bibitem{Bomhof:2006dp}
C.J. Bomhof, P.J. Mulders, and F.~Pijlman.
\newblock {\em Eur.Phys.J.}, C47:147--162, 2006.

\bibitem{Dominguez:2011wm}
Fabio Dominguez, Cyrille Marquet, Bo-Wen Xiao, and Feng Yuan.
\newblock {\em Phys. Rev.}, D83:105005, 2011.

\bibitem{Kotko:2015ura}
P.~Kotko, K.~Kutak, C.~Marquet, E.~Petreska, S.~Sapeta, and A.~van Hameren.
\newblock {\em JHEP}, 09:106, 2015.

\bibitem{Rogers:2010dm}
Ted~C. Rogers and Piet~J. Mulders.
\newblock {\em Phys.Rev.}, D81:094006, 2010.

\bibitem{Catani:1990xk}
S.~Catani, M.~Ciafaloni, and F.~Hautmann.
\newblock {\em Phys. Lett.}, B242:97, 1990.

\bibitem{Catani:1990eg}
S.~Catani, M.~Ciafaloni, and F.~Hautmann.
\newblock {\em Nucl. Phys.}, B366:135--188, 1991.

\bibitem{vanHameren:2012uj}
Andreas van Hameren, Piotr Kotko, and Krzysztof Kutak.
\newblock {\em JHEP}, 12:029, 2012.

\bibitem{vanHameren:2012if}
A.~van Hameren, P.~Kotko, and K.~Kutak.
\newblock {\em JHEP}, 01:078, 2013.

\bibitem{Deak:2009xt}
M.~Deak, F.~Hautmann, H.~Jung, and K.~Kutak.
\newblock {\em JHEP}, 09:121, 2009.

\bibitem{Kutak:2012rf}
Krzysztof Kutak and Sebastian Sapeta.
\newblock {\em Phys. Rev.}, D86:094043, 2012.

\bibitem{Deak:2010gk}
M.~Deak, F.~Hautmann, H.~Jung, and K.~Kutak.
\newblock arXiv:1012.6037, 2010.

\bibitem{vanHameren:2014lna}
A.~van Hameren, P.~Kotko, K.~Kutak, C.~Marquet, and S.~Sapeta.
\newblock {\em Phys. Rev.}, D89(9):094014, 2014.

\bibitem{vanHameren:2014ala}
A.~van Hameren, P.~Kotko, K.~Kutak, and S.~Sapeta.
\newblock {\em Phys. Lett.}, B737:335--340, 2014.

\bibitem{Mangano:1990by}
Michelangelo~L. Mangano and Stephen~J. Parke.
\newblock {\em Phys. Rept.}, 200:301--367, 1991.

\bibitem{Kotko:2016num}
A.~van Hameren, P.~Kotko, K.~Kutak, C.~Marquet, E.~Petreska, and S.~Sapeta.
\newblock arXiv:1607.03121, 2016.

\bibitem{Diehl:2011tt}
Markus Diehl and Andreas Schafer.
\newblock {\em Phys. Lett.}, B698:389--402, 2011.

\bibitem{Diehl:2011yj}
Markus Diehl, Daniel Ostermeier, and Andreas Schafer.
\newblock {\em JHEP}, 03:089, 2012.

\bibitem{Dasgupta:2007wa}
Mrinal Dasgupta, Lorenzo Magnea, and Gavin~P. Salam.
\newblock {\em JHEP}, 02:055, 2008.

\bibitem{Cacciari:2007fd}
Matteo Cacciari and Gavin~P. Salam.
\newblock {\em Phys. Lett.}, B659:119--126, 2008.

\bibitem{Cacciari:2009dp}
Matteo Cacciari, Gavin~P. Salam, and Sebastian Sapeta.
\newblock {\em JHEP}, 04:065, 2010.

\bibitem{Albrow:2006rt}
Michael~G. Albrow et~al.
\newblock volume arXiv:hep-ph/0610012, 2006.

\bibitem{Aad:2010fh}
Georges Aad et~al.
\newblock {\em Phys. Rev.}, D83:112001, 2011.

\bibitem{Aaltonen:2010rm}
T.~Aaltonen et~al.
\newblock {\em Phys. Rev.}, D82:034001, 2010.

\bibitem{Chatrchyan:2012tt}
Serguei Chatrchyan et~al.
\newblock {\em JHEP}, 08:130, 2012.

\bibitem{Aad:2010sp}
G.~Aad et~al.
\newblock {\em Phys. Rev.}, D83:052005, 2011.

\bibitem{Aad:2014eha}
Georges Aad et~al.
\newblock {\em Phys. Rev.}, D90(11):112015, 2014.

\bibitem{Aad:2014hia}
Georges Aad et~al.
\newblock {\em Eur. Phys. J.}, C74(8):2965, 2014.

\bibitem{Khachatryan:2010pv}
Vardan Khachatryan et~al.
\newblock {\em Eur. Phys. J.}, C70:555--572, 2010.

\bibitem{Chatrchyan:2011id}
Serguei Chatrchyan et~al.
\newblock {\em JHEP}, 09:109, 2011.

\bibitem{Aad:2011qe}
Georges Aad et~al.
\newblock {\em Eur. Phys. J.}, C71:1636, 2011.

\bibitem{ATL-PHYS-PUB-2015-019}
Technical Report ATL-PHYS-PUB-2015-019, CERN, Geneva, Jul 2015.

\bibitem{Chatrchyan:2013ala}
Serguei Chatrchyan et~al.
\newblock {\em Eur. Phys. J.}, C73(12):2674, 2013.

\bibitem{Skands:2014pea}
Peter Skands, Stefano Carrazza, and Juan Rojo.
\newblock {\em Eur. Phys. J.}, C74(8):3024, 2014.

\bibitem{Cacciari:2014gra}
Matteo Cacciari, Gavin~P. Salam, and Gregory Soyez.
\newblock {\em Eur.Phys.J.}, C75(2):59, 2015.

\bibitem{Alon:2011xb}
Raz Alon, Ehud Duchovni, Gilad Perez, Aliaksandr~P. Pranko, and Pekka~K.
  Sinervo.
\newblock {\em Phys. Rev.}, D84:114025, 2011.

\bibitem{Jankowiak:2012na}
Martin Jankowiak and Andrew~J. Larkoski.
\newblock {\em JHEP}, 04:039, 2012.

\bibitem{Jankowiak:2011qa}
Martin Jankowiak and Andrew~J. Larkoski.
\newblock {\em JHEP}, 06:057, 2011.

\bibitem{Krohn:2013lba}
David Krohn, Matthew~D. Schwartz, Matthew Low, and Lian-Tao Wang.
\newblock {\em Phys. Rev.}, D90(6):065020, 2014.

\bibitem{Soyez:2012hv}
Gregory Soyez, Gavin~P. Salam, Jihun Kim, Souvik Dutta, and Matteo Cacciari.
\newblock {\em Phys.Rev.Lett.}, 110(16):162001, 2013.

\bibitem{Berta:2014eza}
Peter Berta, Martin Spousta, David~W. Miller, and Rupert Leitner.
\newblock {\em JHEP}, 06:092, 2014.

\bibitem{Dasgupta:2009tm}
Mrinal Dasgupta and Yazid Delenda.
\newblock {\em JHEP}, 07:004, 2009.

\bibitem{Aad:2013tea}
Georges Aad et~al.
\newblock {\em JHEP}, 1405:059, 2014.

\bibitem{Aad:2014rma}
Georges Aad et~al.
\newblock arXiv:1411.1855, 2014.

\bibitem{Aad:2014wha}
Georges Aad et~al.
\newblock {\em Phys. Lett.}, B739:320--342, 2014.

\bibitem{Cacciari:2008gd}
Matteo Cacciari, Juan Rojo, Gavin~P. Salam, and Gregory Soyez.
\newblock {\em JHEP}, 12:032, 2008.

\bibitem{Soyez:2011np}
Gregory Soyez.
\newblock {\em Phys. Lett.}, B698:59--62, 2011.

\bibitem{Chatrchyan:2014gia}
Serguei Chatrchyan et~al.
\newblock {\em Phys. Rev.}, D90(7):072006, 2014.

\bibitem{Dasgupta:2014yra}
Mrinal Dasgupta, Frédéric Dreyer, Gavin~P. Salam, and Gregory Soyez.
\newblock 2014.

\bibitem{Aversa:1989xw}
F.~Aversa, Mario Greco, P.~Chiappetta, and J.~P. Guillet.
\newblock {\em Z. Phys.}, C46:253, 1990.

\bibitem{Ellis:1990ek}
Stephen~D. Ellis, Zoltan Kunszt, and Davison~E. Soper.
\newblock {\em Phys.Rev.Lett.}, 64:2121, 1990.

\bibitem{Ellis:1992en}
Stephen~D. Ellis, Zoltan Kunszt, and Davison~E. Soper.
\newblock {\em Phys.Rev.Lett.}, 69:1496--1499, 1992.

\bibitem{Giele:1994gf}
W.T. Giele, E.W.~Nigel Glover, and David~A. Kosower.
\newblock {\em Phys.Rev.Lett.}, 73:2019--2022, 1994.

\bibitem{Nagy:2001fj}
Zoltan Nagy.
\newblock {\em Phys.Rev.Lett.}, 88:122003, 2002.

\bibitem{Nagy:2003tz}
Zoltan Nagy.
\newblock {\em Phys.Rev.}, D68:094002, 2003.

\bibitem{Bern:2011ep}
Z.~Bern, G.~Diana, L.~J. Dixon, F.~Febres~Cordero, S.~Hoeche, D.~A. Kosower,
  H.~Ita, D.~Maitre, and K.~Ozeren.
\newblock {\em Phys. Rev. Lett.}, 109:042001, 2012.

\bibitem{Badger:2013yda}
Simon Badger, Benedikt Biedermann, Peter Uwer, and Valery Yundin.
\newblock {\em Phys. Rev.}, D89(3):034019, 2014.

\bibitem{Aad:2011tqa}
Georges Aad et~al.
\newblock {\em Eur. Phys. J.}, C71:1763, 2011.

\bibitem{Aad:2015nda}
Georges Aad et~al.
\newblock arXiv:1509.07335, 2015.

\bibitem{Chatrchyan:2011wn}
Serguei Chatrchyan et~al.
\newblock {\em Phys. Lett.}, B702:336--354, 2013.

\bibitem{Khachatryan:2014waa}
Vardan Khachatryan et~al.
\newblock {\em Eur. Phys. J.}, C75(6):288, 2015.

\bibitem{CMS:2014mna}
Vardan Khachatryan et~al.
\newblock {\em Eur. Phys. J.}, C75(5):186, 2015.

\bibitem{Currie:2013dwa}
James Currie, Aude Gehrmann-De~Ridder, E.W.N. Glover, and Joao Pires.
\newblock {\em JHEP}, 1401:110, 2014.

\bibitem{Boughezal:2015dva}
Radja Boughezal, Christfried Focke, Xiaohui Liu, and Frank Petriello.
\newblock {\em Phys. Rev. Lett.}, 115(6):062002, 2015.

\bibitem{GehrmannDeRidder:2005cm}
A.~Gehrmann-De~Ridder, T.~Gehrmann, and E.~W.~Nigel Glover.
\newblock {\em JHEP}, 09:056, 2005.

\bibitem{Ridder:2013mf}
Aude Gehrmann-De~Ridder, Thomas Gehrmann, E.W.N. Glover, and Joao Pires.
\newblock {\em Phys.Rev.Lett.}, 110(16):162003, 2013.

\bibitem{Giele:1993dj}
W.~T. Giele, E.~W.~Nigel Glover, and David~A. Kosower.
\newblock {\em Nucl. Phys.}, B403:633--670, 1993.

\bibitem{Campbell:2002tg}
John~M. Campbell and R.~Keith Ellis.
\newblock {\em Phys. Rev.}, D65:113007, 2002.

\bibitem{Berger:2009zg}
C.~F. Berger, Z.~Bern, Lance~J. Dixon, Fernando Febres~Cordero, D.~Forde,
  T.~Gleisberg, H.~Ita, D.~A. Kosower, and D.~Maitre.
\newblock {\em Phys. Rev. Lett.}, 102:222001, 2009.

\bibitem{KeithEllis:2009bu}
R.~Keith Ellis, Kirill Melnikov, and Giulia Zanderighi.
\newblock {\em Phys. Rev.}, D80:094002, 2009.

\bibitem{Berger:2010zx}
C.~F. Berger, Z.~Bern, Lance~J. Dixon, F.~Febres~Cordero, D.~Forde,
  T.~Gleisberg, H.~Ita, D.~A. Kosower, and D.~Maitre.
\newblock {\em Phys. Rev. Lett.}, 106:092001, 2011.

\bibitem{Bern:2013gka}
Z.~Bern, L.~J. Dixon, F.~Febres~Cordero, S.~Hoeche, H.~Ita, D.~A. Kosower,
  D.~Maître, and K.~J. Ozeren.
\newblock {\em Phys. Rev.}, D88(1):014025, 2013.

\bibitem{Ridder:2015dxa}
A.~Gehrmann-De~Ridder, T.~Gehrmann, E.~W.~N. Glover, A.~Huss, and T.~A. Morgan.
\newblock {\em Phys. Rev. Lett.}, 117(2):022001, 2016.

\bibitem{Stewart:2010tn}
Iain~W. Stewart, Frank~J. Tackmann, and Wouter~J. Waalewijn.
\newblock {\em Phys. Rev. Lett.}, 105:092002, 2010.

\bibitem{Stewart:2009yx}
Iain~W. Stewart, Frank~J. Tackmann, and Wouter~J. Waalewijn.
\newblock {\em Phys. Rev.}, D81:094035, 2010.

\bibitem{Boughezal:2015dra}
Radja Boughezal, Fabrizio Caola, Kirill Melnikov, Frank Petriello, and Markus
  Schulze.
\newblock {\em Phys. Rev. Lett.}, 115(8):082003, 2015.

\bibitem{deFlorian:1999zd}
D.~de~Florian, M.~Grazzini, and Z.~Kunszt.
\newblock {\em Phys. Rev. Lett.}, 82:5209--5212, 1999.

\bibitem{Boughezal:2013uia}
Radja Boughezal, Fabrizio Caola, Kirill Melnikov, Frank Petriello, and Markus
  Schulze.
\newblock {\em JHEP}, 06:072, 2013.

\bibitem{Chen:2014gva}
X.~Chen, T.~Gehrmann, E.~W.~N. Glover, and M.~Jaquier.
\newblock {\em Phys. Lett.}, B740:147--150, 2015.

\bibitem{Boughezal:2015aha}
Radja Boughezal, Christfried Focke, Walter Giele, Xiaohui Liu, and Frank
  Petriello.
\newblock {\em Phys. Lett.}, B748:5--8, 2015.

\bibitem{Czakon:2010td}
M.~Czakon.
\newblock {\em Phys. Lett.}, B693:259--268, 2010.

\bibitem{Campbell:2006xx}
John~M. Campbell, R.~Keith Ellis, and Giulia Zanderighi.
\newblock {\em JHEP}, 10:028, 2006.

\bibitem{Figy:2003nv}
T.~Figy, C.~Oleari, and D.~Zeppenfeld.
\newblock {\em Phys. Rev.}, D68:073005, 2003.

\bibitem{Cacciari:2015jma}
Matteo Cacciari, Frederic~A. Dreyer, Alexander Karlberg, Gavin~P. Salam, and
  Giulia Zanderighi.
\newblock {\em Phys. Rev. Lett.}, 115(8):082002, 2015.

\bibitem{Cullen:2013saa}
G.~Cullen, H.~van Deurzen, N.~Greiner, G.~Luisoni, P.~Mastrolia, E.~Mirabella,
  G.~Ossola, T.~Peraro, and F.~Tramontano.
\newblock {\em Phys. Rev. Lett.}, 111(13):131801, 2013.

\bibitem{Figy:2007kv}
Terrance Figy, Vera Hankele, and Dieter Zeppenfeld.
\newblock {\em JHEP}, 02:076, 2008.

\bibitem{Campanario:2015vqa}
F.~Campanario, M.~Kerner, L.~D. Ninh, M.~Rauch, R.~Roth, and D.~Zeppenfeld.
\newblock In {\em {Proceedings, Advances in Computational Particle Physics:
  Final Meeting (SFB-TR-9)}}, volume 261-262, pages 268--307, 2015.

\bibitem{Campbell:1999ah}
John~M. Campbell and R.~Keith Ellis.
\newblock {\em Phys. Rev.}, D60:113006, 1999.

\bibitem{Arnold:2008rz}
K.~Arnold, M.~Bahr, Giuseppe Bozzi, F.~Campanario, C.~Englert, et~al.
\newblock {\em Comput.Phys.Commun.}, 180:1661--1670, 2009.

\bibitem{Baglio:2014uba}
J.~Baglio, J.~Bellm, F.~Campanario, B.~Feigl, J.~Frank, et~al.
\newblock arXiv:1404.3940, 2014.

\bibitem{Gleisberg:2008ta}
T.~Gleisberg, Stefan. Hoeche, F.~Krauss, M.~Schonherr, S.~Schumann, F.~Siegert,
  and J.~Winter.
\newblock {\em JHEP}, 02:007, 2009.

\bibitem{Badger:2012pg}
Simon Badger, Benedikt Biedermann, Peter Uwer, and Valery Yundin.
\newblock {\em Comput. Phys. Commun.}, 184:1981--1998, 2013.

\bibitem{Alwall:2014hca}
J.~Alwall, R.~Frederix, S.~Frixione, V.~Hirschi, F.~Maltoni, O.~Mattelaer,
  H.~S. Shao, T.~Stelzer, P.~Torrielli, and M.~Zaro.
\newblock {\em JHEP}, 07:079, 2014.

\bibitem{Cullen:2014yla}
Gavin Cullen et~al.
\newblock {\em Eur. Phys. J.}, C74(8):3001, 2014.

\bibitem{Cascioli:2011va}
Fabio Cascioli, Philipp Maierhofer, and Stefano Pozzorini.
\newblock {\em Phys. Rev. Lett.}, 108:111601, 2012.

\bibitem{Dittmaier:2012kx}
Stefan Dittmaier, Alexander Huss, and Christian Speckner.
\newblock {\em JHEP}, 11:095, 2012.

\bibitem{Denner:2009gj}
Ansgar Denner, Stefan Dittmaier, Tobias Kasprzik, and Alexander Muck.
\newblock {\em JHEP}, 08:075, 2009.

\bibitem{Denner:2014ina}
Ansgar Denner, Lars Hofer, Andreas Scharf, and Sandro Uccirati.
\newblock {\em JHEP}, 01:094, 2015.

\bibitem{Baur:2006sn}
U.~Baur.
\newblock {\em Phys. Rev.}, D75:013005, 2007.

\bibitem{Kallweit:2014xda}
Stefan Kallweit, Jonas~M. Lindert, Philipp Maierhofer, Stefano Pozzorini, and
  Marek Schonherr.
\newblock {\em JHEP}, 04:012, 2015.

\bibitem{Aad:2013ysa}
Georges Aad et~al.
\newblock {\em JHEP}, 07:032, 2013.

\bibitem{Khachatryan:2014zya}
Vardan Khachatryan et~al.
\newblock {\em Phys. Rev.}, D91(5):052008, 2015.

\bibitem{Aad:2014qxa}
Georges Aad et~al.
\newblock {\em Eur. Phys. J.}, C75(2):82, 2015.

\bibitem{Khachatryan:2014uva}
Vardan Khachatryan et~al.
\newblock {\em Phys. Lett.}, B741:12--37, 2015.

\bibitem{Bengtsson:1986hr}
Mats Bengtsson and Torbjorn Sjostrand.
\newblock {\em Phys. Lett.}, B185:435, 1987.

\bibitem{Baer:1991caa}
Howard Baer and Mary~Hall Reno.
\newblock {\em Phys. Rev.}, D45:1503--1511, 1992.

\bibitem{Catani:1996vz}
S.~Catani and M.~H. Seymour.
\newblock {\em Nucl. Phys.}, B485:291--419, 1997.
\newblock [Erratum: Nucl. Phys.B510,503(1998)].

\bibitem{Herwig7}
Johannes Bellm et~al.
\newblock {\em Eur. Phys. J.}, C76(4):196, 2016.

\bibitem{Nagy:2014mqa}
Zoltan Nagy and Davison~E. Soper.
\newblock {\em JHEP}, 06:097, 2014.

\bibitem{Czakon:2015cla}
M.~Czakon, H.~B. Hartanto, M.~Kraus, and M.~Worek.
\newblock {\em JHEP}, 06:033, 2015.

\bibitem{Bevilacqua:2011xh}
G.~Bevilacqua, M.~Czakon, M.~V. Garzelli, A.~van Hameren, A.~Kardos, C.~G.
  Papadopoulos, R.~Pittau, and M.~Worek.
\newblock {\em Comput. Phys. Commun.}, 184:986--997, 2013.

\bibitem{Hoche:2012wh}
Stefan Hoeche and Marek Schonherr.
\newblock {\em Phys. Rev.}, D86:094042, 2012.

\bibitem{Hoeche:2012ft}
Stefan Hoeche, Frank Krauss, Marek Schonherr, and Frank Siegert.
\newblock {\em Phys. Rev. Lett.}, 110(5):052001, 2013.

\bibitem{Nason:2012pr}
Paolo Nason and Bryan Webber.
\newblock {\em Ann. Rev. Nucl. Part. Sci.}, 62:187--213, 2012.

\bibitem{Hoeche:2011fd}
Stefan Hoeche, Frank Krauss, Marek Schonherr, and Frank Siegert.
\newblock {\em JHEP}, 09:049, 2012.

\bibitem{Alioli:2010xd}
Simone Alioli, Paolo Nason, Carlo Oleari, and Emanuele Re.
\newblock {\em JHEP}, 06:043, 2010.

\bibitem{Alioli:2010xa}
Simone Alioli, Keith Hamilton, Paolo Nason, Carlo Oleari, and Emanuele Re.
\newblock {\em JHEP}, 04:081, 2011.

\bibitem{Kardos:2014dua}
Adam Kardos, Paolo Nason, and Carlo Oleari.
\newblock {\em JHEP}, 04:043, 2014.

\bibitem{Alioli:2010qp}
Simone Alioli, Paolo Nason, Carlo Oleari, and Emanuele Re.
\newblock {\em JHEP}, 01:095, 2011.

\bibitem{Re:2012zi}
Emanuele Re.
\newblock {\em JHEP}, 10:031, 2012.

\bibitem{Campbell:2013vha}
John~M. Campbell, R.~Keith Ellis, Paolo Nason, and Giulia Zanderighi.
\newblock {\em JHEP}, 08:005, 2013.

\bibitem{Alioli:2011as}
Simone Alioli, Sven-Olaf Moch, and Peter Uwer.
\newblock {\em JHEP}, 01:137, 2012.

\bibitem{Campbell:2012am}
John~M. Campbell, R.~Keith Ellis, Rikkert Frederix, Paolo Nason, Carlo Oleari,
  and Ciaran Williams.
\newblock {\em JHEP}, 07:092, 2012.

\bibitem{Alioli:2013hqa}
Simone Alioli, Christian~W. Bauer, Calvin Berggren, Frank~J. Tackmann,
  Jonathan~R. Walsh, and Saba Zuberi.
\newblock {\em JHEP}, 06:089, 2014.

\bibitem{Alioli:2015toa}
Simone Alioli, Christian~W. Bauer, Calvin Berggren, Frank~J. Tackmann, and
  Jonathan~R. Walsh.
\newblock {\em Phys. Rev.}, D92(9):094020, 2015.

\bibitem{Becher:2014oda}
Thomas Becher, Alessandro Broggio, and Andrea Ferroglia.
\newblock arXiv:1410.1892, 2014.

\bibitem{Rubin:2010xp}
Mathieu Rubin, Gavin~P. Salam, and Sebastian Sapeta.
\newblock {\em JHEP}, 1009:084, 2010.

\bibitem{Maitre:2013wha}
Daniel Maitre and Sebastian Sapeta.
\newblock {\em Eur. Phys. J.}, C73(12):2663, 2013.

\bibitem{Campanario:2012fk}
Francisco Campanario and Sebastian Sapeta.
\newblock {\em Phys.Lett.}, B718:100--104, 2012.

\bibitem{Campanario:2013wta}
Francisco Campanario, Michael Rauch, and Sebastian Sapeta.
\newblock {\em Nucl.Phys.}, B879:65--79, 2014.

\bibitem{Campanario:2015nha}
Francisco Campanario, Michael Rauch, and Sebastian Sapeta.
\newblock {\em JHEP}, 08:070, 2015.

\bibitem{Hamilton:2012np}
Keith Hamilton, Paolo Nason, and Giulia Zanderighi.
\newblock {\em JHEP}, 10:155, 2012.

\bibitem{Hamilton:2012rf}
Keith Hamilton, Paolo Nason, Carlo Oleari, and Giulia Zanderighi.
\newblock {\em JHEP}, 05:082, 2013.

\bibitem{Catani:2001cc}
S.~Catani, F.~Krauss, R.~Kuhn, and B.~R. Webber.
\newblock {\em JHEP}, 11:063, 2001.

\bibitem{Luisoni:2013kna}
Gionata Luisoni, Paolo Nason, Carlo Oleari, and Francesco Tramontano.
\newblock {\em JHEP}, 10:083, 2013.

\bibitem{Hamilton:2013fea}
Keith Hamilton, Paolo Nason, Emanuele Re, and Giulia Zanderighi.
\newblock {\em JHEP}, 10:222, 2013.

\bibitem{Karlberg:2014qua}
Alexander Karlberg, Emanuele Re, and Giulia Zanderighi.
\newblock {\em JHEP}, 09:134, 2014.

\bibitem{Hoeche:2012yf}
Stefan Hoeche, Frank Krauss, Marek Schonherr, and Frank Siegert.
\newblock {\em JHEP}, 04:027, 2013.

\bibitem{Cascioli:2013gfa}
F.~Cascioli, S.~Hoche, F.~Krauss, P.~Maierhofer, S.~Pozzorini, and F.~Siegert.
\newblock {\em JHEP}, 01:046, 2014.

\bibitem{Hoeche:2014lxa}
Stefan Hoeche, Frank Krauss, and Marek Schonherr.
\newblock {\em Phys. Rev.}, D90(1):014012, 2014.

\bibitem{Hoeche:2014qda}
Stefan Hoeche, Frank Krauss, Philipp Maierhoefer, Stefano Pozzorini, Marek
  Schonherr, and Frank Siegert.
\newblock {\em Phys. Lett.}, B748:74--78, 2015.

\bibitem{Hoeche:2014rya}
S.~Hoeche, F.~Krauss, S.~Pozzorini, M.~Schoenherr, J.~M. Thompson, and K.~C.
  Zapp.
\newblock {\em Phys. Rev.}, D89(9):093015, 2014.

\bibitem{Lonnblad:2012ix}
Leif Lonnblad and Stefan Prestel.
\newblock {\em JHEP}, 03:166, 2013.

\bibitem{Platzer:2012bs}
Simon Plätzer.
\newblock {\em JHEP}, 08:114, 2013.

\bibitem{Lonnblad:2012ng}
Leif Lonnblad and Stefan Prestel.
\newblock {\em JHEP}, 02:094, 2013.

\bibitem{Hoeche:2014aia}
Stefan Hoeche, Ye~Li, and Stefan Prestel.
\newblock {\em Phys. Rev.}, D91(7):074015, 2015.

\bibitem{Hoche:2014dla}
Stefan Hoeche, Ye~Li, and Stefan Prestel.
\newblock {\em Phys. Rev.}, D90(5):054011, 2014.

\bibitem{Catani:2007vq}
Stefano Catani and Massimiliano Grazzini.
\newblock {\em Phys.Rev.Lett.}, 98:222002, 2007.

\bibitem{Frederix:2012ps}
Rikkert Frederix and Stefano Frixione.
\newblock {\em JHEP}, 12:061, 2012.

\bibitem{Frederix:2015eii}
Rikkert Frederix, Stefano Frixione, Andreas Papaefstathiou, Stefan Prestel, and
  Paolo Torrielli.
\newblock {\em JHEP}, 02:131, 2016.

\bibitem{Christiansen:2015jpa}
Jesper~Roy Christiansen and Stefan Prestel.
\newblock {\em Eur. Phys. J.}, C76(1):39, 2016.

\bibitem{Mangano:2002ea}
Michelangelo~L. Mangano, Mauro Moretti, Fulvio Piccinini, Roberto Pittau, and
  Antonio~D. Polosa.
\newblock {\em JHEP}, 07:001, 2003.

\bibitem{AlcarazMaestre:2012vp}
J.~Alcaraz~Maestre et~al.
\newblock In {\em {Proceedings, 7th Les Houches Workshop on Physics at TeV
  Colliders}}, pages 1--220, 2012.

\bibitem{Chatrchyan:2011tz}
Serguei Chatrchyan et~al.
\newblock {\em Phys.Lett.}, B699:25--47, 2011.

\bibitem{Banfi:2012jm}
Andrea Banfi, Pier~Francesco Monni, Gavin~P. Salam, and Giulia Zanderighi.
\newblock {\em Phys.Rev.Lett.}, 109:202001, 2012.

\bibitem{Banfi:2015pju}
Andrea Banfi, Fabrizio Caola, Frédéric~A. Dreyer, Pier~F. Monni, Gavin~P.
  Salam, Giulia Zanderighi, and Falko Dulat.
\newblock {\em JHEP}, 04:049, 2016.

\bibitem{Anastasiou:2005qj}
Charalampos Anastasiou, Kirill Melnikov, and Frank Petriello.
\newblock {\em Nucl.Phys.}, B724:197--246, 2005.

\bibitem{Grazzini:2008tf}
Massimiliano Grazzini.
\newblock {\em JHEP}, 0802:043, 2008.

\bibitem{Catani:2009sm}
Stefano Catani, Leandro Cieri, Giancarlo Ferrera, Daniel de~Florian, and
  Massimiliano Grazzini.
\newblock {\em Phys.Rev.Lett.}, 103:082001, 2009.

\bibitem{Anastasiou:2015ema}
Charalampos Anastasiou, Claude Duhr, Falko Dulat, Franz Herzog, and Bernhard
  Mistlberger.
\newblock {\em Phys. Rev. Lett.}, 114:212001, 2015.

\bibitem{Banfi:2013eda}
Andrea Banfi, Pier~Francesco Monni, and Giulia Zanderighi.
\newblock {\em JHEP}, 01:097, 2014.

\bibitem{Banfi:2012yh}
Andrea Banfi, Gavin~P. Salam, and Giulia Zanderighi.
\newblock {\em JHEP}, 1206:159, 2012.

\bibitem{Banfi:2004yd}
Andrea Banfi, Gavin~P. Salam, and Giulia Zanderighi.
\newblock {\em JHEP}, 0503:073, 2005.

\bibitem{Andersen:2014efa}
J.~R. Andersen et~al.
\newblock arXiv:1405.1067, 2014.

\bibitem{Becher:2012qa}
Thomas Becher and Matthias Neubert.
\newblock {\em JHEP}, 07:108, 2012.

\bibitem{Becher:2013xia}
Thomas Becher, Matthias Neubert, and Lorena Rothen.
\newblock {\em JHEP}, 10:125, 2013.

\bibitem{Alioli:2013hba}
Simone Alioli and Jonathan~R. Walsh.
\newblock {\em JHEP}, 03:119, 2014.

\bibitem{Campanario:2010xn}
Francisco Campanario, Christoph Englert, and Michael Spannowsky.
\newblock {\em Phys.Rev.}, D82:054015, 2010.

\bibitem{Campanario:2014lza}
Francisco Campanario, Robin Roth, and Dieter Zeppenfeld.
\newblock {\em Phys.Rev.}, D91(5):054039, 2015.

\bibitem{Stewart:2011cf}
Iain~W. Stewart and Frank~J. Tackmann.
\newblock {\em Phys. Rev.}, D85:034011, 2012.

\bibitem{Lipatov:1976zz}
L.~N. Lipatov.
\newblock {\em Sov. J. Nucl. Phys.}, 23:338--345, 1976.
\newblock [Yad. Fiz.23,642(1976)].

\bibitem{Kuraev:1977fs}
E.~A. Kuraev, L.~N. Lipatov, and Victor~S. Fadin.
\newblock {\em Sov. Phys. JETP}, 45:199--204, 1977.
\newblock [Zh. Eksp. Teor. Fiz.72,377(1977)].

\bibitem{Balitsky:1978ic}
I.~I. Balitsky and L.~N. Lipatov.
\newblock {\em Sov. J. Nucl. Phys.}, 28:822--829, 1978.
\newblock [Yad. Fiz.28,1597(1978)].

\bibitem{Ciafaloni:1987ur}
Marcello Ciafaloni.
\newblock {\em Nucl. Phys.}, B296:49, 1988.

\bibitem{Catani:1989sg}
S.~Catani, F.~Fiorani, and G.~Marchesini.
\newblock {\em Nucl. Phys.}, B336:18, 1990.

\bibitem{Catani:1989yc}
S.~Catani, F.~Fiorani, and G.~Marchesini.
\newblock {\em Phys. Lett.}, B234:339, 1990.

\bibitem{Mueller:1986ey}
Alfred~H. Mueller and H.~Navelet.
\newblock {\em Nucl. Phys.}, B282:727, 1987.

\bibitem{Andersen:2009nu}
Jeppe~R. Andersen and Jennifer~M. Smillie.
\newblock {\em JHEP}, 01:039, 2010.

\bibitem{Andersen:2009he}
Jeppe~R. Andersen and Jennifer~M. Smillie.
\newblock {\em Phys. Rev.}, D81:114021, 2010.

\bibitem{Bury:2016cue}
Marcin Bury, Michal Deak, Krzysztof Kutak, and Sebastian Sapeta.
\newblock {\em Phys. Lett.}, B760:594--601, 2016.

\bibitem{Jung:2010si}
H.~Jung et~al.
\newblock {\em Eur. Phys. J.}, C70:1237--1249, 2010.

\bibitem{Bury:2015dla}
M.~Bury and A.~van Hameren.
\newblock {\em Comput. Phys. Commun.}, 196:592--598, 2015.

\bibitem{vanHameren:2015uia}
A.~van Hameren, P.~Kotko, and K.~Kutak.
\newblock {\em Phys. Rev.}, D92(5):054007, 2015.

\bibitem{Kimber:2001sc}
M.~A. Kimber, Alan~D. Martin, and M.~G. Ryskin.
\newblock {\em Phys. Rev.}, D63:114027, 2001.

\bibitem{Andersen:2011hs}
Jeppe~R. Andersen and Jennifer~M. Smillie.
\newblock {\em JHEP}, 06:010, 2011.

\bibitem{Andersen:2011zd}
Jeppe~R. Andersen, Leif Lonnblad, and Jennifer~M. Smillie.
\newblock {\em JHEP}, 07:110, 2011.

\bibitem{Andersen:2012gk}
Jeppe~R. Andersen, Tuomas Hapola, and Jennifer~M. Smillie.
\newblock {\em JHEP}, 09:047, 2012.

\bibitem{Bartels:2001ge}
J.~Bartels, D.~Colferai, and G.~P. Vacca.
\newblock {\em Eur. Phys. J.}, C24:83--99, 2002.

\bibitem{Bartels:2002yj}
J.~Bartels, D.~Colferai, and G.~P. Vacca.
\newblock {\em Eur. Phys. J.}, C29:235--249, 2003.

\bibitem{Ducloue:2013wmi}
B.~Ducloue, L.~Szymanowski, and S.~Wallon.
\newblock {\em JHEP}, 05:096, 2013.

\bibitem{Colferai:2010wu}
D.~Colferai, F.~Schwennsen, L.~Szymanowski, and S.~Wallon.
\newblock {\em JHEP}, 12:026, 2010.

\bibitem{Caporale:2011cc}
F.~Caporale, D.~{\relax Yu}. Ivanov, B.~Murdaca, A.~Papa, and A.~Perri.
\newblock {\em JHEP}, 02:101, 2012.

\bibitem{Caporale:2013uva}
F.~Caporale, B.~Murdaca, A.~Sabio~Vera, and C.~Salas.
\newblock {\em Nucl. Phys.}, B875:134--151, 2013.

\bibitem{Ducloue:2013bva}
B.~Ducloué, L.~Szymanowski, and S.~Wallon.
\newblock {\em Phys. Rev. Lett.}, 112:082003, 2014.

\bibitem{Caporale:2014gpa}
Francesco Caporale, Dmitry~{\relax Yu}. Ivanov, Beatrice Murdaca, and
  Alessandro Papa.
\newblock {\em Eur. Phys. J.}, C74(10):3084, 2014.
\newblock [Erratum: Eur. Phys. J.C75,no.11,535(2015)].

\bibitem{Kutak:2014wga}
Krzysztof Kutak.
\newblock {\em Phys. Rev.}, D91(3):034021, 2015.

\bibitem{Gribov:1984tu}
L.~V. Gribov, E.~M. Levin, and M.~G. Ryskin.
\newblock {\em Phys. Rept.}, 100:1--150, 1983.

\bibitem{Balitsky:1995ub}
I.~Balitsky.
\newblock {\em Nucl. Phys.}, B463:99--160, 1996.

\bibitem{Kovchegov:1999yj}
Yuri~V. Kovchegov.
\newblock {\em Phys. Rev.}, D60:034008, 1999.

\bibitem{Albacete:2010sy}
Javier~L. Albacete, Nestor Armesto, Jose~Guilherme Milhano, Paloma
  Quiroga-Arias, and Carlos~A. Salgado.
\newblock {\em Eur. Phys. J.}, C71:1705, 2011.

\bibitem{Aad:2014pua}
Georges Aad et~al.
\newblock {\em Eur. Phys. J.}, C74(11):3117, 2014.

\bibitem{CMS:2014oma}
CMS Collaboration.
\newblock CMS-PAS-FSQ-12-008, 2014.

\bibitem{CMS:2013eda}
CMS Collaboration.
\newblock CMS-PAS-FSQ-12-002, 2013.

\bibitem{Aad:2011jz}
Georges Aad et~al.
\newblock {\em JHEP}, 09:053, 2011.

\bibitem{Binosi:2008ig}
D.~Binosi, J.~Collins, C.~Kaufhold, and L.~Theussl.
\newblock {\em Comput. Phys. Commun.}, 180:1709--1715, 2009.

\end{thebibliography}
\end{document}